\documentclass[twocolumn]{aastex63}
\usepackage{diagbox}
\usepackage{multirow}
\usepackage{amsmath}
\usepackage{threeparttable}
\usepackage{lineno}
\graphicspath{{./}{figures/}}

\begin{document}

\title{Magnetar-powered long gamma-ray bursts and connection to superluminous supernovae and fast radio bursts}

\author{Yu-Qi Zhou}
\affiliation{School of Physics and Physical Engineering, Qufu Normal University, Qufu 273165, China; yisx2015@qfnu.edu.cn}
\affiliation{College of Physics and Electronic information, Dezhou University, Dezhou 253023, China}

\author[0000-0003-0672-5646]{Shuang-Xi Yi}
\affiliation{School of Physics and Physical Engineering, Qufu Normal University, Qufu 273165, China; yisx2015@qfnu.edu.cn}

\author{Yu-Peng Yang}
\affiliation{School of Physics and Physical Engineering, Qufu Normal University, Qufu 273165, China; yisx2015@qfnu.edu.cn}

\author{Yan-Kun Qu}
\affiliation{School of Physics and Physical Engineering, Qufu Normal University, Qufu 273165, China; yisx2015@qfnu.edu.cn}

\author{Ning Gai}
\affiliation{College of Physics and Electronic information, Dezhou University, Dezhou 253023, China}

\author{Yanke Tang}
\affiliation{College of Physics and Electronic information, Dezhou University, Dezhou 253023, China}

\author[0000-0003-4157-7714]{Fa-Yin Wang}
\affiliation{School of Astronomy and Space Science, Nanjing University, Nanjing 210023, China; fayinwang@nju.edu.cn}

\begin{abstract}
Based on X-ray afterglow observations from the \textit{Swift} satellite, we construct a sample of 169 long gamma-ray bursts (LGRBs) exhibiting the canonical magnetar plateau signature, i.e., a plateau followed by a $t^{-2}$ decay. We derive the plateau luminosity $L_0$ and break time $t_b$ for each burst by performing Markov Chain Monte Carlo (MCMC) fits to the light curves, and estimate pseudo-redshifts for bursts lacking known redshifts via the Amati relation. The fundamental magnetar parameters are subsequently inferred: the surface polar magnetic field strength $B_p \in [0.39,\ 23.08] \times 10^{15}$~G and the initial spin period $P_0 \in [0.95,\ 13.79]$~ms. Statistical analysis shows that both the known-redshift subsample and the full sample follow the Dainotti correlation between $L_0$ and $t_b$ with a slope close to $-1$, supporting a constant energy injection rate during the plateau phase. Furthermore, we identify a significant correlation between $B_p$ and $P_0$: $B_p \propto P_0^{0.83 \pm 0.09}$ for the full sample and $B_p \propto P_0^{0.80 \pm 0.16}$ for the known-redshift subsample, with both slopes consistent within uncertainties. Compared to magnetars powering superluminous supernovae (SLSNe), GRB magnetars possess systematically stronger magnetic fields (by approximately one order of magnitude), suggesting fundamental differences in their progenitor systems or collapse conditions; while their magnetic field strengths show no significant difference from those powering fast radio bursts (FRBs), suggesting a possible common evolutionary pathway. This study provides a physics-motivated, model-consistent sample of magnetar-candidate GRBs, offering a robust foundation for statistical investigations within the magnetar central engine model and placing new observational constraints on the birth properties of these extreme compact objects.
\end{abstract}


\section{Introduction}
\label{section:1}

Gamma-ray bursts (GRBs) are the most violent transient astrophysical phenomena in the universe and represent one of the core frontiers in contemporary high-energy astrophysics\citep{2004RvMP...76.1143P,2015PhR...561....1K}. They serve as natural laboratories for studying relativistic jets, offering an unparalleled probe of extreme relativistic shocks, particle acceleration mechanisms, and physical processes in strong gravitational fields and extreme magnetic fields \citep{2006ARA&A..44..507W, 2013ApJ...776..120Y}. Furthermore, with the highest detected redshift reaching $z \sim 9.4$ \citep{2011ApJ...736....7C} and their enormous isotropic-equivalent energies (reaching $\sim 10^{55}$ erg, \citealp{2009ARA&A..47..567G, 2023SciA....9I1405O}), GRBs have become crucial probes for studying cosmic evolution and the high-redshift universe \citep{2015NewAR..67....1W,Bargiacchi2025}.

GRBs are traditionally classified into two types based on their observed duration $T_{90}$: long GRBs (LGRBs) with $T_{90} > 2 s$ and short GRBs (SGRBs) with $T_{90} < 2 s$ \citep{1993ApJ...413L.101K}. It is generally accepted that LGRBs are predominantly associated with core-collapse supernovae of massive stars \citep{2002AJ....123.1111B, 2003Natur.423..847H}, whereas SGRBs originate from the mergers of binary compact systems, such as neutron star-neutron star or neutron star-black hole binaries \citep{1989Natur.340..126E, 1992ApJ...395L..83N, 2005Natur.437..845F, 2013ApJ...774L..23B, 2017ApJ...848L..12A}. Although this classification framework provides a basis for understanding the progenitor systems of GRBs, the physical mechanisms of the central engine that power these extreme explosions remain a central and active area of current research \citep{1994MNRAS.270..480T, 1999ApJ...524..262M, 2008MNRAS.385.1455M, 2011ApJ...726...90Z}.

It is theorized that the final outcome of a massive stellar collapse or a binary compact star merger is a newborn stellar-mass black hole surrounded by a rapidly rotating accretion disk \citep{1993ApJ...405..273W}. Within this framework, the energy powering the relativistic jets is thought to originate primarily from the rotational energy of the black hole (extracted via the Blandford-Znajek mechanism; \citealp{1977MNRAS.179..433B, 2017JHEAp..13....1Y, 2021MNRAS.507.1047Y, 2021ApJ...908..242D}) or from neutrino-antineutrino annihilation and viscous dissipation in the hyper-accretion disk \citep{1997A&A...319..122R, 1999ApJ...518..356P, 2007ApJ...657..383C, 2013ApJ...765..125L, 2009ApJ...700.1970L, 2017NewAR..79....1L}.

However, since the launch of the Neil Gehrels Swift Observatory (Swift), a wealth of observational data, particularly the persistent plateaus and ubiquitous flares in the afterglow light curves, has challenged the traditional black hole-accretion disk model \citep{2006ApJ...642..354Z, 2006ApJ...642..389N, 2007ApJ...662.1093W,Wang2013, 2016ApJS..224...20Y}. These plateaus exhibit long-lasting and stable energy injection characteristics that are difficult to fully explain through the instabilities or intermittent energy supply mechanisms commonly associated with accretion processes.

Various models have been developed  to address the question of the origin of the long-duration plateau. Among these, the jet structure model \citep{2020ApJ...893...88O, 2020A&A...641A..61A} and the dynamical model proposed by \citet{2022NatCo..13.5611D} can account for the long plateau duration without invoking additional energy sources. An alternative class of models, collectively referred to as energy injection models, attributes the plateau formation to continuous energy supply from the central engine to the external shock. The magnetar model is the most representative one in this category.

This model \citep{1992Natur.357..472U, 1998A&A...333L..87D, 1998PhRvL..81.4301D, 2001ApJ...552L..35Z, 2009ApJ...702.1171C, 2011A&A...526A.121D, 2014MNRAS.439.3916M} suggests that, instead of directly forming a black hole, the collapse may produce a newborn magnetar --- an extremely rapidly rotating neutron star (with a spin period on the order of milliseconds) possessing an ultra-strong magnetic field (with a surface dipole field strength of $\sim 10^{14}-10^{15} G$). Such a magnetar can continuously convert its enormous rotational energy into electromagnetic emission via magnetic dipole radiation, thereby providing a stable and predictable energy source for the gamma-ray burst and its afterglow --- particularly the plateau features \citep{2001ApJ...552L..35Z, 2010MNRAS.402..705L, 2011MNRAS.413.2031M, 2013MNRAS.430.1061R}. The specific details and physical mechanisms of this theoretical model will be systematically discussed in Section \ref{section:2}.

Although the magnetar model provides a compelling theoretical explanation for the plateau phase and a pathway to derive central engine parameters ($P_0$, $B_p$) from observables ($L_0$, $t_b$), statistical studies in this field still face challenges. For instance, the identification of the plateau and the extraction of the initial observables rely on empirical functions used for fitting, such as the BPL model \citep{2006ApJ...642..354Z, 2009ApJ...698...43R, 2019ApJS..245....1T, 2026arXiv260101586L} or the W07 model \citep{2007ApJ...662.1093W}. The choice of different empirical functions leads to variations in the derived physical parameters. Therefore, constructing a pure, homogeneous sample of 'magnetar candidates' screened based on a physical model is crucial for reliably testing the magnetar model and investigating its physical universality in a statistical sense.

The magnetar central engine model predicts a distinct temporal signature in the X-ray afterglow: an extended plateau phase of nearly constant luminosity, followed by a steep decay scaling as $t^{-2}$. This signature provides a key observational criterion for identifying magnetar-driven GRBs. Guided by this physical criterion, we systematically analyze the Swift GRB sample to construct a clean selection of events exhibiting clear magnetar plateau signatures.

For each GRB in this magnetar candidate sample, we estimate the plateau observables ($L_0$, $t_b$) and subsequently derive the underlying magnetar parameters ($B_p$, $P_0$), thereby testing the model's predictions. Furthermore, we investigate the correlation between $L_0$ and $t_b$ within our sample. Known as the Dainotti correlation, it follows a power-law form $L_0 \propto t_b^{-\alpha}$ \citep{2008MNRAS.391L..79D, 2010ApJ...722L.215D, 2013MNRAS.436...82D, 2017ApJ...848...88D}. Through this luminosity correlation, GRBs can be calibrated as standard candles and thus serve as independent cosmological probes at high redshifts \citep{2022ApJ...924...97W, 2025ApJ...991..145Z}. Recent simulations indicate that with high-quality X-ray GRBs, the 2D Dainotti relation can achieve constraints on dark energy comparable to those from the Planck satellite \citep{2026arXiv260318223D}.

In addition to the 2D relation, the 3D Dainotti correlation has also been widely employed to constrain cosmological parameters. \cite{2022MNRAS.514.1828D} used GRBs with both optical and X-ray data to constrain cosmological parameters, and found that the 3D relation in the optical band is equally effective for cosmological studies. Moreover, they showed that after correcting for redshift evolution, the intrinsic scatter $\sigma_{\mathrm{int}}$ of the $L_a$--$T_a$--$L_{\mathrm{peak}}$ correlation is significantly reduced, thereby improving the precision of cosmological parameter constraints \citep{2023MNRAS.518.2201D}.

In this paper, we try to carry out such a work with a sample of 169 GRBs, in order to reveal some statistical properties of magnetar-powered long GRBs and the corresponding magnetars, and compare the results with statistics of superluminous supernovae (SLSNe) and fast radio bursts (FRBs). The structure of this paper is organized as follows. Section \ref{section:2} describes the magnetar model, the data selection and analysis methods. Section \ref{section:3} presents the derived magnetar parameters and the Dainotti and $B_p-P_0$ correlations. Sections \ref{section:4} and \ref{section:5} summarize the results and discuss their implications, respectively. We conclude in Section \ref{section:6}. A flat $\Lambda$CDM cosmology with Planck parameters is adopted throughout \citep{2020A&A...641A...6P}.

\section{Data and Methods}
\label{section:2}
\subsection{Magnetar Model}
\label{section:2.1}

The energy reservoir of a newly born millisecond magnetar is the total rotational energy, which can be estimated as
\begin{equation}\label{eq:1}
E_{\mathrm{rot}}=\frac{1}{2}I\Omega_0^2\simeq2 \times 10^{52} \mathrm{erg} M_{1.4}R_6^2P_{0,-3}^{-2}
\end{equation}
where $I$ is the stellar moment of inertia, $\Omega_0 = 2\pi/P_0$ is the initial angular frequency of the magnetar (with $P_0$ being the initial period), and $R$ is the stellar radius. The mass is expressed in units of $1.4 M_\odot$ via $M_{1.4} \equiv M/(1.4 M_\odot)$. Throughout this paper, we adopt cgs units and follow the convention $Q = 10^x Q_x$ for all other dimensional quantities unless otherwise noted.

During the spin-down process, if the rotational energy is predominantly released via electromagnetic dipole radiation, the energy-loss rate is $\dot{E}_{\mathrm{rot}} = -L_{\mathrm{EM}}$, where $L_{\mathrm{EM}}$ is the dipole radiation luminosity. Assuming a constant magnetic inclination angle, its temporal evolution can be approximated as \citep{1998A&A...333L..87D, 2001ApJ...552L..35Z}
\begin{equation}\label{eq:2}
L_{\mathrm{EM}}(t) = \frac{L_{\mathrm{EM},0}}{\left(1 + t / t_{b}\right)^2}.
\end{equation}
Here, $L_{\mathrm{EM},0} \equiv L_{\mathrm{EM}}(t=0)$ is the theoretical initial dipole luminosity, and $t_b$ is the characteristic spin-down timescale. This equation predicts the observed afterglow pattern: a nearly constant plateau ($L_{\mathrm{EM}} \approx L_{\mathrm{EM},0}$ for $t \ll t_b$) followed by a steep $t^{-2}$ decay ($L_{\mathrm{EM}} \propto t^{-2}$ for $t \gg t_b$).

The observed X-ray afterglow luminosity, $L_X(t)$, is related to the dipole luminosity by a radiative efficiency factor $\eta_x$:
\begin{equation}\label{eq:3}
L_X(t) = \eta_x \, L_{\mathrm{EM}}(t) = \eta_x \frac{L_{\mathrm{EM},0}}{\left(1 + t / t_{b}\right)^2}.
\end{equation}

We fit the X-ray light curves with the following functional form:
\begin{equation}\label{eq:4}
L_X(t) = \frac{L_0}{\left(1 + t / t_{b}\right)^2}.
\end{equation}
where $L_0$ and $t_b$ are the free parameters obtained from the fit. The observed X-ray plateau luminosity is $L_0 = \eta_x L_{\mathrm{EM},0}$. $\eta_x$ represents the conversion efficiency from the magnetar's electromagnetic energy to X-ray emission. In previous studies, there has been no strict consensus regarding the specific value of the radiative efficiency \( \eta_x \) \citep{2019ApJ...878...62X, 2014MNRAS.443.1779R,  2024ApJ...963L..26Z}. To avoid theoretical scenarios where an excessively low efficiency value (e.g., \( \eta_x < 0.1 \)) would lead to a derived magnetar spin velocity exceeding the Keplerian limit \citep{2025ApJS..280...45L}, we adopt \( \eta_x = 0.5 \) as the fiducial value for calculation \citep{2011MNRAS.413.2031M, 2016PhRvD..93d4065G, 2022ApJ...934..125X, 2025ApJS..280...45L, 2025arXiv251122149L}. The systematic dependence of the derived magnetar parameters on this assumption is elaborated in Section \ref{section:3.1}.

The theoretical parameters $L_{\mathrm{EM},0}$ and $t_b$ are determined by the magnetar's fundamental properties:
\begin{equation}\label{eq:5}
\begin{aligned}
L_{\mathrm{EM},0} &= \frac{B_p^2 R^6 \Omega_0^4}{6c^3} \\
      &\simeq 1.0 \times 10^{49} \ \mathrm{erg \ s^{-1}} \ B_{p,15}^{2} P_{0,-3}^{-4} R_6^{6},
\end{aligned}
\end{equation}
\begin{equation}\label{eq:6}
\begin{aligned}
t_b &= \frac{3c^3 I}{B_p^2 R^6 \Omega_0^2} \\
    &\simeq 2.05 \times 10^{3} \ \mathrm{s} \ I_{45} B_{p,15}^{-2} P_{0,-3}^{2} R_6^{-6},
\end{aligned}
\end{equation}
where $B_p$ is the dipole magnetic field strength at the pole. In this work, we adopt the canonical neutron star parameters \( I_{45} = I/(10^{45}\,\mathrm{g \cdot cm^2}) \) and \( R_6 = R/(10^{6}\,\mathrm{cm}) \).

\subsection{GRB Sample Selection and Light Curve Fitting}
\label{section:2.2}

Prior to sample selection, we visually inspected each light curve and excluded any potential early steep decay phase---typically identified by its occurrence at early times ($<10^3$ s) and a slope significantly steeper than $-2$ (often in the range of $-3$ to $-5$)---to ensure that the subsequent analysis focuses on the plateau emission powered by magnetar spin-down. Only data after this phase were retained for fitting the plateau emission.

Guided by the theoretical plateau signature described above, we conducted a systematic search for LGRBs exhibiting such features from the Swift GRB catalog of the UK Swift Science Data Centre (\url{https://www.swift.ac.uk/xrt_curves/}; \citealt{2007A&A...469..379E, 2009MNRAS.397.1177E, 2010A&A...519A.102E}). The sample selection was carried out in two steps. First, we visually inspected the X-ray afterglow light curves within the Swift GRB catalog and selected candidates that displayed a clear plateau phase (with a slope near 0) followed by a decay with a slope of approximately $-2$. Subsequently, we performed model fitting using Equation (\ref{eq:4}) on these candidates to finalize our sample.

To accurately select magnetar-powered LGRBs from the Swift GRB catalog, we adopted a set of five selection criteria, following a methodology similar to that of \citet{2022ApJ...924...97W}:
\begin{itemize}
    \item The plateau duration criterion requires $T_p = (t_b - t_1) / t_b > 0.75$, where $t_1$ and $t_b$ represent the start time and end time of the plateau, respectively.
    \item The decay phase following the plateau must satisfy $T_{decay} = (t_{last} - t_b) / t_b > 5$, where $t_{last}$ represents the time of the last data point in the light curve.
    \item There are no weak flares.
    \item To ensure continuity, a sufficient number of non-clustered data points are required.
    \item The reduced chi-square ($\chi^2$) is required to be close to 1, preferably within (0.7, 2).
\end{itemize}
These five criteria collectively ensure that the light curve of the plateau phase is well-defined, complete, and reliably fitted, thereby providing a high-quality, pure sample of magnetar candidates for our subsequent physical analysis. Applying these criteria to the Swift catalog, our selection process yields a final sample of 169 GRBs, of which 78 have measured redshifts. This compilation incorporates the 30 candidates identified by \citet{2022ApJ...924...97W}, from which we have excluded GRB 090205 due to its potential classification as a SGRB \citep{2010A&A...522A..20D}.

\subsection{Pseudo-redshift Estimation}
\label{section:2.3}

For GRBs without spectroscopic redshifts, we estimate pseudo-redshifts using empirical correlations. Several methods have been proposed for this purpose based on observed GRB properties \citep[e.g.,][]{2004ApJ...609..935Y, 2007ApJ...660...16S, 2025A&A...698A..92N}. In this study, we apply the Amati relation \citep{2002A&A...390...81A,Wang2016} for estimation. The Amati relation is a robust empirical correlation connecting the isotropic-equivalent radiated energy ($E_{\gamma,\mathrm{iso}}$) and the rest-frame peak energy ($E_p$) of GRBs, and is expressed as follows \citep{2002A&A...390...81A}:
\begin{equation}\label{eq:7}
\log \left(\frac{E_{\gamma,\mathrm{iso}}}{\ {\mathrm{erg}}}\right) =a+b\log \left(\frac{E_p}{\ {\mathrm{keV}}} \right)
\end{equation}
The rest-frame peak energy is
\begin{equation}\label{eq:8}
E_p=E_{p,\mathrm{obs}} \times (1+z)
\end{equation}
The isotropic energy of the prompt emission is
\begin{equation}\label{eq:9}
E_{\gamma,\mathrm{iso}}=\frac{4\pi d_L^2S_{\mathrm{bolo}}}{1+z}
\end{equation}
where $S_{\mathrm{bolo}}=K \times S_{\mathrm{obs}}$ is the observed bolometric fluence. More detailed discussions about the K-correction process can be found in, e.g., \citet{1993ApJ...413..281B, Wang2011,2023ApJ...953...58L} and  \citet{2023ApJ...958...74T}.

The parameters $a$ and $b$ in this relation are determined through sample fitting. For this work, we adopted the values derived by \citet{2021JCAP...09..042K} based on the A118 sample set, specifically $a = 49.830$ and $b = 1.207$, obtained under the flat $\Lambda$CDM cosmological model.

For those GRBs requiring pseudo-redshift estimation, we directly extract their observed peak energy $E_{p,\mathrm{obs}}$ and observed fluence $S_{\mathrm{obs}}$ from the circulars of the Gamma-ray Coordinates Network (GCN; \url{https://gcn.nasa.gov/}). Using Equations (\ref{eq:8}) and (\ref{eq:9}), we derive the corresponding rest-frame quantities: the peak energy $E_p$ and the isotropic energy $E_{\gamma,\mathrm{iso}}$. These derived values are then used, within the framework of the Amati relation (Equation \ref{eq:7}), to sample the posterior probability distribution of the redshift $z$ using the Markov Chain Monte Carlo (MCMC) method \citep{2013PASP..125..306F}. The final pseudo-redshift is taken as the median of the posterior distribution. GRBs with redshifts determined through this method are marked with an asterisk (*) in Table \ref{tab:grb_data_1}.

While the Amati relation has intrinsic scatter and may be subject to selection effects, its application allows us to extend the statistical analysis to a larger sample of magnetar-candidate GRBs, providing a more complete view of the population despite the associated uncertainties. We will further discuss the potential impact of these pseudo-redshifts on our results in Section \ref{section:5}.

It is worth noting that some empirical correlations used for redshift estimation directly involve luminosity (e.g., the Yonetoku relation, $L_{\gamma,\mathrm{iso}}-E_p$), which could introduce circularity when subsequently studying luminosity-related correlations such as the Dainotti relation. Other relations, such as the Ghirlanda relation ($E_{\gamma,\mathrm{iso}}-E_{p,\mathrm{jet}}$), require the jet opening angle $\theta_j$, which is typically inferred from the jet break phase---an evolutionary stage distinct from the plateau phase analyzed in this work. To avoid these issues, we adopt the Amati relation ($E_{\gamma,\mathrm{iso}}-E_p$), which connects two quantities that are both independent of luminosity after redshift correction.

\section{Magnetar Parameters}
\label{section:3}
\subsection{Magnetar Parameters for the GRB Sample}
\label{section:3.1}

From the fitted plateau luminosity $L_0$ and the characteristic timescale $t_b$, combined with the relation $L_0 = \eta_x L_{\mathrm{EM},0}$ as well as Equation (\ref{eq:5}) and (\ref{eq:6}), the initial period $P_0$ and the polar magnetic field $B_p$ can be derived through inversion:
\begin{equation}
P_{0,-3}=1.42\ \mathrm{s} \left( I_{45}^\frac{1}{2}L_{EM,49}^{-\frac{1}{2} }t_{3}^{-\frac{1}{2}} \right )
\label{eq:10}
\end{equation}
and
\begin{equation}
B_{p,15}=2.05\ \mathrm{G}\left( I_{45}^\frac{1}{2}L_{EM,49}^{-1/2 }t_{3}^{-1}R_6^{-3} \right )
\label{eq:11}
\end{equation}

We employ the MCMC method, implemented via the public code \textbf{emcee} \citep{2013PASP..125..306F}, to derive the best-fit physical parameters of the nascent magnetar---specifically, its initial spin period \(P_0\) and surface dipole magnetic field strength \(B_p\)---along with their associated \(1\sigma\) uncertainties. Uniform priors are adopted over broad ranges: \(P_0 \in [0.5, 20]\,\text{ms}\) and \(B_p \in [10^{13}, 10^{17}]\,\text{G}\). The best-fit parameters are listed in Table~\ref{tab:grb_data_1}.

In Figure \ref{fig:distribution}, we present the histograms of the distributions of $P_0$ and $B_p$ for the full sample, along with the results of fitting with a Gaussian distribution. All derived magnetar parameters are obtained under the assumption of a fixed X-ray radiative efficiency $\eta_x = 0.5$. The resulting distributions of $P_0$ and $B_p$ are shown in Figure \ref{fig:distribution}, and their characteristic values are summarized in Section \ref{section:4}.

In addition, we performed the Dainotti correlation \citep{2008MNRAS.391L..79D, 2013MNRAS.436...82D, 2018ApJ...863...50S} analysis for both the redshift-measured sample and the full sample. This correlation can be expressed as
\begin{equation}
\log \left(\frac{ L_0}{10^{47} \rm erg/s} \right) =k\times \log\left(\frac{t_b}{10^3(1+z) \rm s}  \right)  +m.
\label{eq:12}
\end{equation}

Some selection effects (i.e.,the redshift dependence of $t_b$ and $L_0$, the threshold of the detector) would affect the Dainotti relation. Fortunately, this correlation has been tested against selection bias robustly \citep{2013MNRAS.436...82D}. The fitted correlations are shown in Figure \ref{fig:Dainotti}, with the detailed best-fit parameters reported in Section \ref{section:4}. The slopes are approximately $-1$, consistent with the magnetar model prediction.

\subsection{The $B_p-P_0$ Correlation in GRB Magnetars}
\label{section:3.2}

Figure \ref{fig:BP} presents the $B_p-P_0$ correlation for the GRB samples. Both the redshift-measured subsample and the full sample exhibit a proportional relationship, with the fitted slope of approximately 0.8.

Apart from GRBs, magnetars could also potentially serve as the central engine for superluminous supernovae (SLSNe)\citep{2017ApJ...840...12Y}. The study by \citet{2023ApJ...943...42C} investigated 70 SLSNe and derived the magnetar parameters for these events. The distributions of the initial spin period $P_0$ and the polar magnetic field strength $B_p$ were found to span the following ranges: $P_0 \in [1.96, 4.22]$ ms and $B_p \in [0.35, 1.86] \times 10^{14}$ G.

In Figure \ref{fig:SLSN}, we overlay the data from these 70 SLSNe (red points) onto the $B_p-P_0$ diagram of the full GRB sample, presenting the distribution of both GRB and SLSNe populations in the parameter space.

We note that the derivation of the above magnetar parameters depends on the assumed X-ray radiative efficiency $\eta_x = 0.5$. From Equations (\ref{eq:10}) and (\ref{eq:11}) together with $L_0 = \eta_x L_{\mathrm{EM},0}$, both $P_0$ and $B_p$ are proportional to $\eta_x^{1/2}$, and they decrease as $\eta_x$ decreases. Adopting a different value of $\eta_x$ would shift the inferred $P_0$ and $B_p$ for all GRBs globally, thereby changing the intercept of the $B_p$--$P_0$ correlation, while leaving its slope unaffected. Thus, the choice of $\eta_x$ does not alter the main conclusion of this work regarding the positive $B_p$--$P_0$ correlation.

\section{Results}
\label{section:4}

Following the selection criteria outlined in Section \ref{section:2.2}, we obtained the full sample of 169 magnetar-candidate LGRBs. Among these, 78 GRBs have spectroscopically confirmed redshifts (the redshift-measured sample), with a redshift distribution spanning $0.54 < z < 5.91$. For the remaining 91 GRBs lacking direct redshift measurements, we estimated their pseudo-redshifts using the Amati relation, yielding a distribution in the range of $0.5 < z < 9.95$.

Assuming $\eta_x=0.5$, we calculated the magnetar parameters for all 169 GRBs using Equations (\ref{eq:10}) and (\ref{eq:11}). The distribution of the surface polar magnetic field strength $B_p$ across the full sample spans $[0.39,\ 23.08] \times 10^{15}$ G. A Gaussian fit to $\log B_p$ yields the $\mu_{\log B_p} = 15.41$ with the standard deviation $\sigma_{\log B_p} = 0.36$. Meanwhile, the distribution of the initial spin period $P_0$ ranges from 0.95 ms to 13.79 ms. A Gaussian fit to $\log P_0$ gives $\mu_{\log P_0} = 0.60$ and $\sigma_{\log P_0} = 0.24$. This corresponds to a characteristic magnetic field strength of $\sim 2.57 \times 10^{15} \ \mathrm{G}$ and a characteristic initial spin period of $\sim 3.98\ \mathrm{ms}$ for the population.

We investigated the Dainotti correlation within this sample. For the subsample of 78 GRBs with spectroscopic redshifts, the fit yields the slope $k=-1.10^{+0.11}_{-0.11}$, the intercept $m=1.55^{+0.14}_{-0.14}$, and the intrinsic scatter $\sigma_{\mathrm{int}} = 0.58$. For the full sample of 169 GRBs, which includes the 91 GRBs with pseudo-redshifts, the fit gives the slope $k = -1.07^{+0.09}_{-0.09}$, the intercept $m = 1.24^{+0.11}_{-0.12}$, and the intrinsic scatter $\sigma_{\mathrm{int}} = 0.66$. The fitting results are presented in Figure \ref{fig:Dainotti} and Table \ref{tab:addlabe2}.

Based on the derived magnetar parameters ($B_p$, $P_0$) for all 169 candidates, we identify a positive correlation between $B_p$ and $P_0$. For the full sample, the best-fit power-law relation is $B_p \propto P_0^{\,0.83 \pm 0.09}$. For the redshift-measured subsample, the relation is $B_p \propto P_0^{\,0.80 \pm 0.16}$. The distributions of the magnetar parameters for our GRB sample and the fitting results for both correlations are summarized in Figure \ref{fig:distribution}, Figure \ref{fig:BP}, and Table \ref{tab:addlabe2}.

\section{Discussion}
\label{section:5}

\subsection{GRBs Powered by Magnetar Central Engines}
\label{subsection:5.1}
Based on the pure sample constructed in this work, we derived the fundamental magnetar parameters ($B_p$, $P_0$) for all 169 candidates. We find that GRB magnetars in the sample predominantly occupy the high magnetic field regime, with a mean $B_p$ of $2.6 \times 10^{15}$ G and a mean initial spin period $P_0 \sim 4.0$ ms. A significant positive correlation between $B_p$ and $P_0$ is identified. For the full sample, $B_p \propto P_0^{0.8}$. The derived $B_p$ values span a range of $0.4 \times 10^{15}$ to $23 \times 10^{15}$ G, while $P_0$ ranges from $1$ to $14$ ms.

The $B_p-P_0$ positive correlation we found is qualitatively consistent with several previous studies, jointly supporting the view of a possible intrinsic connection between the parameters of newborn magnetars. The slope reported by \citet{2018ApJ...869..155S} ($B \propto P^{0.83\pm0.17}$) is almost identical to ours. \citet{2015ApJ...813...92R} also derived $B_p$ and $P_0$ from GRB X-ray plateaus and presented a correlation in the form $\log P_0 = -6.2 + 0.22\log B_0$, which corresponds to $B_p \propto P_0^{4.55}$. Although this slope is steeper than ours and that of \citet{2018ApJ...869..155S}, it still indicates a positive correlation between $B_p$ and $P_0$. This discrepancy likely arises from two main factors: first, differences in sample composition¡ªthe sample of \citet{2015ApJ...813...92R} includes short GRBs and X-ray flashes, whose $B_p$ and $P_0$ distributions may differ from those of long GRBs; second, different treatments of redshift evolution¡ªwhile that study did not apply evolutionary corrections to the light curves, we have adopted the method of \citet{2013MNRAS.436...82D} to remove selection effects and redshift evolution, allowing the derived parameters to better reflect the intrinsic physics. Among other studies, \citet{2025ApJS..280...45L} reported a slope of approximately 0.8 for the $B_p-P_0$ correlation in their sample, after accounting for the evolution of neutron star radius ($R$) and moment of inertia ($I$), and assuming a baryonic mass $M_b = 2.0 M_{\odot}$ and a radiative efficiency $\eta_x = 0.5$. \citet{2015ApJ...805...89L} also found a positive trend between $B_p$ and $P_0$ when analyzing a set of GRBs with X-ray plateaus, although their sample and selection criteria differ from this work. Furthermore, studies by \citet{2013MNRAS.430.1061R, 2014MNRAS.443.1779R, 2022ApJ...934..125X} also provide support for this trend. \citet{2017ApJ...840...12Y} re-fitted the sample selected by \citet{2014ApJ...785...74L} and obtained a positive $B_p-P_0$ correlation with a slope of $\sim 0.8$. The convergence of these studies points to an intrinsic $B_p-P_0$ correlation in the newborn magnetar population.

It should be noted that the $P_0$ derived in this work corresponds to the effective initial spin period at the birth of the magnetar within the framework of the standard magnetic dipole radiation braking model. However, this interpretation is not unique. \citet{2025arXiv251122149L} pointed out that if significant fallback accretion and its interaction with the magnetosphere (i.e., the ``propeller" effect) are considered, the observed plateau may correspond to a state where the magnetar has reached spin equilibrium. In this scenario, the fitted $P_0$ should be interpreted as an equilibrium spin period, whose value is jointly determined by the accretion rate, magnetic field strength, and magnetospheric radius. Under this picture, the $B_p-P_0$ correlation may reflect a connection between the properties of the accretion flow and the magnetic field strength, rather than originating directly from the initial conditions at the moment of collapse.

The physical origin of the $B_p-P_0$ correlation remains an open question. Using different physical models also reveals similar correlations. For instance, \citet{2025arXiv251122149L}, employing the propeller model, found $B_p \propto P^{7/6}$ and suggested that the $P_0$ inferred from X-ray plateau data may represent an equilibrium spin period achieved via fallback accretion rather than the true initial value. The fact that a similar correlation appears in both the standard dipole spin-down model (this work) and the propeller model suggests that this correlation is a robust observational feature. However, whether it reflects an intrinsic birth property of magnetars or arises from the parameter inversion process under different model assumptions requires further theoretical investigation.

\subsection{Other Transients Possibly Powered by Magnetar Engines}
\label{subsection:5.2}

Apart from GRBs, magnetars could also potentially serve as the central engine for SLSNe \citep{Woosley2010,Kasen2010}. The study by \citet{2023ApJ...943...42C} investigated 70 SLSNe and derived the magnetar parameters for these events. The distributions of the initial spin period $P_0$ and the polar magnetic field strength $B_p$ were found to span the following ranges: $P_0 \in [1.96, 4.22]$ ms and $B_p \in [0.35, 1.86] \times 10^{14}$ G.

Placing the GRB magnetar parameters from this work and SLSNe magnetar parameters on the same $B_p-P_0$ plane (Figure~\ref{fig:SLSN}), it can be seen that the magnetic field strength of GRB magnetars ($B_p \sim 10^{15}$ G) is generally about an order of magnitude higher than that of SLSNe magnetars ($B_p \sim 10^{14}$ G), while their initial spin period ranges show substantial overlap. This observational fact is qualitatively consistent with theoretical expectations: a stronger magnetic field may be more critical for efficiently extracting energy from a rapidly rotating compact star and forming a highly collimated relativistic jet. For SLSNe, their energy needs to be deposited into the surrounding non-relativistic ejecta, where a moderately strong magnetic field combined with appropriate rotation can already achieve effective energy injection through dipole radiation. This hints at fundamental physical differences in the progenitor systems or collapse processes that give rise to these two types of transients.

In addition to SLSNe, magnetars are also considered promising central engines for fast radio bursts (FRBs) \citep{2019PhR...821....1P, 2020Natur.587...59B, 2020Natur.587...54C, 2020Natur.587...45Z, Wang2022, 2025MNRAS.tmp.2057B}. Detailed modeling shows that the quasi-steady radio emission from such systems can be explained by synchrotron radiation from a magnetar wind nebula, which may be powered by the magnetar's rotational energy \citep{Murase2016} or internal magnetic field reservoir \citep{2018ApJ...868L...4M}. Recent studies of FRB 20121102A, FRB 20190520B, and FRB 20240114A suggest their persistent radio sources can be powered by magnetar wind nebulae with energy injection by internal magnetic field decay in magnetars with strong magnetic fields ($B_{\mathrm{int}} \sim 10^{16}$ G) \citep{2021ApJ...923L..17Z, 2025ApJ...988..276R}. We use the magnetic fields derived by fitting the spectra and light curves of persistent radio sources (Zhao et al. in prep.). The initial spin period is hard to constrain. We assume it randomly distributes in 1 ms to 10 ms. The results are shown as blue rectangles in Figure \ref{fig:SLSN}. These inferred magnetic field strengths are notably similar as those of GRB sample, suggesting a common evolutionary may be at play.

Despite the observational differences among GRBs, SLSNe, and FRBs, our statistical analysis of their magnetar parameters reveals several common features, suggesting that they may share a similar physical origin and central engine mechanism. First, all three types of transients can be reasonably explained by the millisecond magnetar model, with their initial spin periods concentrated in the range of about 1-10 ms, implying a similar angular momentum structure in their progenitor cores. Second, their observed properties are consistent with energy injection processes dominated by magnetar spin-down---for instance, the X-ray plateau lasting $10^3$--$10^4$~s in GRBs, the optical light curves evolving over several weeks in SLSNe, and the sustained radio emission over decades exhibited by repeating FRBs. Furthermore, \citet{2017ApJ...840...12Y} and \citet{2017ApJ...843...84N} found that the host galaxies of all three phenomena tend to be dwarf galaxies with low metallicity and high star-formation rates, indicating that they likely originate from similar massive, metal-poor stars, whose evolutionary environment provides favorable conditions for the formation of rapidly rotating magnetars upon core collapse. However, it should be emphasized that these comparisons are qualitative and subject to various uncertainties, as the magnetar parameters for different transients are derived from different emission bands and model assumptions. A more quantitative and unified assessment requires further observational and theoretical efforts.

\subsection{Implications of the Dainotti Correlation}
\label{subsection:5.3}

We note that the intrinsic scatter $\sigma_{\mathrm{int}}$ of the Dainotti correlation is approximately $0.6$ for both the known-redshift subsample and the full sample. This value is larger than the scatter reported by \cite{2022ApJ...924...97W} ($\sigma_{\mathrm{int}} = 0.22$) for a gold sample of $10$ GRBs selected under similar physical criteria. This difference likely arises because our sample is substantially larger (expanded from $10$ to $169$ GRBs) and our selection criteria are designed to identify physically self-consistent magnetar candidates rather than to minimize the intrinsic scatter; an increase in scatter is therefore expected.

We note that the intrinsic scatter parameter $\sigma_{int}$ of the Dainotti correlation is approximately 0.6 for both the known-redshift subsample and the full sample. In contrast, \cite{2020ApJ...904...97D} constructed a "platinum sample" of 47 gamma-ray bursts (GRBs), for which the intrinsic scatter of the three-dimensional X-ray fundamental plane relation is $\sigma_{int} = 0.22$. \cite{2024JHEAp..44..323F} used a platinum sample of 20 GRBs calibrated with cosmological clocks and obtained $\sigma_{int} = 0.20$ for the two-dimensional relation. \cite{2022ApJ...924...97W}, based on physical criteria similar to those used in this study, obtained $\sigma_{int} = 0.22$ for the two-dimensional Dainotti relation using a Gold Standard sample of 10 GRBs. The differences between these results and those presented in this study may stem from variations in selection criteria, sample size, and the choice of correlation. Furthermore, our selection criteria are designed to identify physically consistent magnetar candidates rather than to minimize intrinsic scatter; therefore, an increase in scatter is to be expected.

Nevertheless, the relatively large scatter implies that the two-dimensional correlation in this sample has limited power for constraining cosmological parameters. \cite{2022MNRAS.516.1386C} found that GRB data favor a three-parameter correlation over a two-parameter one, offering a viable alternative for future cosmological applications of this sample. We emphasize that our primary purpose in examining the two-dimensional Dainotti correlation is to verify whether its slope is consistent with the prediction of the magnetar model (i.e., $L_0 \propto t_b^{-1}$).

As described in Section \ref{section:2.2}, our sample selection requires the light curve to exhibit a plateau followed by a $t^{-2}$ decay. This morphological feature also serves to rule out the alternative central engine models mentioned in the Introduction. The structured jet model predicts high-latitude emission with a considerably shallower decay, typically in the range of $-0.5$ to $-1$. The dynamical model with a low initial Lorentz factor, on the other hand, asymptotically approaches the decay slope of a standard external afterglow, roughly between $-1$ and $-1.2$. Neither of these alternative models can simultaneously reproduce both a clear plateau and the precise $t^{-2}$ decay that follows it.

Recently, \cite{2025JHEAp..4700384L} proposed that the GRB plateau can be powered by the spin-down of a Kerr black hole via the Blandford--Znajek mechanism. This model also predicts an anti-correlation between X-ray luminosity and time, and thus exhibits a degeneracy with the magnetar model in terms of the Dainotti slope. \cite{2025JHEAp..4700384L} pointed out that when the energy budget of a GRB exceeds the theoretical limit of a magnetar ($\sim 2\times 10^{52}$ erg), the black hole spin-down model may be a more natural choice. However, the two models differ in the detailed shape of the light curve: the magnetar dipole radiation naturally and robustly predicts a post-plateau decay slope of $t^{-2}$, whereas the decay slope in the black hole spin-down model depends on the accretion rate and magnetic field evolution and is not necessarily $-2$. Since our sample selection explicitly requires a $t^{-2}$ decay, the magnetar model remains the most natural explanation for this particular class of GRBs. This does not imply that all GRB central engines are magnetars; the model of \cite{2025JHEAp..4700384L} provides a valuable complement for bursts whose energy budgets exceed the magnetar limit or whose decay slopes deviate from $-2$. In addition, \cite{2022ApJ...925...15L} found a similar luminosity versus time anti-correlation in the radio band, with a corrected slope also around $-1$, suggesting that such plateau phenomena are common across multiple wavelengths and that the underlying energy injection mechanism may have a unified physical origin.

\subsection{Implications of pseudo-redshift}
\label{subsection:5.4}

A substantial fraction of the GRBs in our full sample lack spectroscopic redshifts; we therefore estimated their distances using the Amati relation. To assess the potential impact of these pseudo-redshifts, we compared the results obtained from the subsample of $78$ GRBs with known redshifts with those from the full sample that includes pseudo-redshifts. As shown in Table \ref{tab:addlabe2}, the slopes of the $L_0$-$t_b$ and $B_p$-$P_0$ correlations derived from the subsample are consistent within $1\sigma$ uncertainties with those obtained from the full sample. This consistency indicates that the inclusion of GRBs with pseudo-redshifts estimated via the Amati relation does not significantly bias the observed trends of these correlations. Nevertheless, subsequent studies utilizing the sample containing pseudo-redshifts presented in this work should carefully evaluate the potential effects introduced by these estimates.

In recent years, machine learning-based methods have also been employed to estimate GRB redshifts \citep{2024ApJ...967L..30D, 2024ApJS..271...22D}. These approaches can incorporate multiple observational features without relying on a single empirical correlation, offering advantages in both precision and generality. In comparison, empirical methods such as the Amati relation are computationally inexpensive and straightforward to implement, making them a reasonable choice for constructing large samples in exploratory studies.

However, systematic uncertainties inherent to the Amati relation---including its intrinsic scatter and the choice of calibration parameters---may affect the reliability of individual redshift estimates. For the statistical trends investigated in this work, the consistency between the subsample and the full sample suggests that these uncertainties do not significantly bias our main conclusions. For future high-precision cosmological applications, adopting machine learning techniques to obtain more accurate redshift estimates would be a valuable improvement.

\section{Conclusion}
\label{section:6}

We present a systematic analysis of LGRBs exhibiting the canonical magnetar plateau signature. By directly fitting X-ray afterglow light curves with the theoretical magnetar spin-down model, we construct a model-consistent sample of 169 magnetar-candidate GRBs. From this sample, and under the assumption of $\eta_x=0.5$, we derive characteristic magnetar parameters of $B_p \sim 2.6 \times 10^{15}$ G and $P_0 \sim 4.0$ ms, confirm the Dainotti correlation with slope $\approx -1$, and identify a significant positive correlation $B_p \propto P_0^{0.8}$. These results provide new observational constraints on magnetar birth properties and support constant energy injection via dipole spin-down during the plateau phase.

The comparison of our GRB-derived parameters with those inferred for SLSNe and FRBs reveals a continuum of magnetar properties, suggesting a unified framework where a highly magnetized, rapidly rotating neutron star powers different transients through distinct energy extraction channels.

We emphasize that the results presented in this work may have important implications for future studies exploring the application of the Dainotti relation as a cosmological probe. The large, homogeneous sample of magnetar-candidate GRBs constructed here, together with the confirmation that the slope of the $L_0$-$t_b$ correlation is consistent with the magnetar spin-down model, provides a physically motivated foundation for using this correlation in cosmological contexts. As demonstrated in recent works \citep{2022ApJ...924...97W, 2023ApJ...951...63D}, the Dainotti relation shows potential for constraining cosmological parameters when properly calibrated. Future observations from missions like \textit{Einstein Probe} \citep{2015arXiv150607735Y} and \textit{SVOM}\citep{Wei2016} will provide a larger sample of GRBs with spectroscopic redshifts, offering critical data to refine the magnetar model and further enhancing the utility of this correlation as a standard candle for cosmology.

\section*{Acknowledgments}
We would like to thank the anonymous referee for helpful comments. We acknowledge the use of public data from the \textit{Swift} data archives. This work is supported by the Natural Science Foundation of China (NSFC) grants 12494575 and 12273009, the Shandong Provincial Natural Science Foundation (grants ZR2025MS47 and ZR2025MS16), the Natural Science Foundation of Jiangxi Province of China (grant No. 20242BAB26012), and the China Postdoctoral Science Foundation (grant No. 2025M783226).

\clearpage

\begin{figure*}[h!]
\centering

\noindent
\resizebox{55mm}{!}{\includegraphics[]{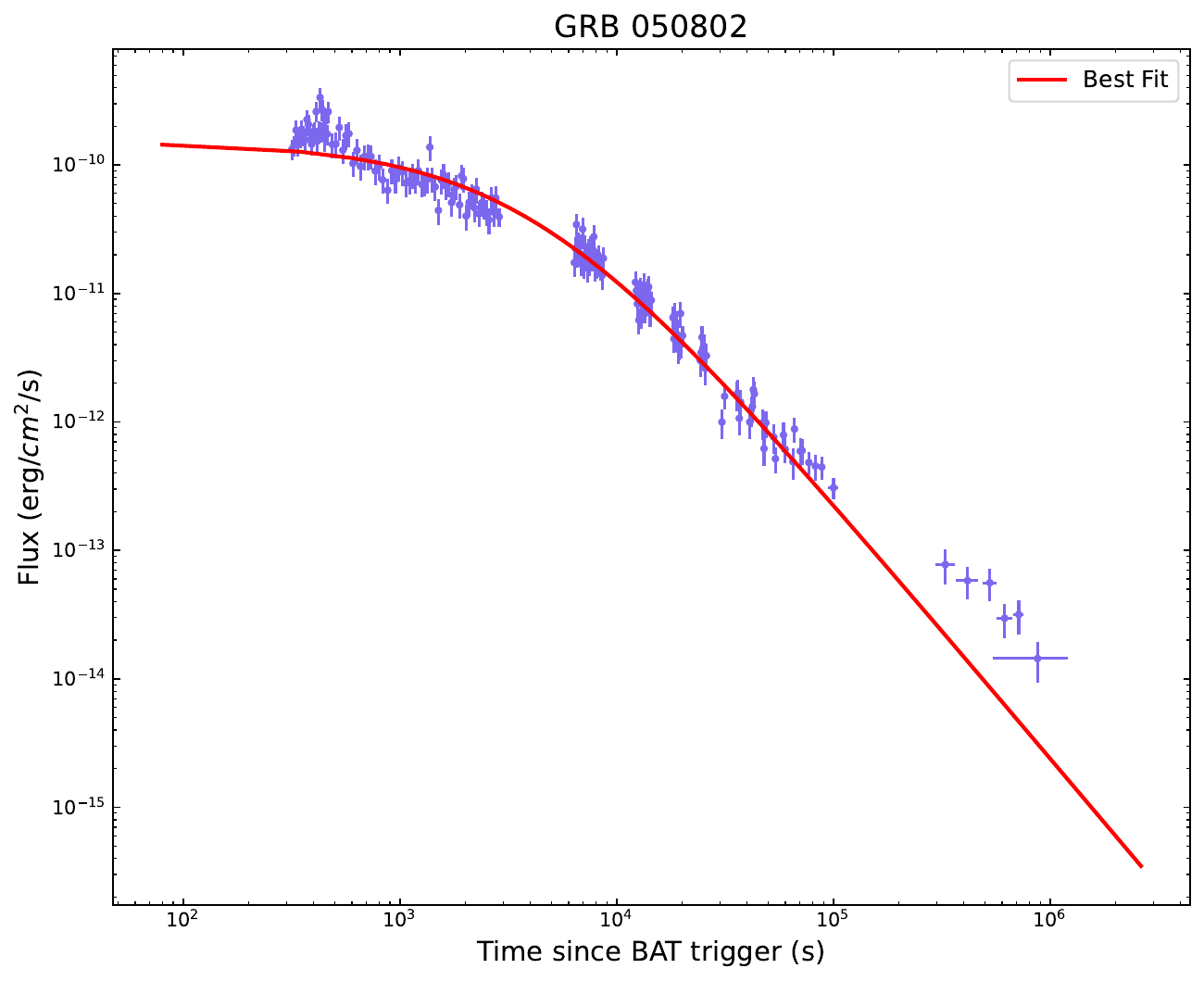}}%
\resizebox{55mm}{!}{\includegraphics[]{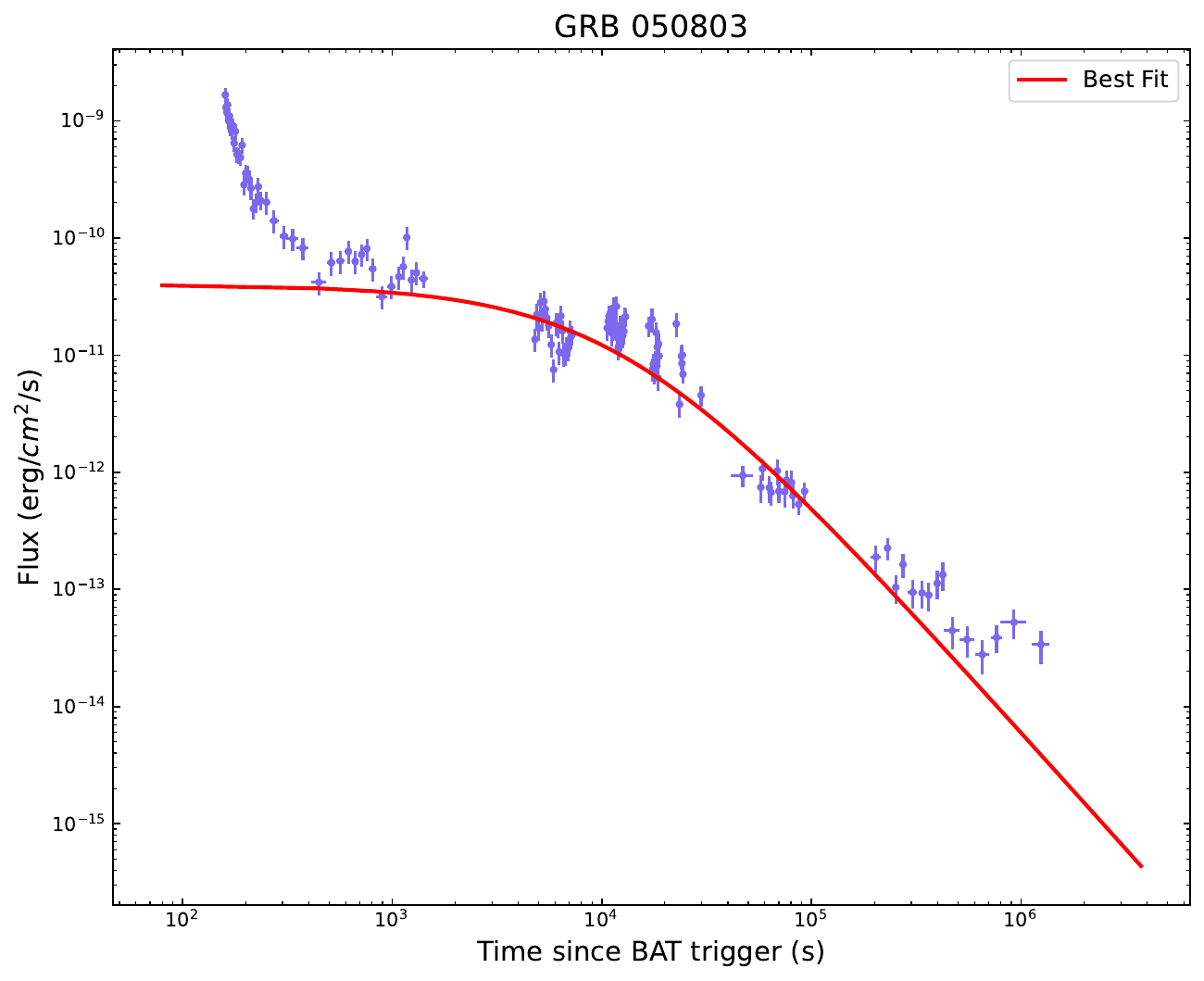}}%
\resizebox{55mm}{!}{\includegraphics[]{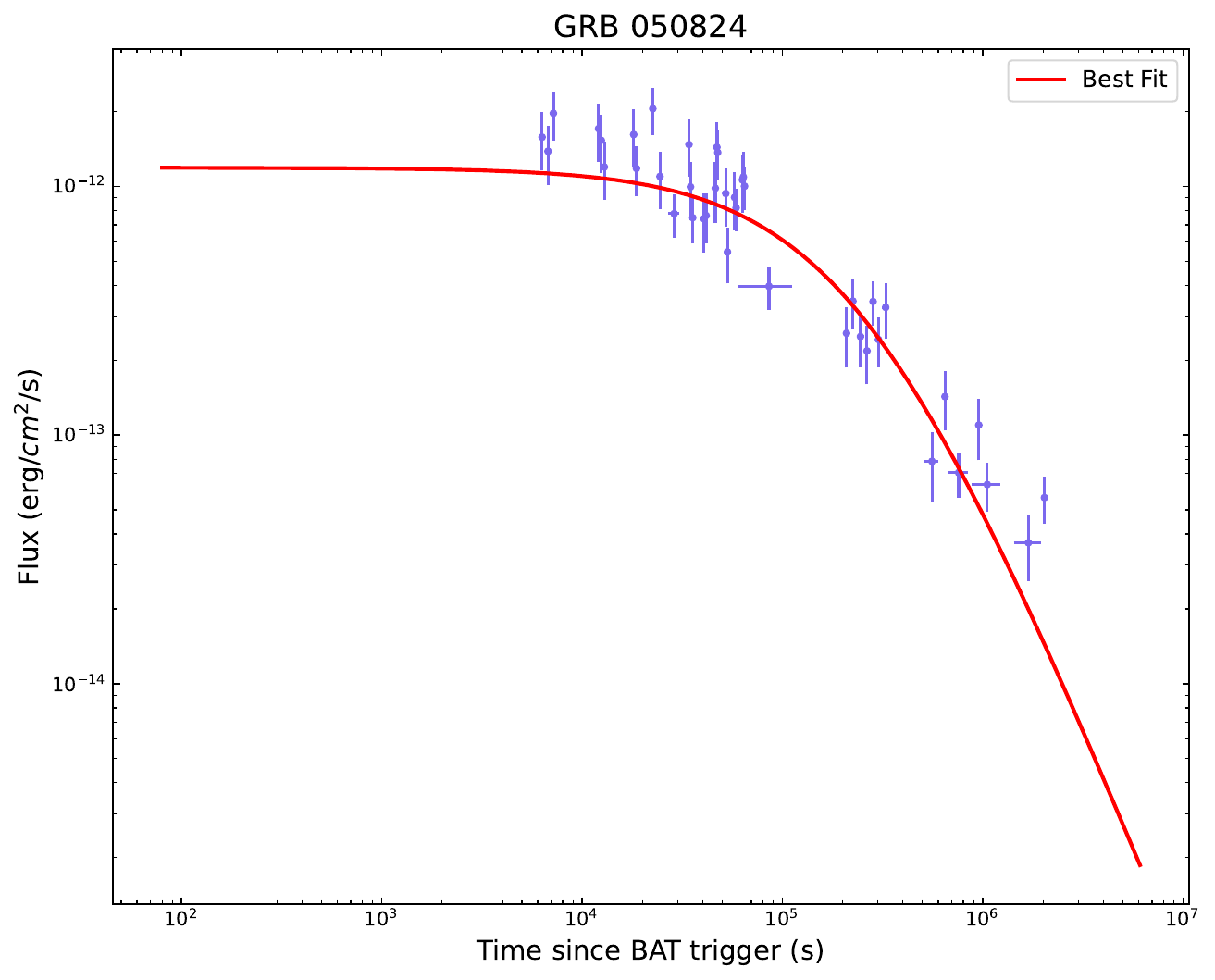}}%

\noindent
\resizebox{55mm}{!}{\includegraphics[]{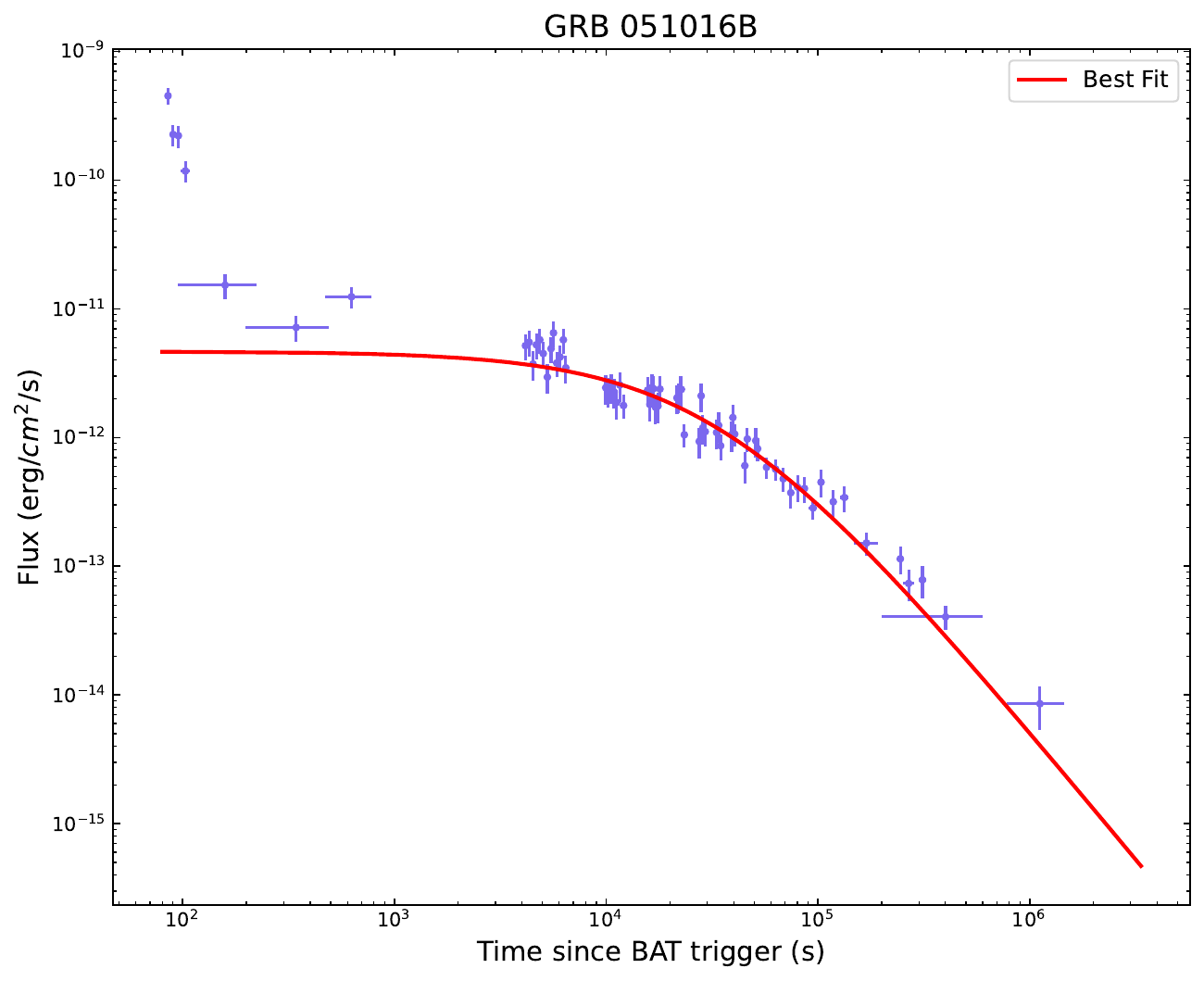}}%
\resizebox{55mm}{!}{\includegraphics[]{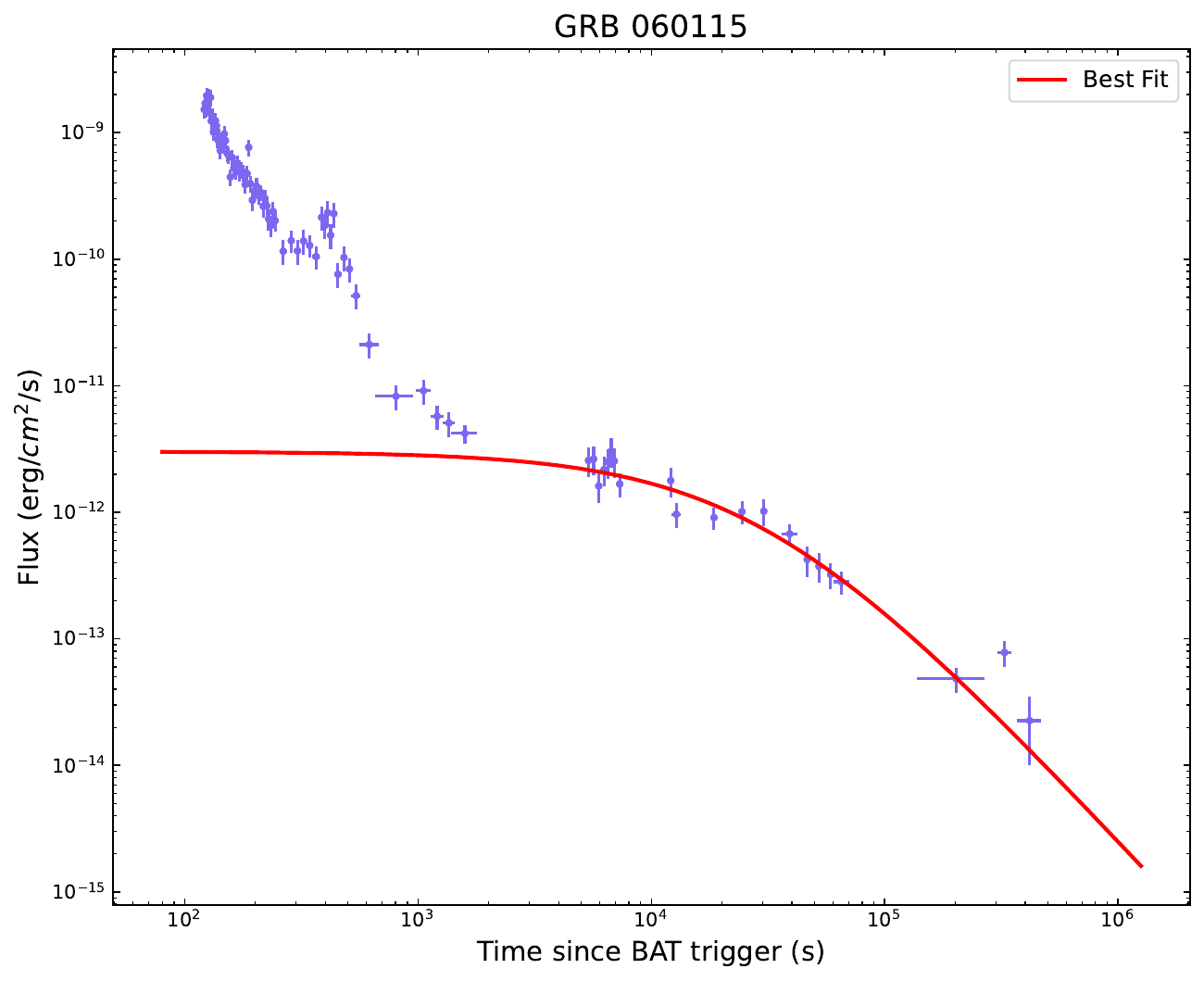}}%
\resizebox{55mm}{!}{\includegraphics[]{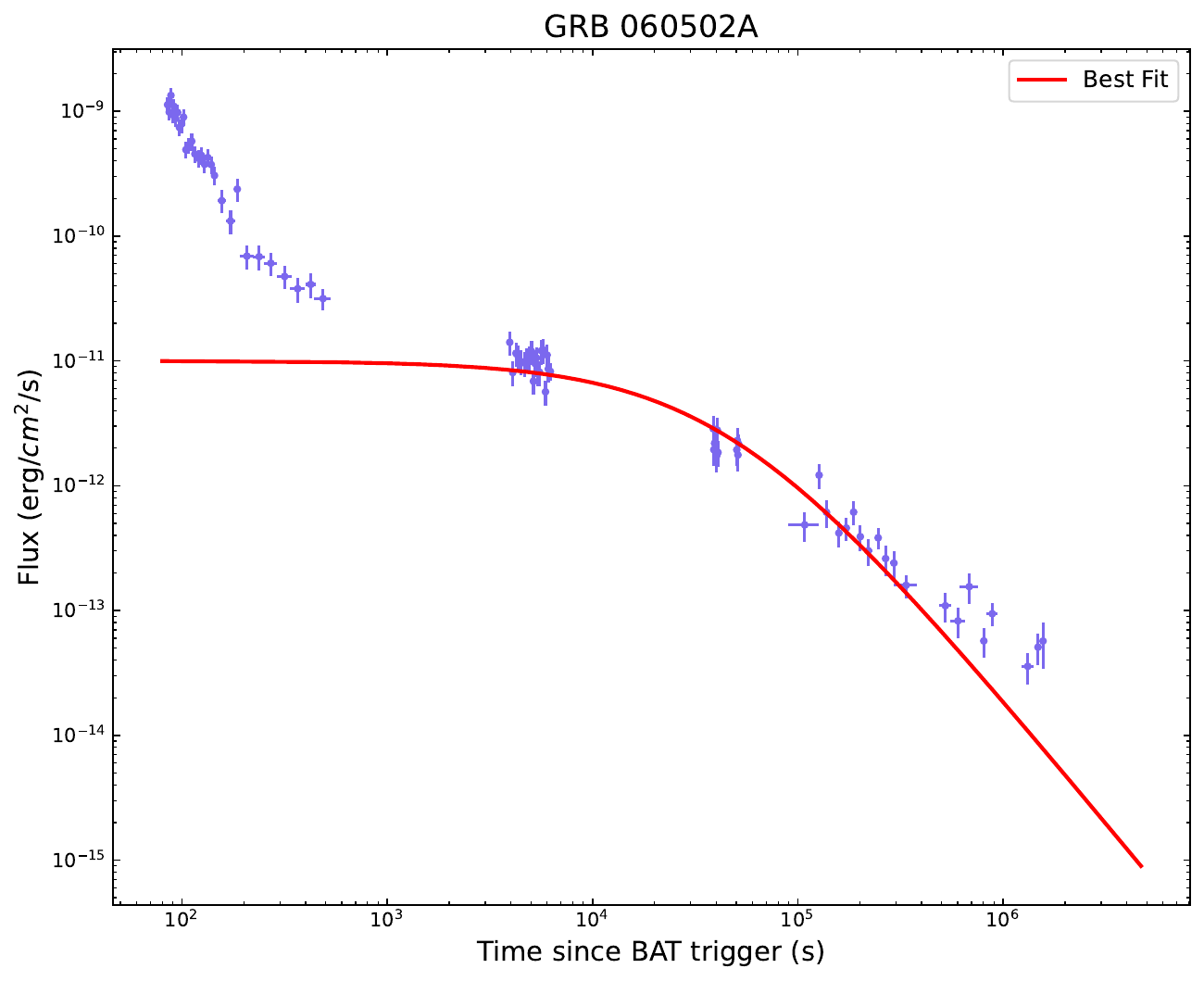}}%

\noindent
\resizebox{55mm}{!}{\includegraphics[]{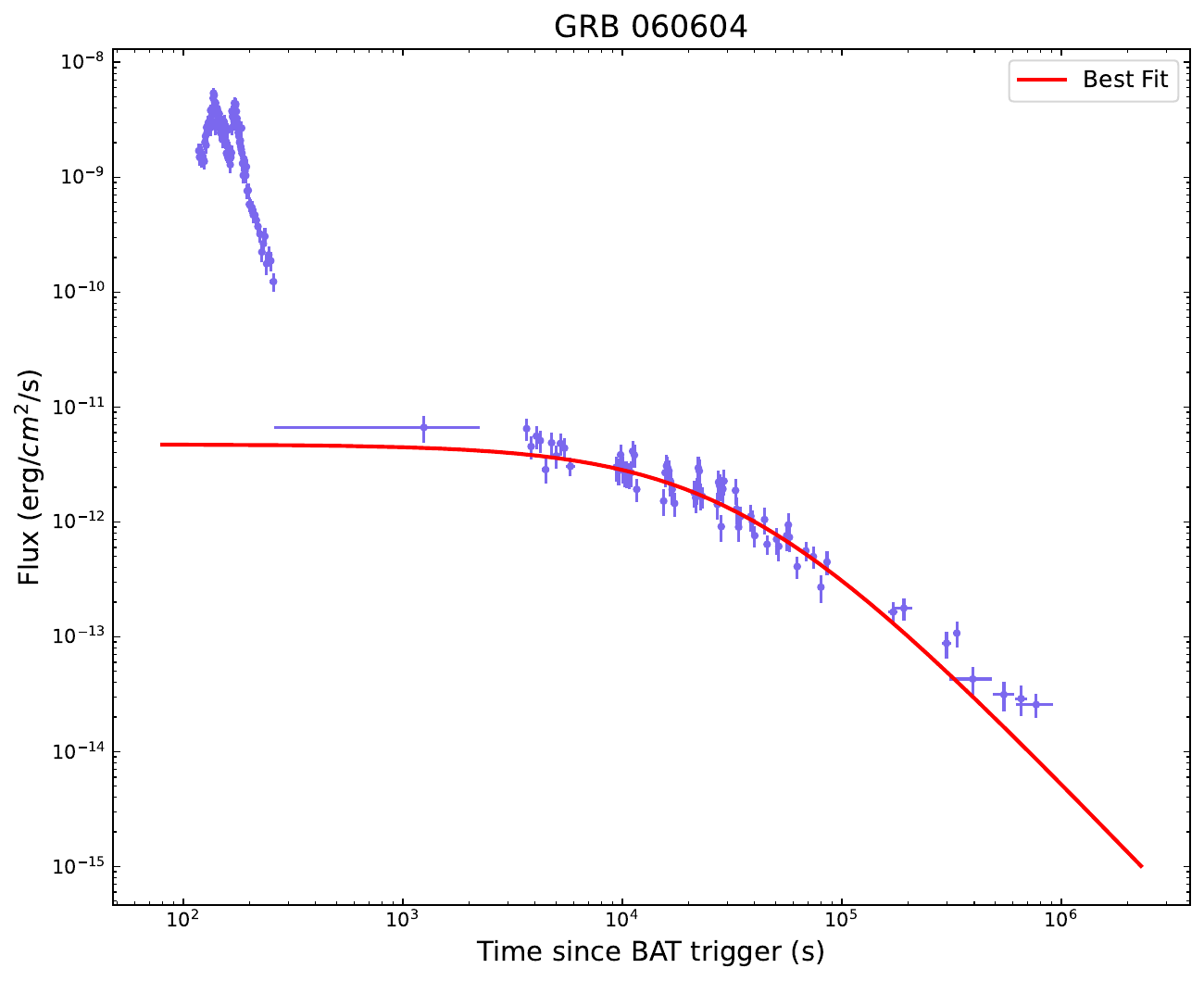}}%
\resizebox{55mm}{!}{\includegraphics[]{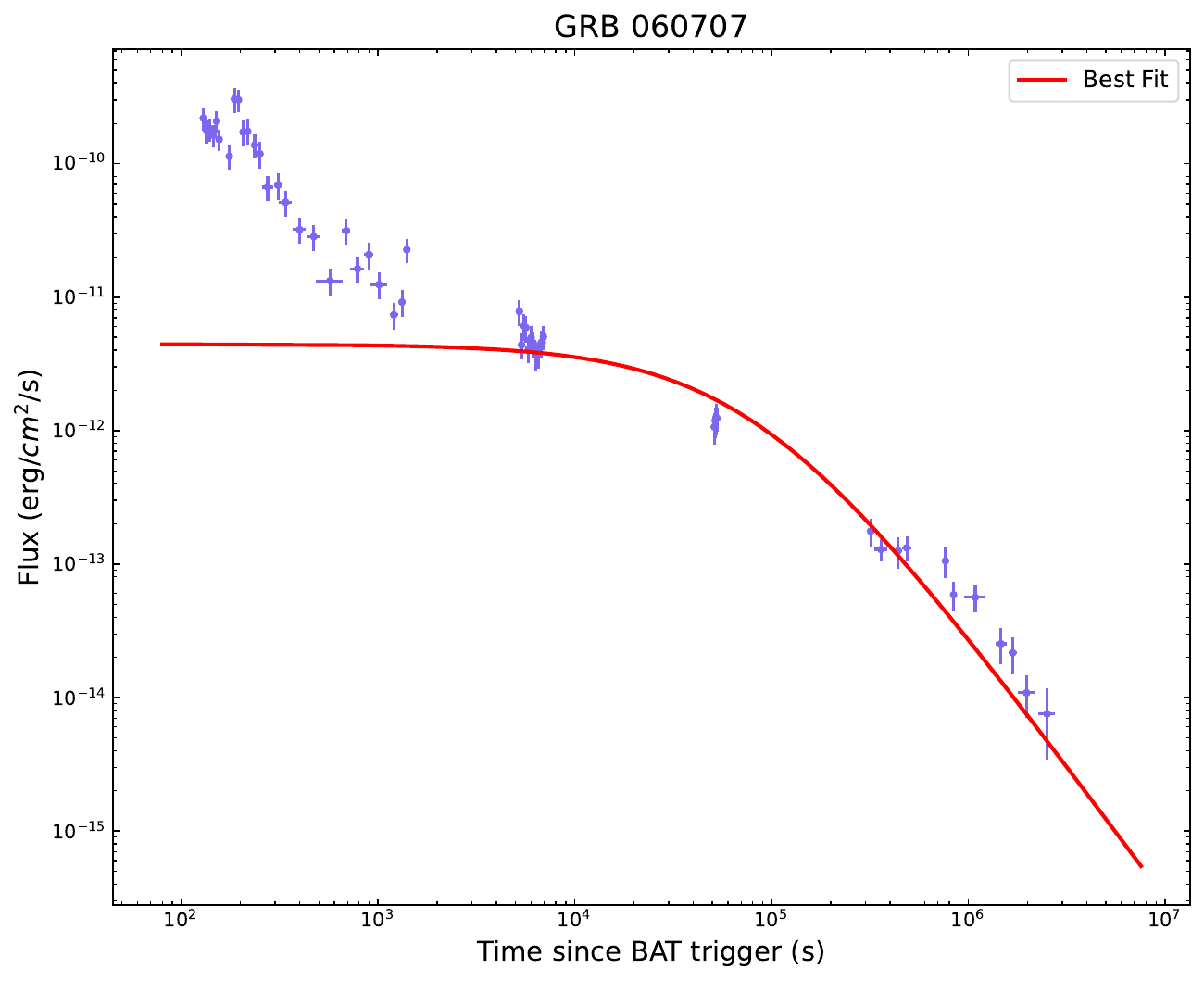}}%
\resizebox{55mm}{!}{\includegraphics[]{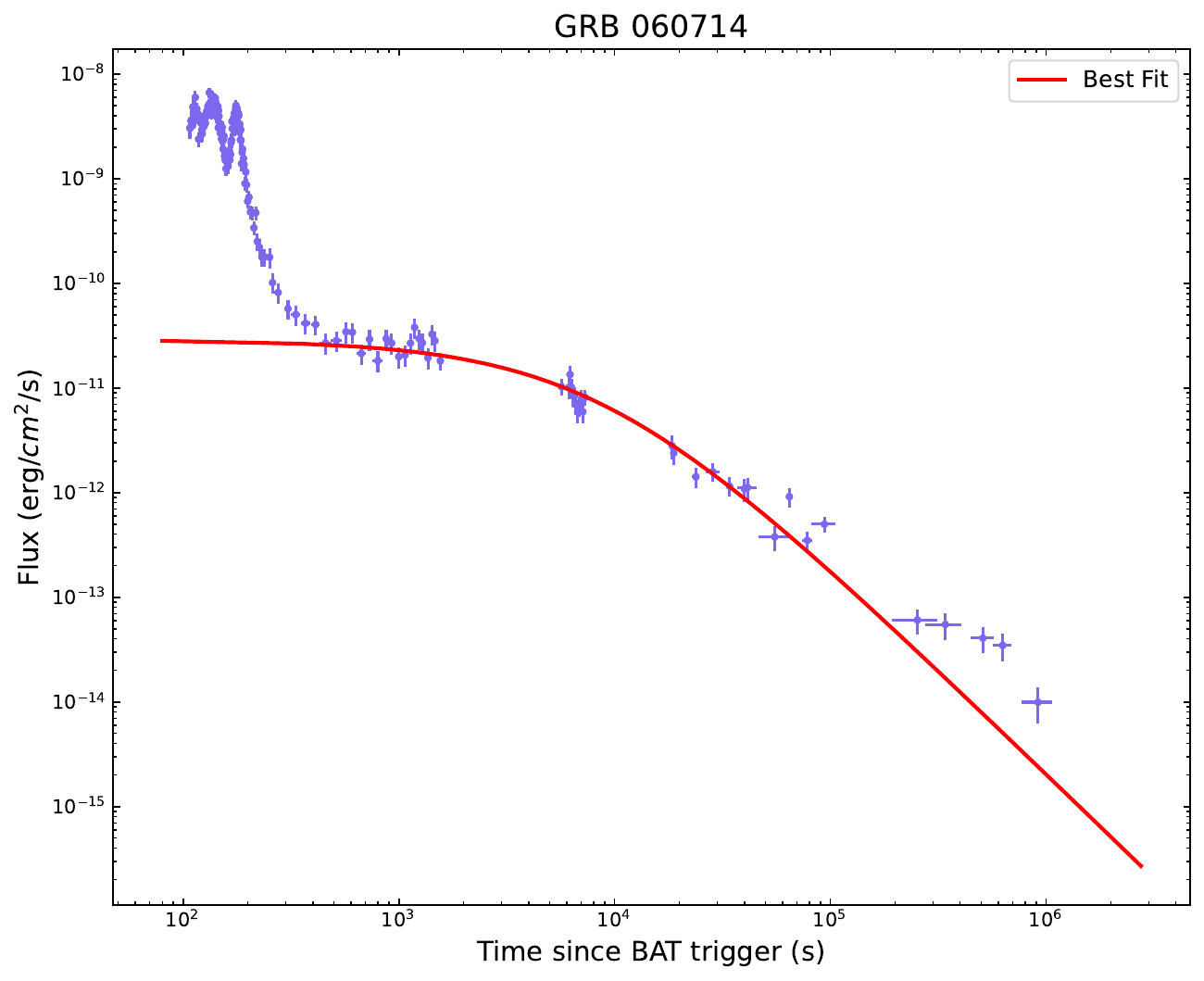}}%

\noindent
\resizebox{55mm}{!}{\includegraphics[]{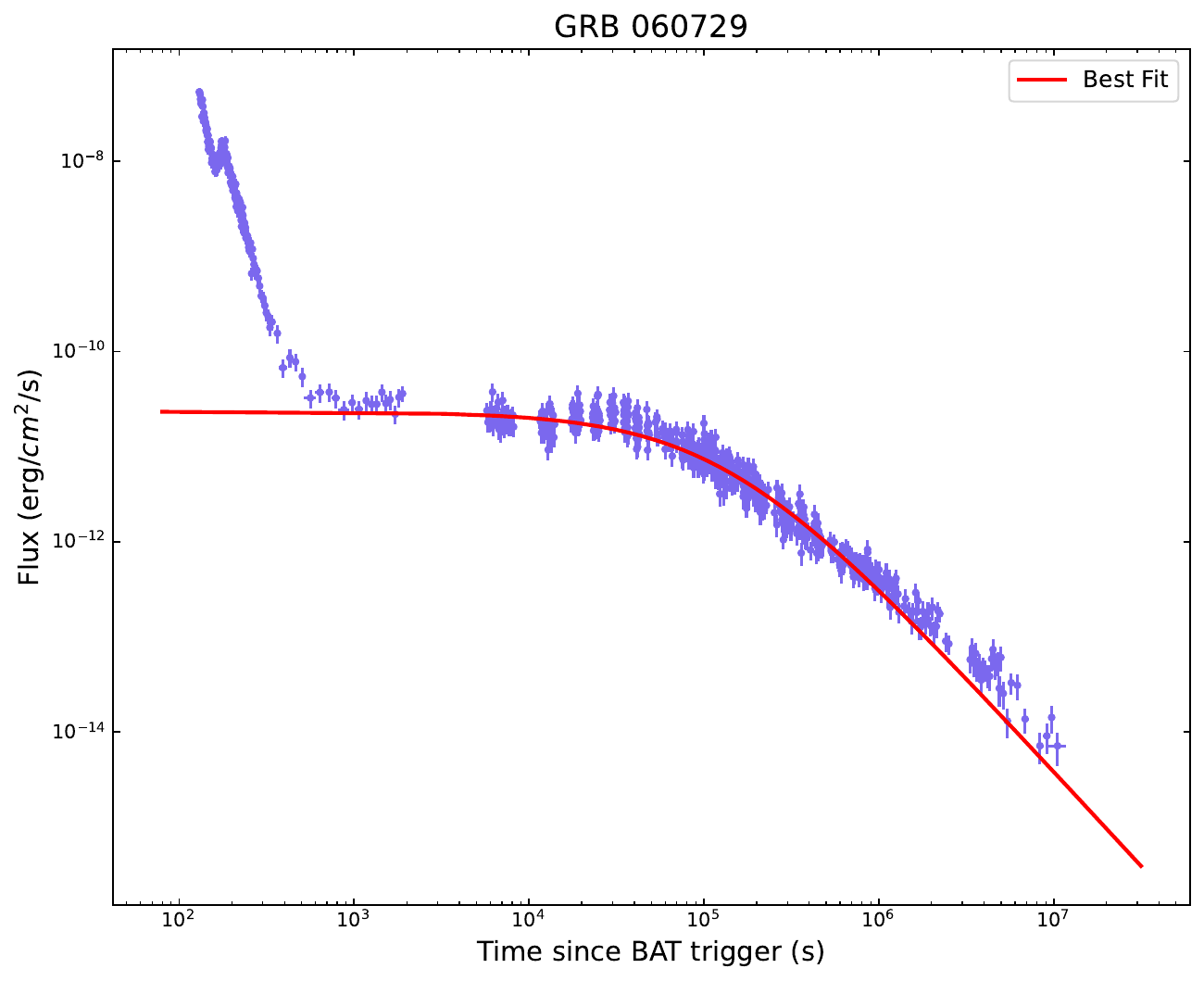}}%
\resizebox{55mm}{!}{\includegraphics[]{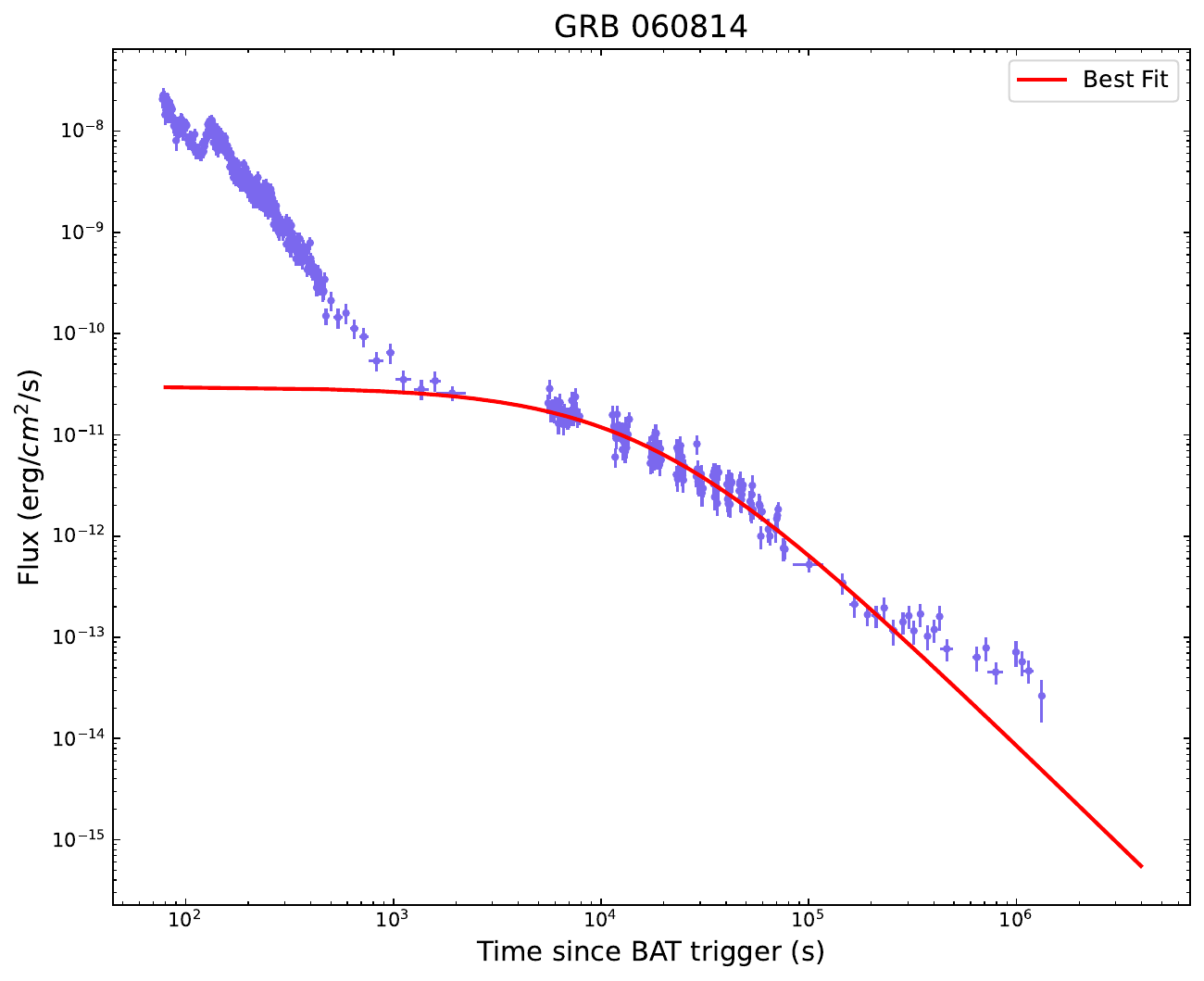}}%
\resizebox{55mm}{!}{\includegraphics[]{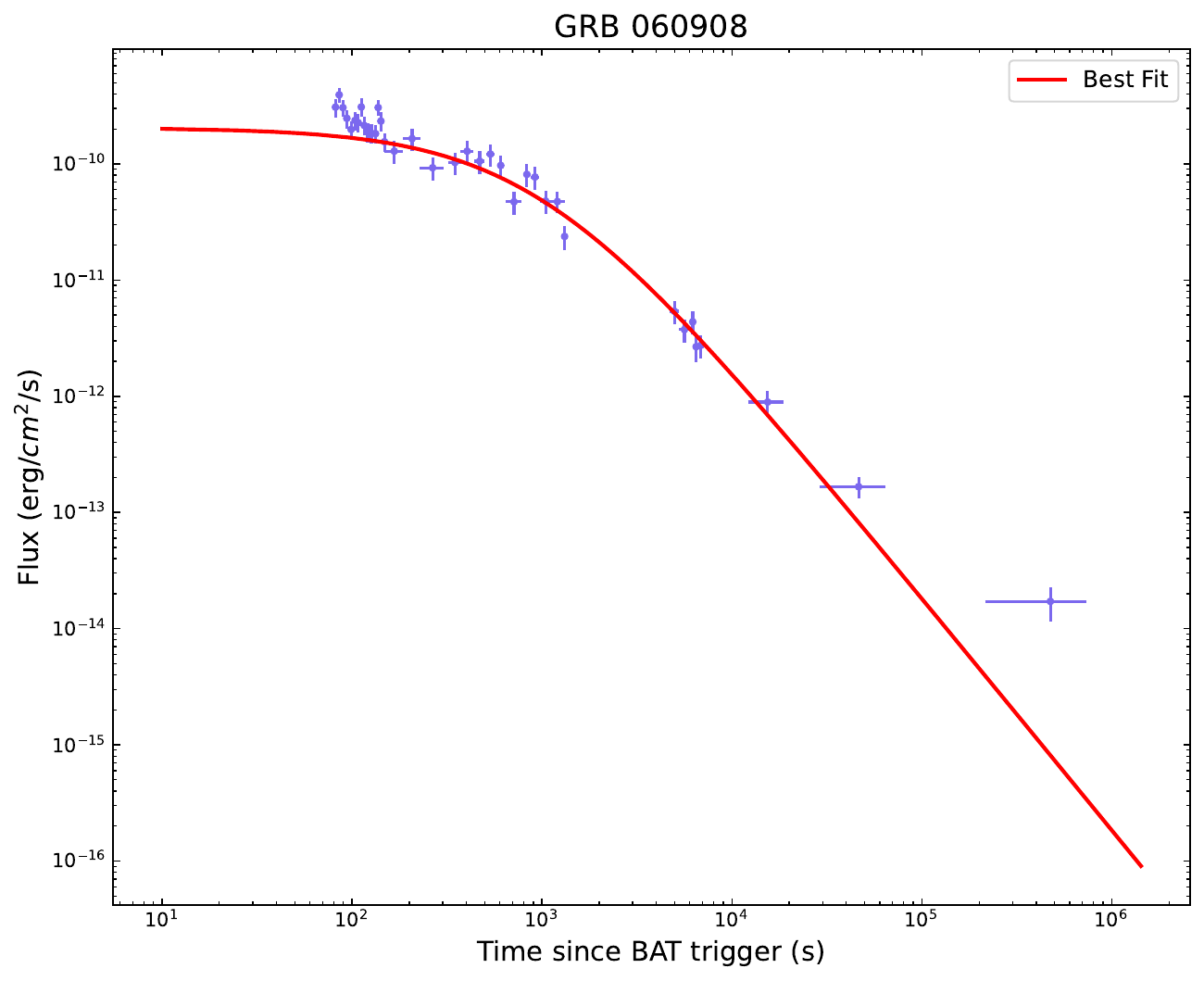}}%

\noindent
\resizebox{55mm}{!}{\includegraphics[]{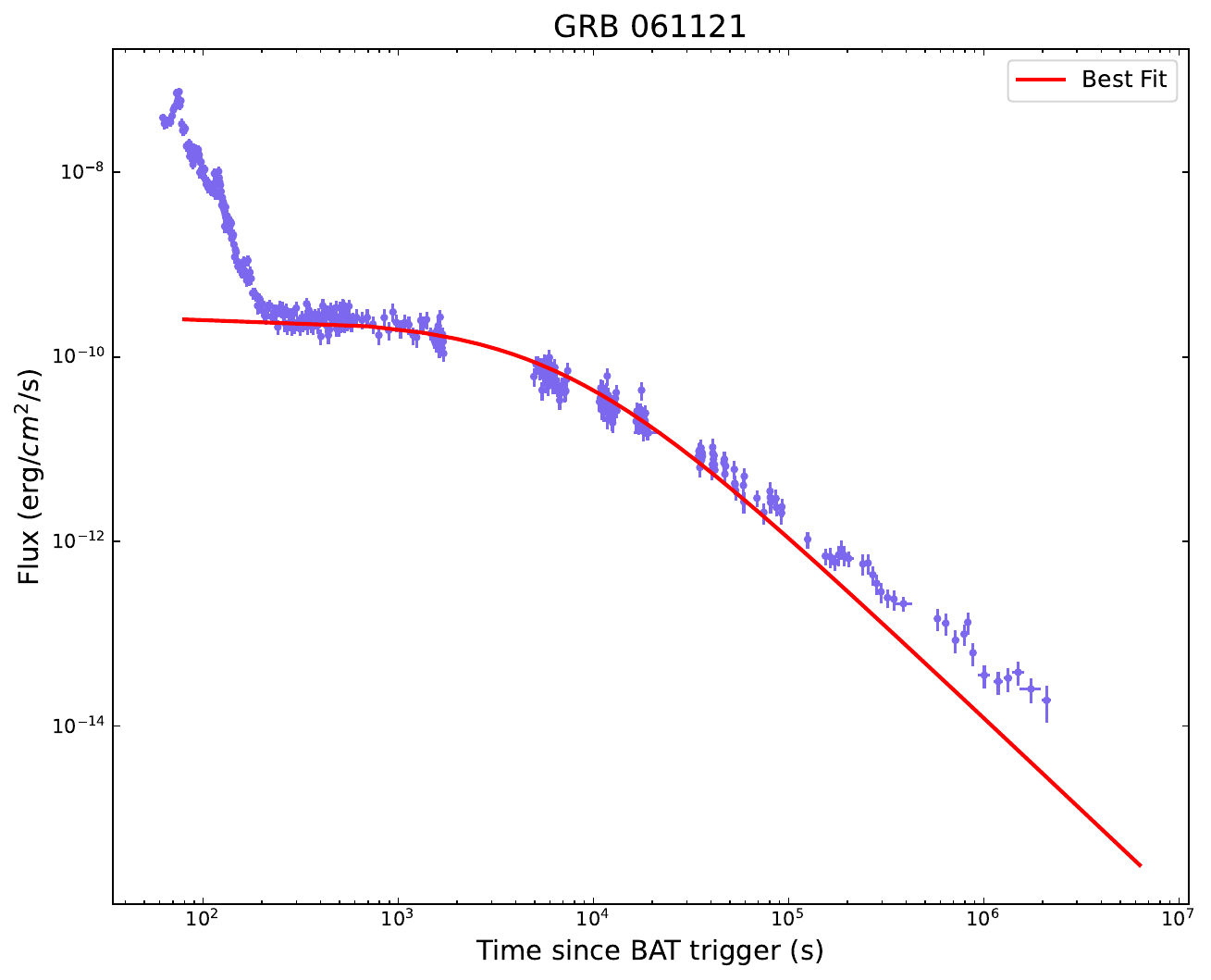}}%
\resizebox{55mm}{!}{\includegraphics[]{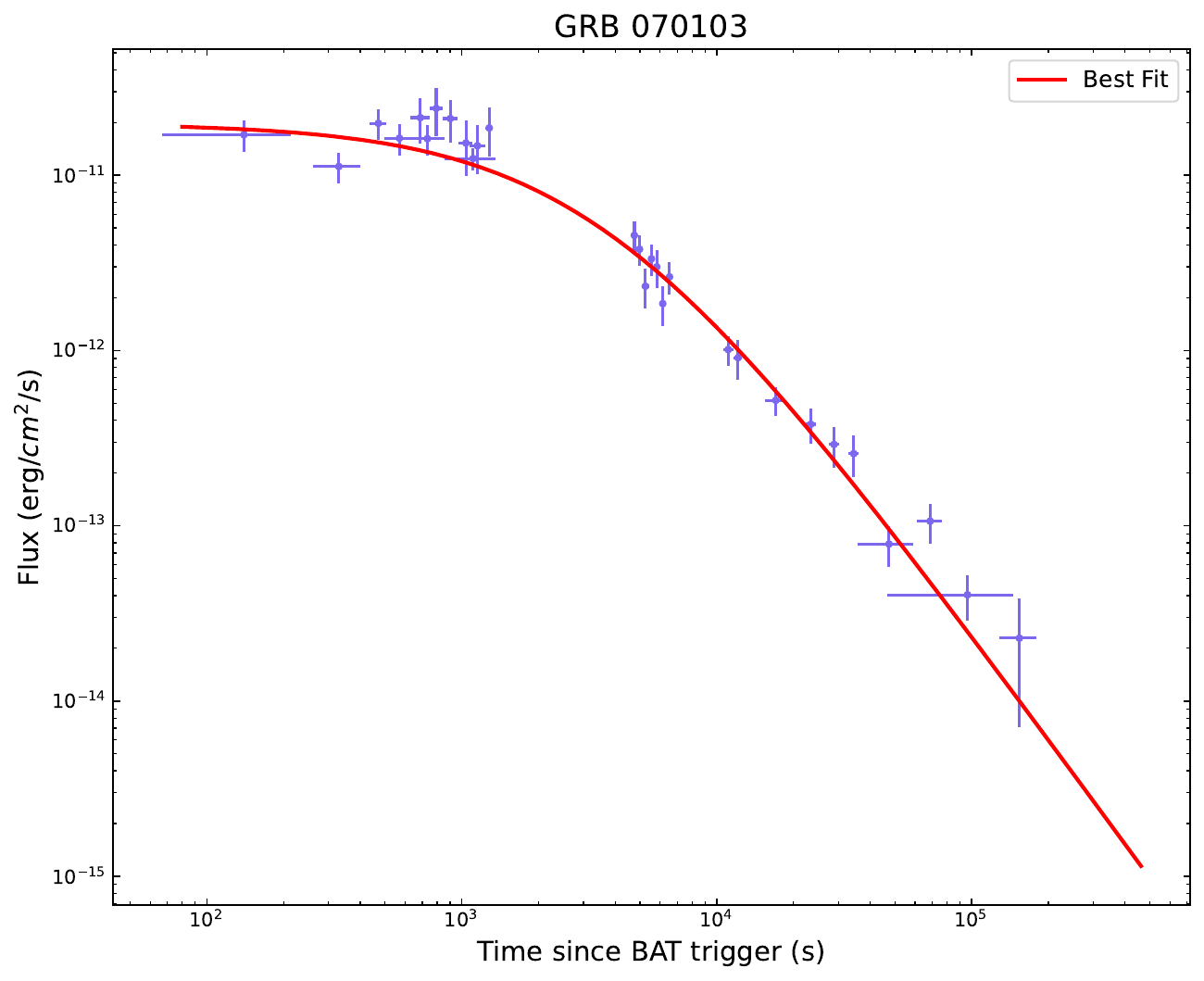}}%
\resizebox{55mm}{!}{\includegraphics[]{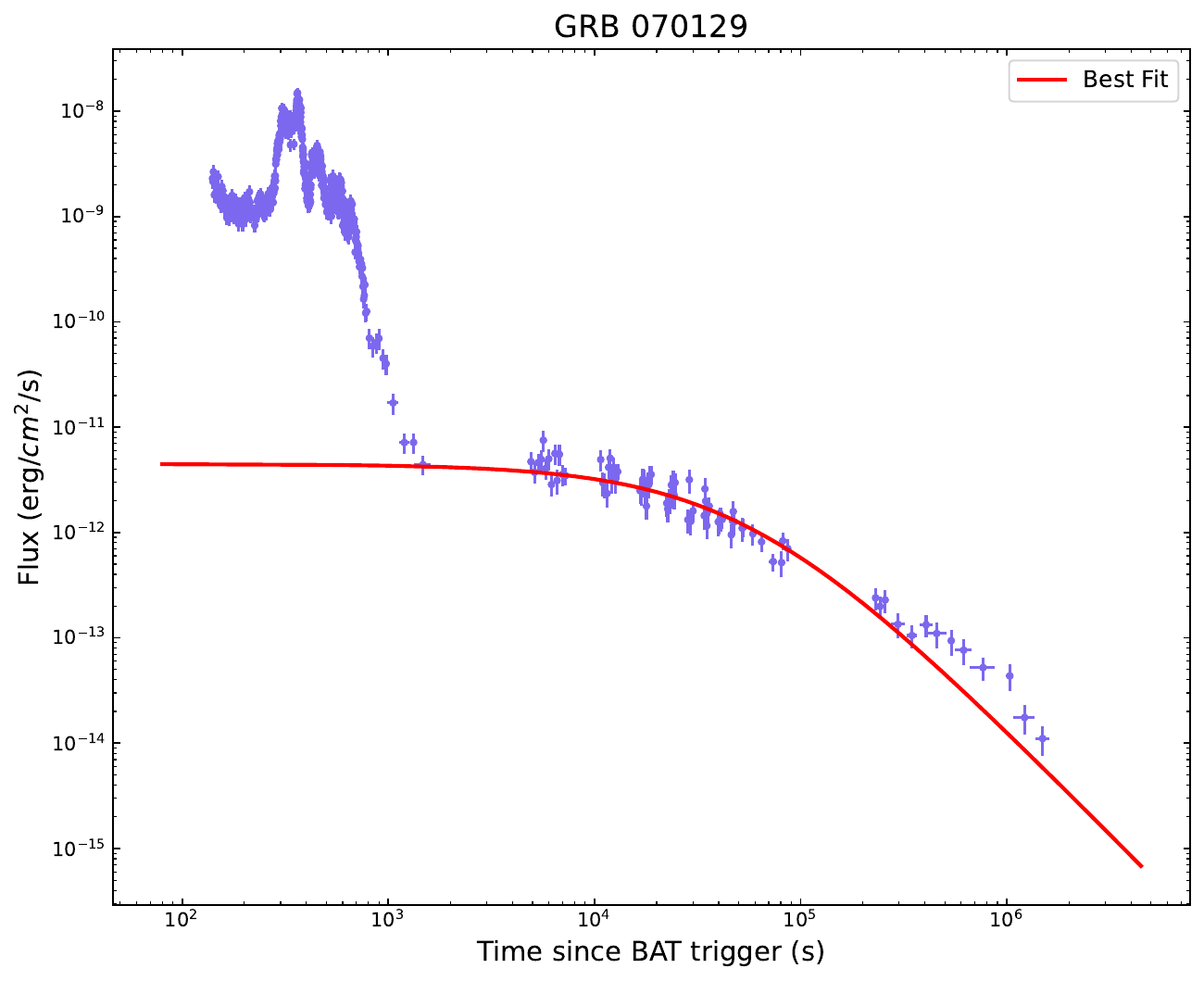}}%

\caption{The X-ray light curves of 139 GRBs, excluding the 30 from \cite{2022ApJ...924...97W} (GRB 090205 may be a SGRB \citep{2010A&A...522A..20D}). The first 51 are GRBs with redshifts. The remaining 88 are GRBs with pseudo-redshifts.}
\label{fig:figure1}
\end{figure*}

\addtocounter{figure}{-1}
\begin{figure*}[ht!]

\noindent
\resizebox{55mm}{!}{\includegraphics[]{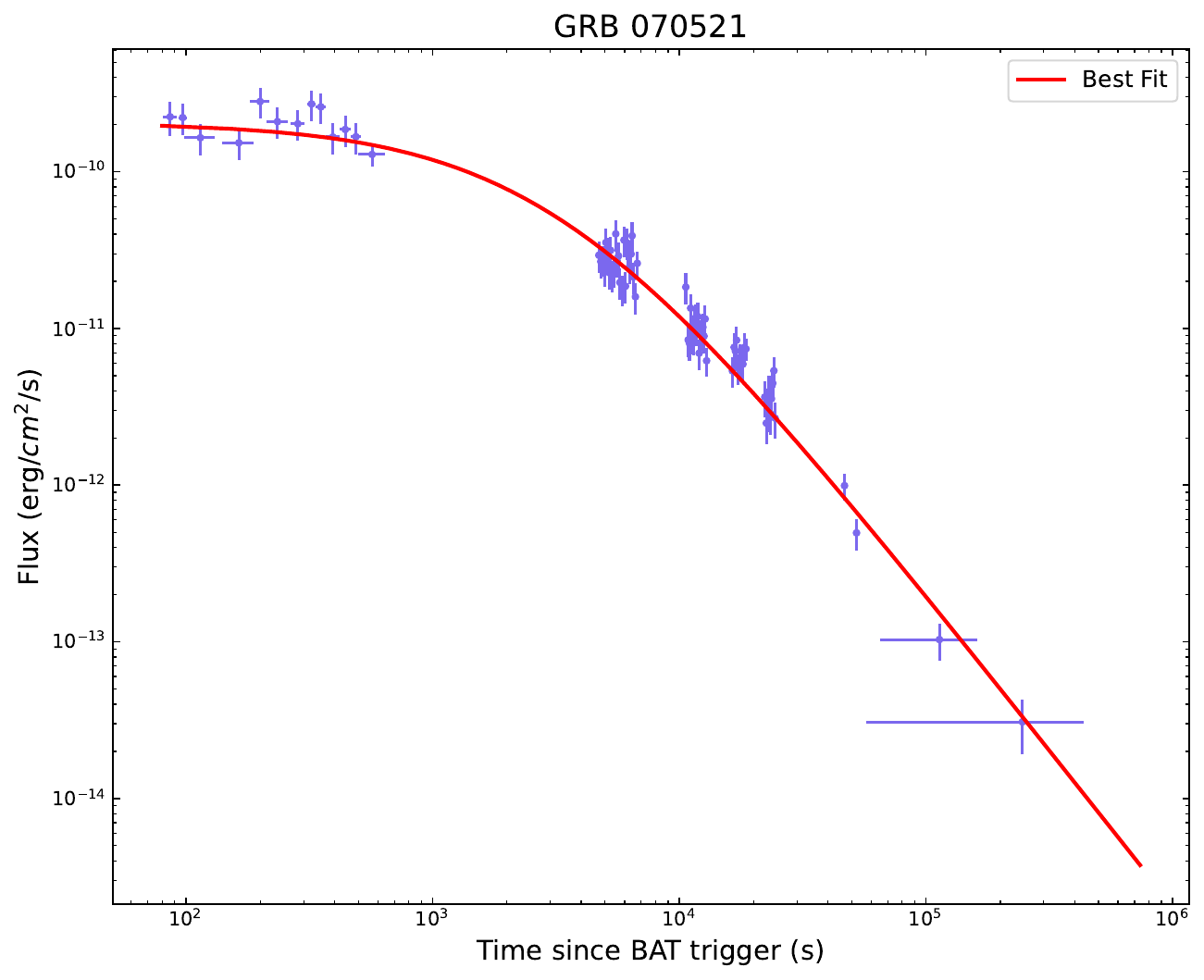}}%
\resizebox{55mm}{!}{\includegraphics[]{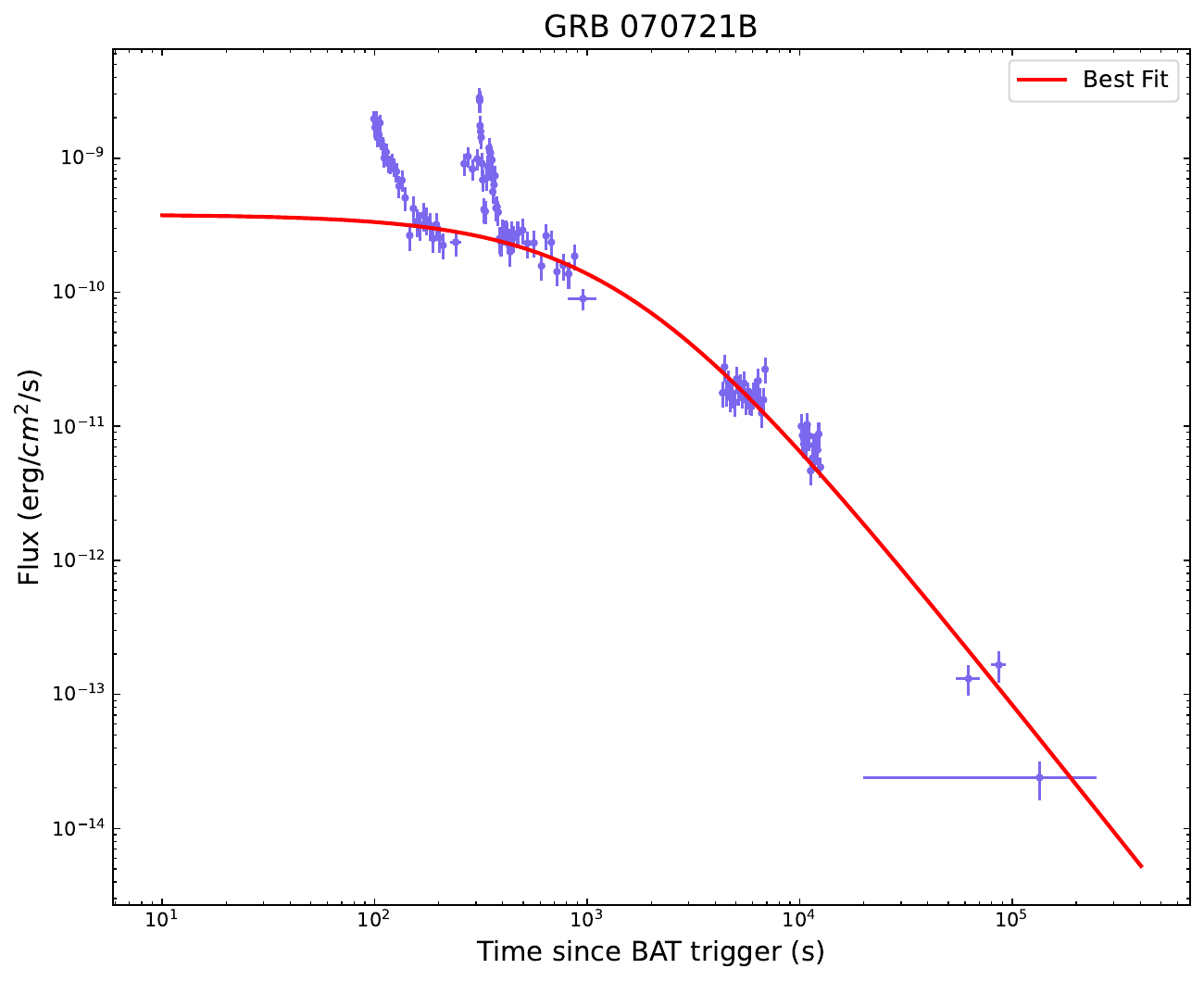}}%
\resizebox{55mm}{!}{\includegraphics[]{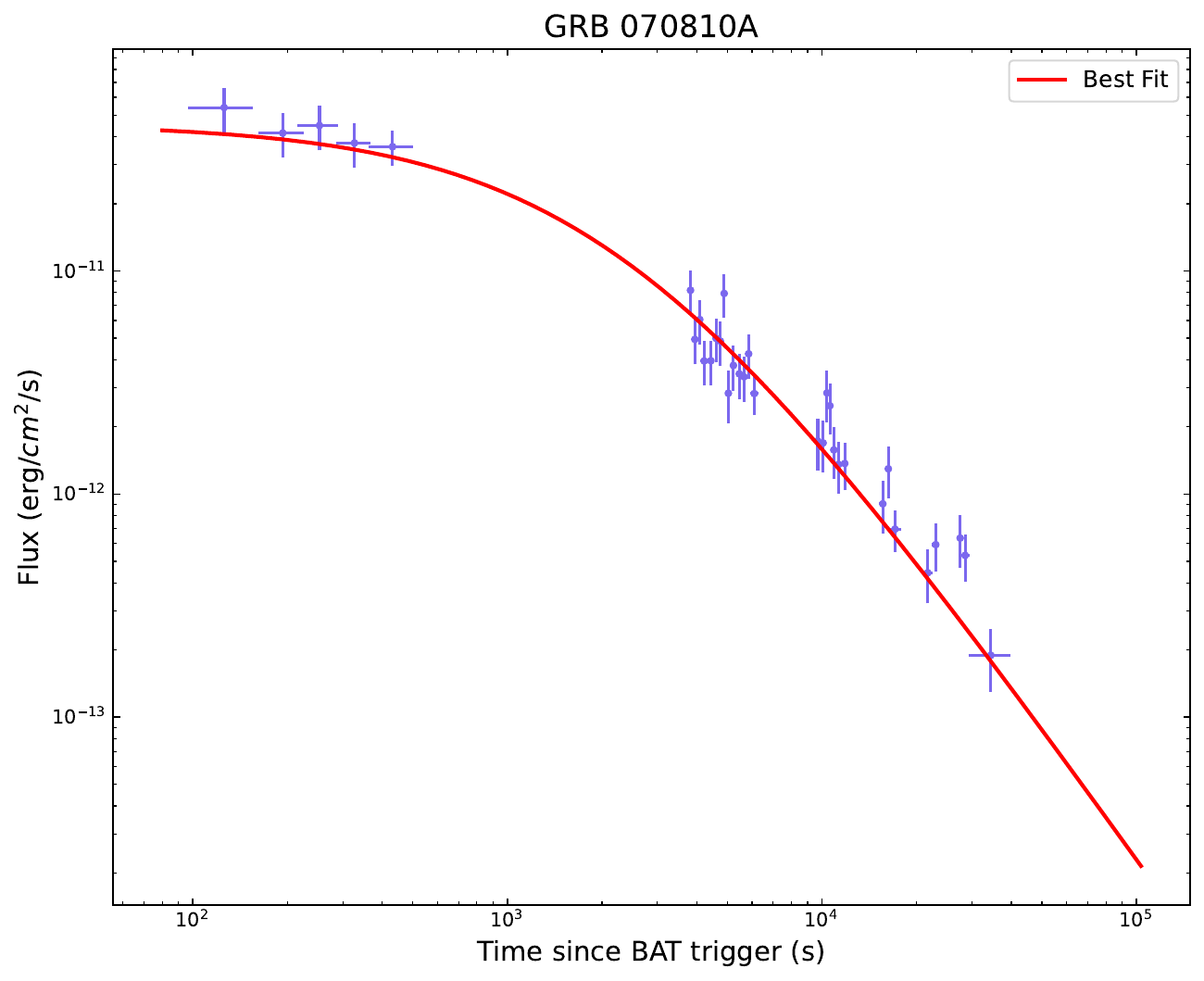}}%

\noindent
\resizebox{55mm}{!}{\includegraphics[]{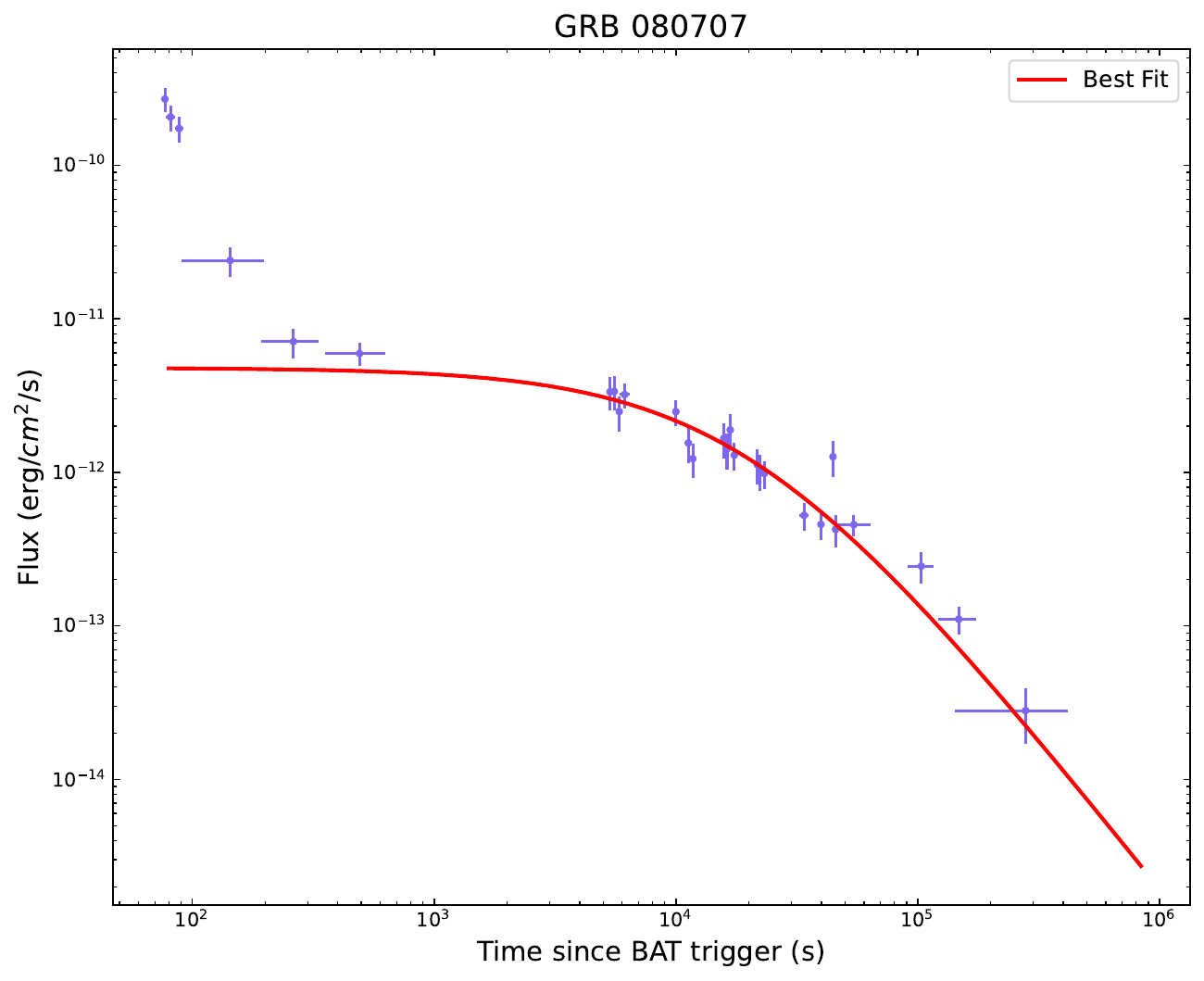}}%
\resizebox{55mm}{!}{\includegraphics[]{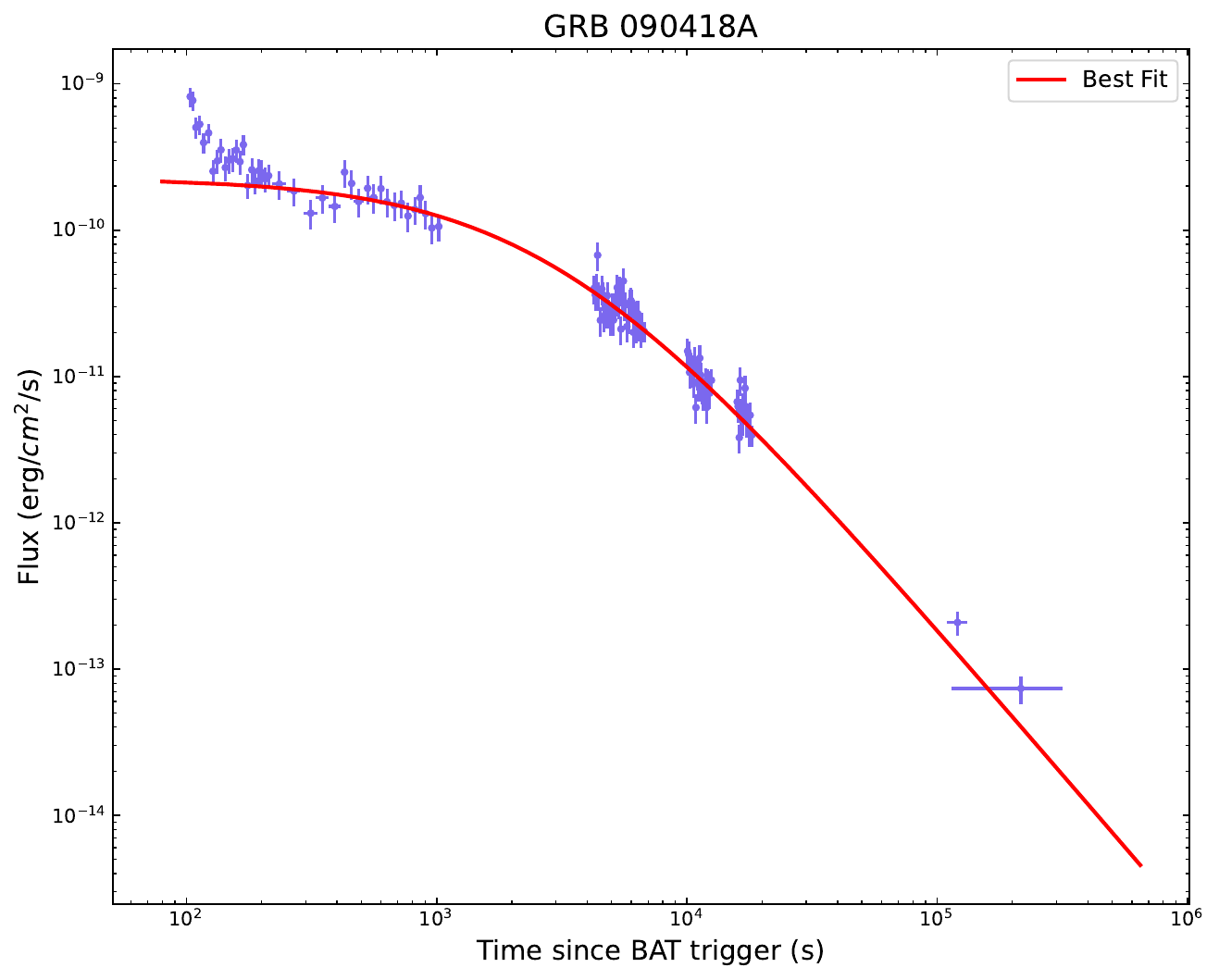}}%
\resizebox{55mm}{!}{\includegraphics[]{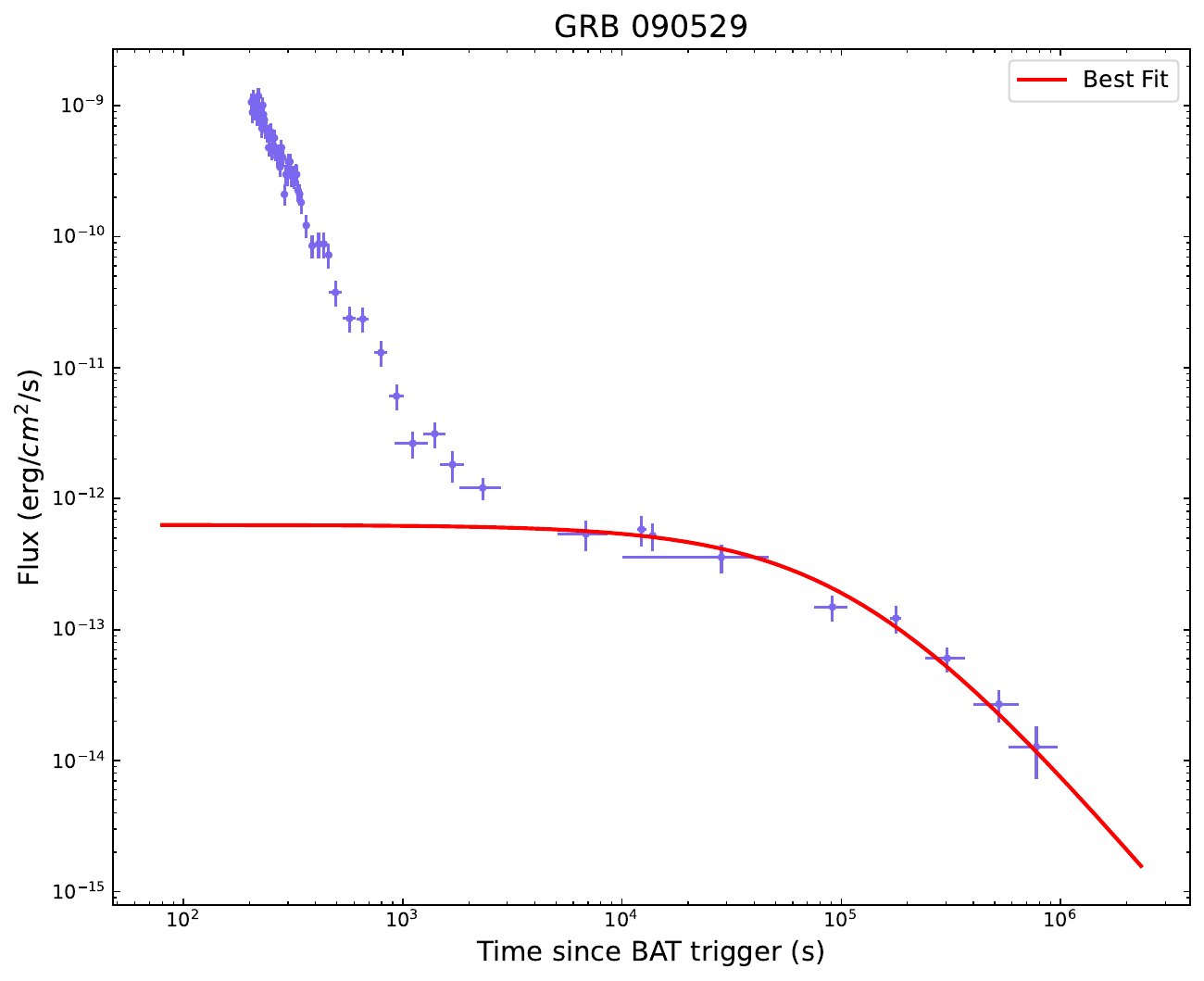}}%

\noindent
\resizebox{55mm}{!}{\includegraphics[]{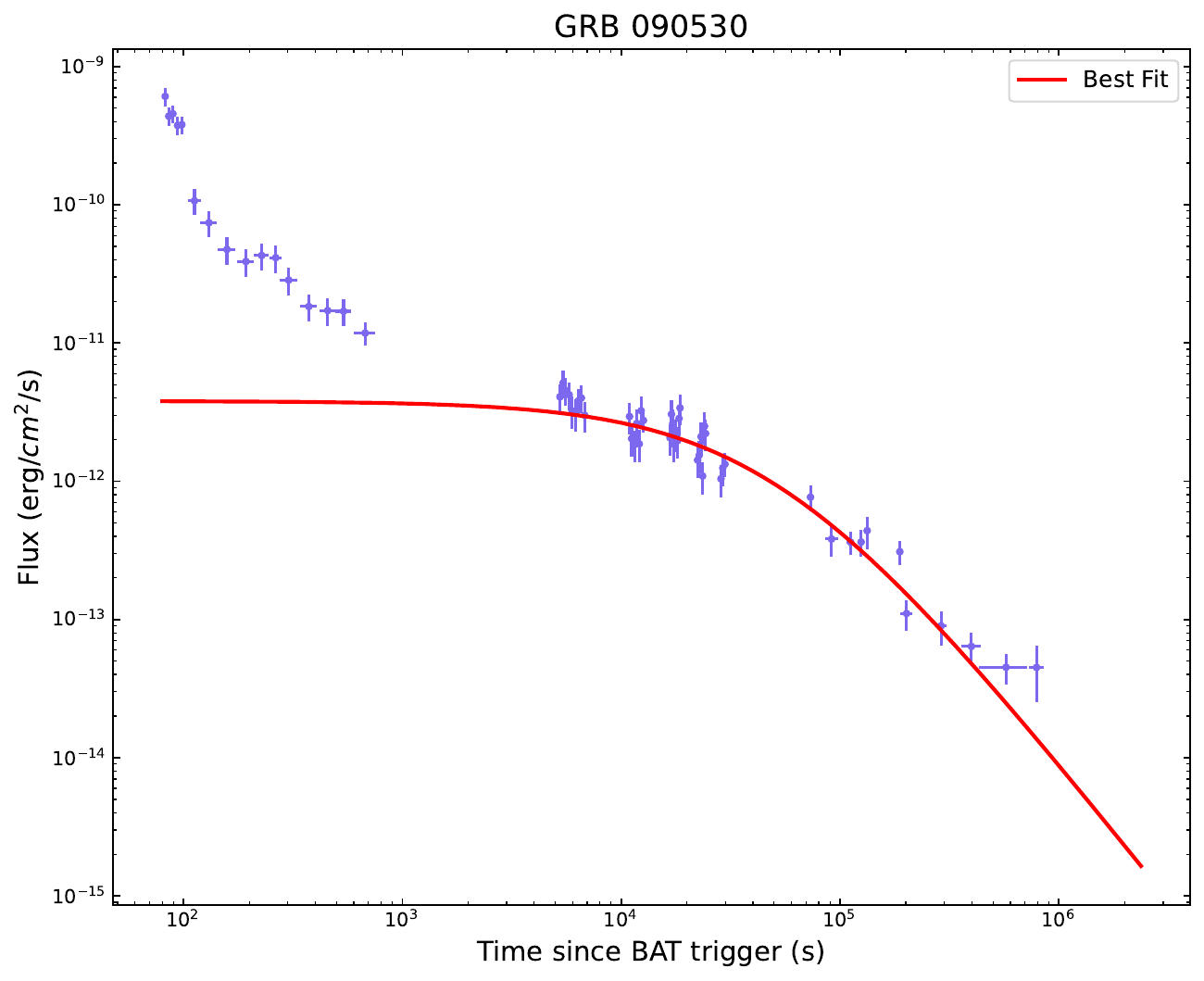}}%
\resizebox{55mm}{!}{\includegraphics[]{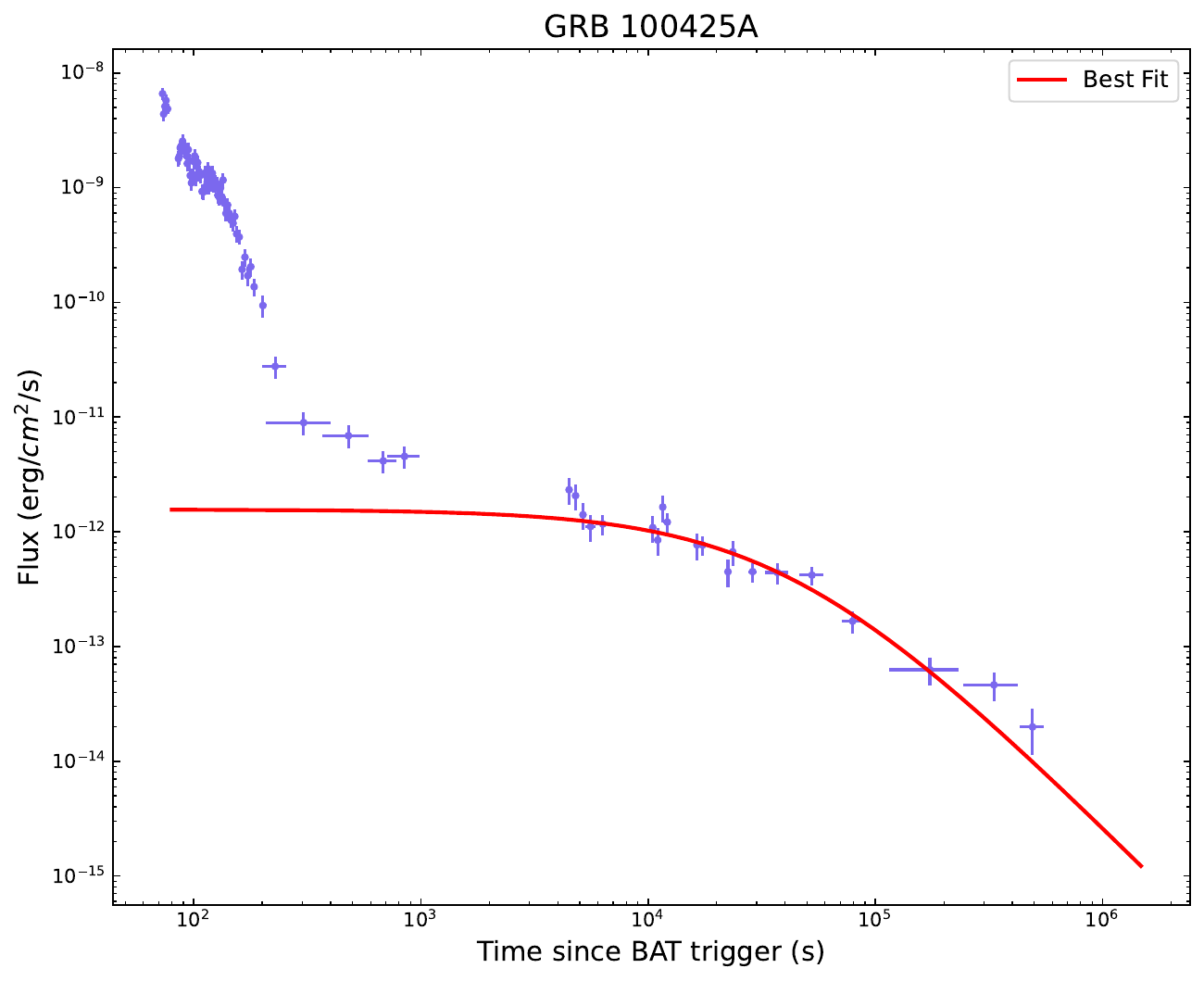}}%
\resizebox{55mm}{!}{\includegraphics[]{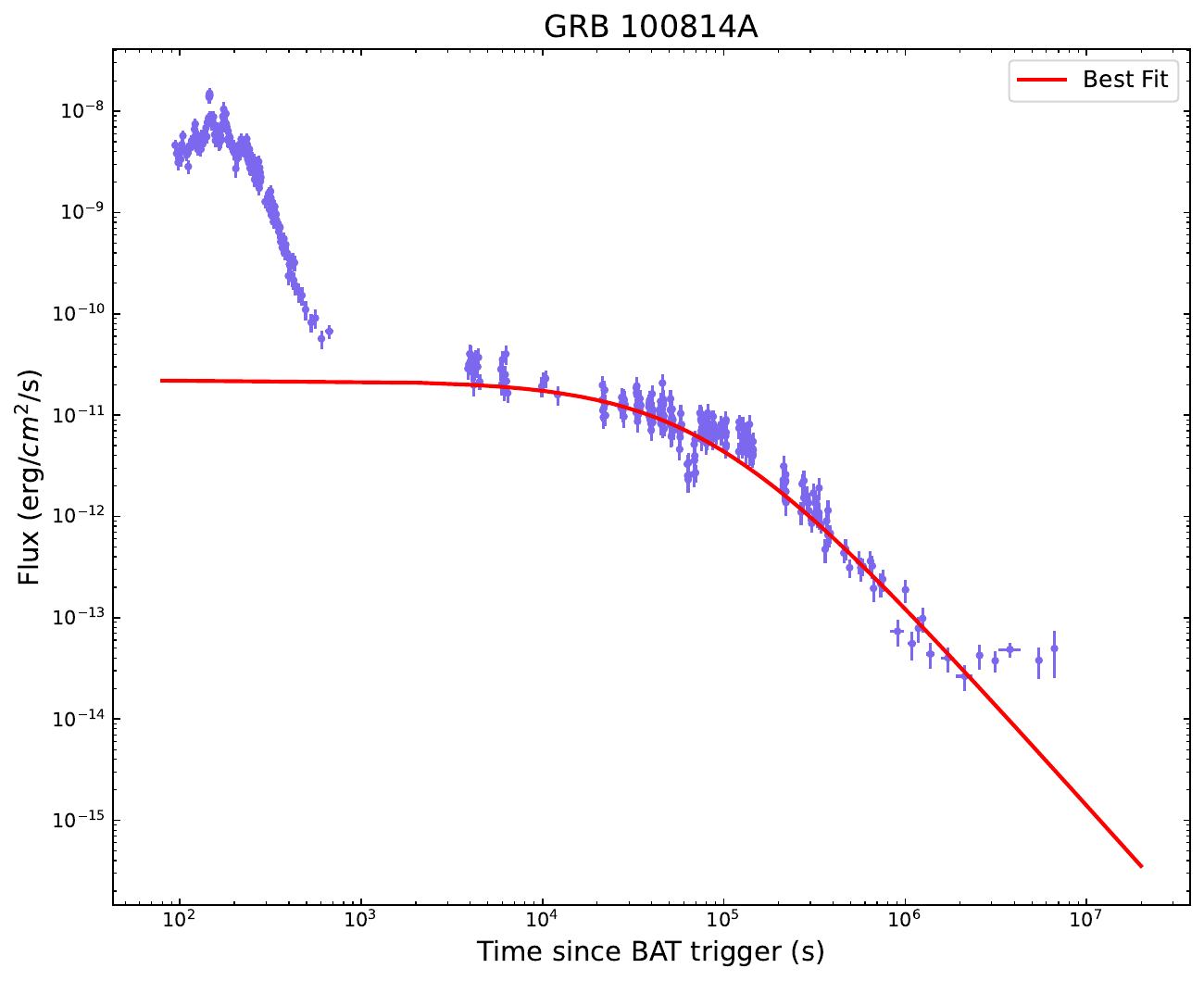}}%

\noindent
\resizebox{55mm}{!}{\includegraphics[]{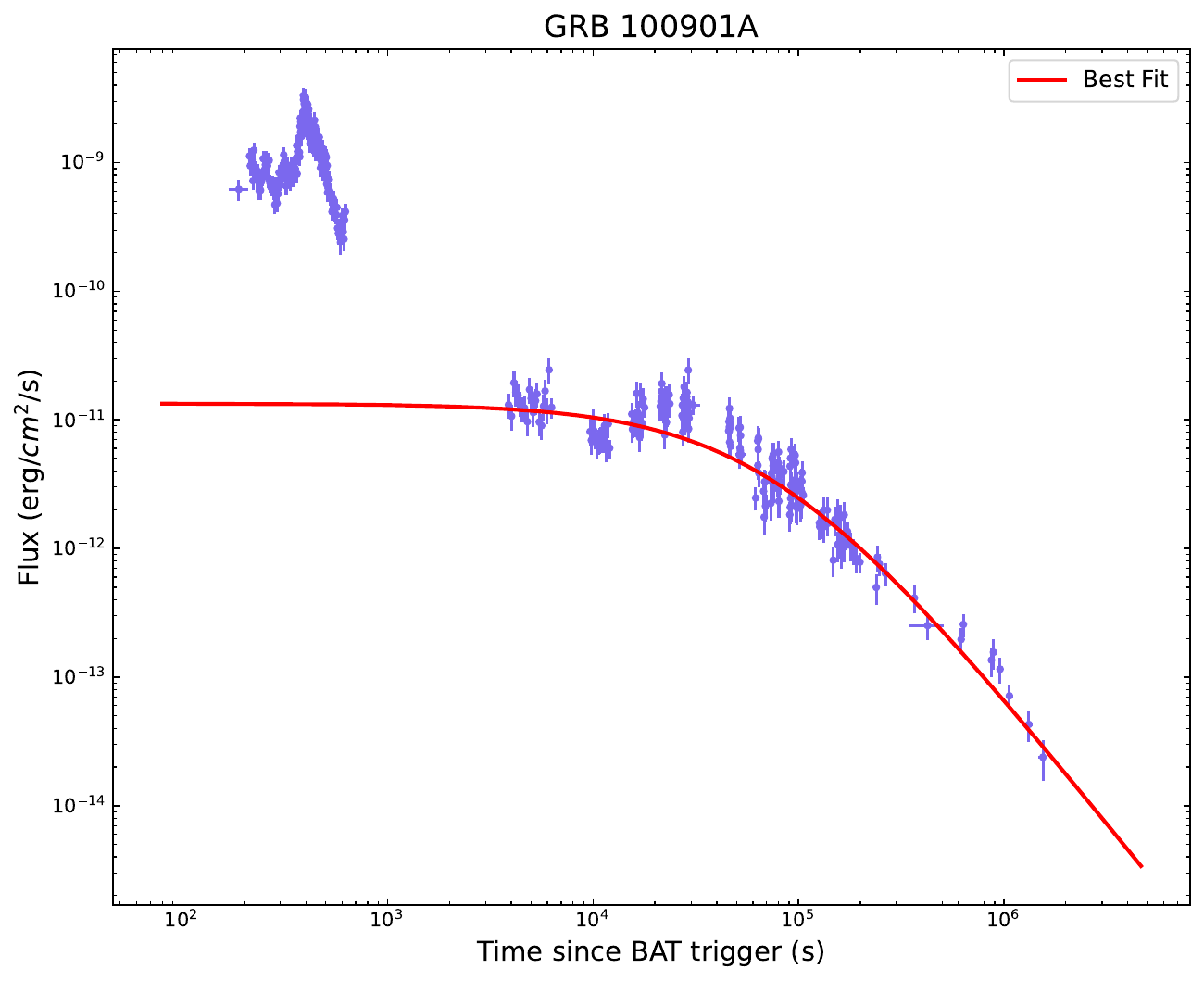}}%
\resizebox{55mm}{!}{\includegraphics[]{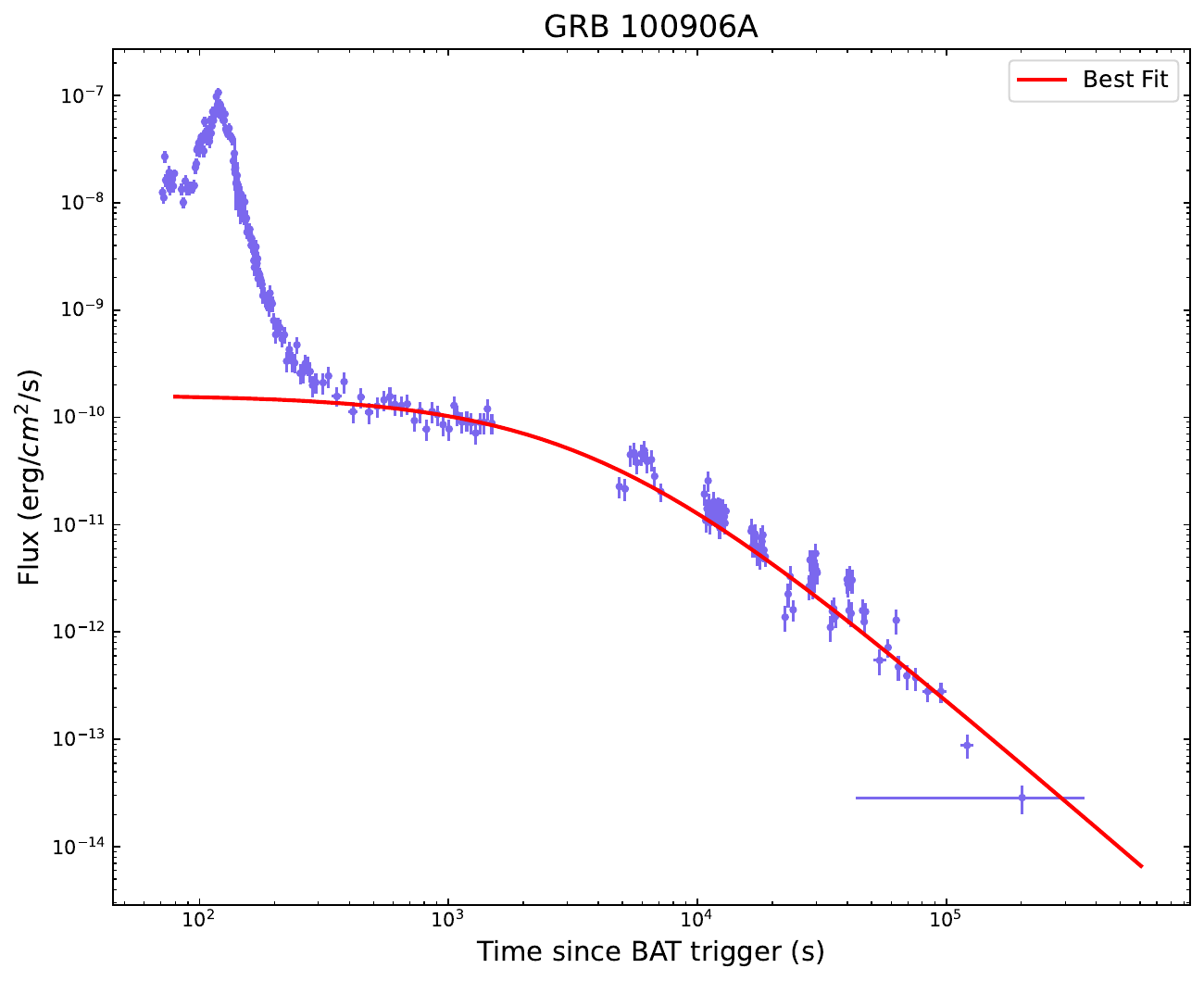}}%
\resizebox{55mm}{!}{\includegraphics[]{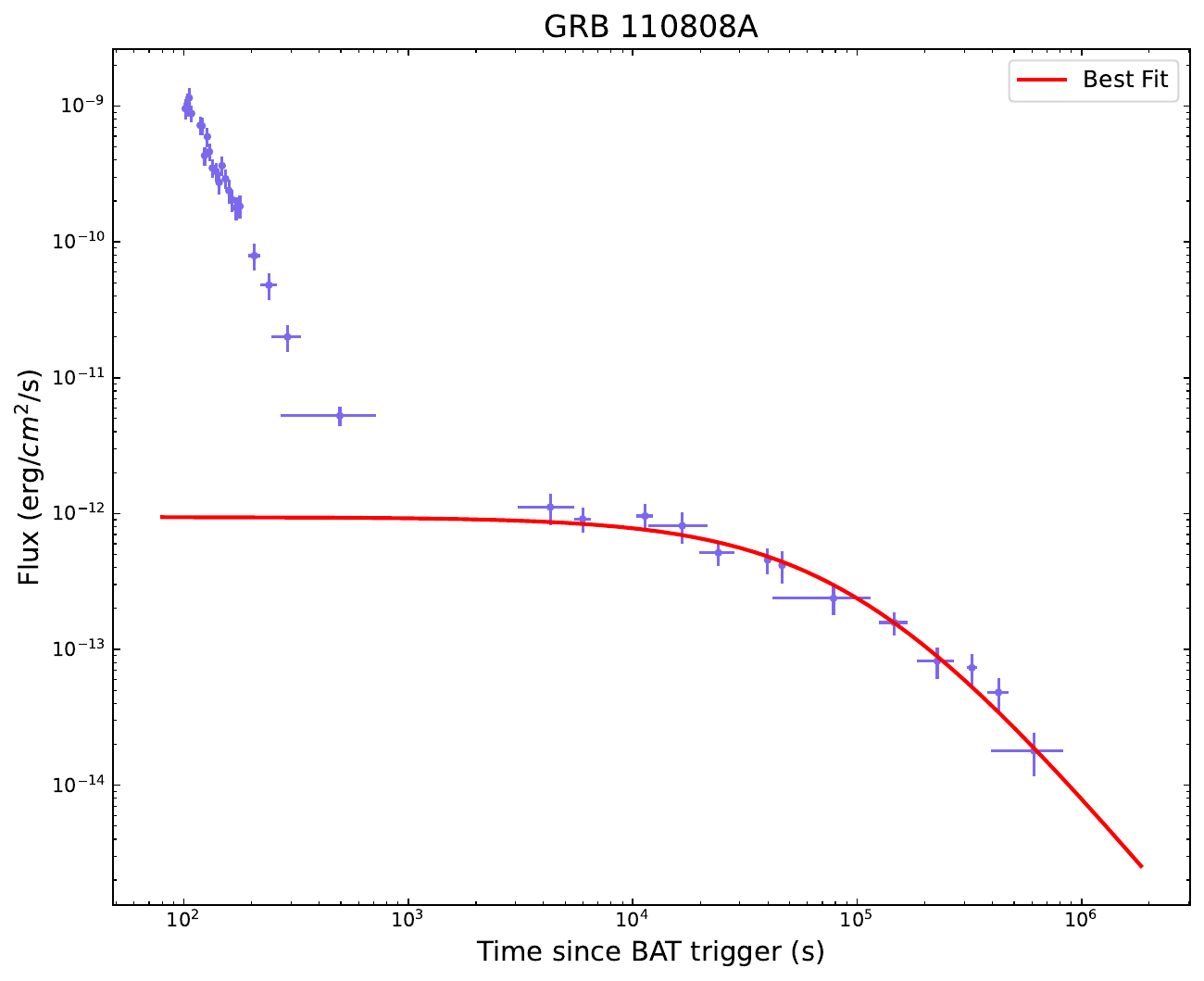}}%

\noindent
\resizebox{55mm}{!}{\includegraphics[]{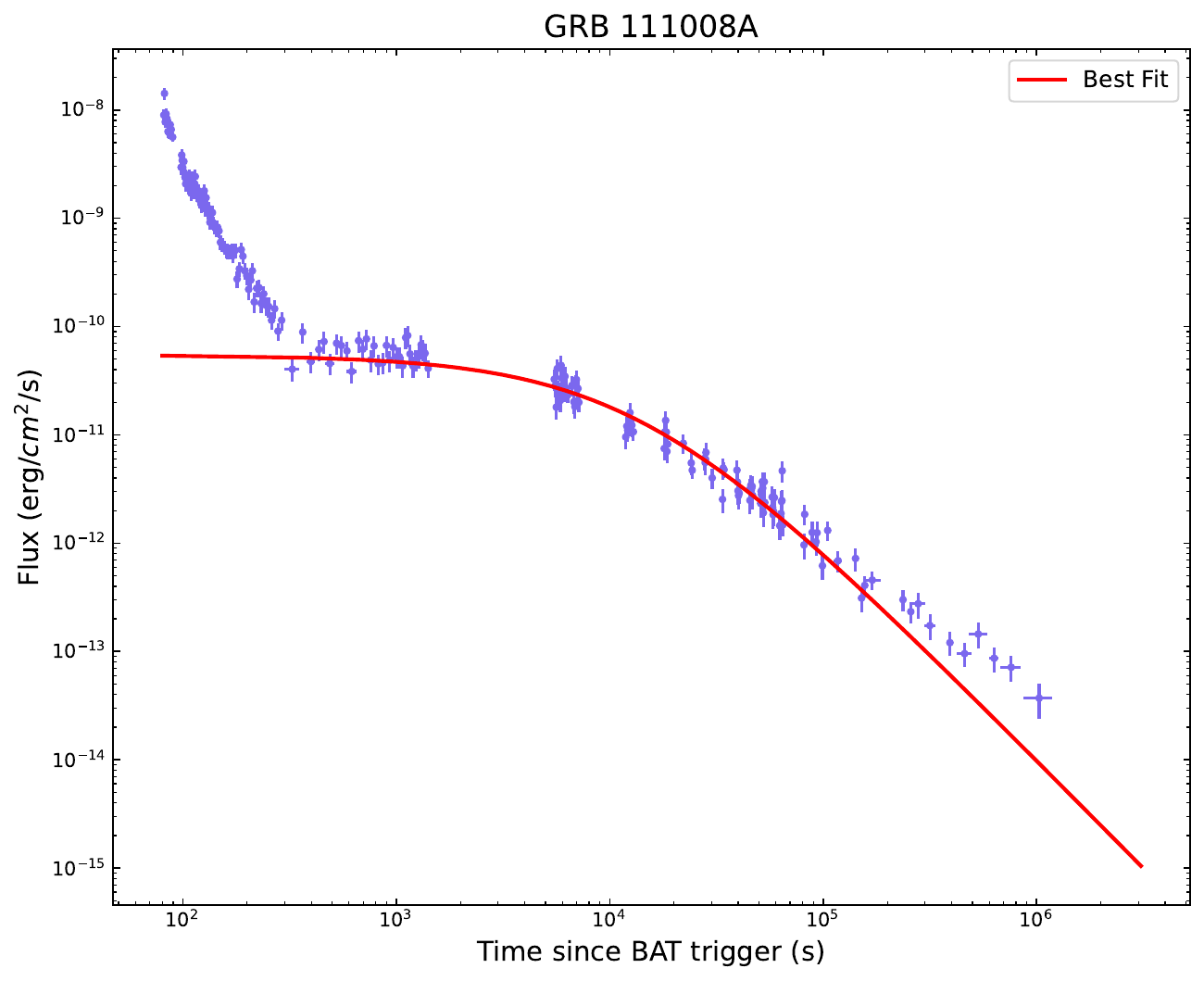}}%
\resizebox{55mm}{!}{\includegraphics[]{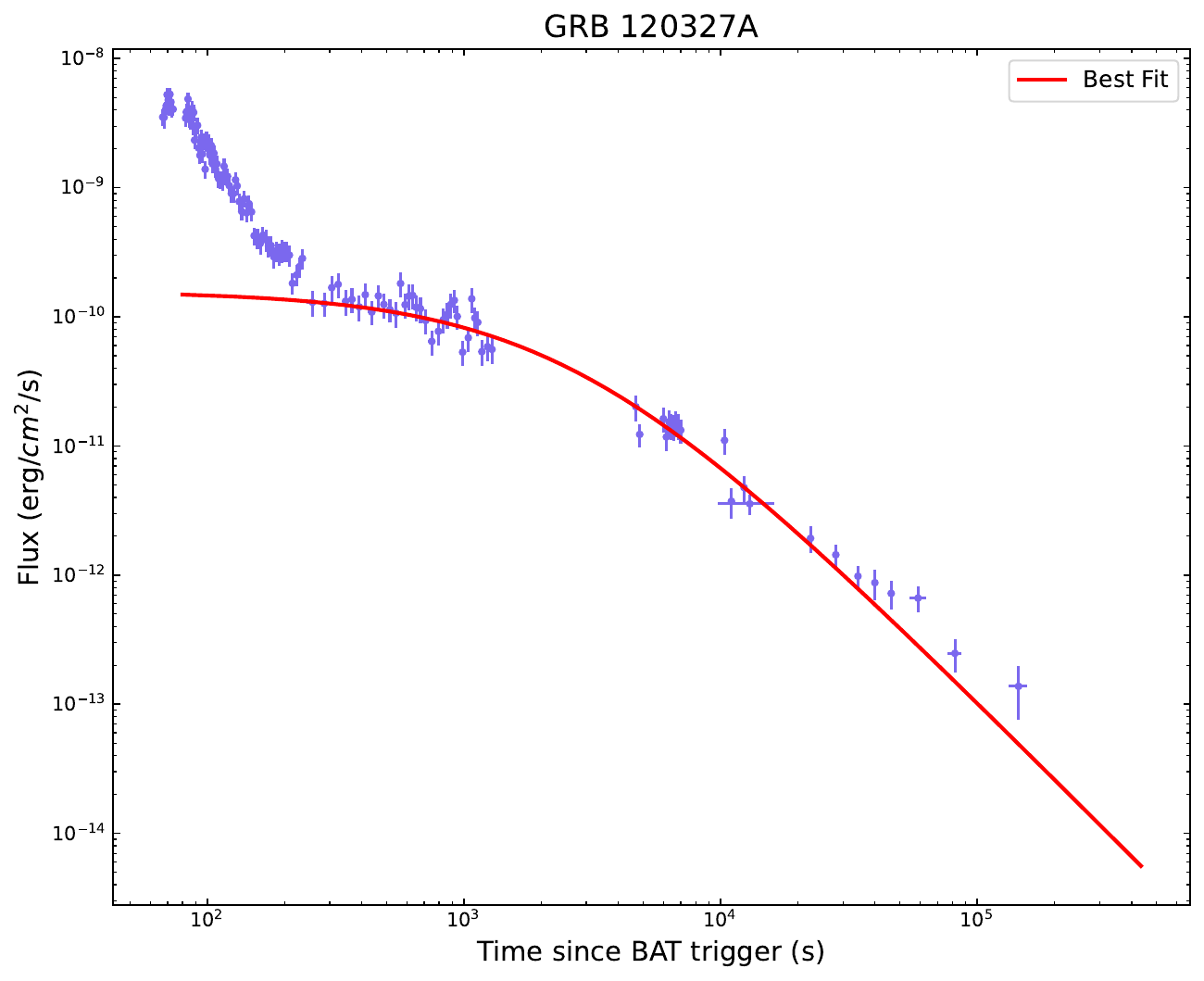}}%
\resizebox{55mm}{!}{\includegraphics[]{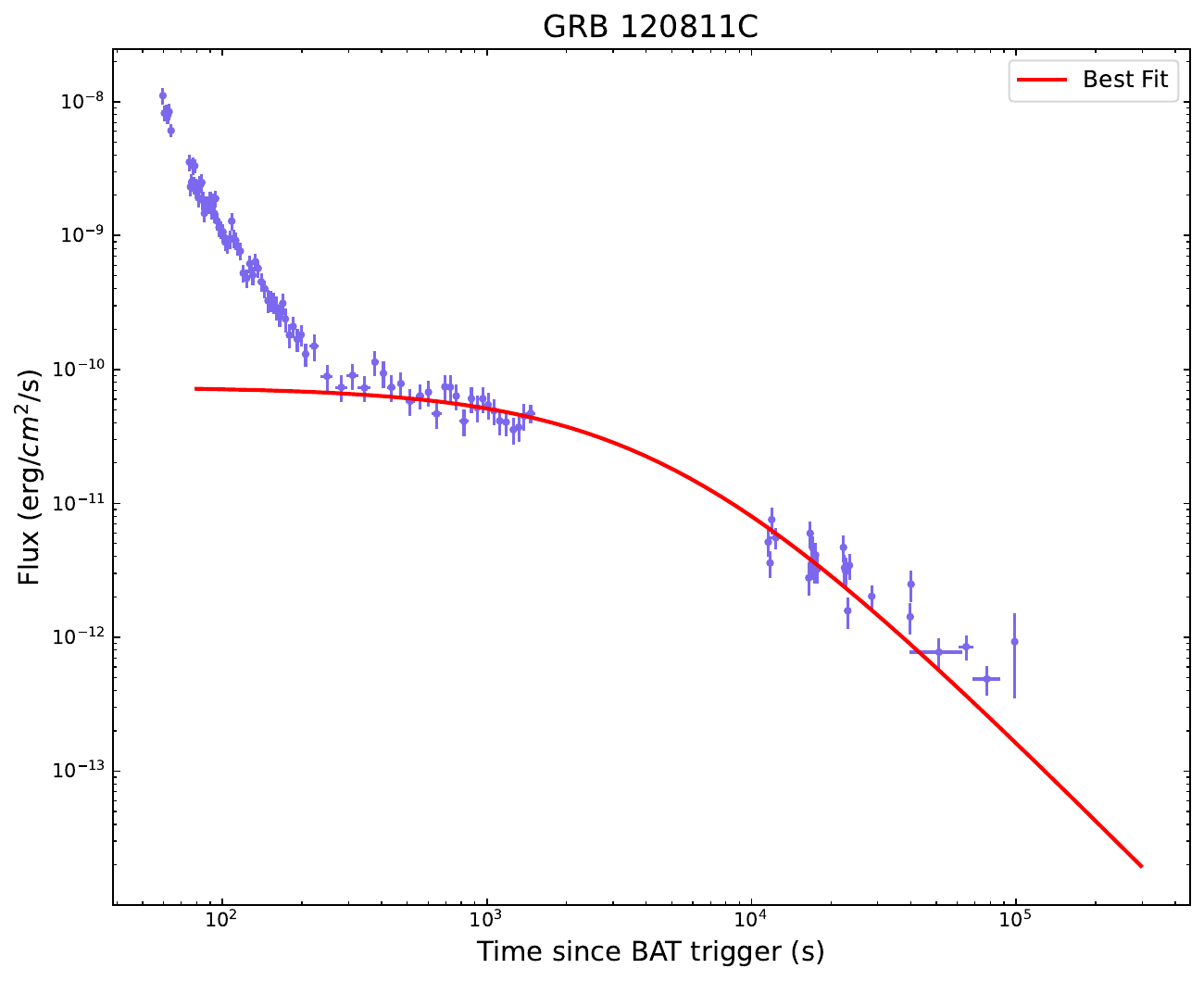}}%

\caption{(Continued)}
\end{figure*}

\addtocounter{figure}{-1}
\begin{figure*}[ht!]

\noindent
\resizebox{55mm}{!}{\includegraphics[]{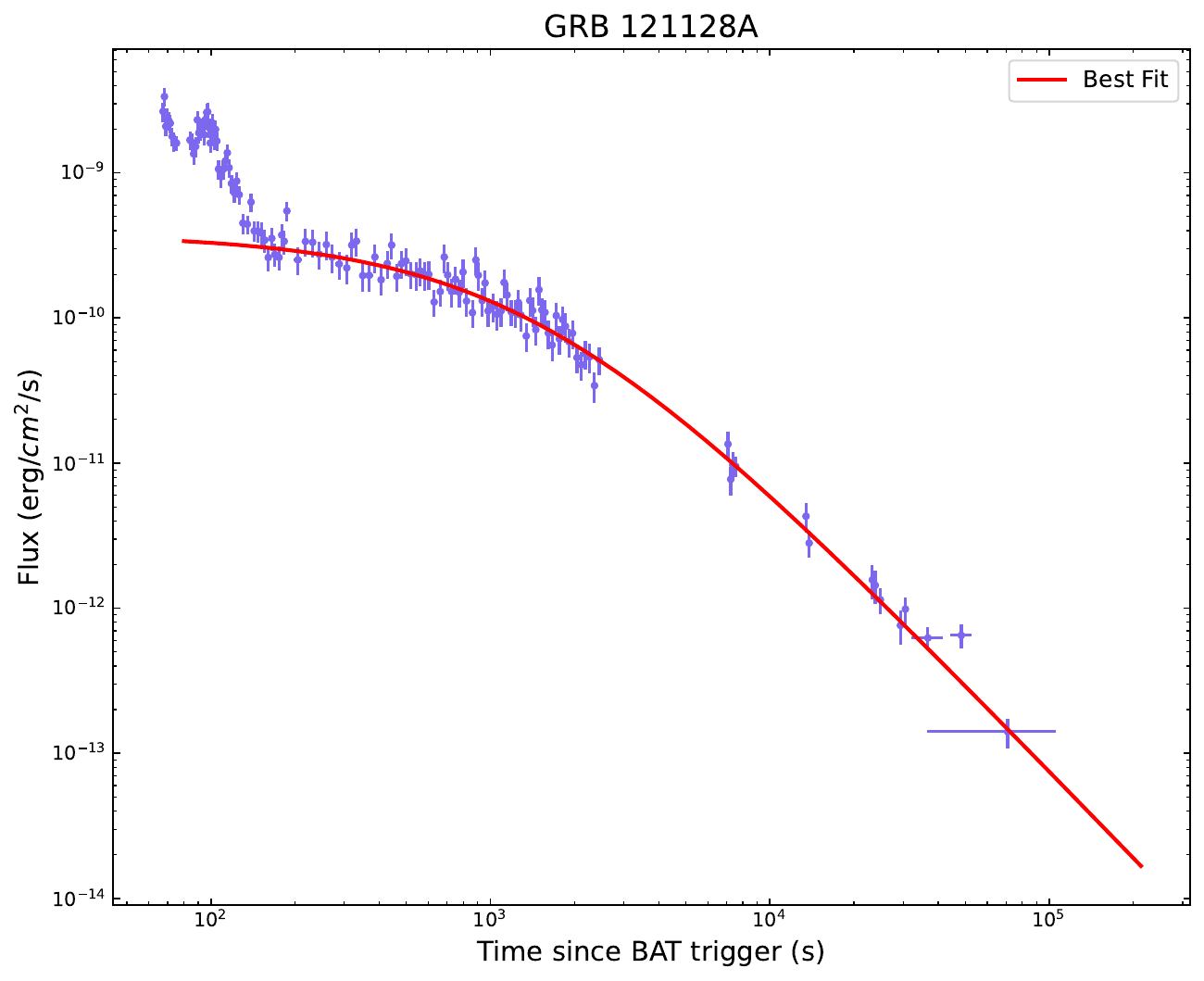}}%
\resizebox{55mm}{!}{\includegraphics[]{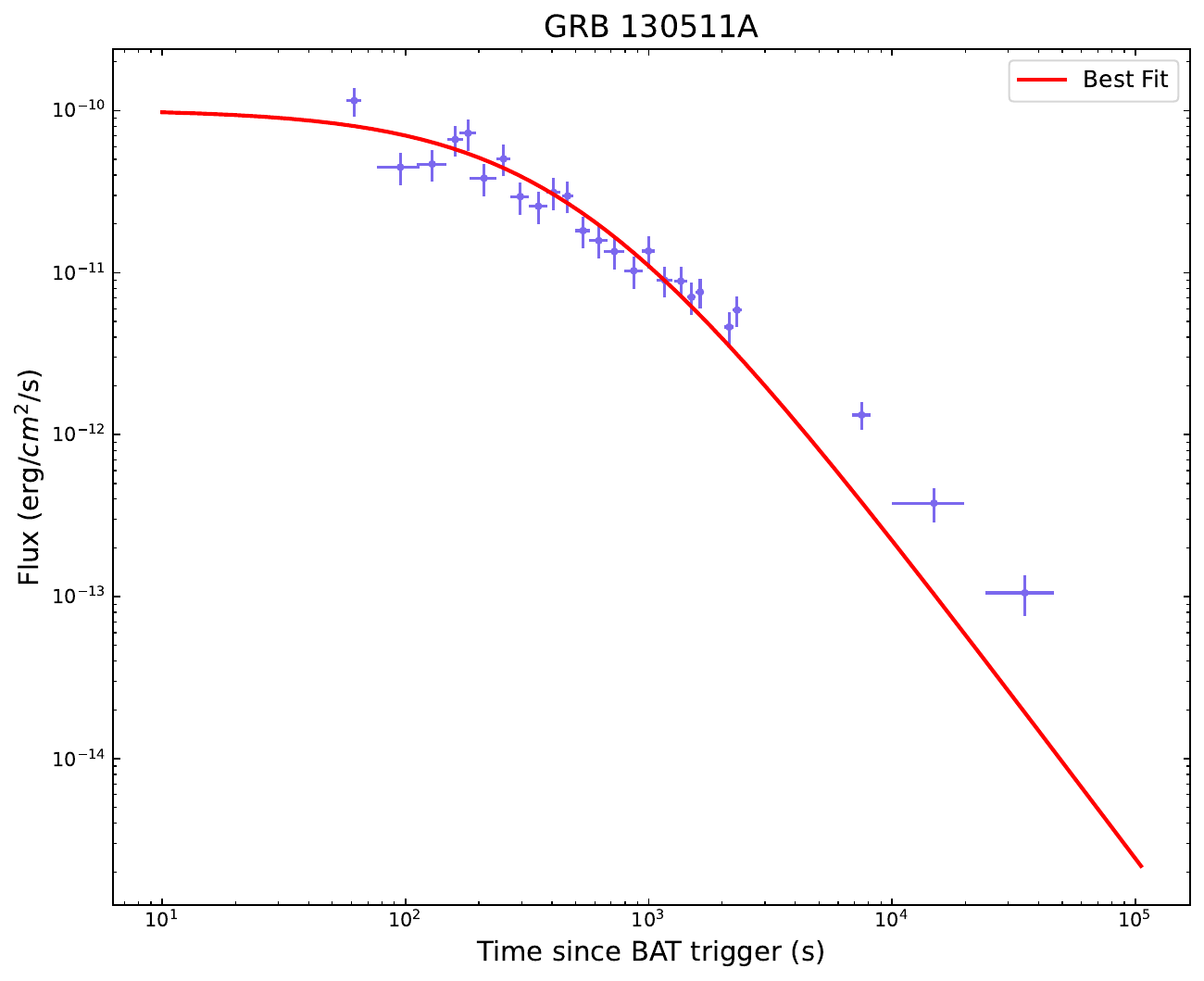}}%
\resizebox{55mm}{!}{\includegraphics[]{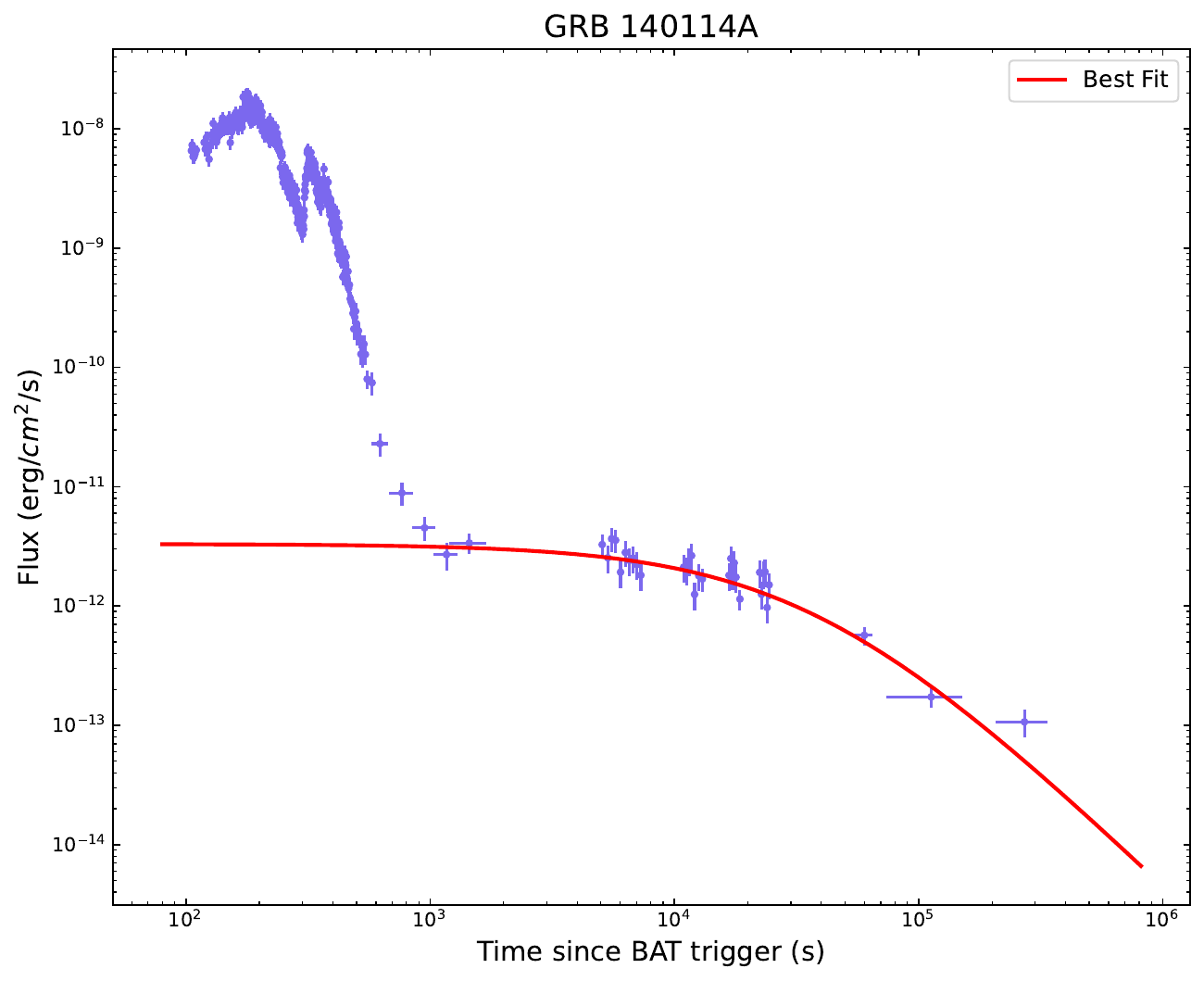}}%

\noindent
\resizebox{55mm}{!}{\includegraphics[]{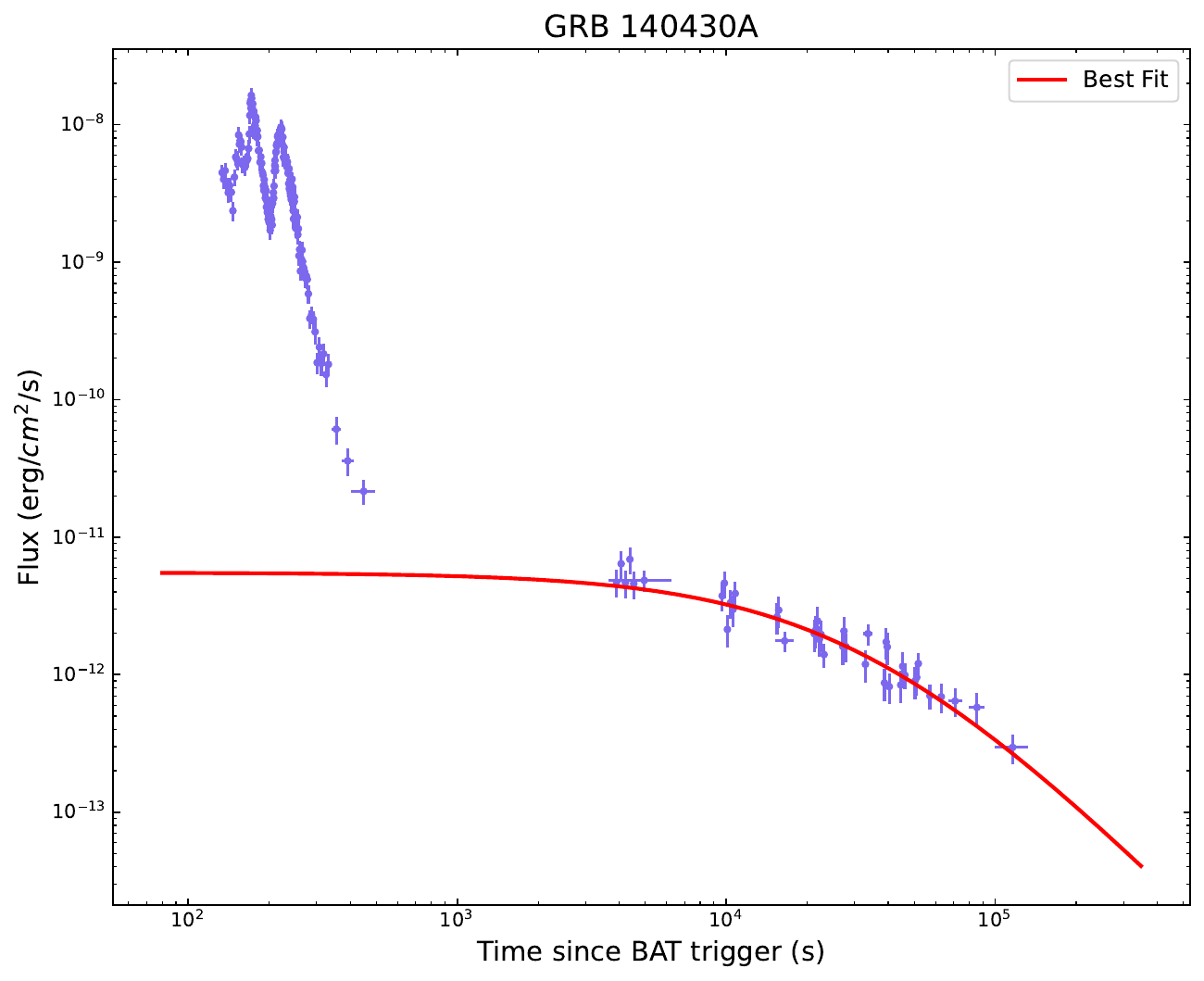}}%
\resizebox{55mm}{!}{\includegraphics[]{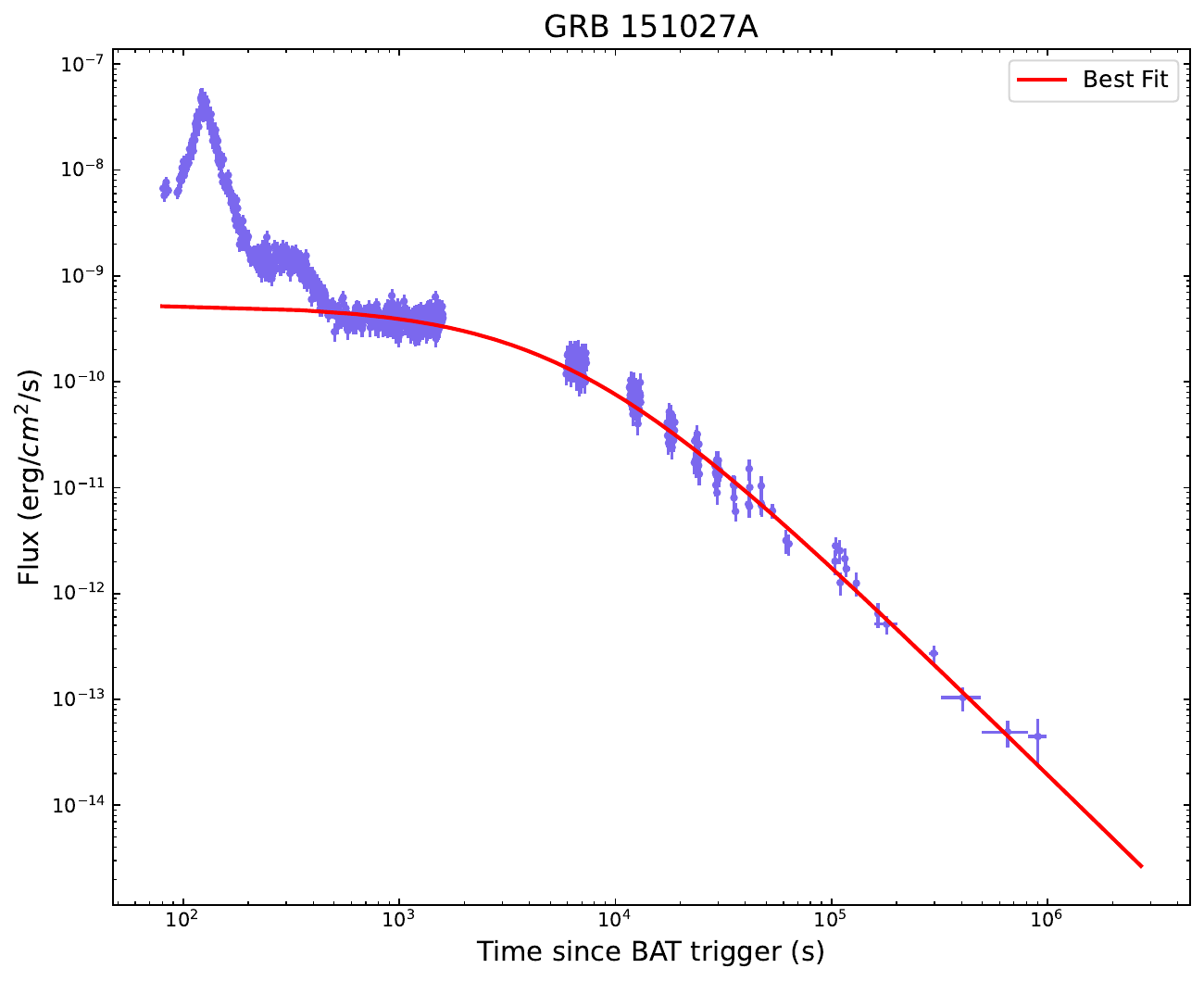}}%
\resizebox{55mm}{!}{\includegraphics[]{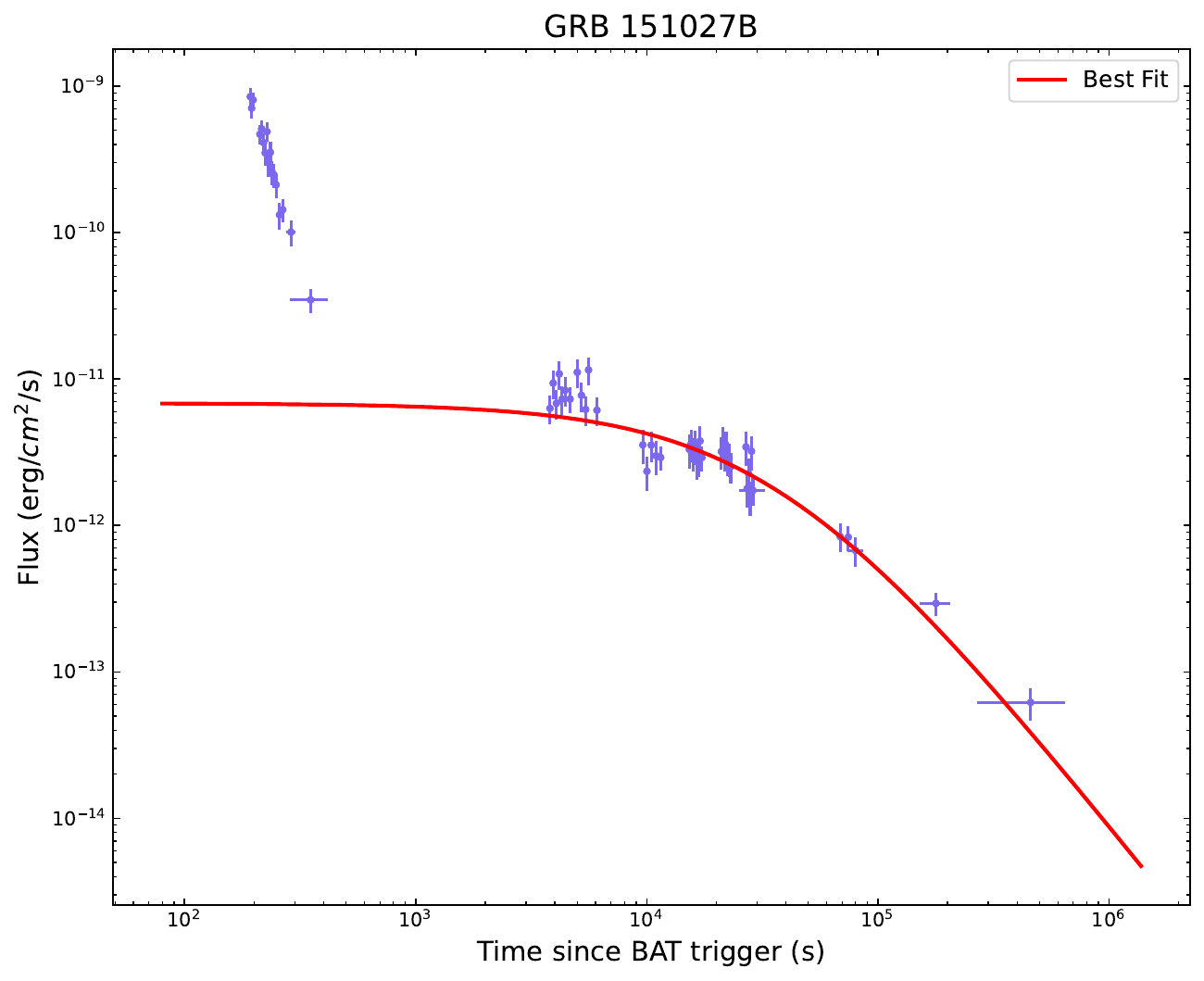}}%

\noindent
\resizebox{55mm}{!}{\includegraphics[]{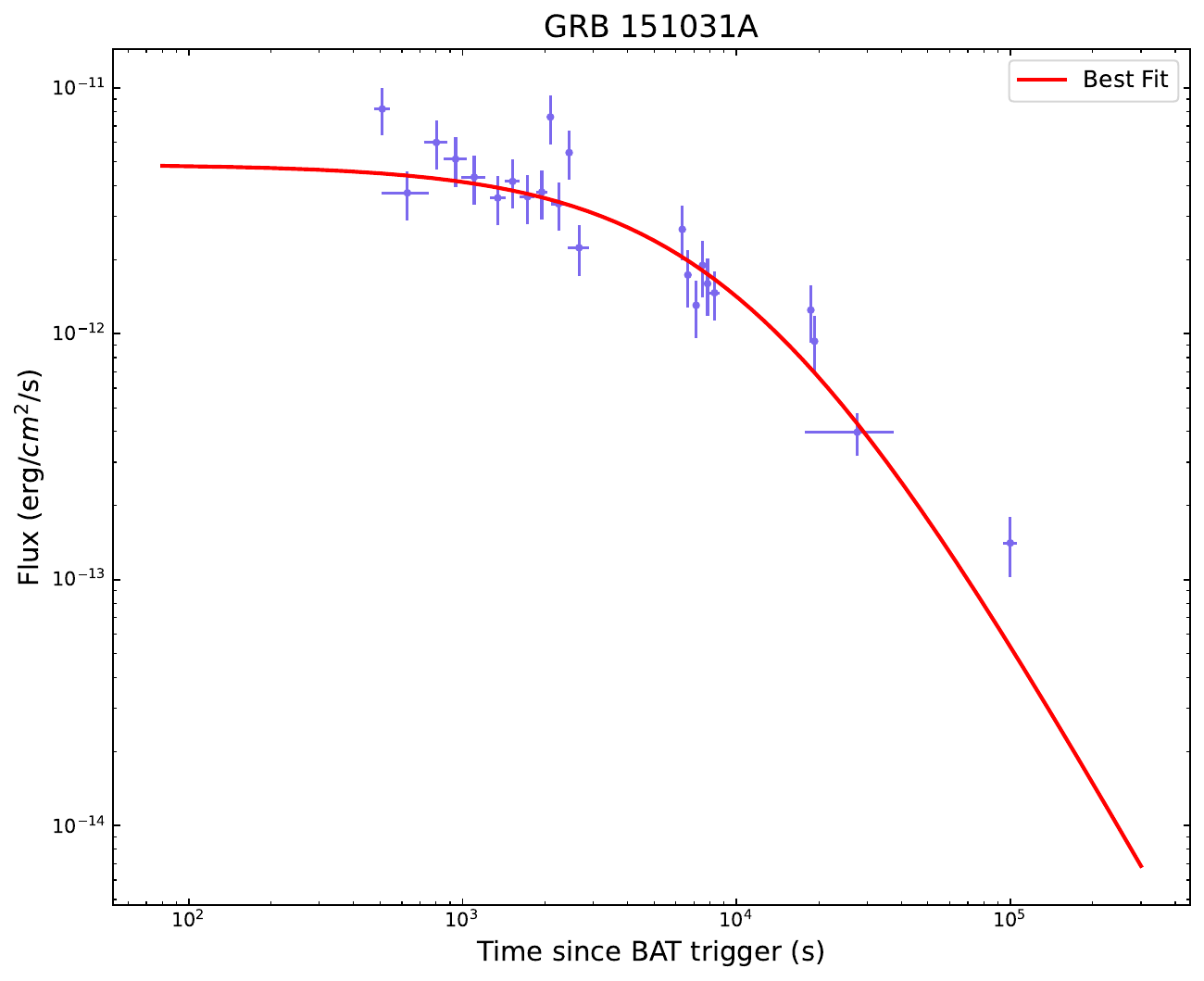}}%
\resizebox{55mm}{!}{\includegraphics[]{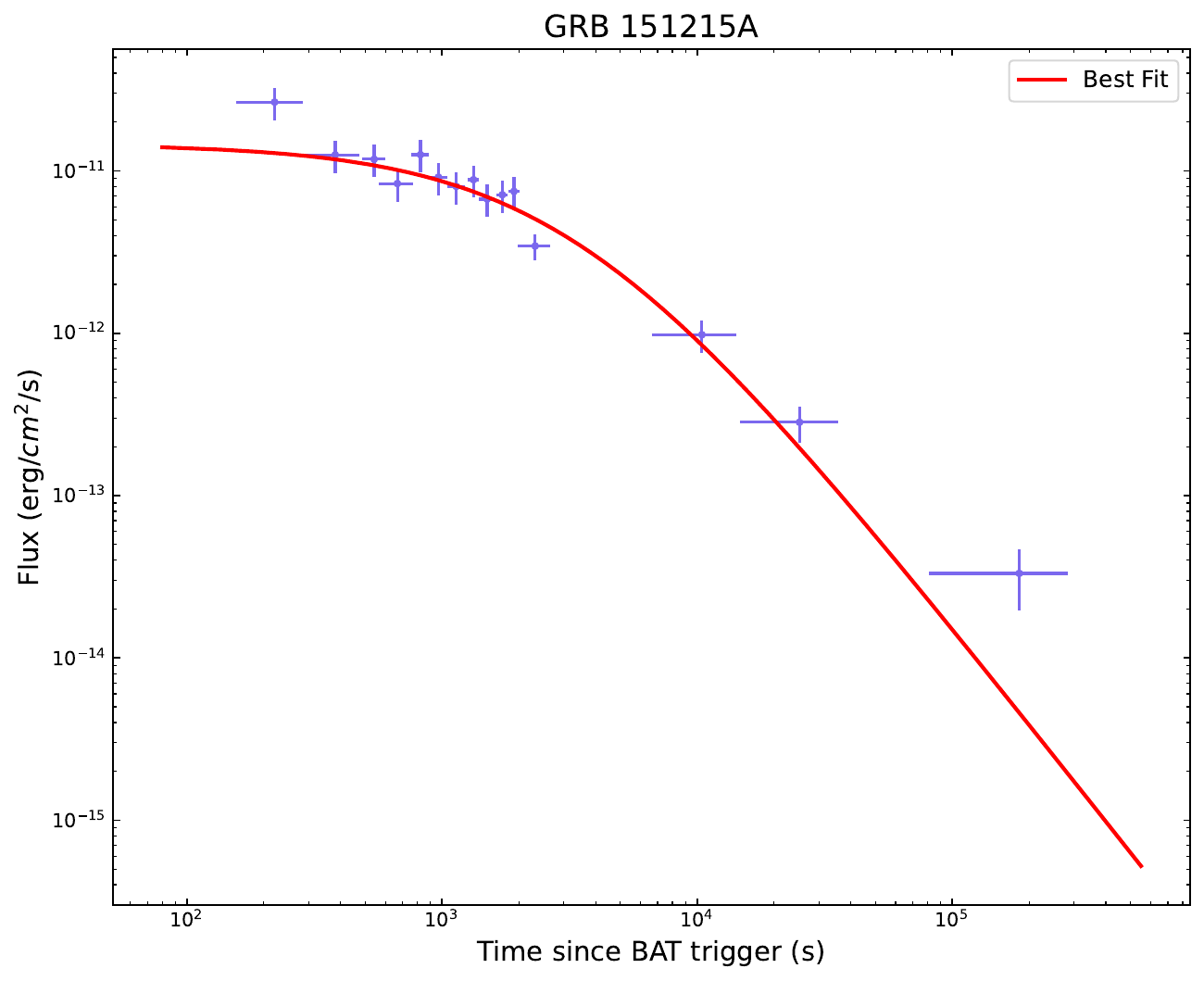}}%
\resizebox{55mm}{!}{\includegraphics[]{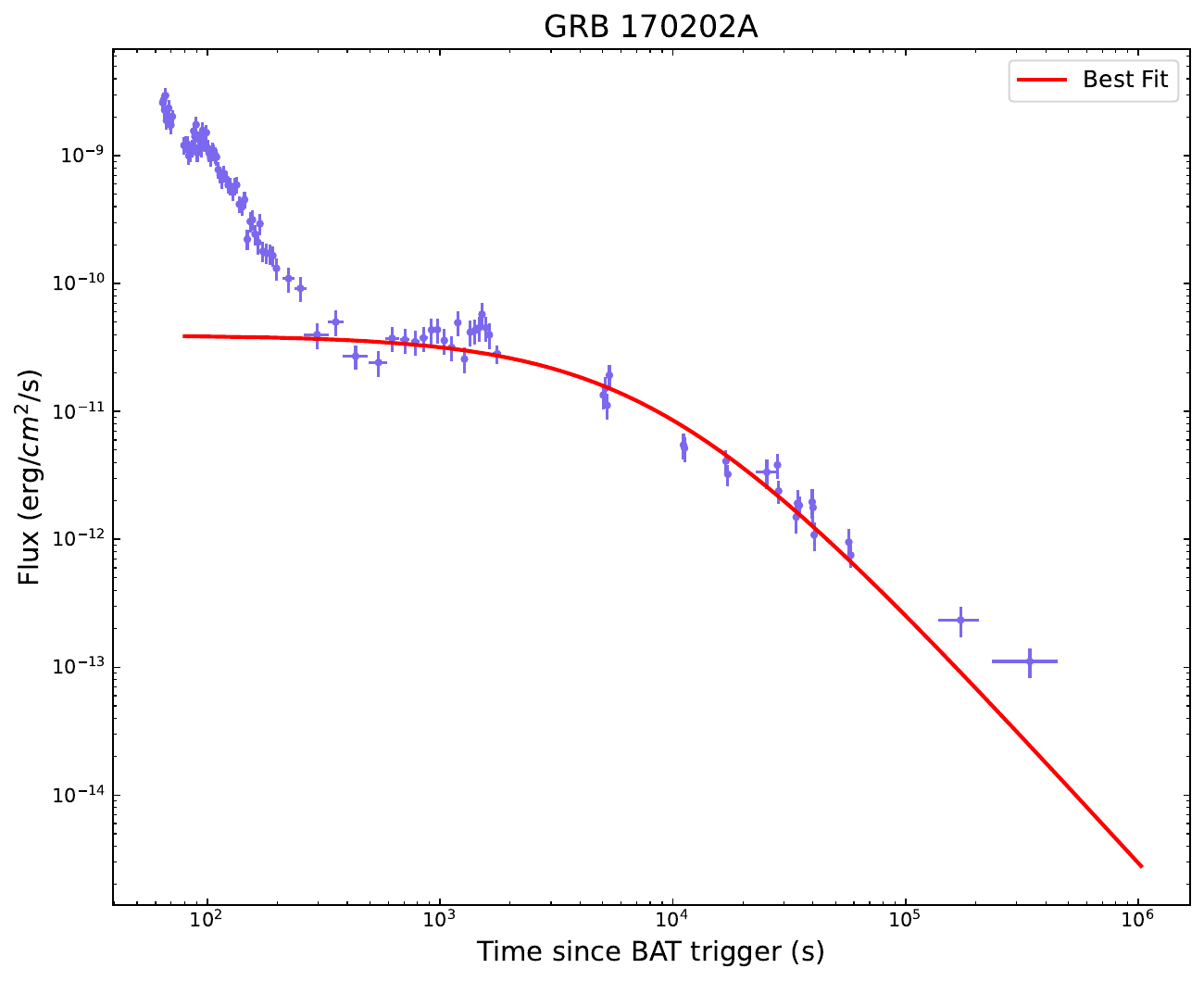}}%

\noindent
\resizebox{55mm}{!}{\includegraphics[]{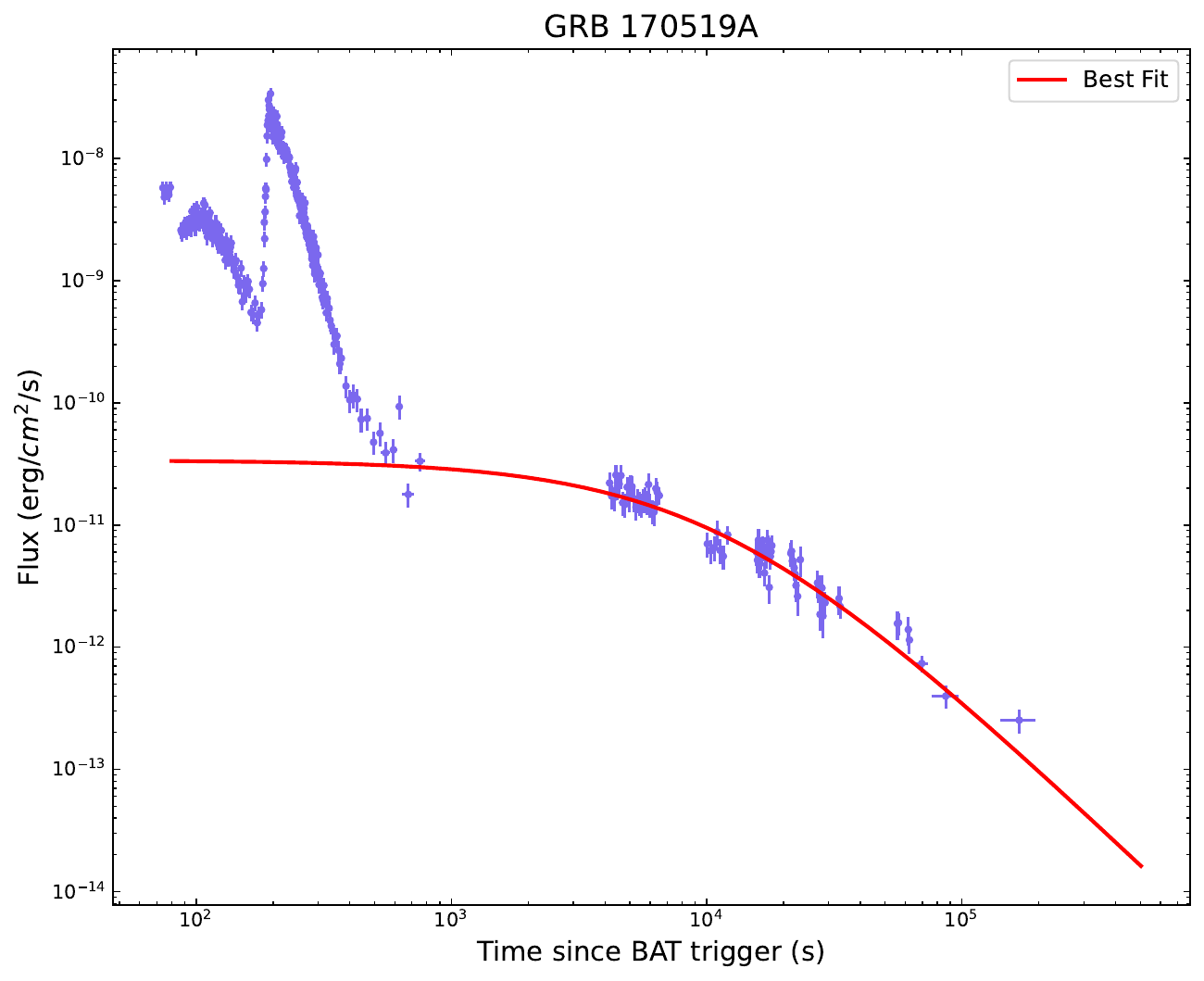}}%
\resizebox{55mm}{!}{\includegraphics[]{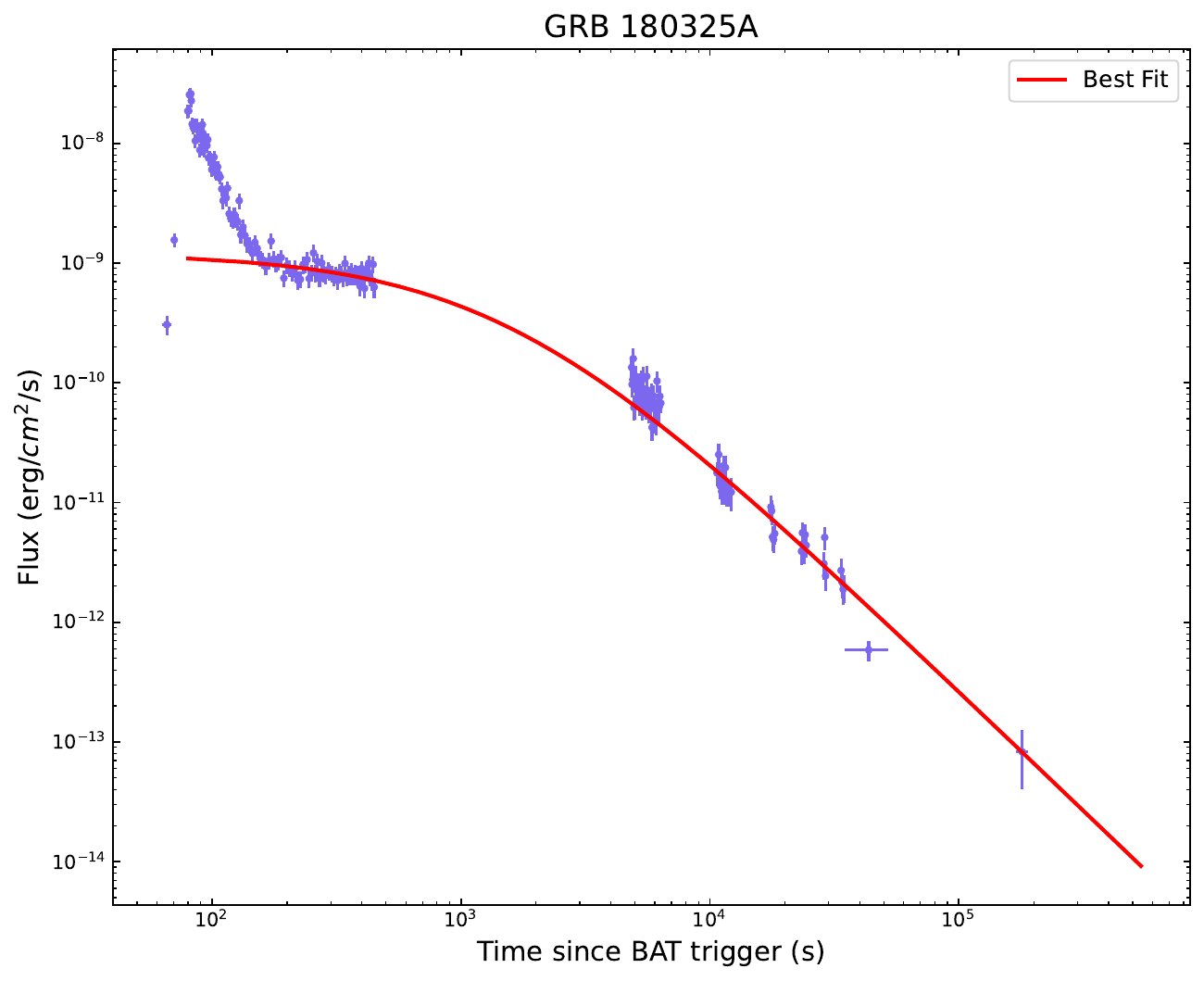}}%
\resizebox{55mm}{!}{\includegraphics[]{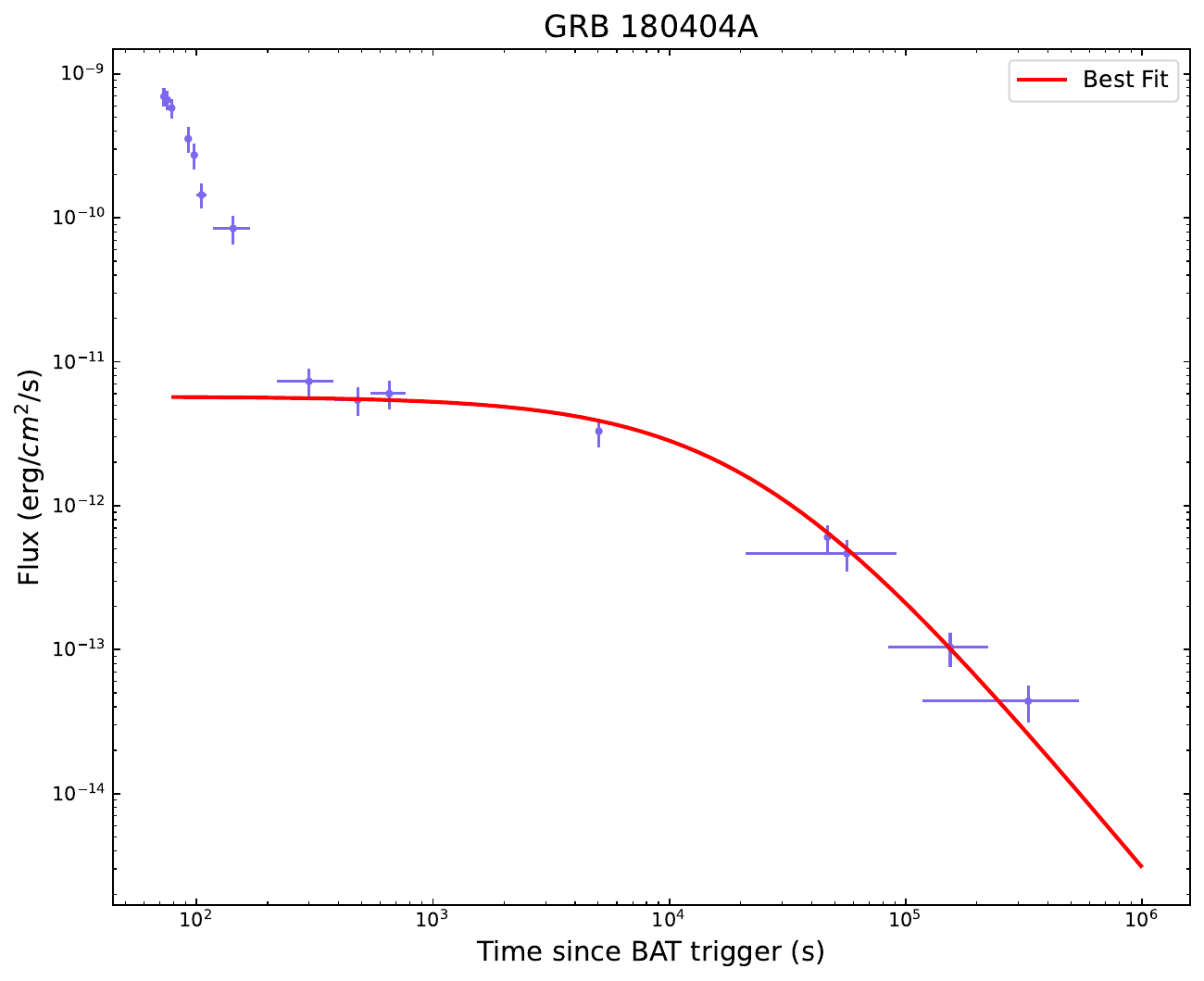}}%

\noindent
\resizebox{55mm}{!}{\includegraphics[]{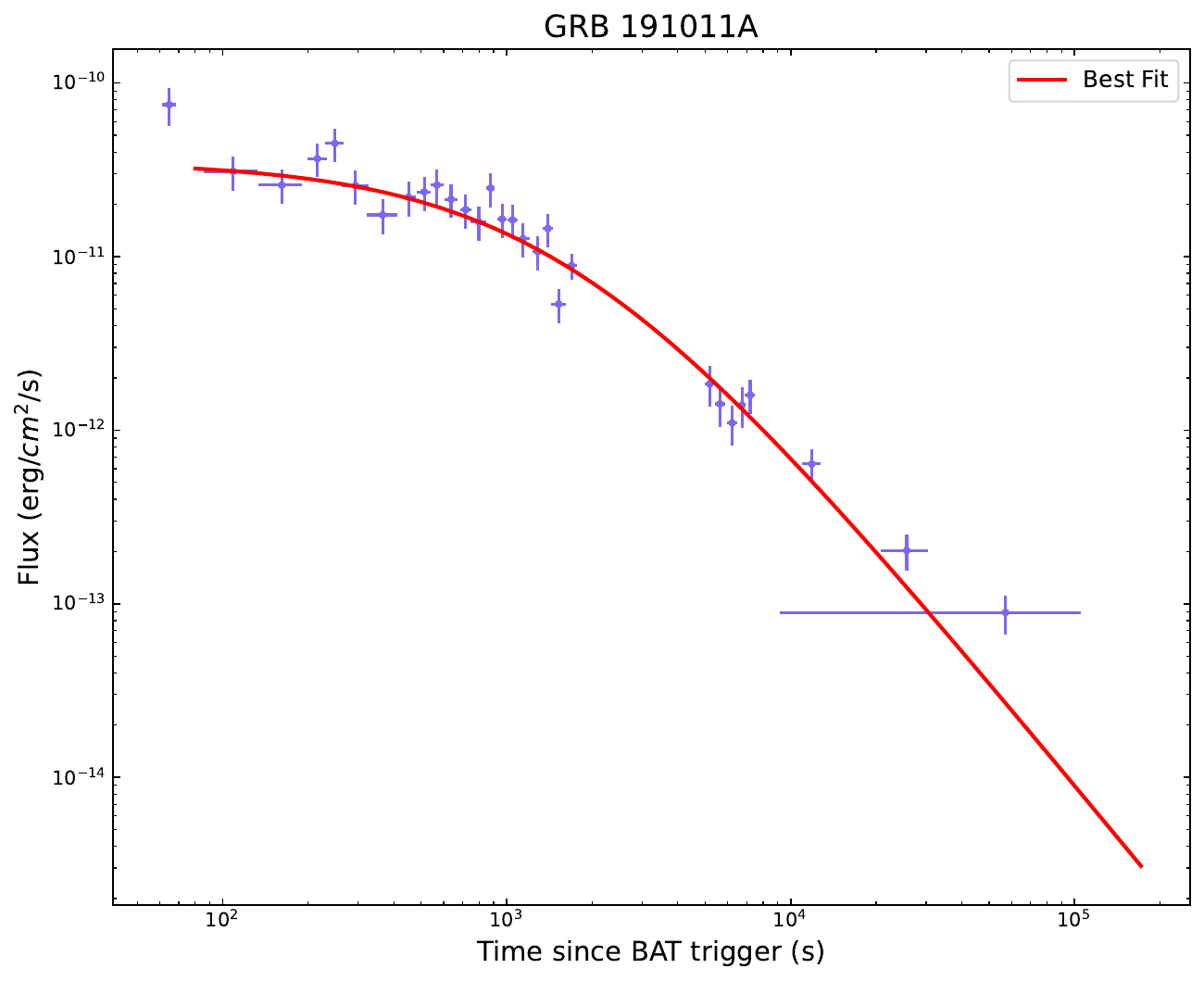}}%
\resizebox{55mm}{!}{\includegraphics[]{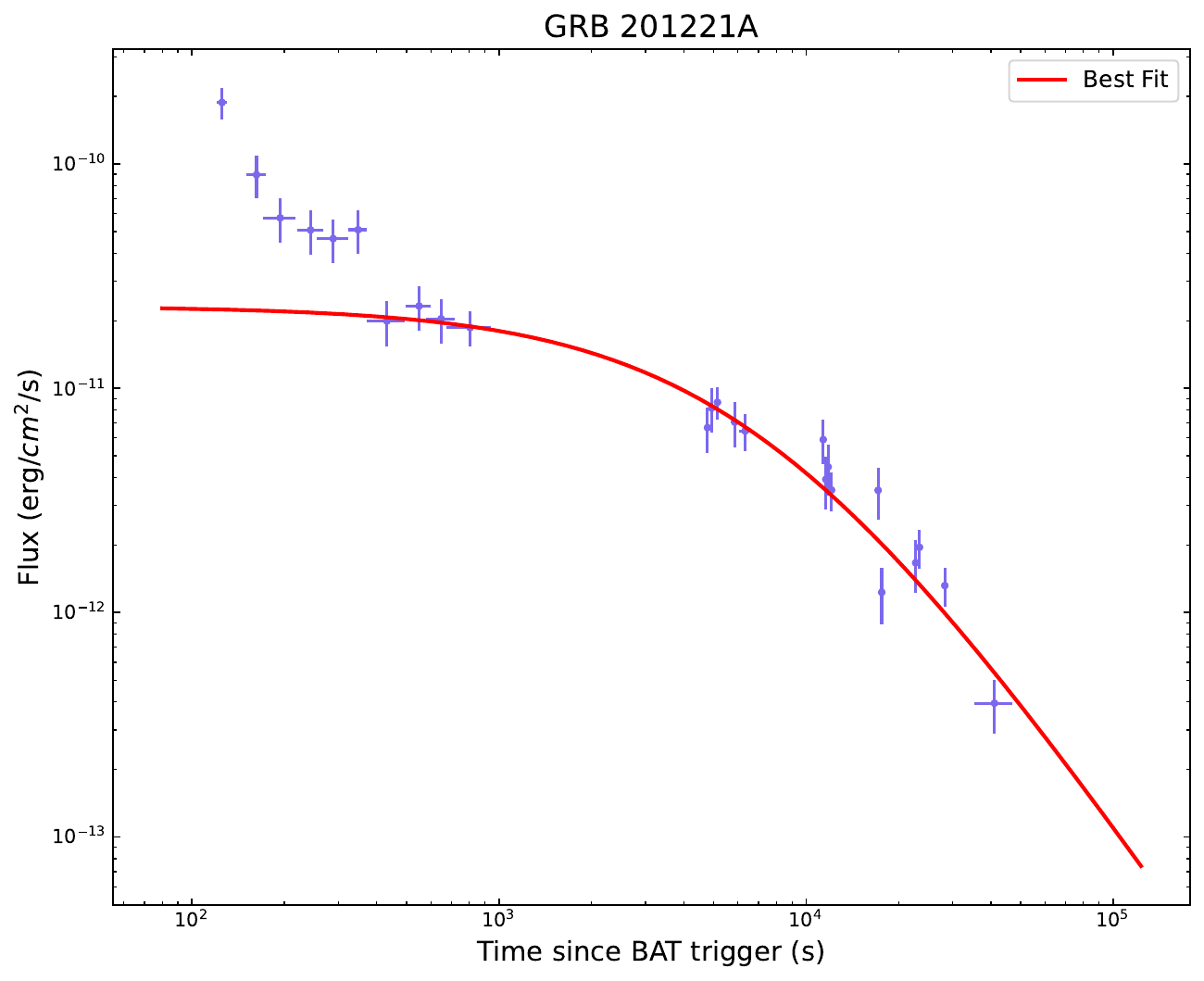}}%
\resizebox{55mm}{!}{\includegraphics[]{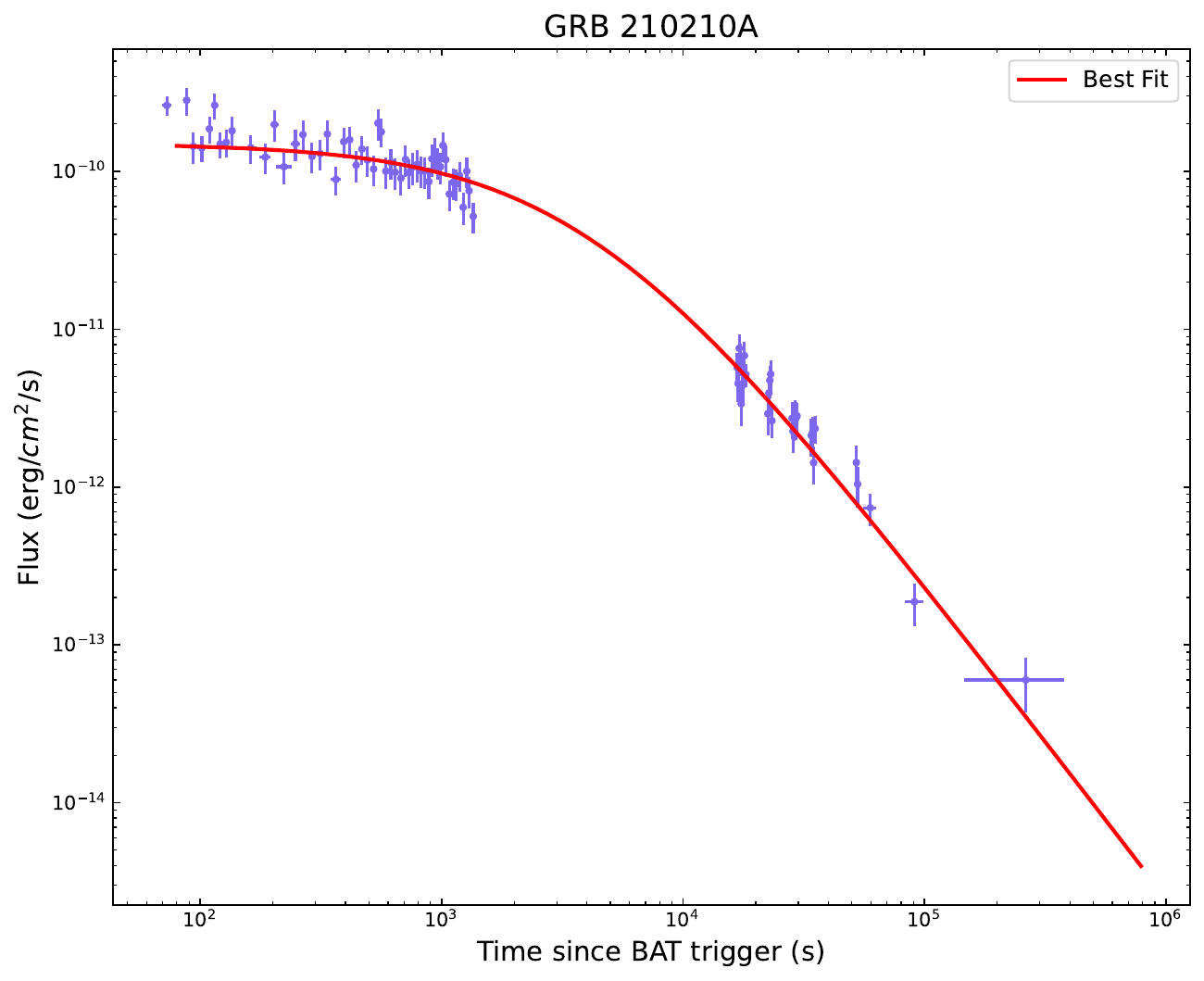}}%
\caption{(Continued)}
\end{figure*}

\addtocounter{figure}{-1}
\begin{figure*}[ht!]

\noindent
\resizebox{55mm}{!}{\includegraphics[]{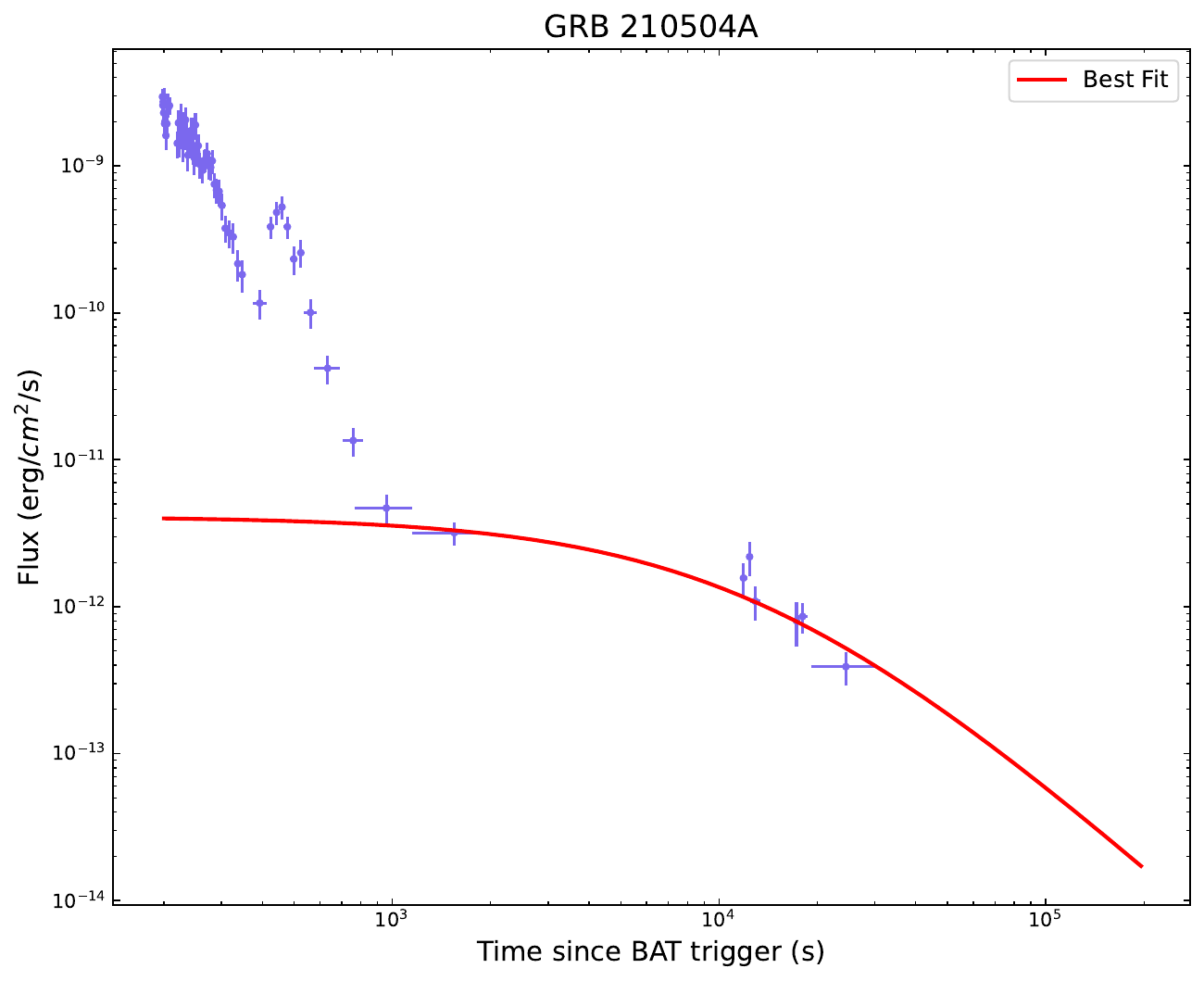}}%
\resizebox{55mm}{!}{\includegraphics[]{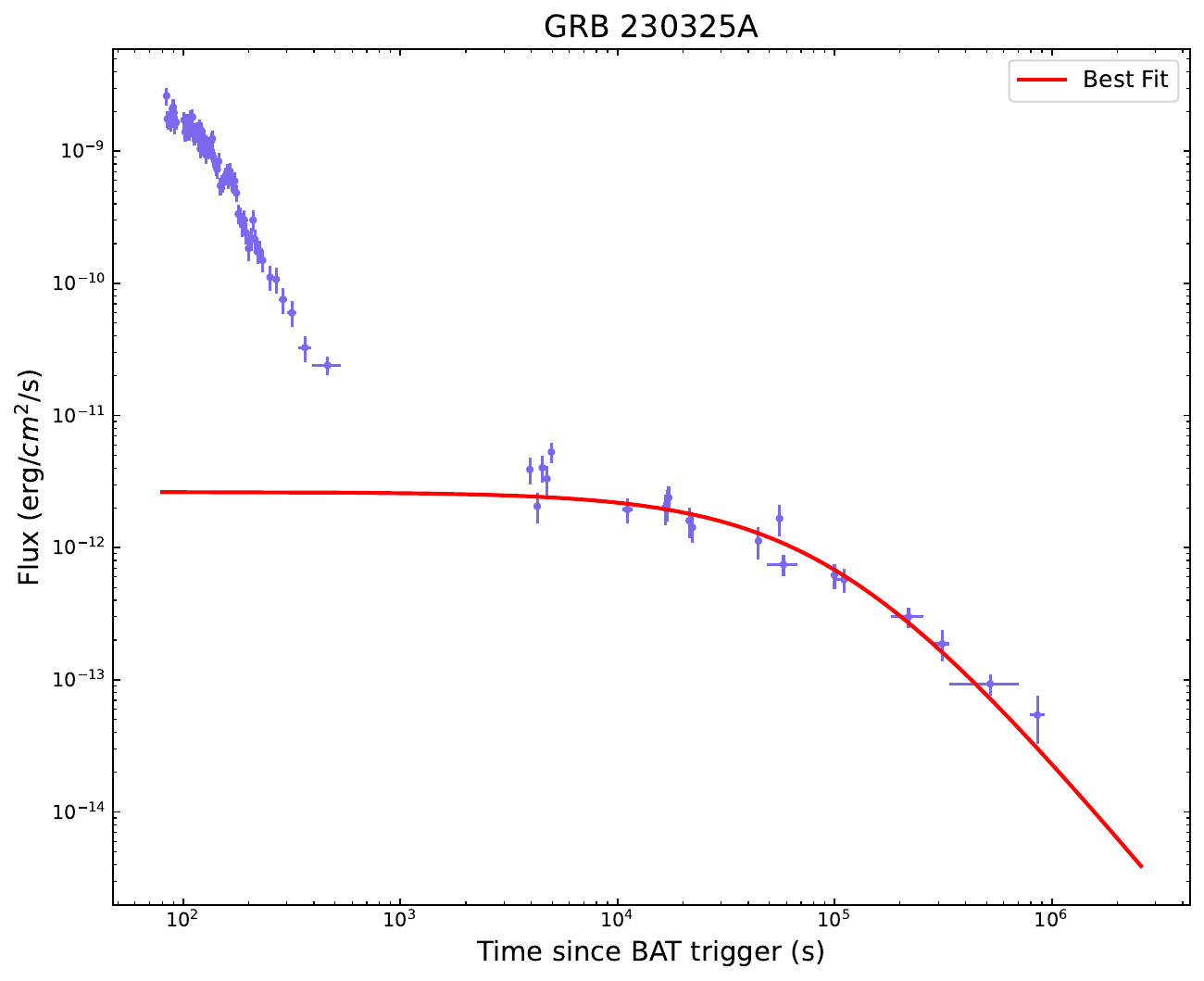}}%
\resizebox{55mm}{!}{\includegraphics[]{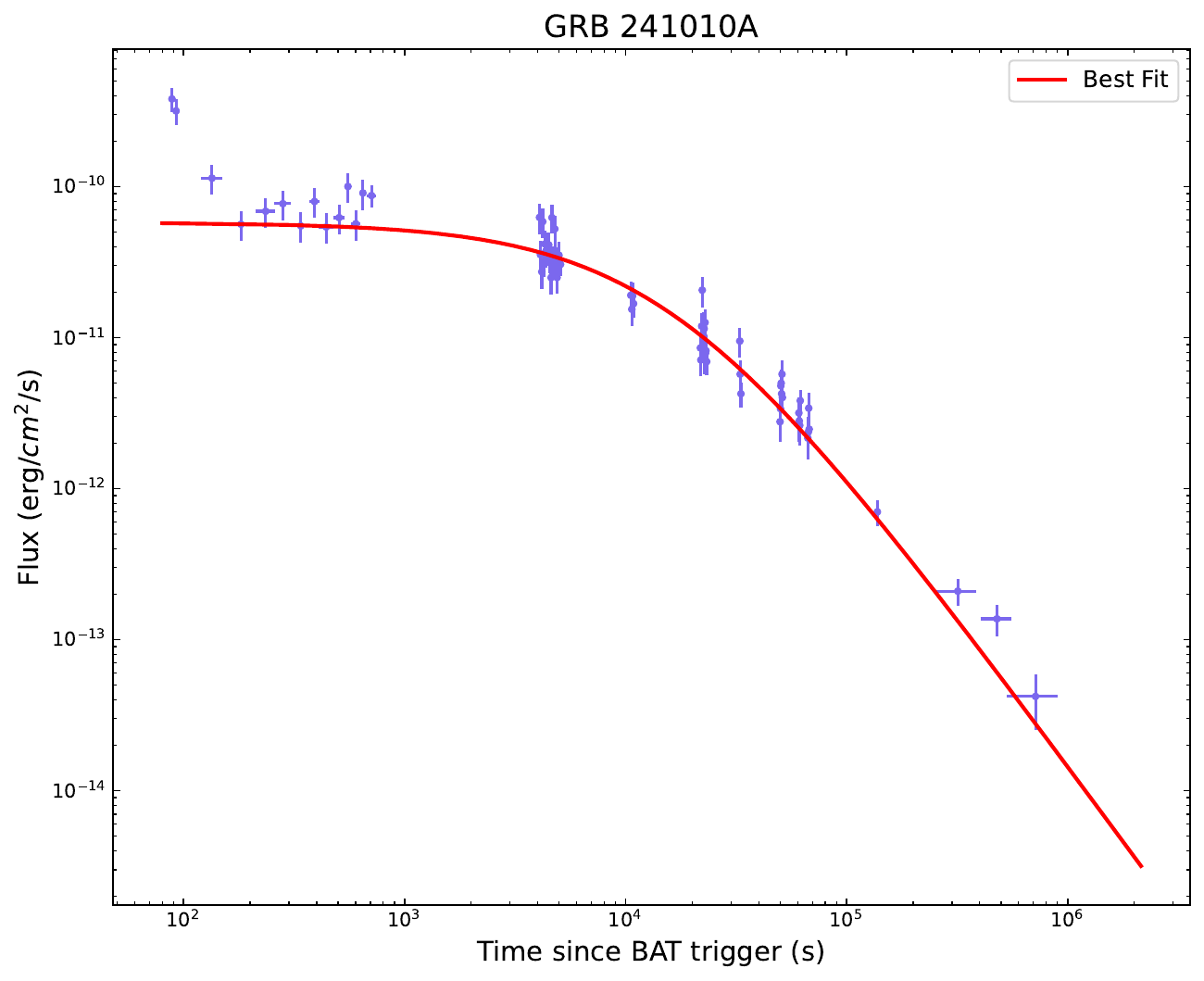}}%

\noindent
\resizebox{55mm}{!}{\includegraphics[]{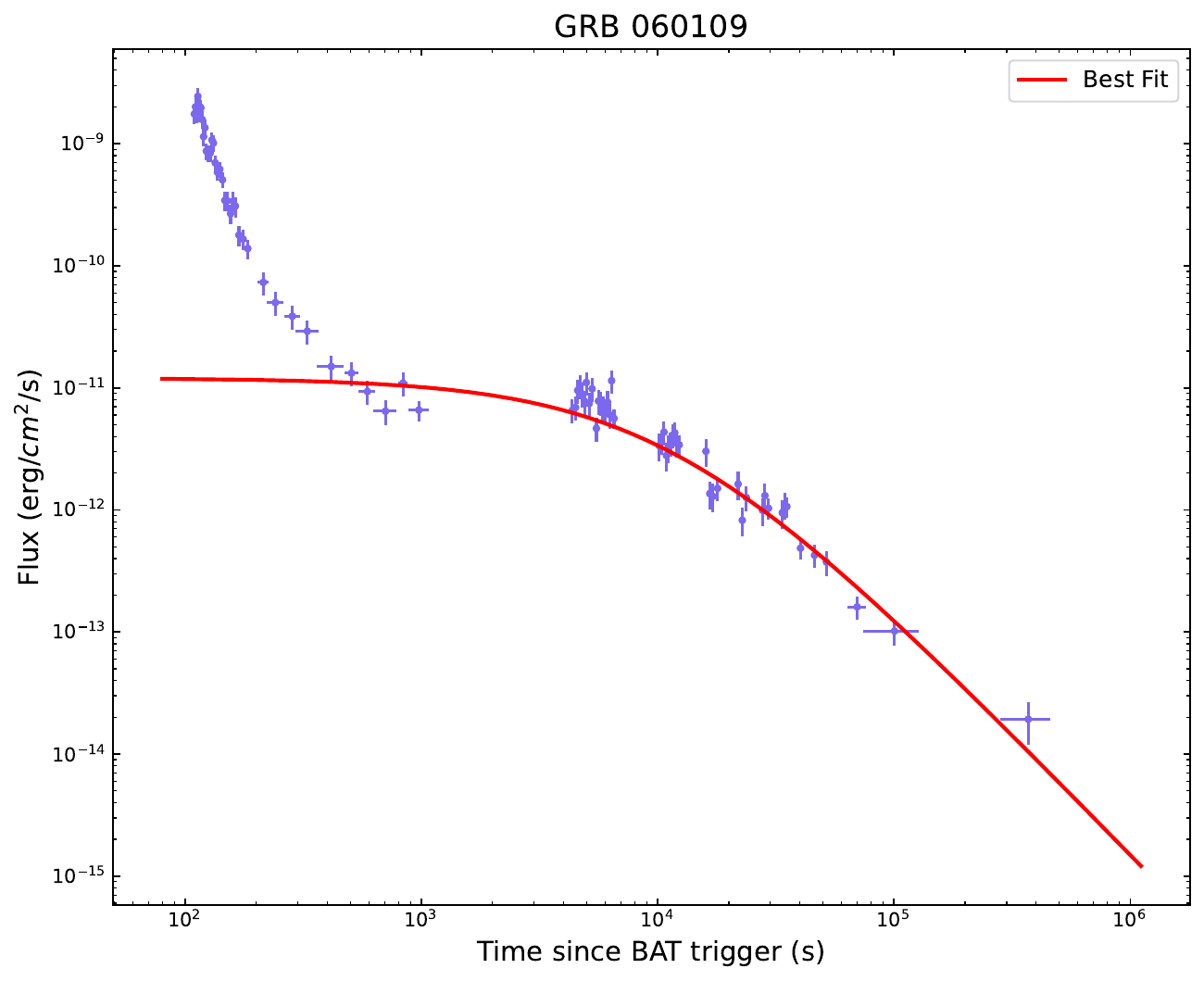}}%
\resizebox{55mm}{!}{\includegraphics[]{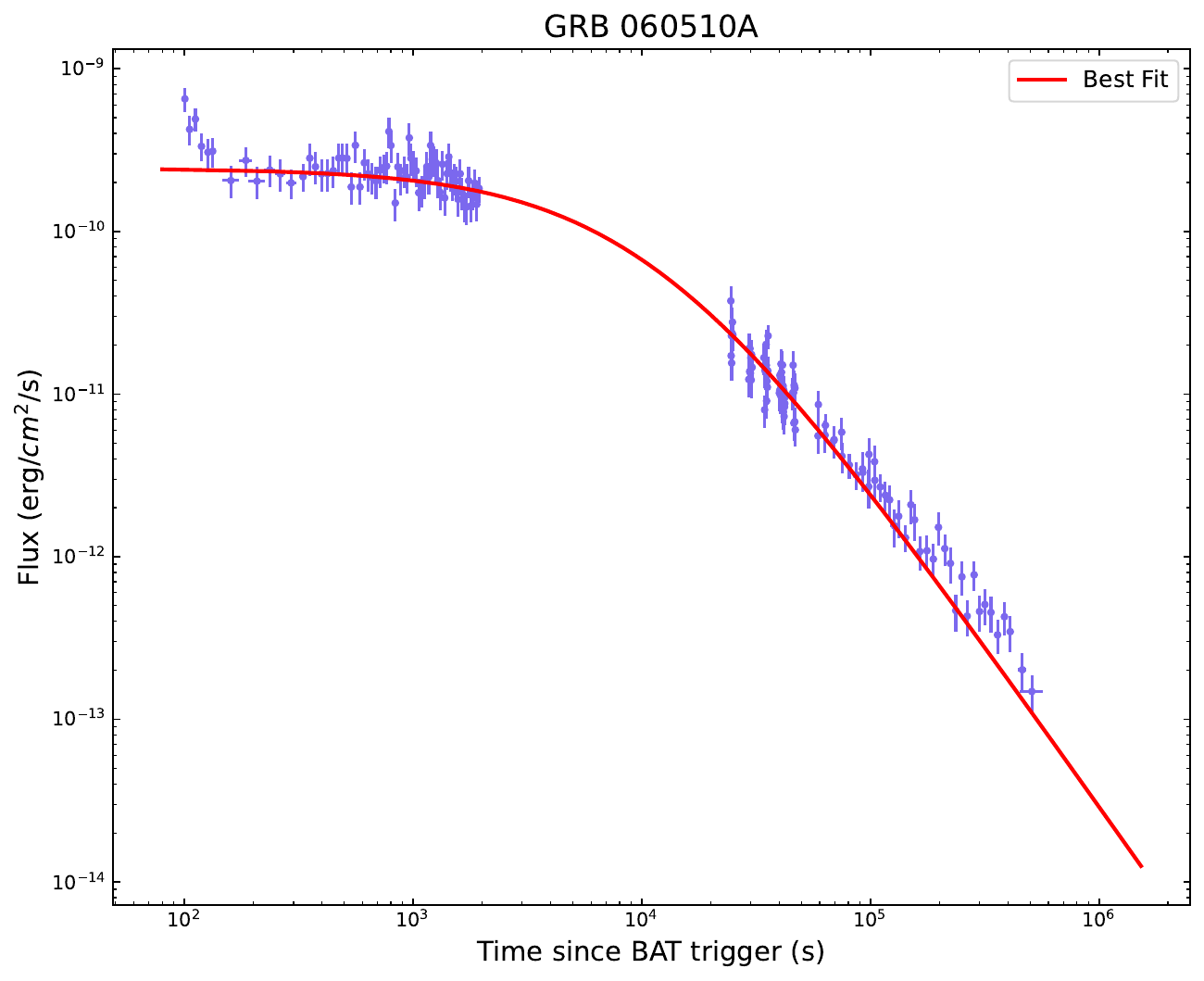}}%
\resizebox{55mm}{!}{\includegraphics[]{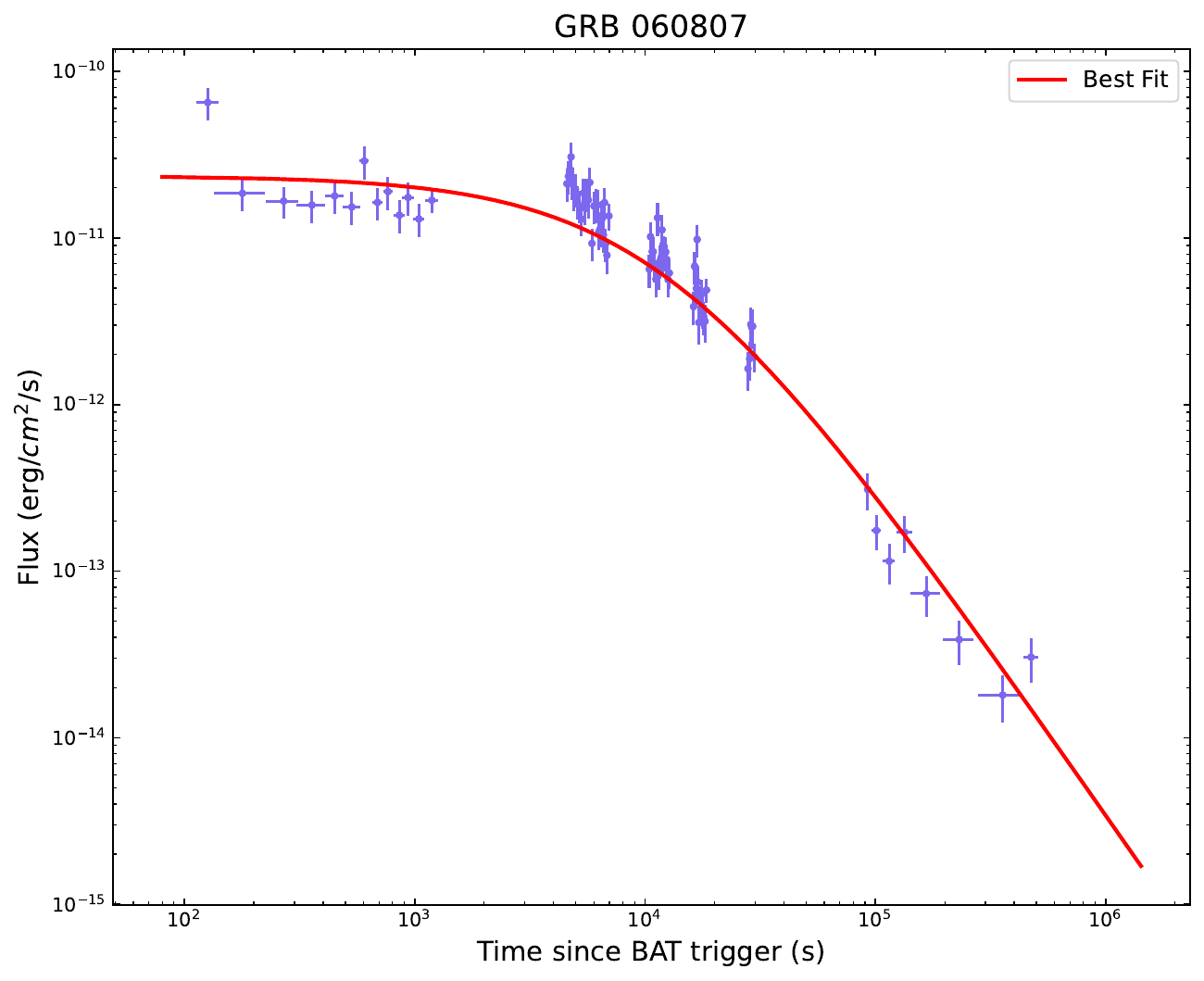}}%

\noindent
\resizebox{55mm}{!}{\includegraphics[]{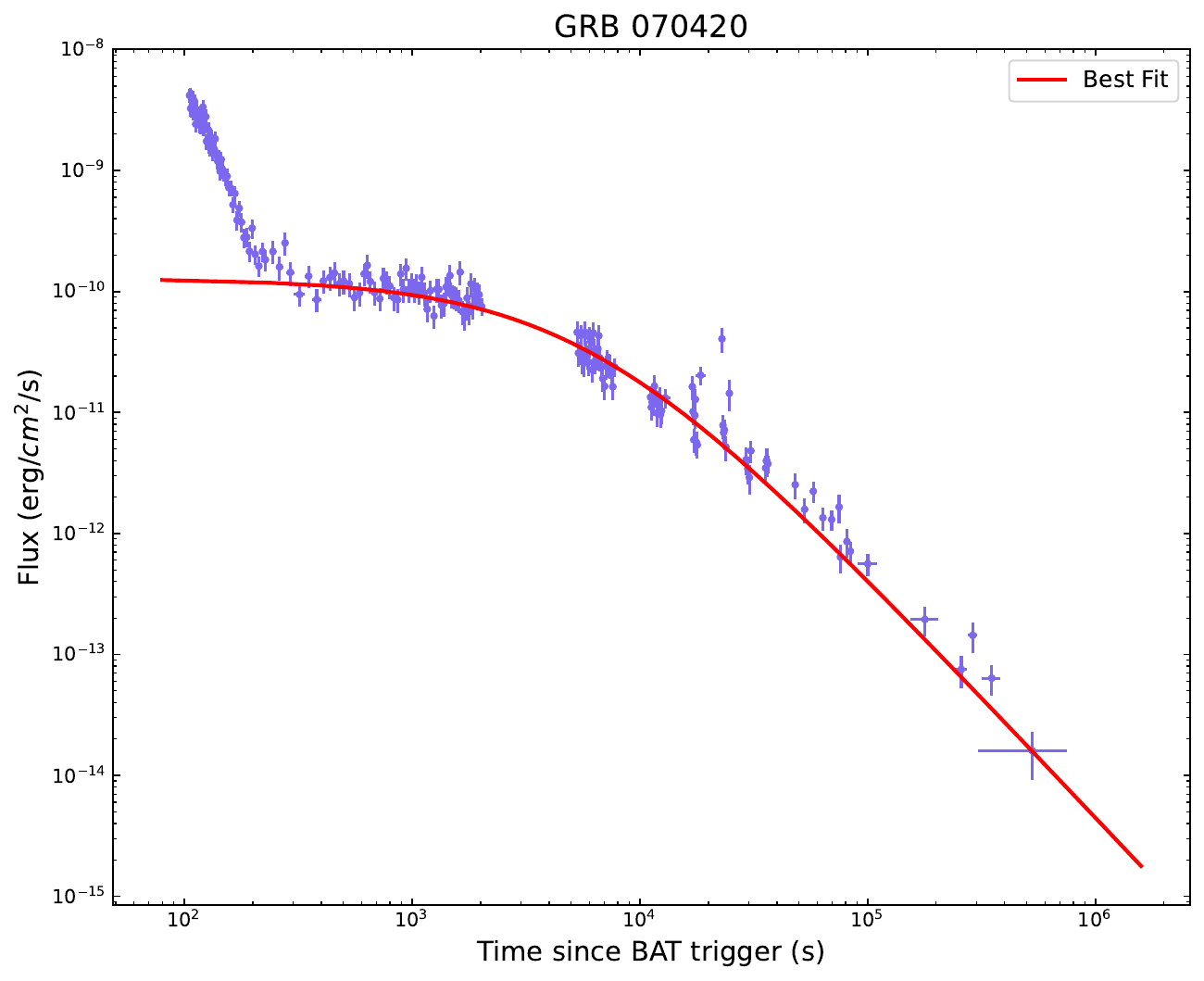}}%
\resizebox{55mm}{!}{\includegraphics[]{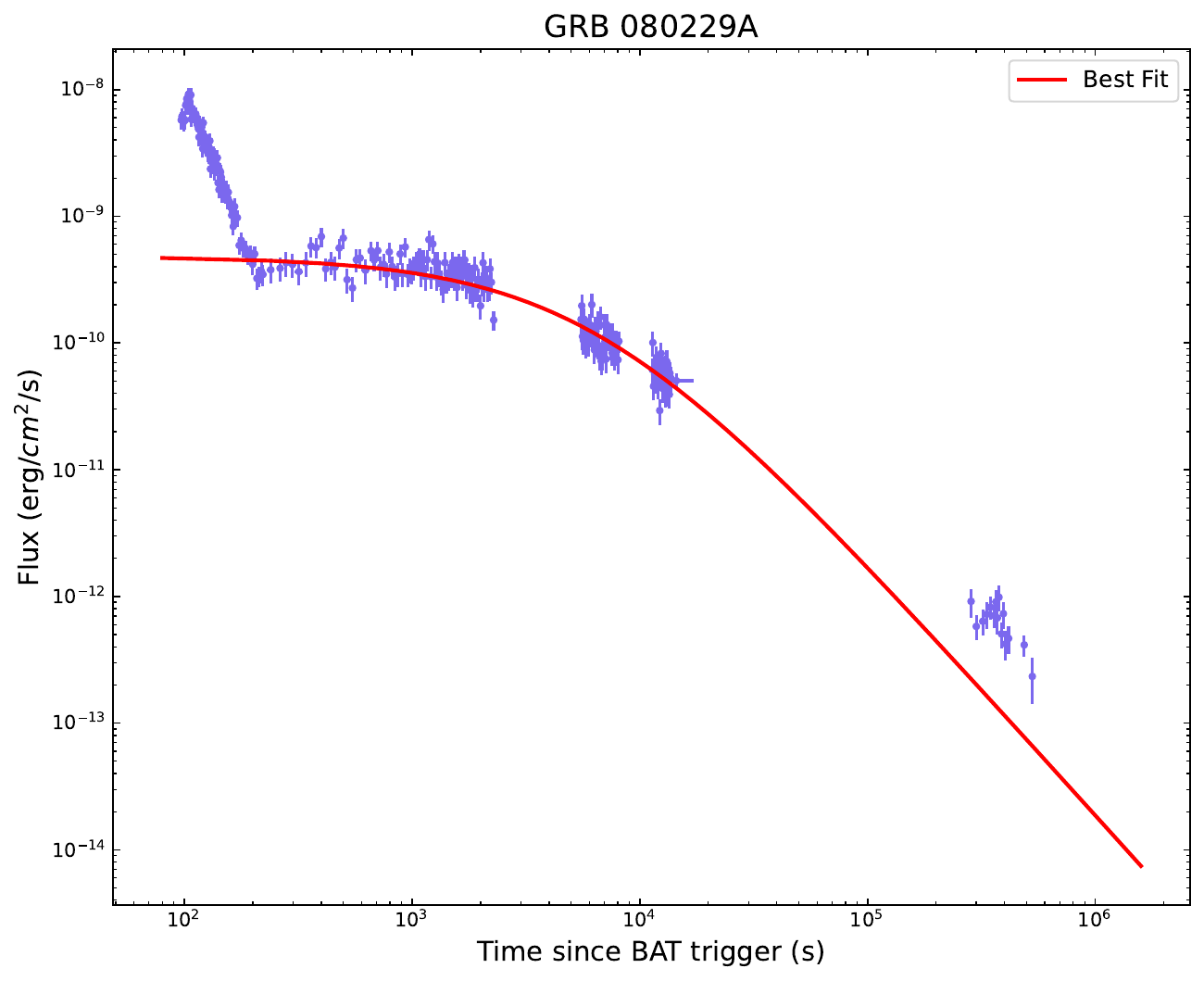}}%
\resizebox{55mm}{!}{\includegraphics[]{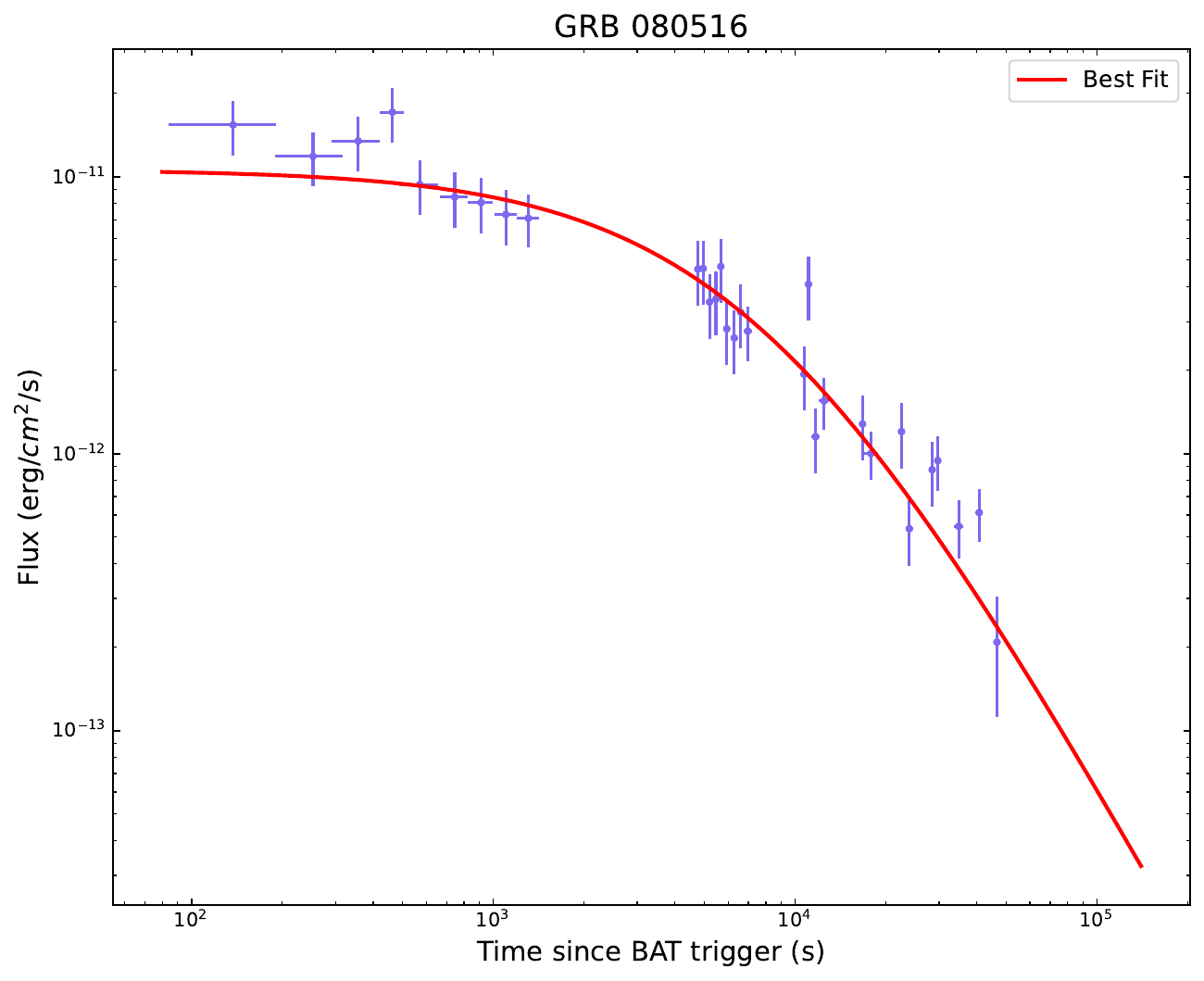}}%

\noindent
\resizebox{55mm}{!}{\includegraphics[]{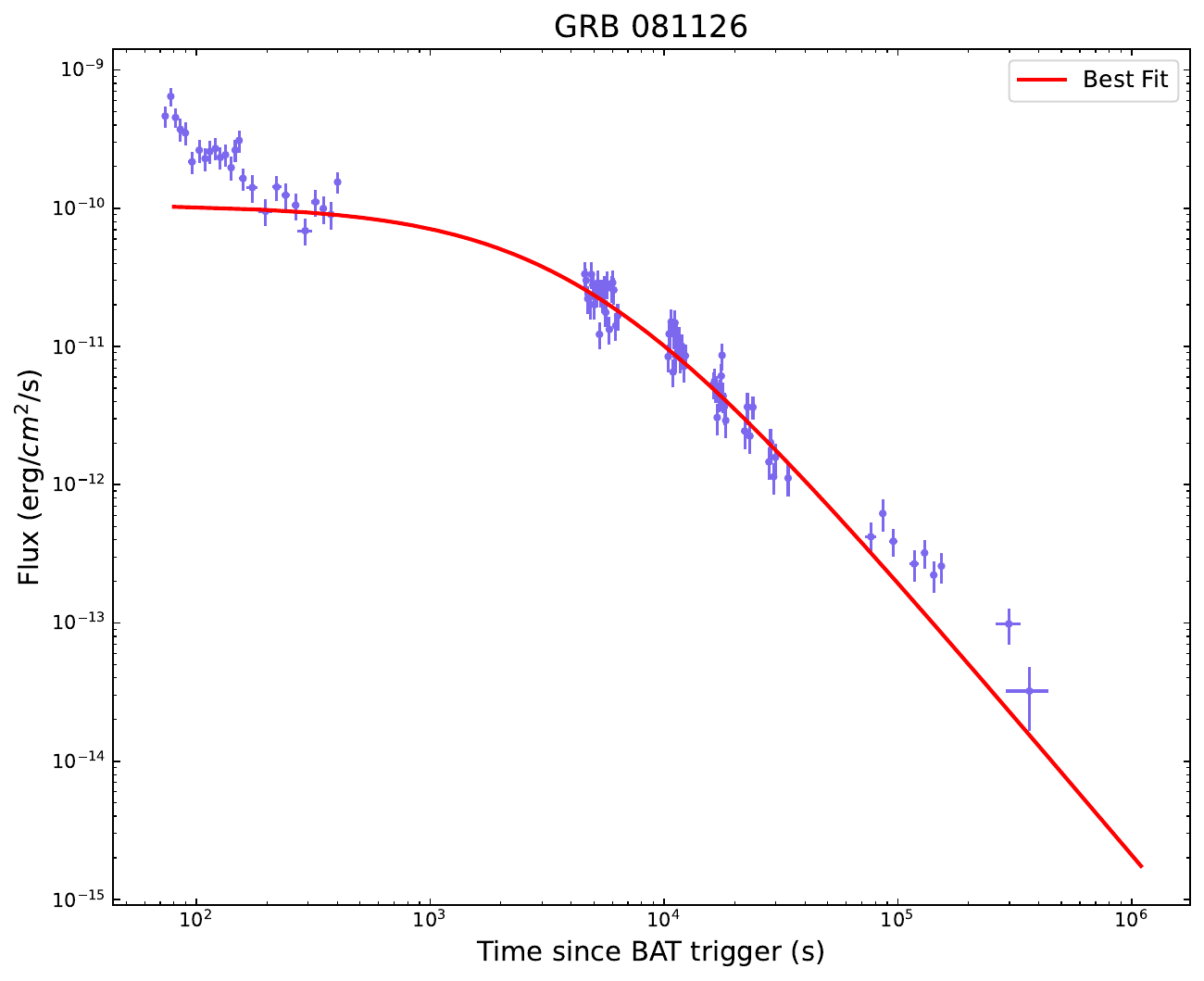}}%
\resizebox{55mm}{!}{\includegraphics[]{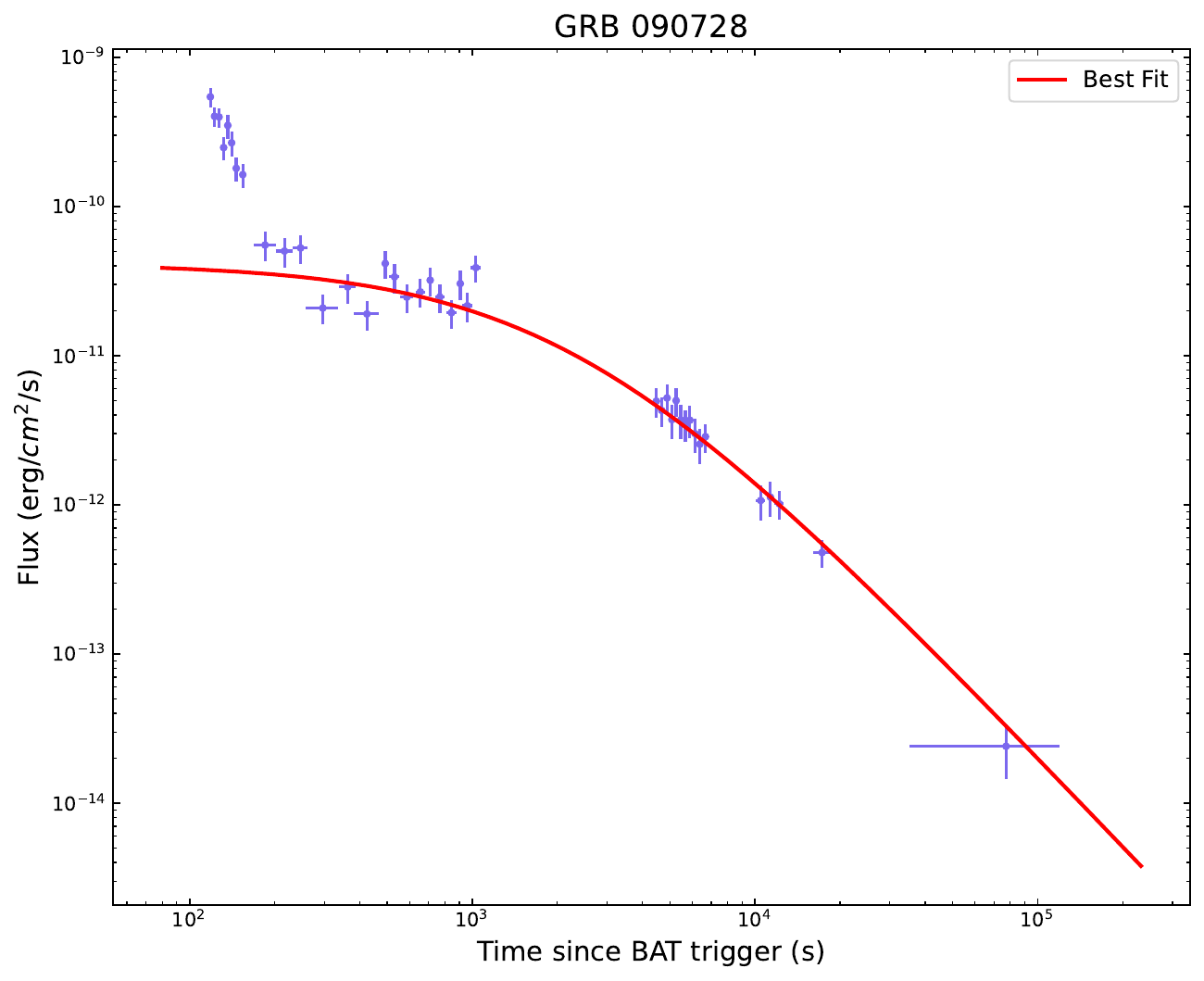}}%
\resizebox{55mm}{!}{\includegraphics[]{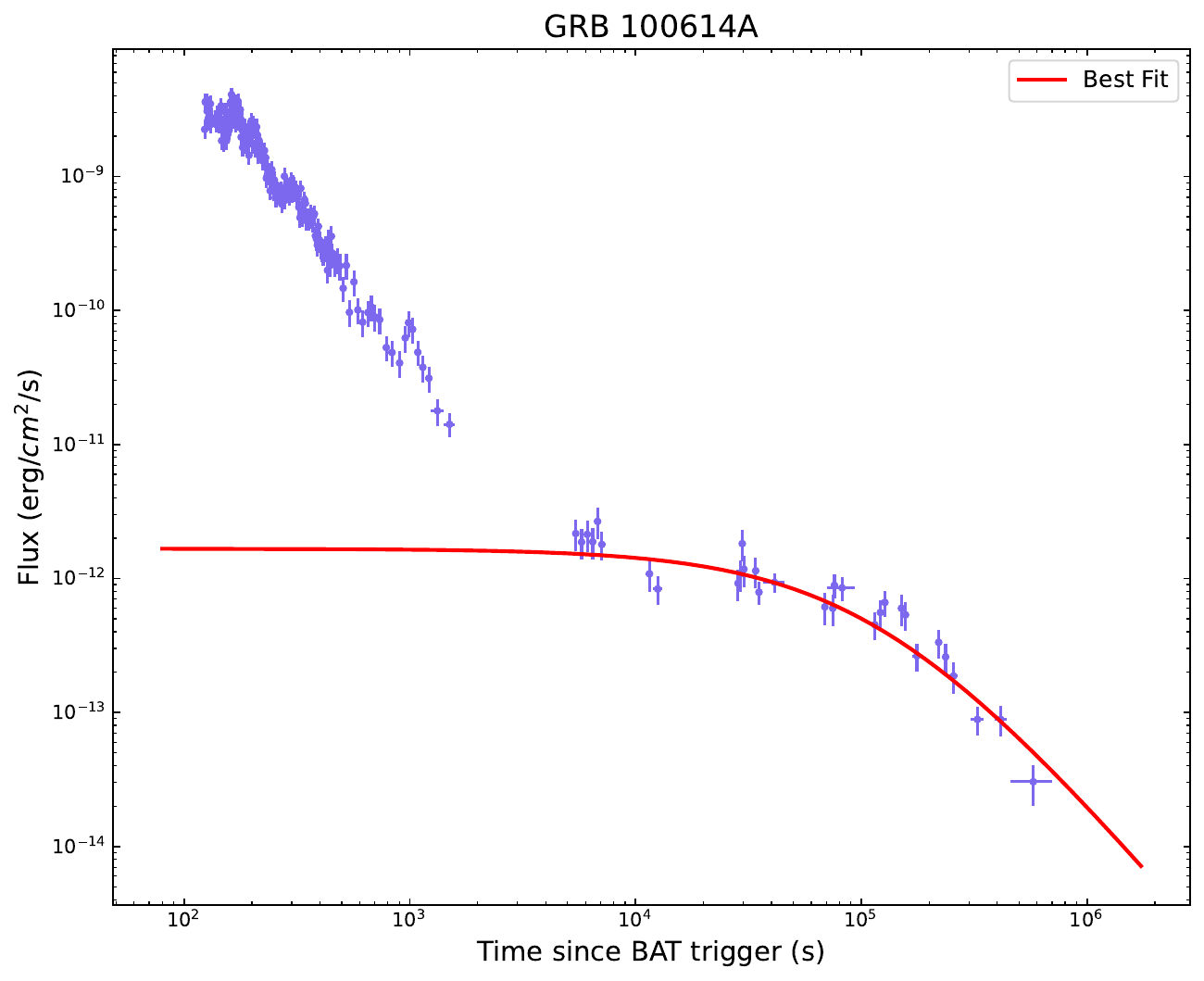}}%

\noindent
\resizebox{55mm}{!}{\includegraphics[]{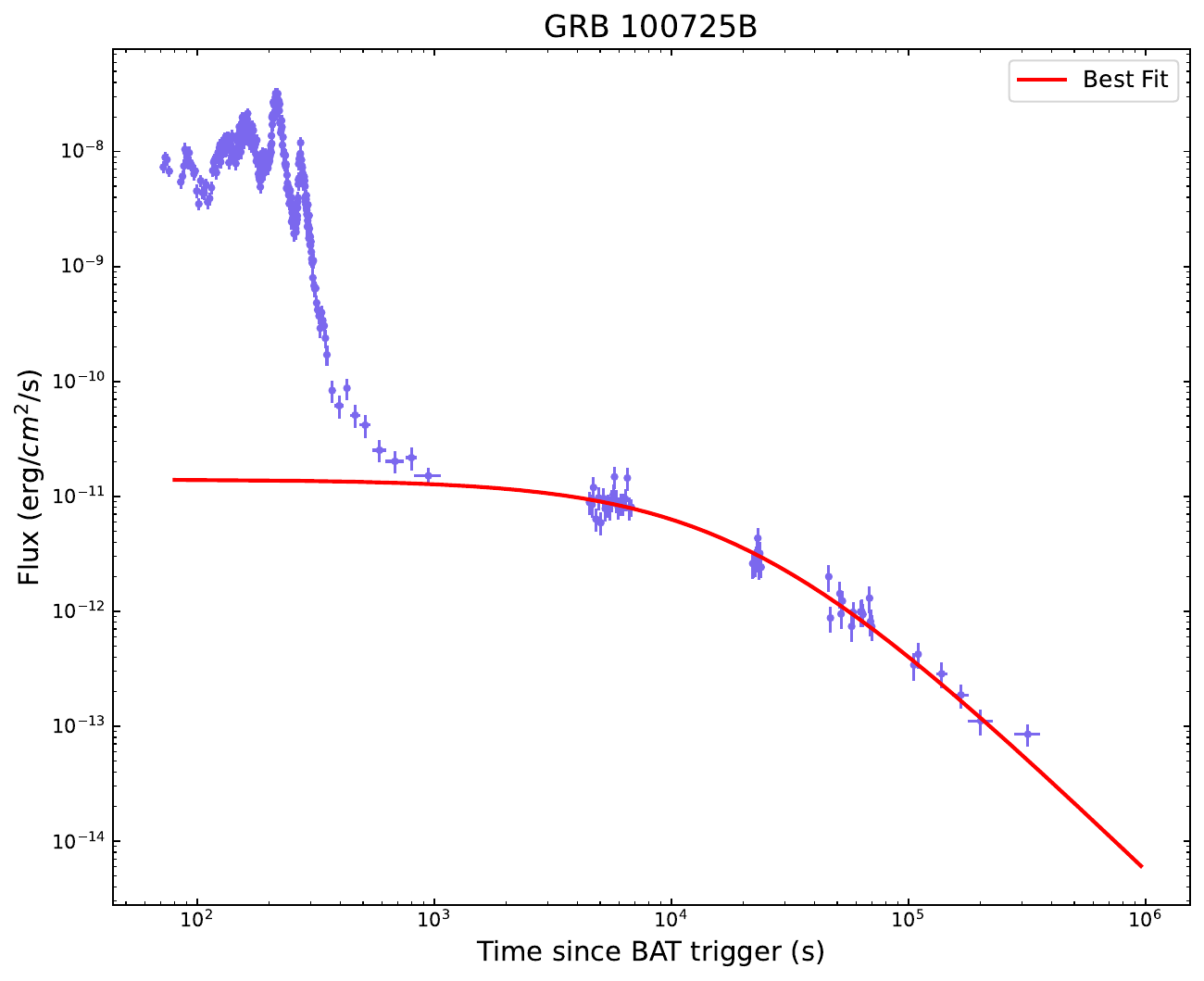}}%
\resizebox{55mm}{!}{\includegraphics[]{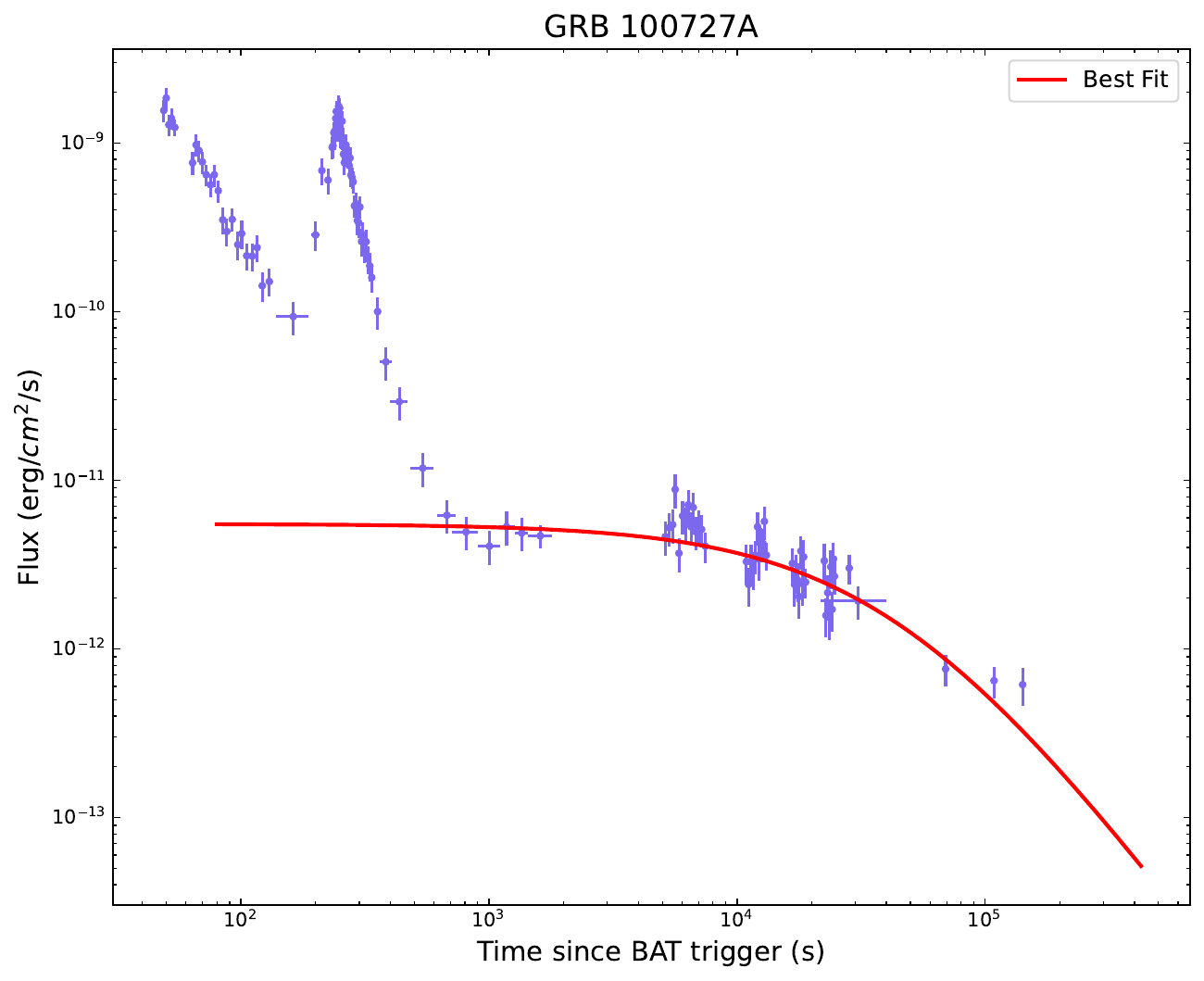}}%
\resizebox{55mm}{!}{\includegraphics[]{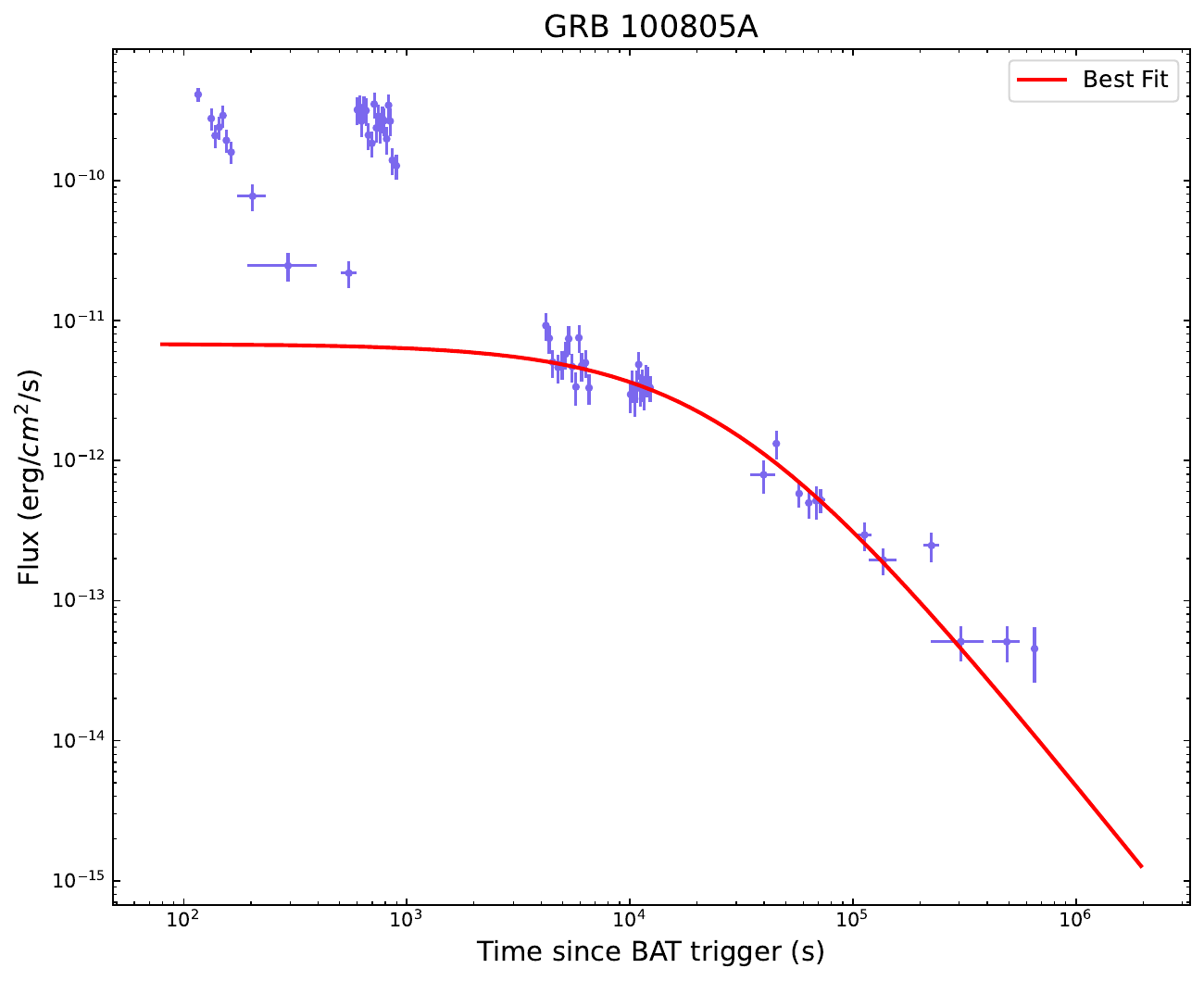}}%
\caption{(Continued)}
\end{figure*}

\addtocounter{figure}{-1}
\begin{figure*}[ht!]

\noindent
\resizebox{55mm}{!}{\includegraphics[]{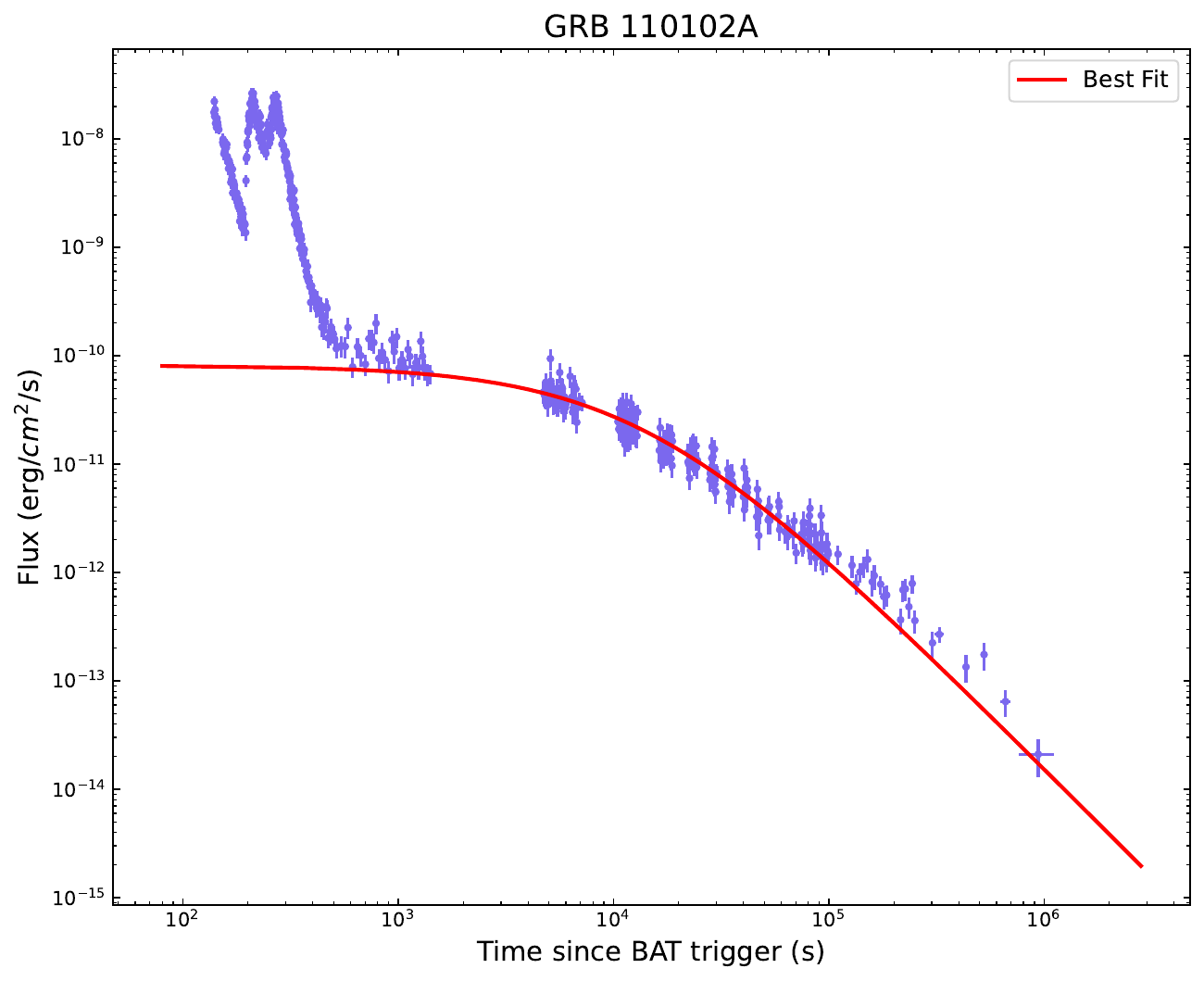}}%
\resizebox{55mm}{!}{\includegraphics[]{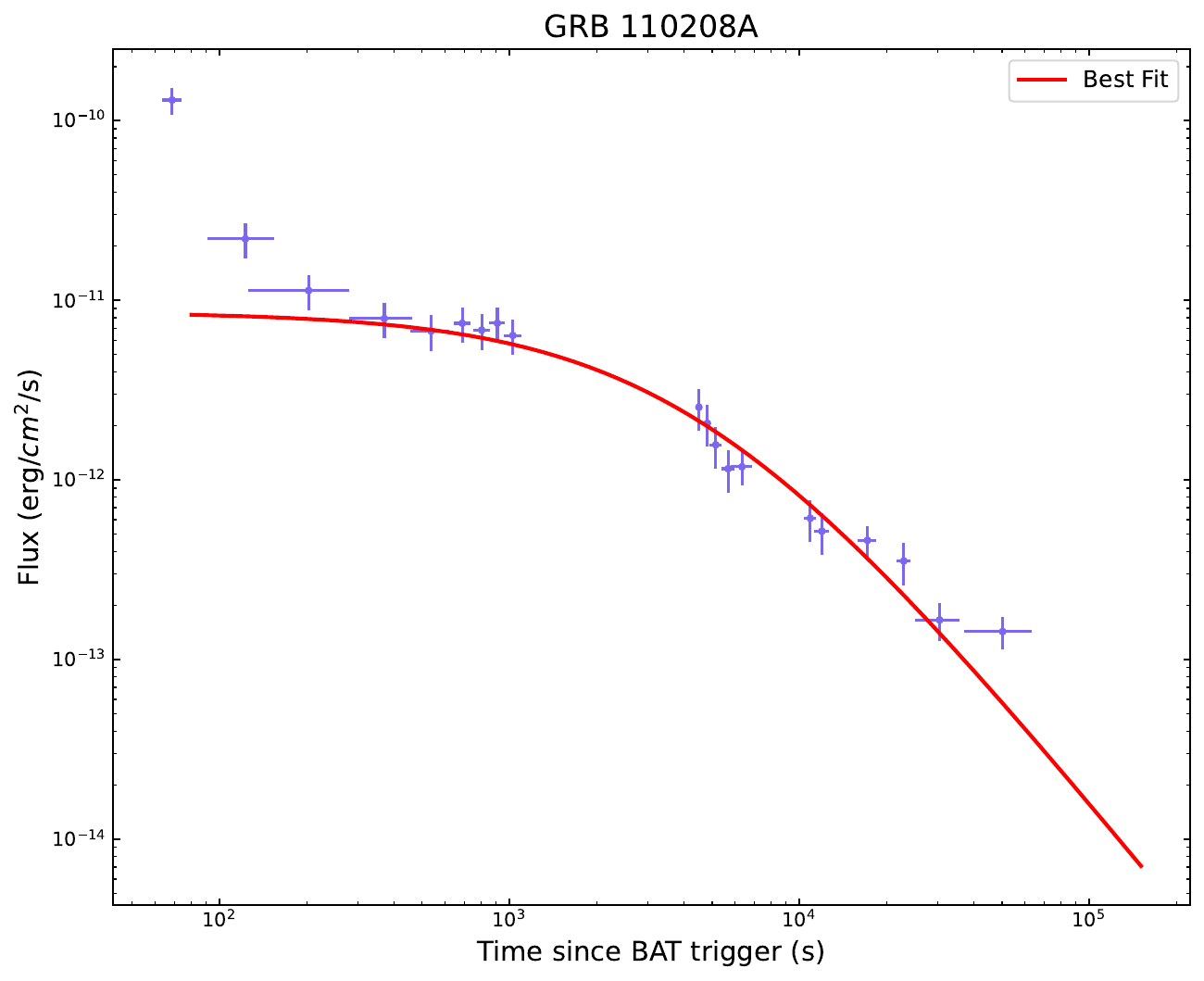}}%
\resizebox{55mm}{!}{\includegraphics[]{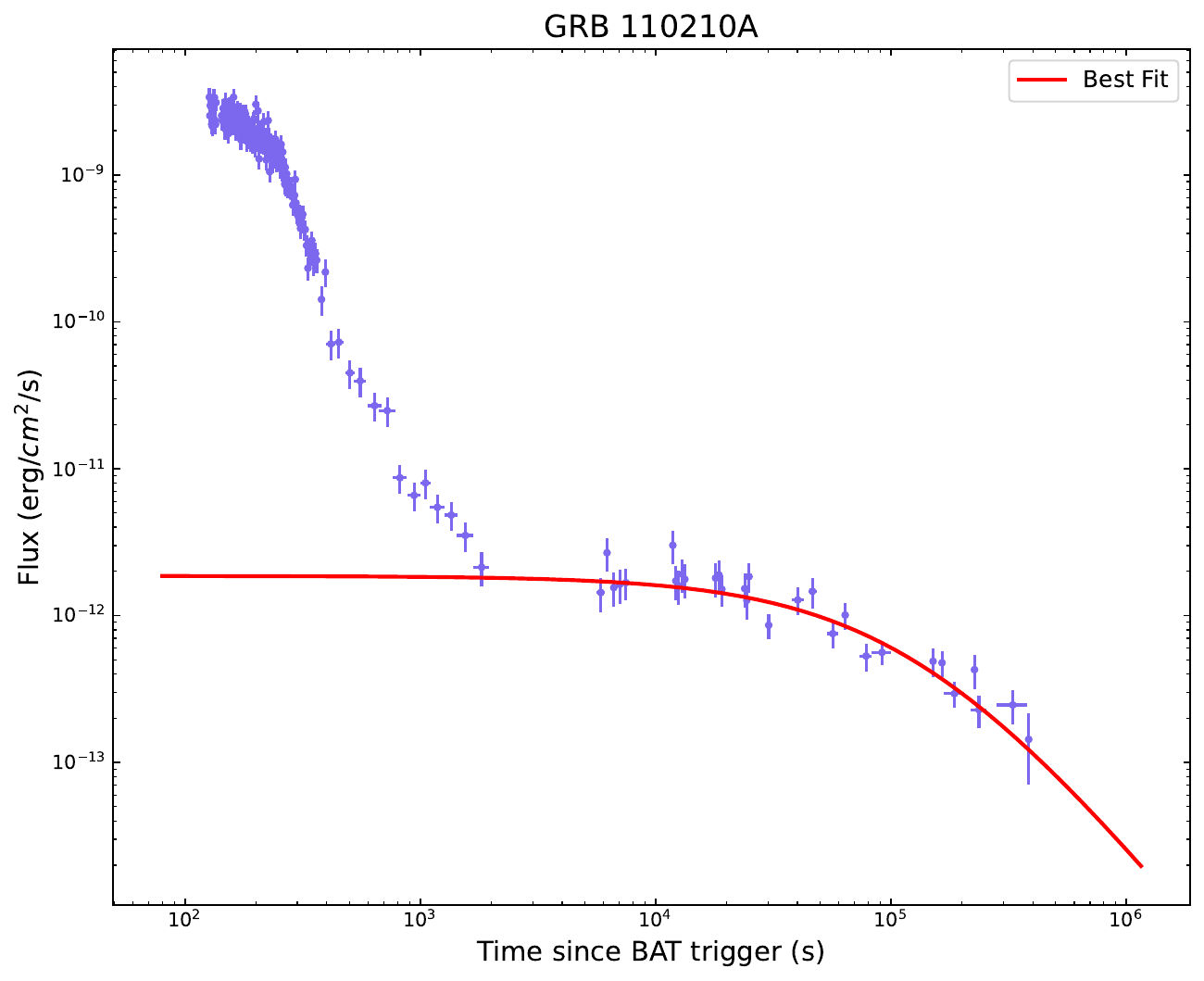}}%

\noindent
\resizebox{55mm}{!}{\includegraphics[]{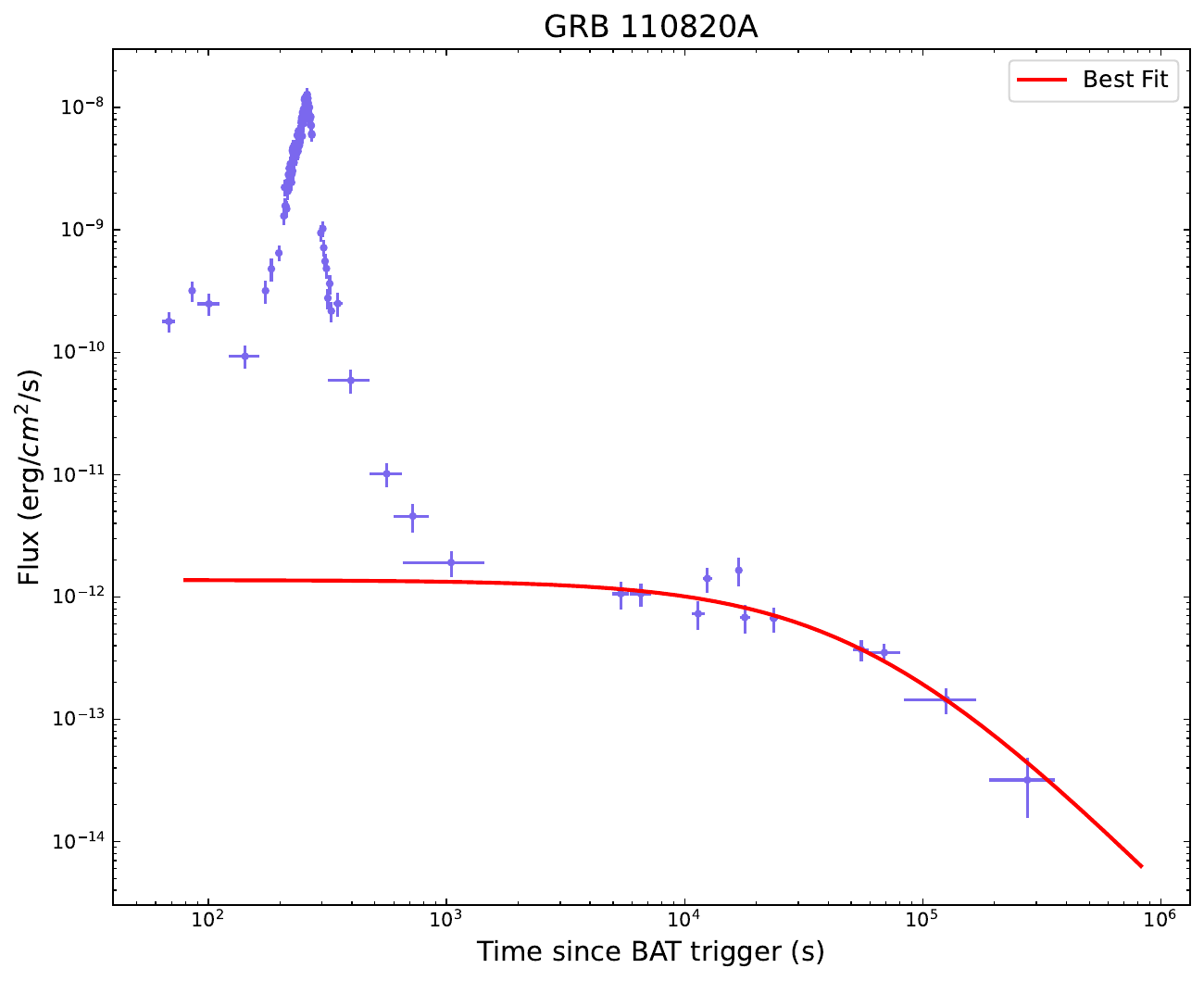}}%
\resizebox{55mm}{!}{\includegraphics[]{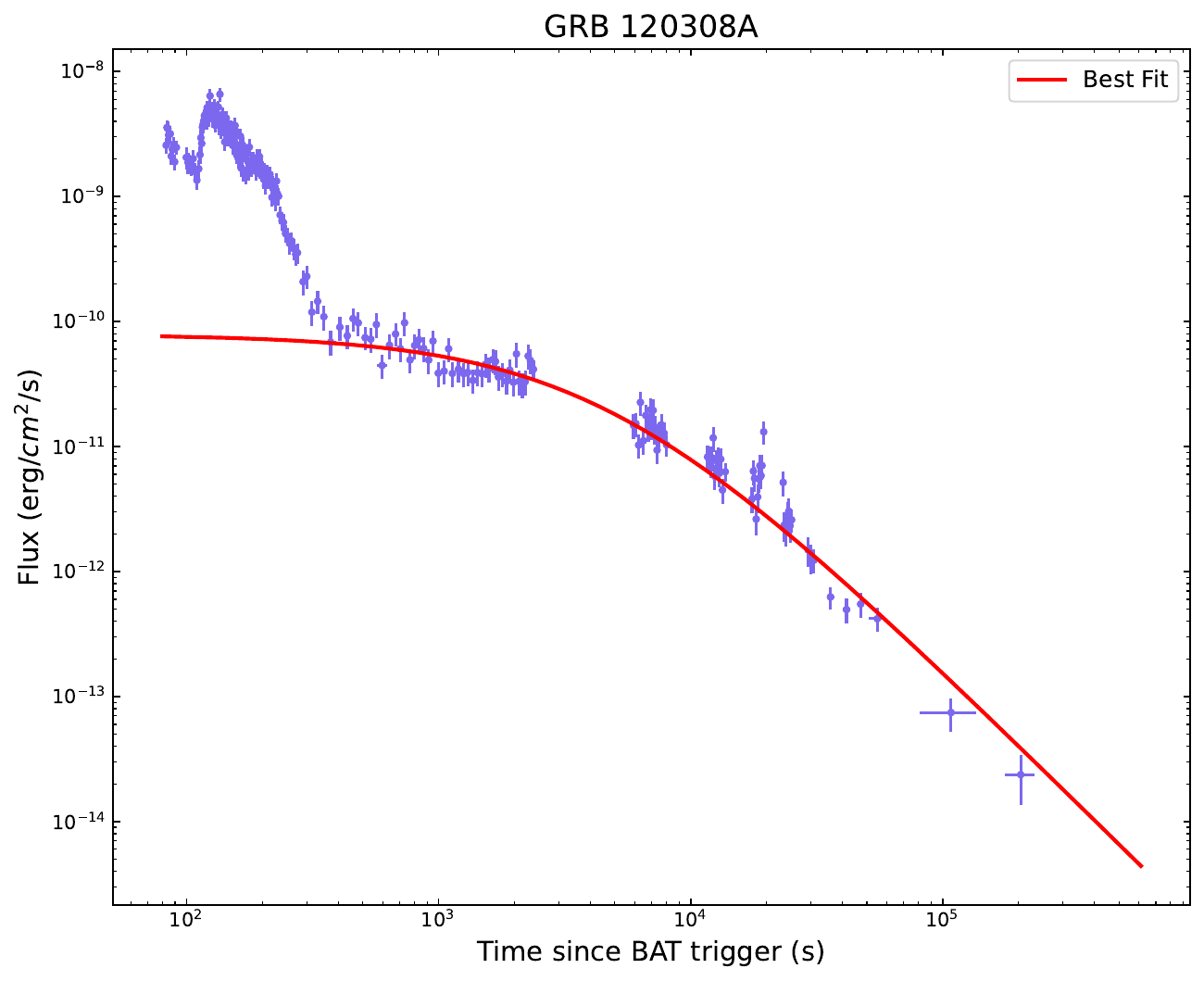}}%
\resizebox{55mm}{!}{\includegraphics[]{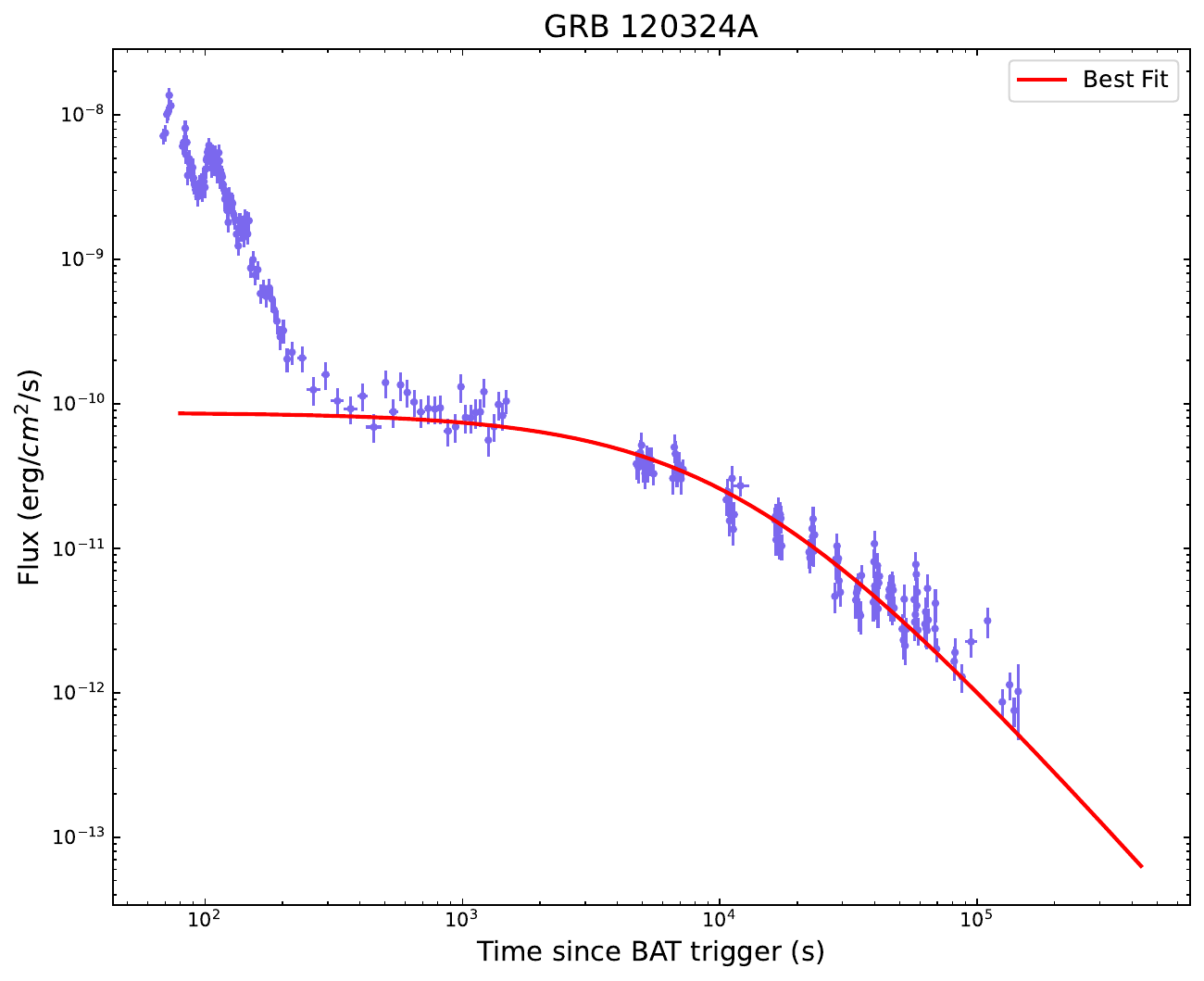}}%

\noindent
\resizebox{55mm}{!}{\includegraphics[]{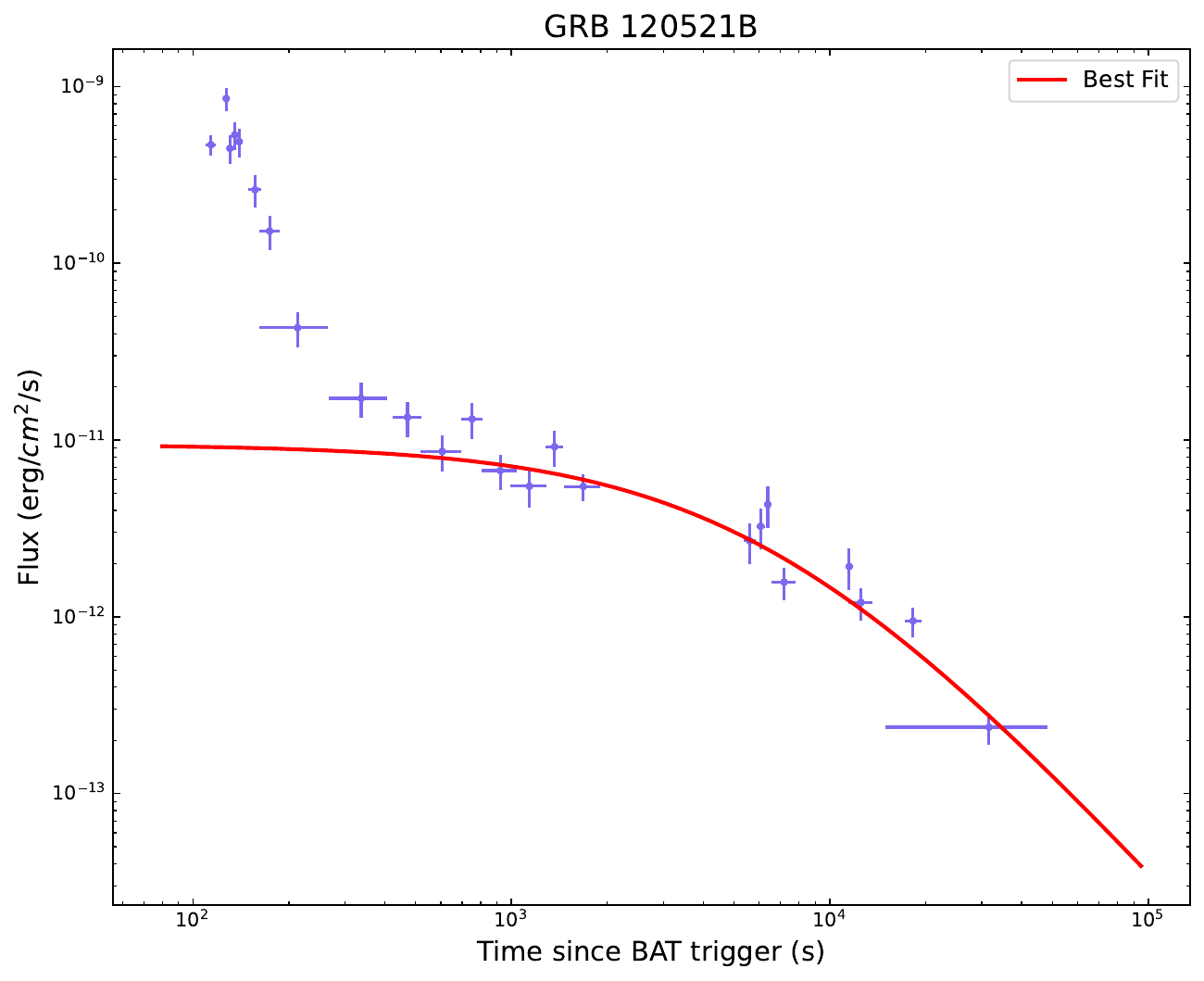}}%
\resizebox{55mm}{!}{\includegraphics[]{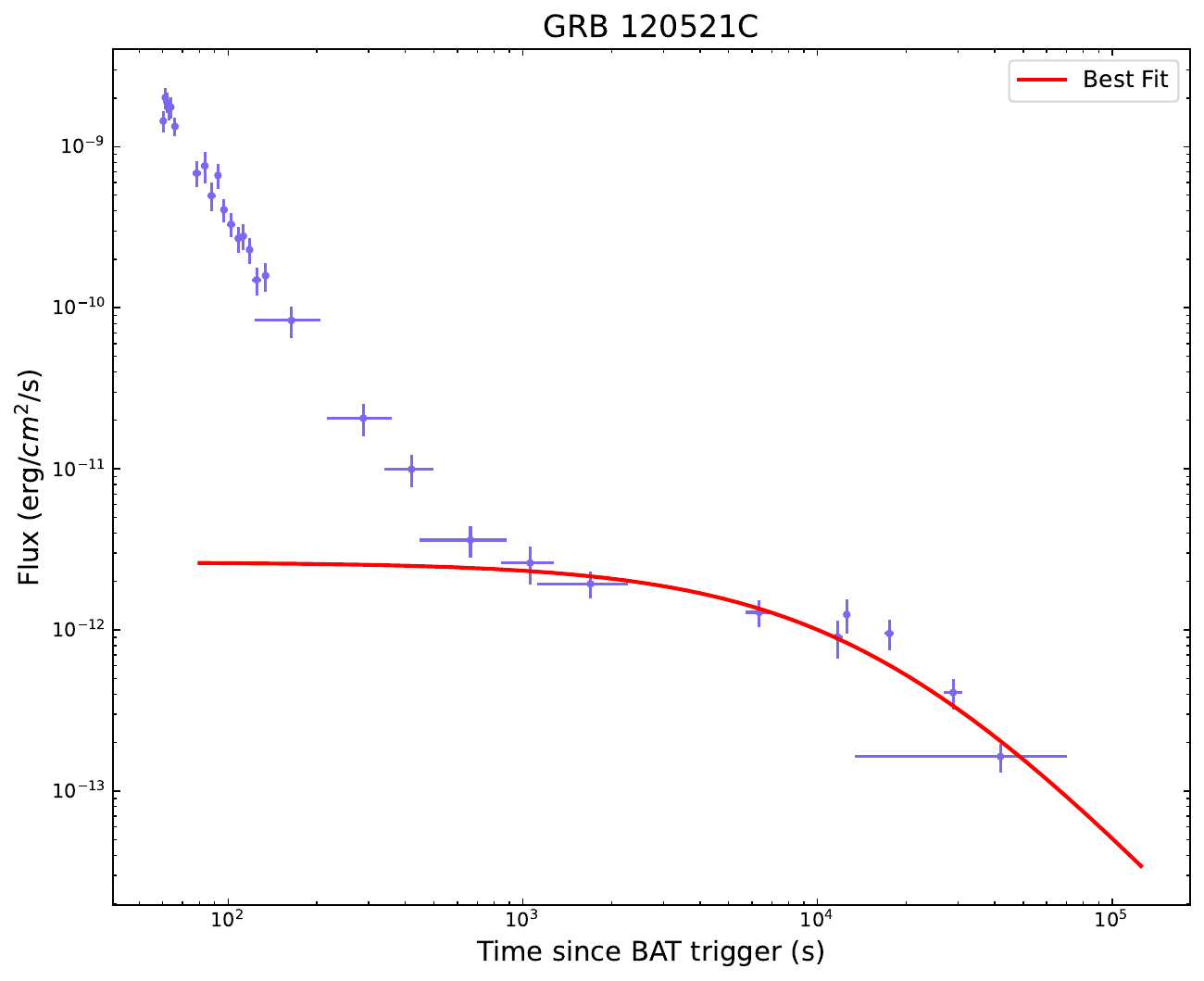}}%
\resizebox{55mm}{!}{\includegraphics[]{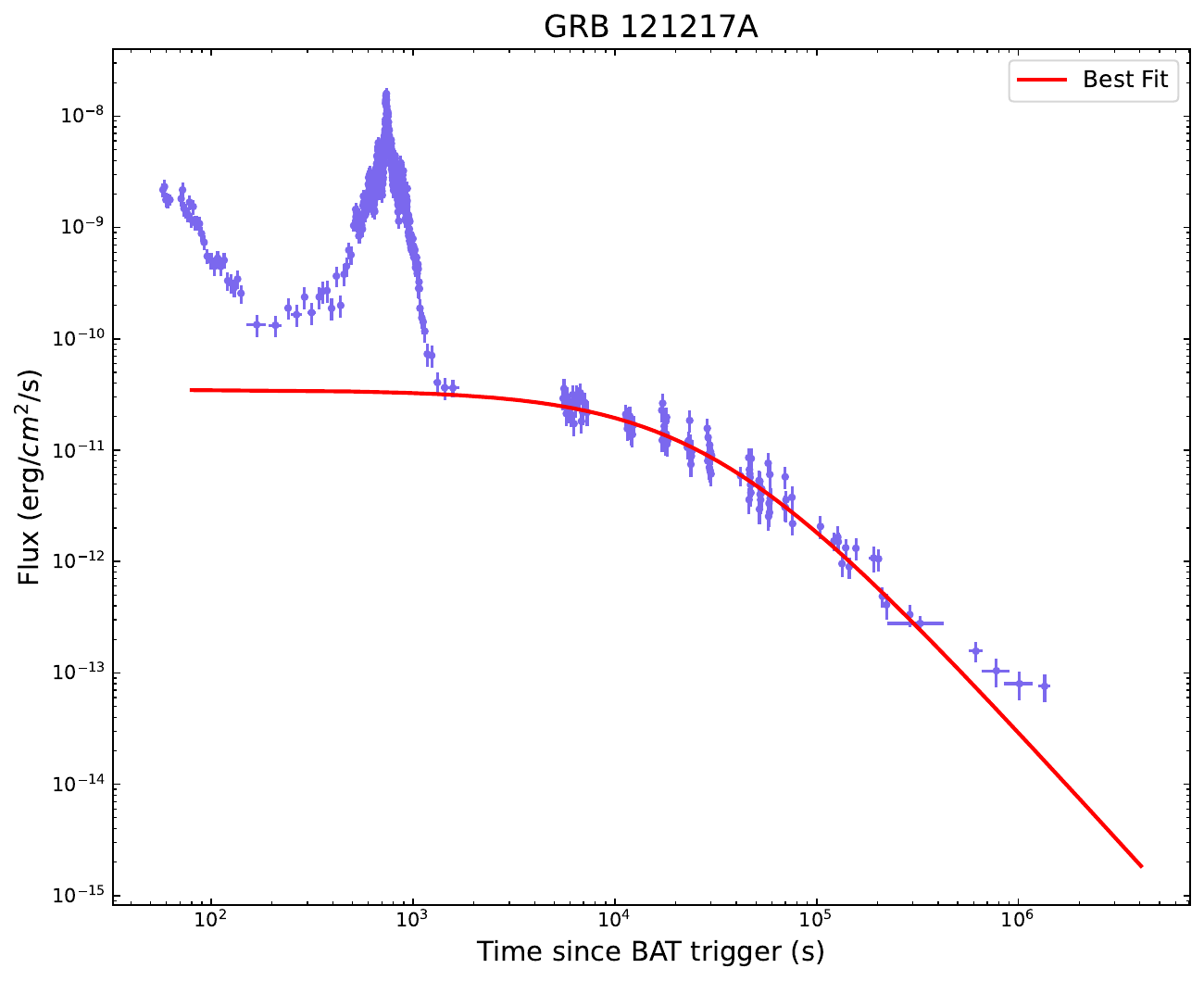}}%

\noindent
\resizebox{55mm}{!}{\includegraphics[]{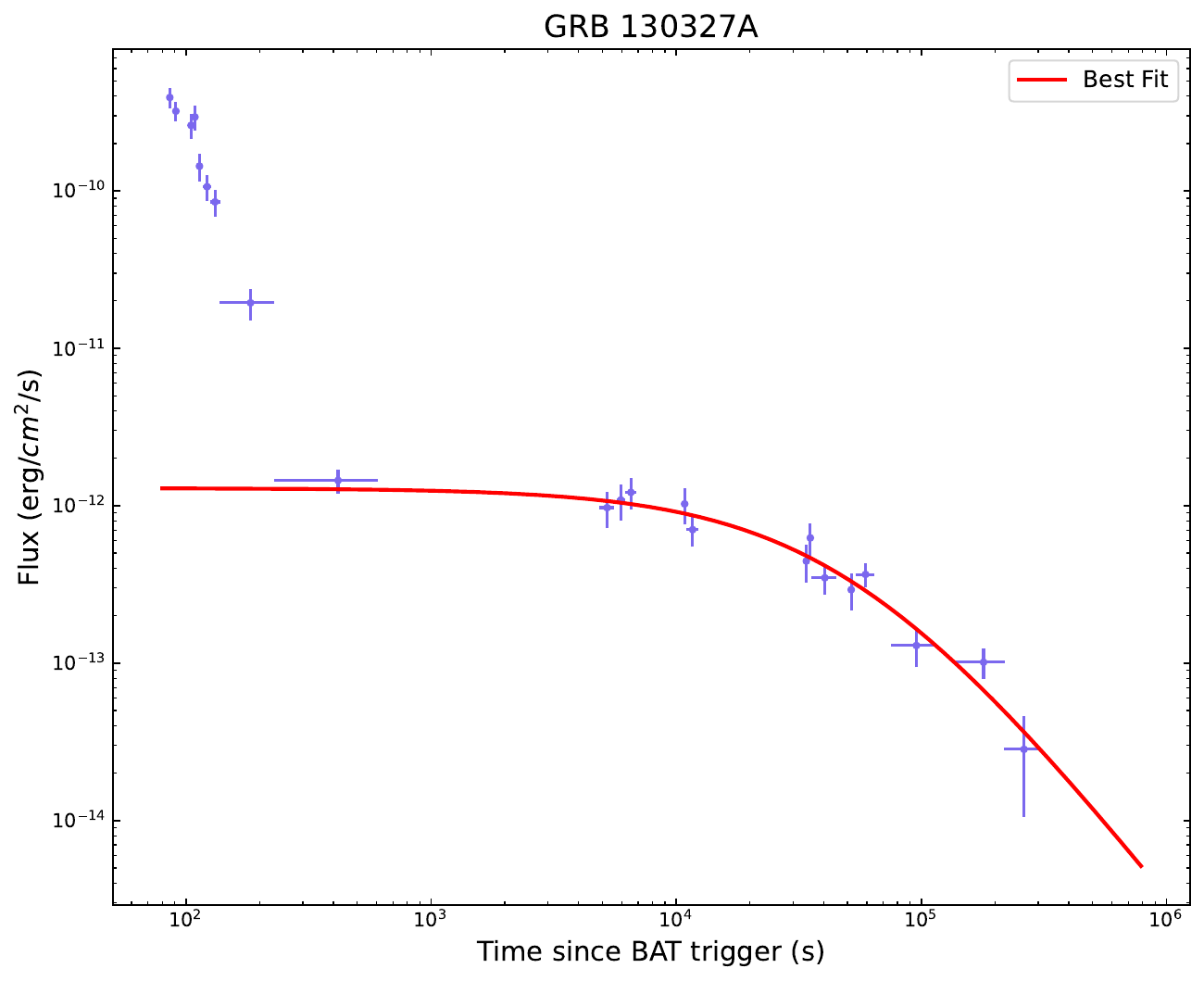}}%
\resizebox{55mm}{!}{\includegraphics[]{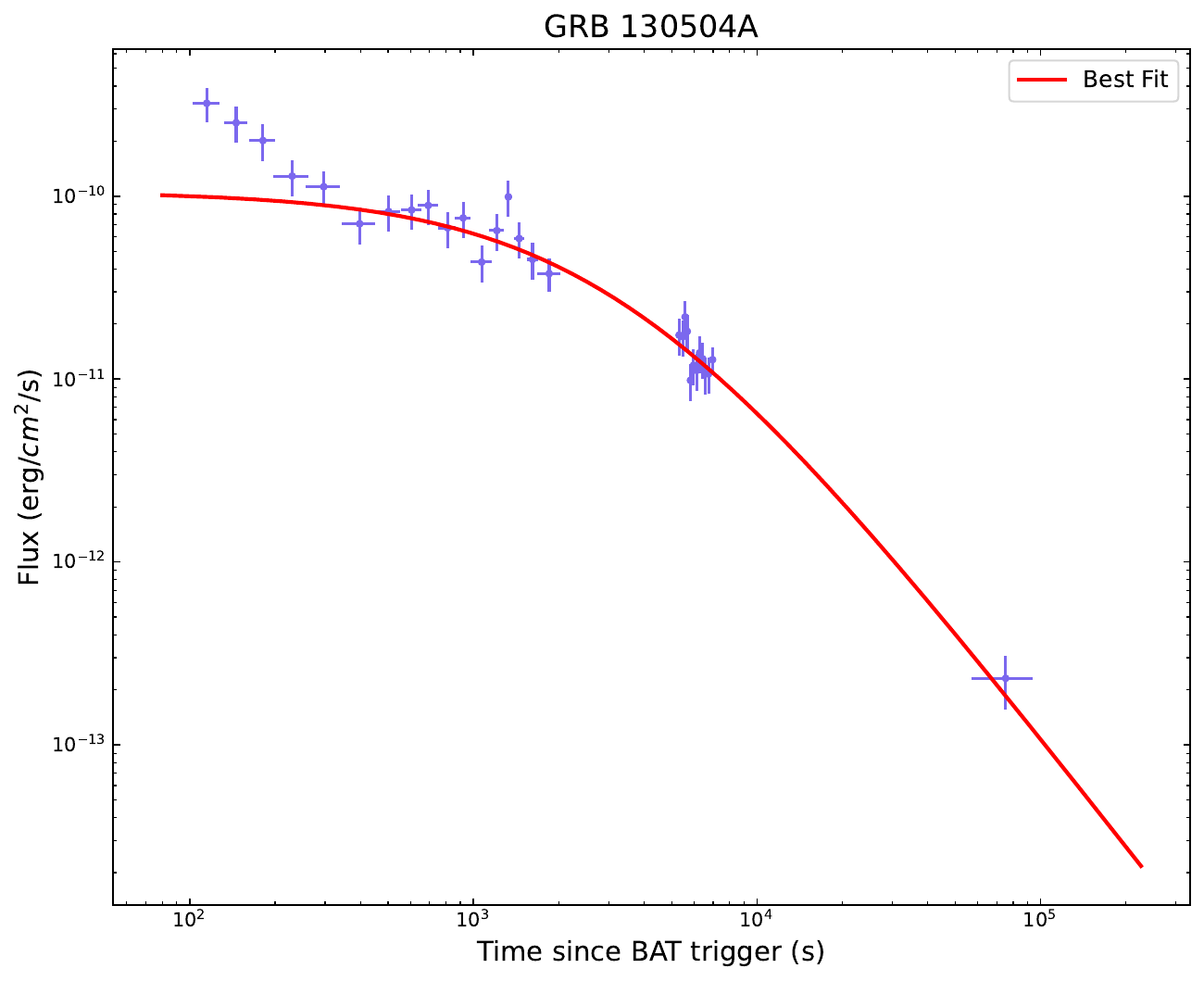}}%
\resizebox{55mm}{!}{\includegraphics[]{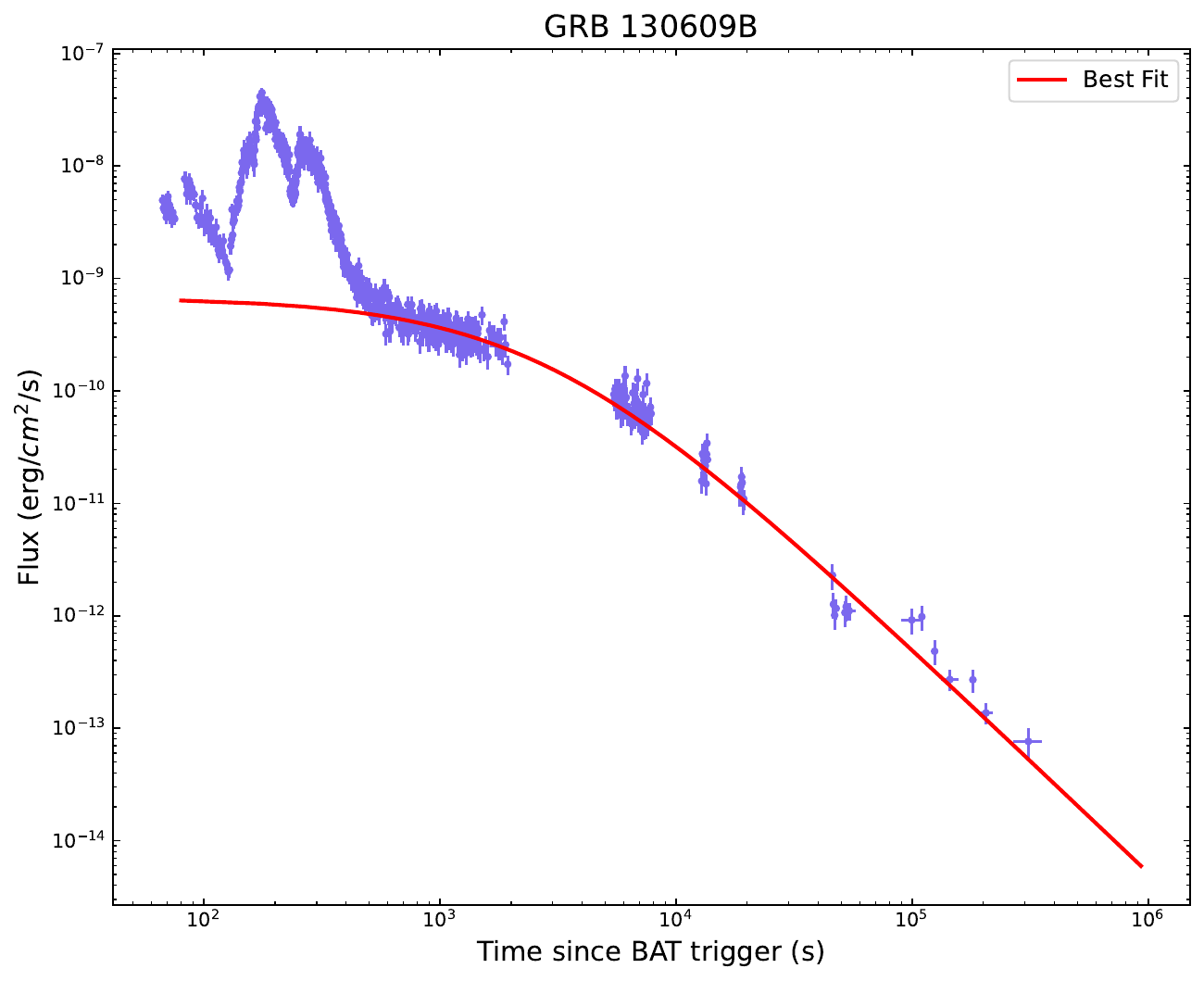}}%

\noindent
\resizebox{55mm}{!}{\includegraphics[]{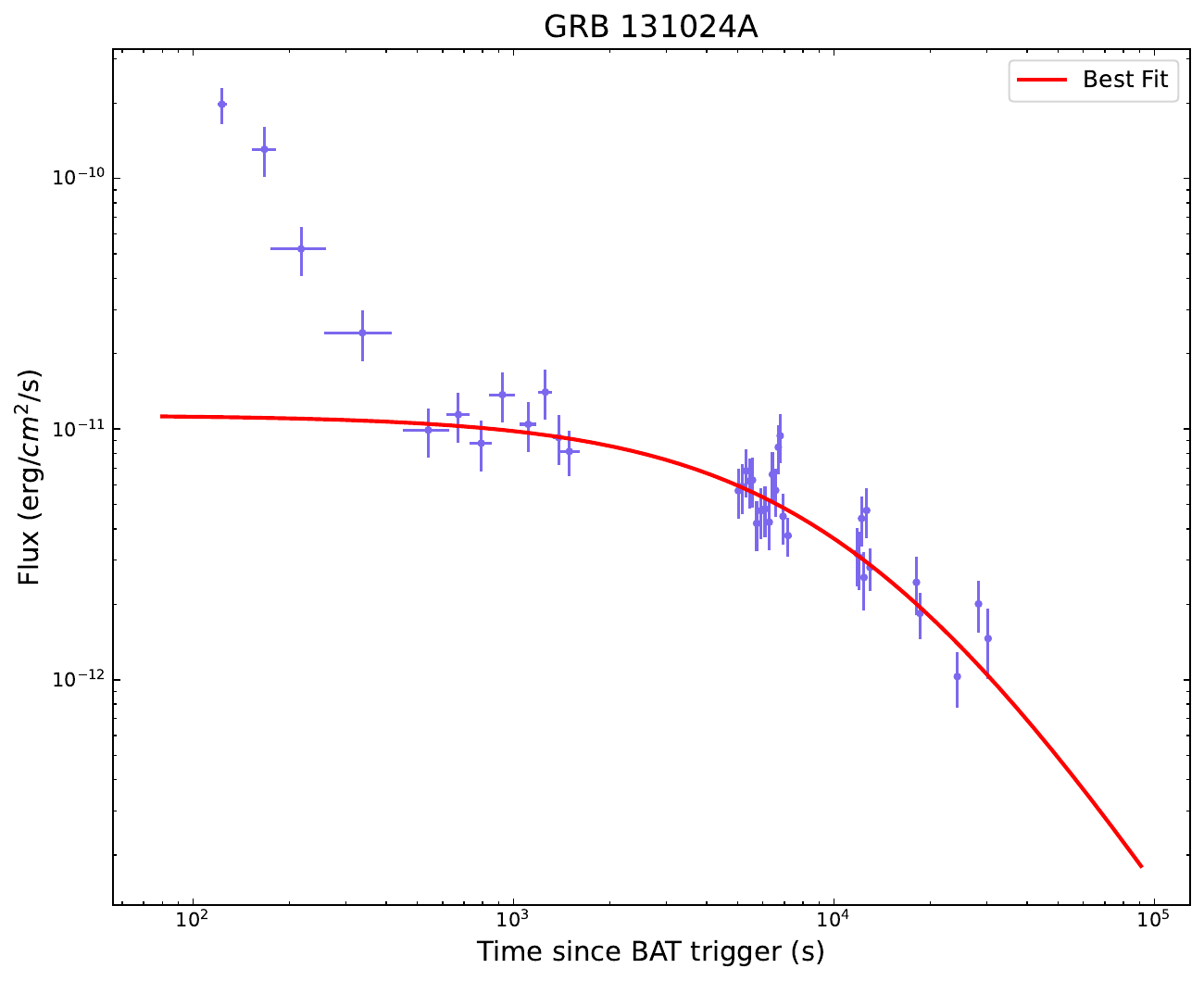}}%
\resizebox{55mm}{!}{\includegraphics[]{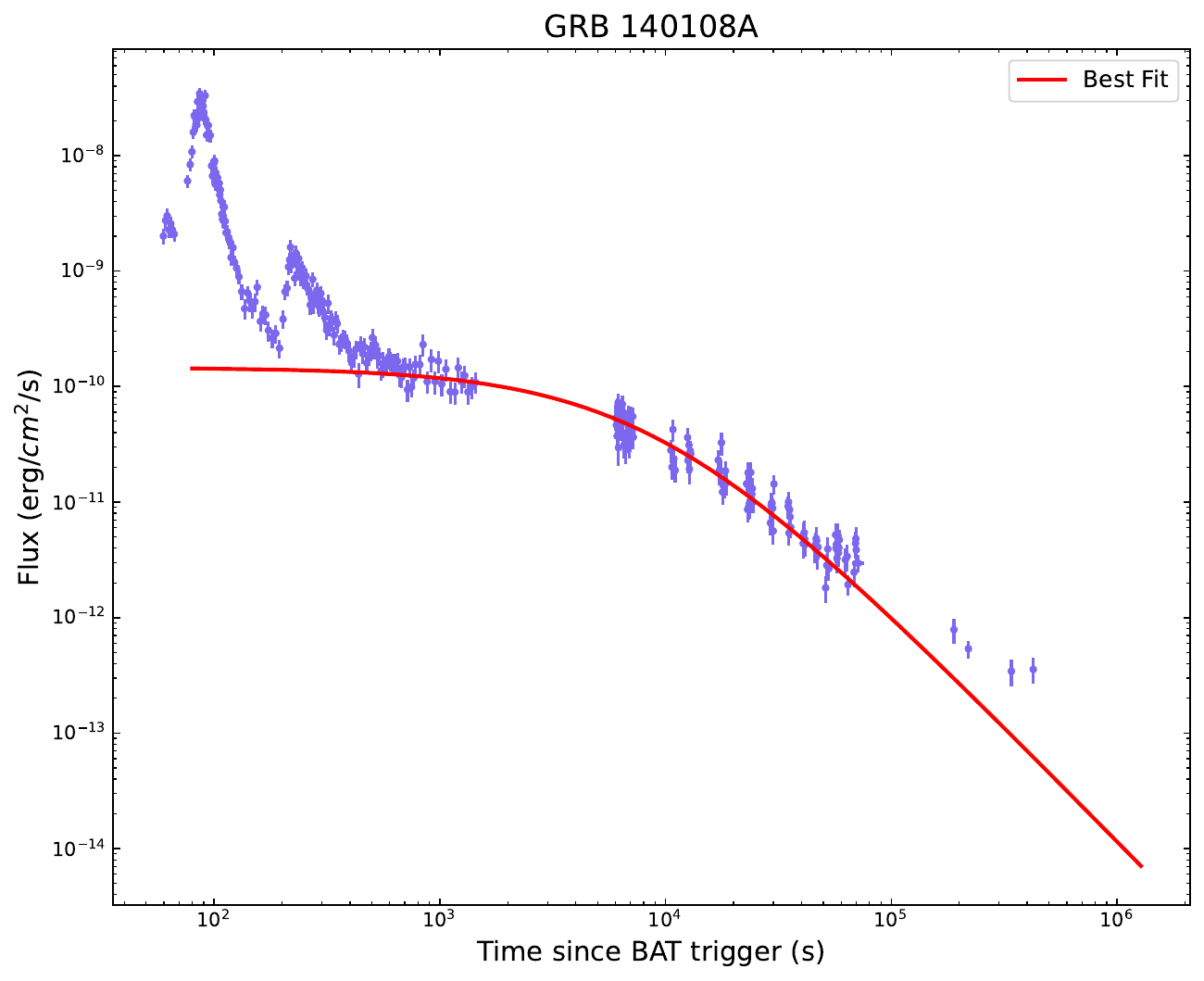}}%
\resizebox{55mm}{!}{\includegraphics[]{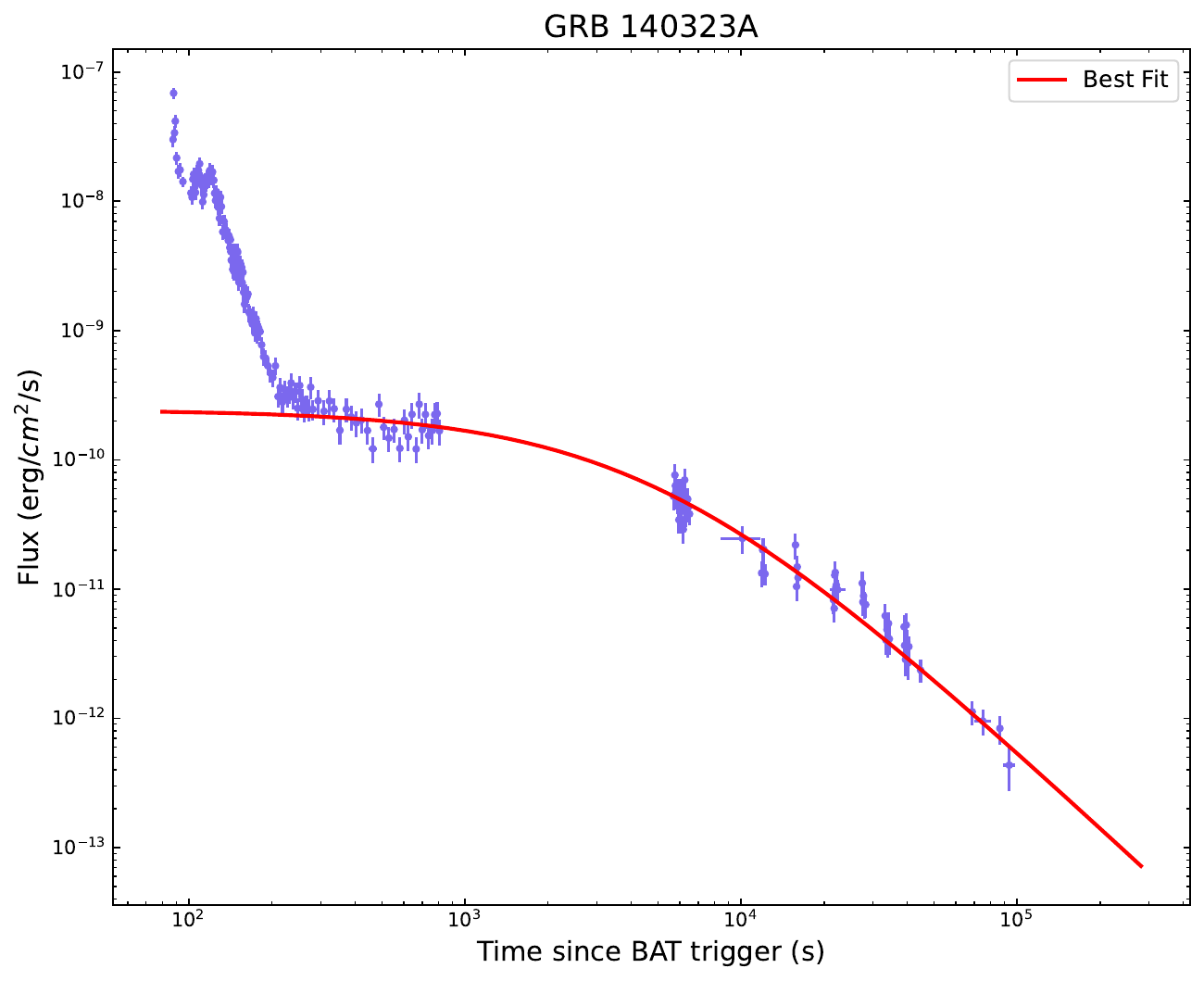}}%

\caption{(Continued)}
\end{figure*}

\addtocounter{figure}{-1}
\begin{figure*}[ht!]

\noindent
\resizebox{55mm}{!}{\includegraphics[]{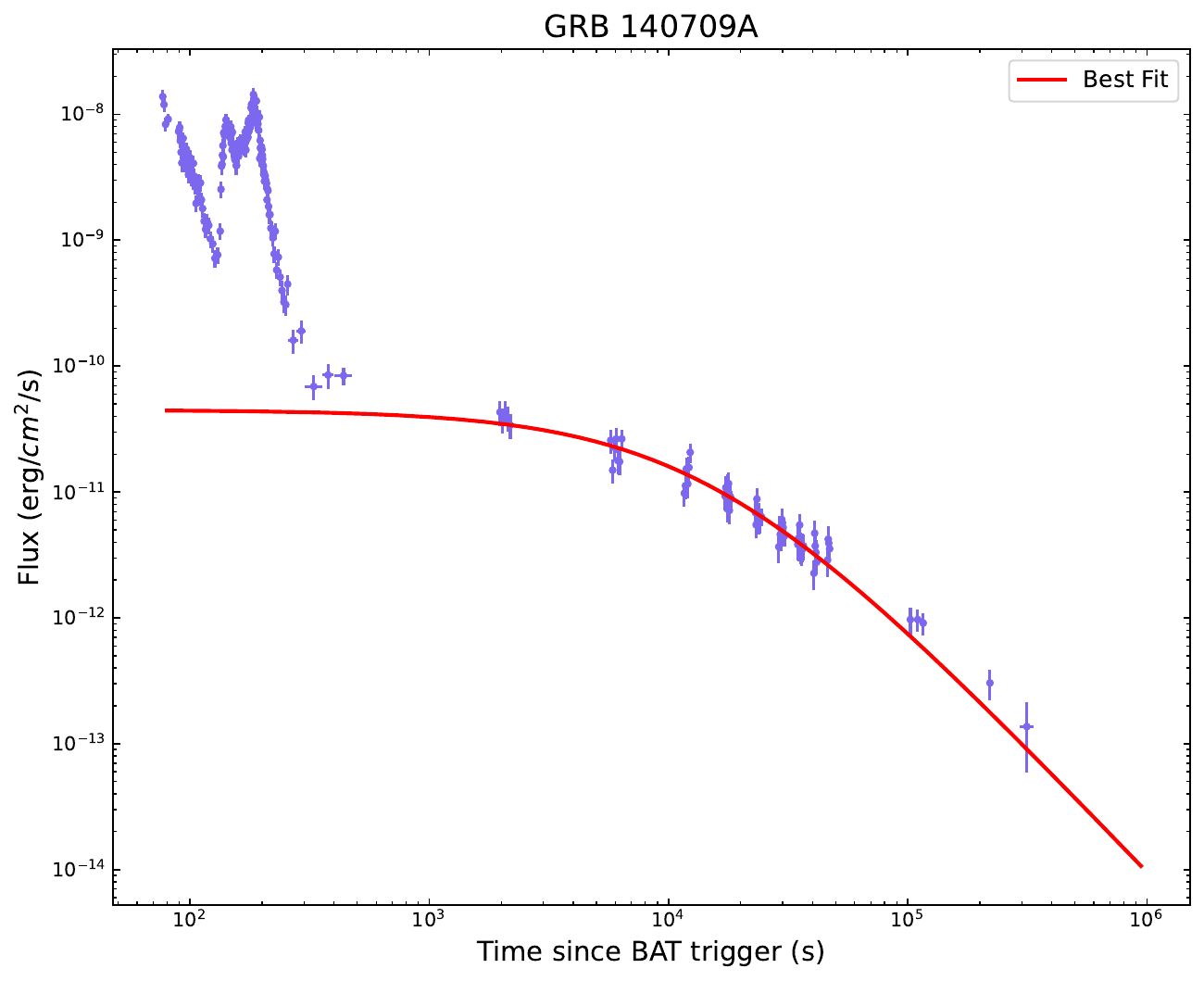}}%
\resizebox{55mm}{!}{\includegraphics[]{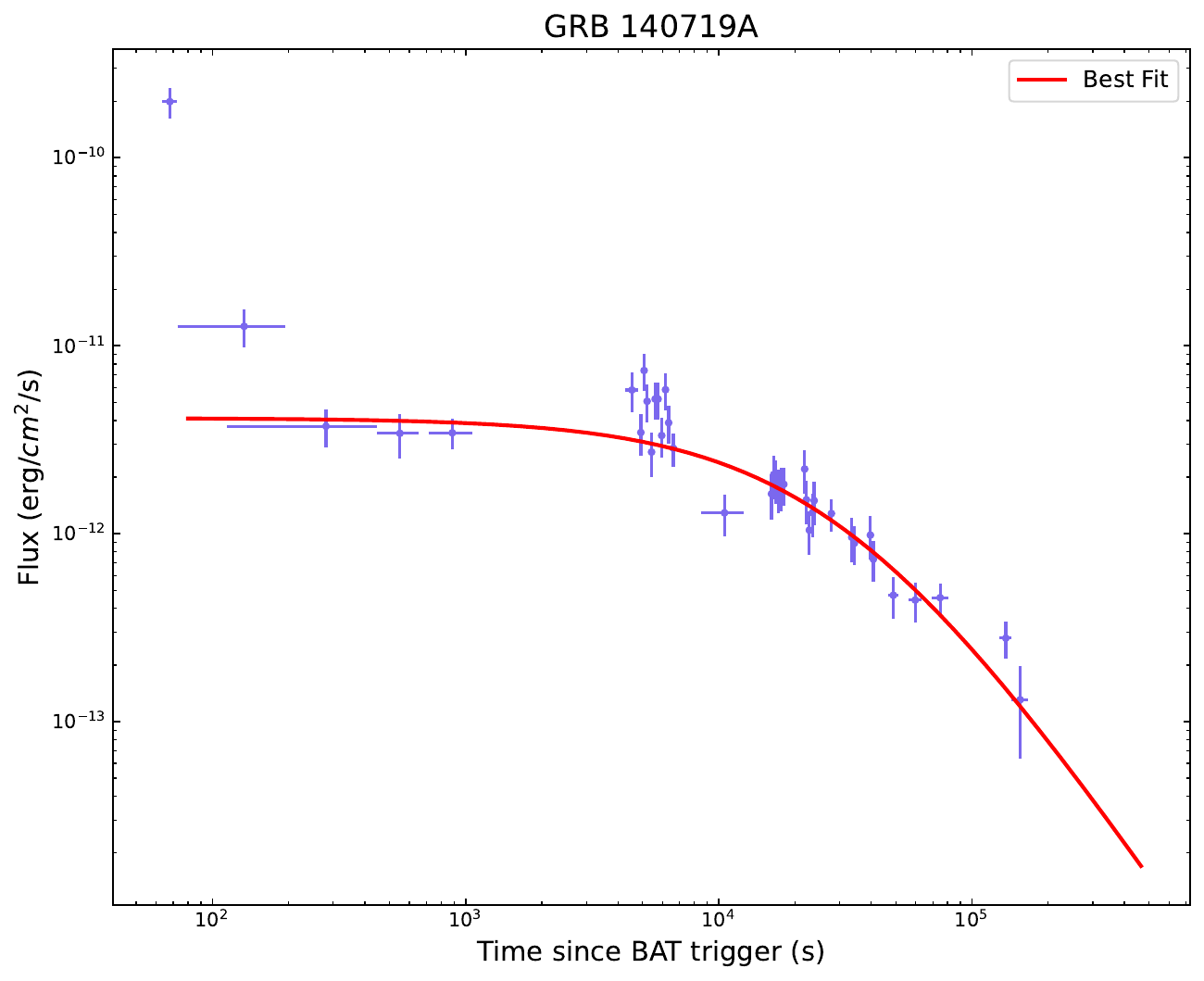}}%
\resizebox{55mm}{!}{\includegraphics[]{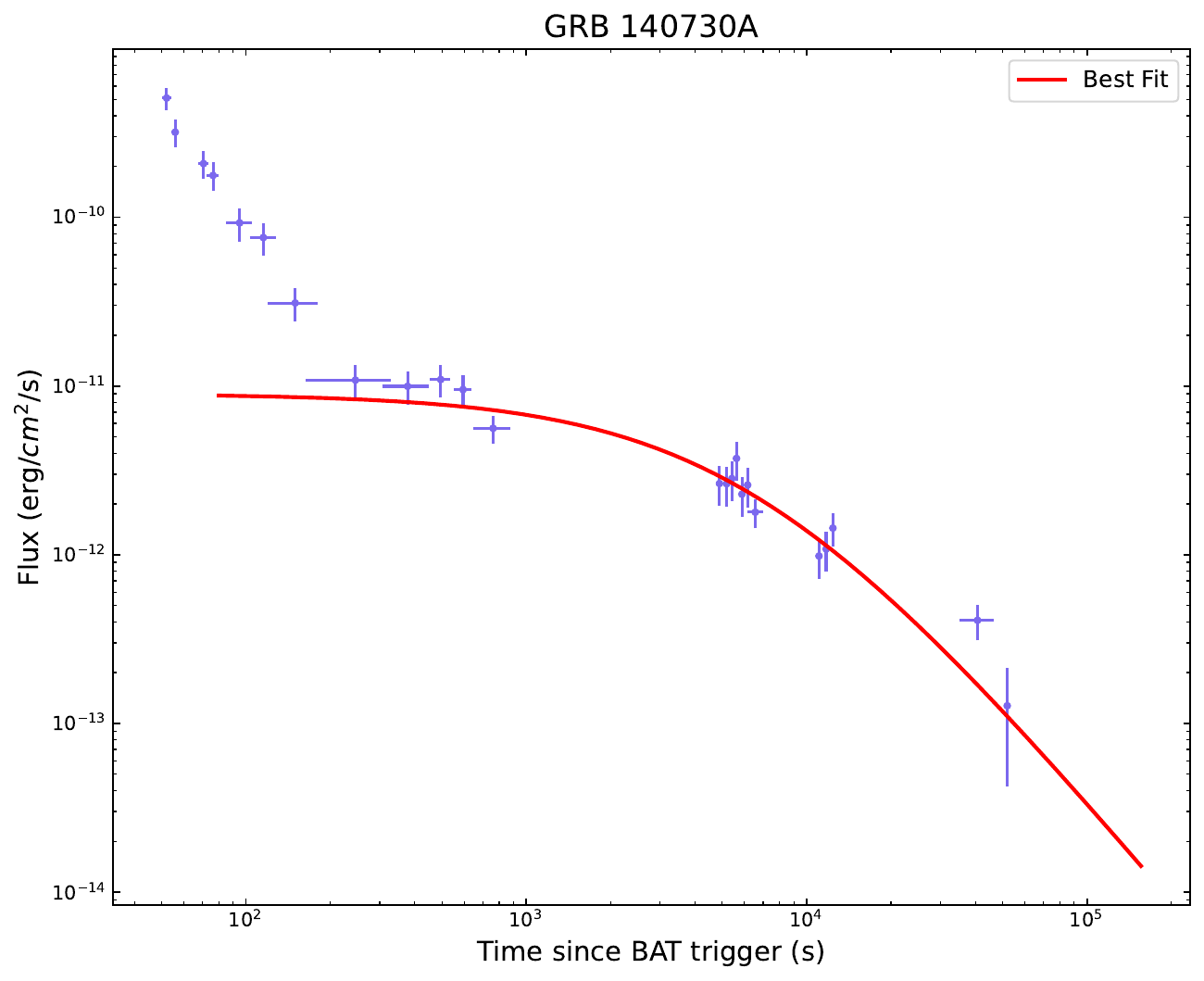}}%

\noindent
\resizebox{55mm}{!}{\includegraphics[]{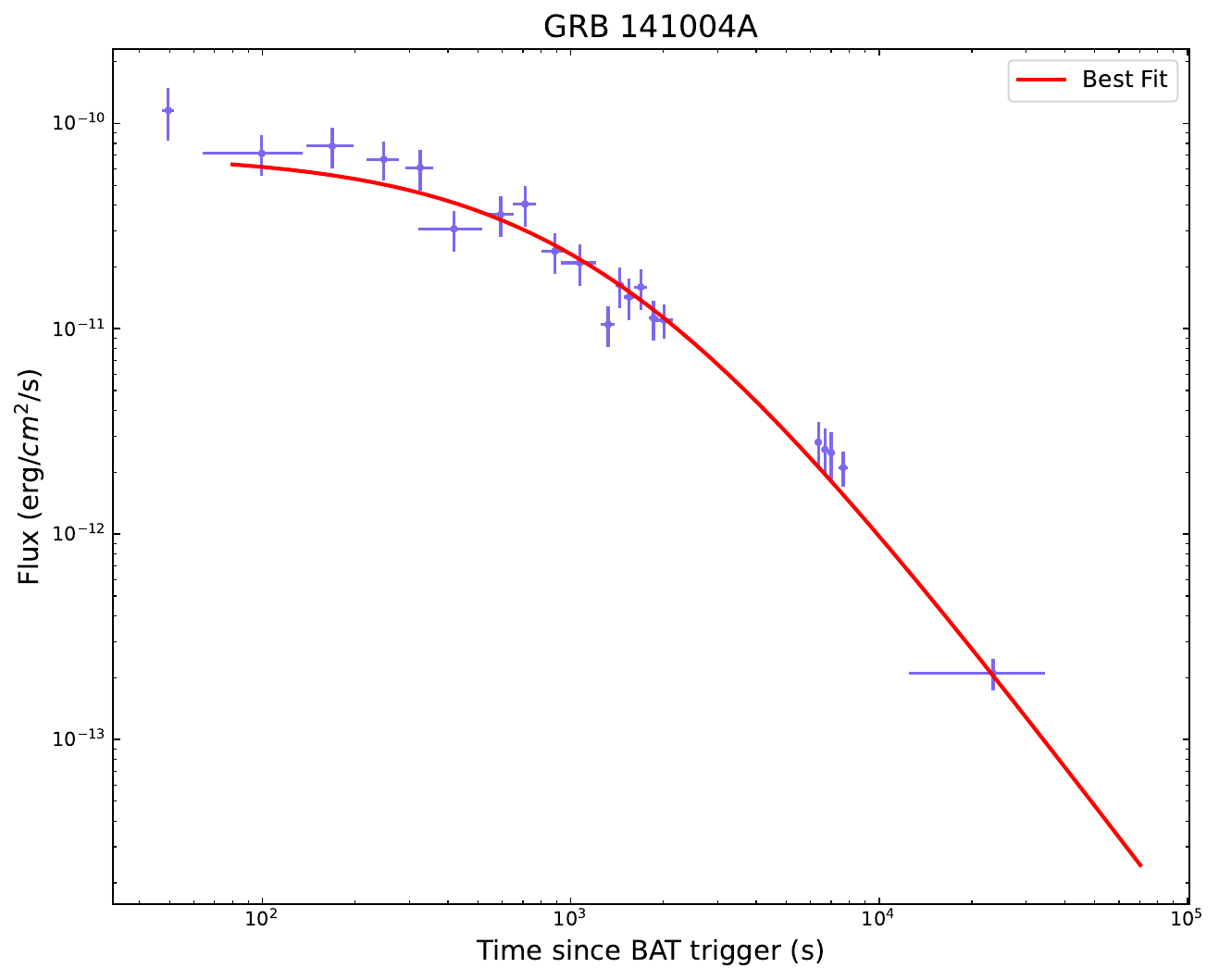}}%
\resizebox{55mm}{!}{\includegraphics[]{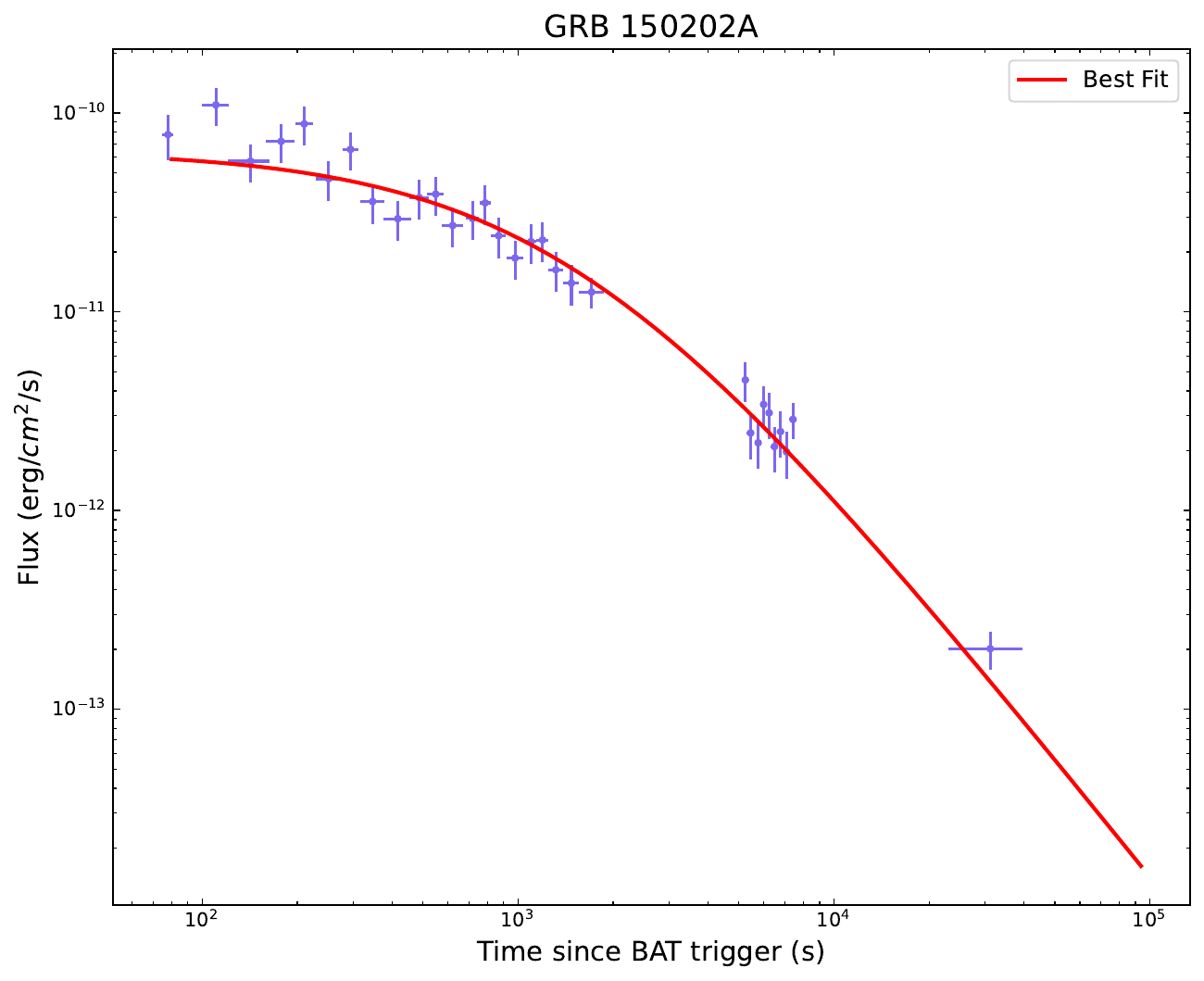}}%
\resizebox{55mm}{!}{\includegraphics[]{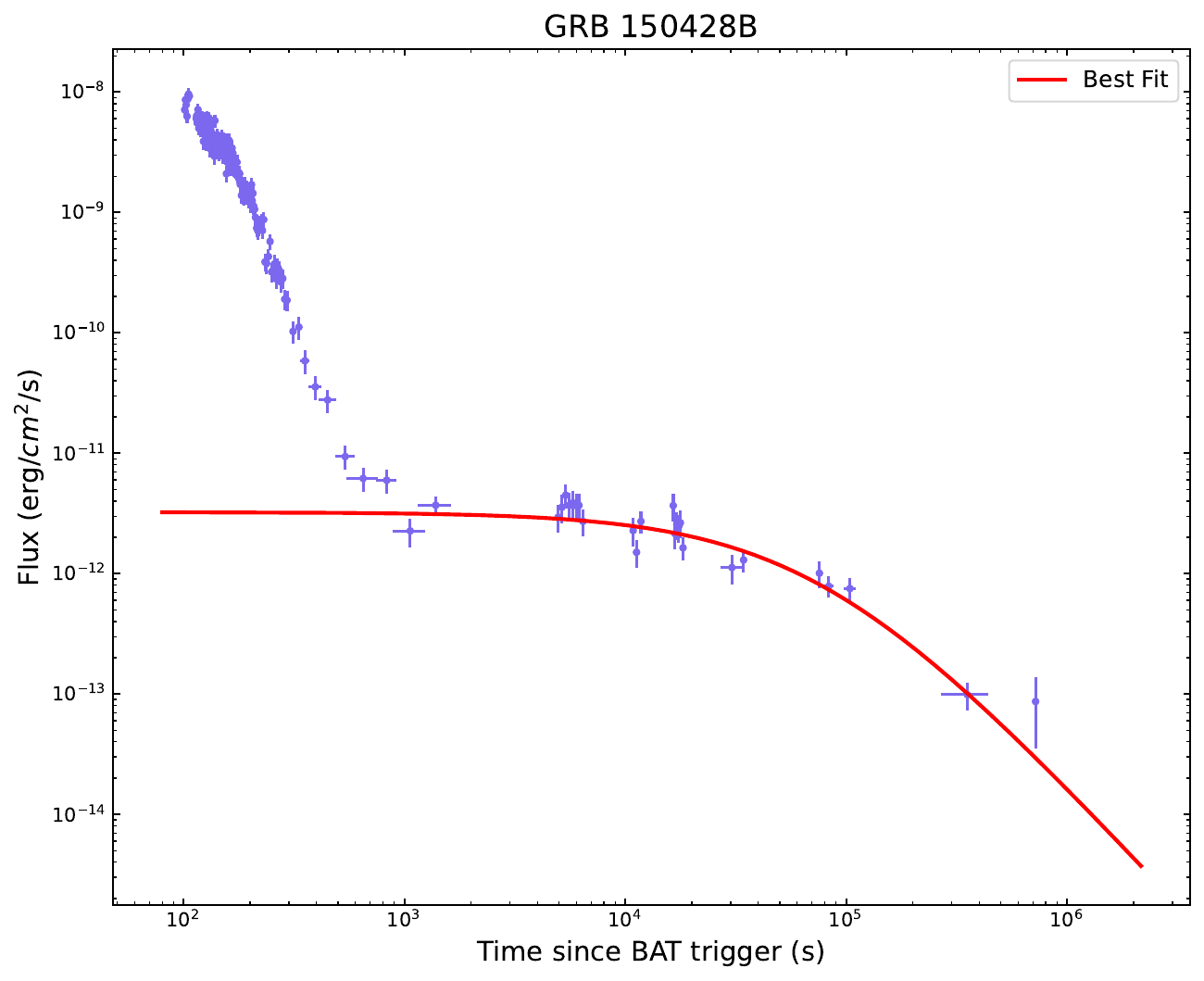}}%

\noindent
\resizebox{55mm}{!}{\includegraphics[]{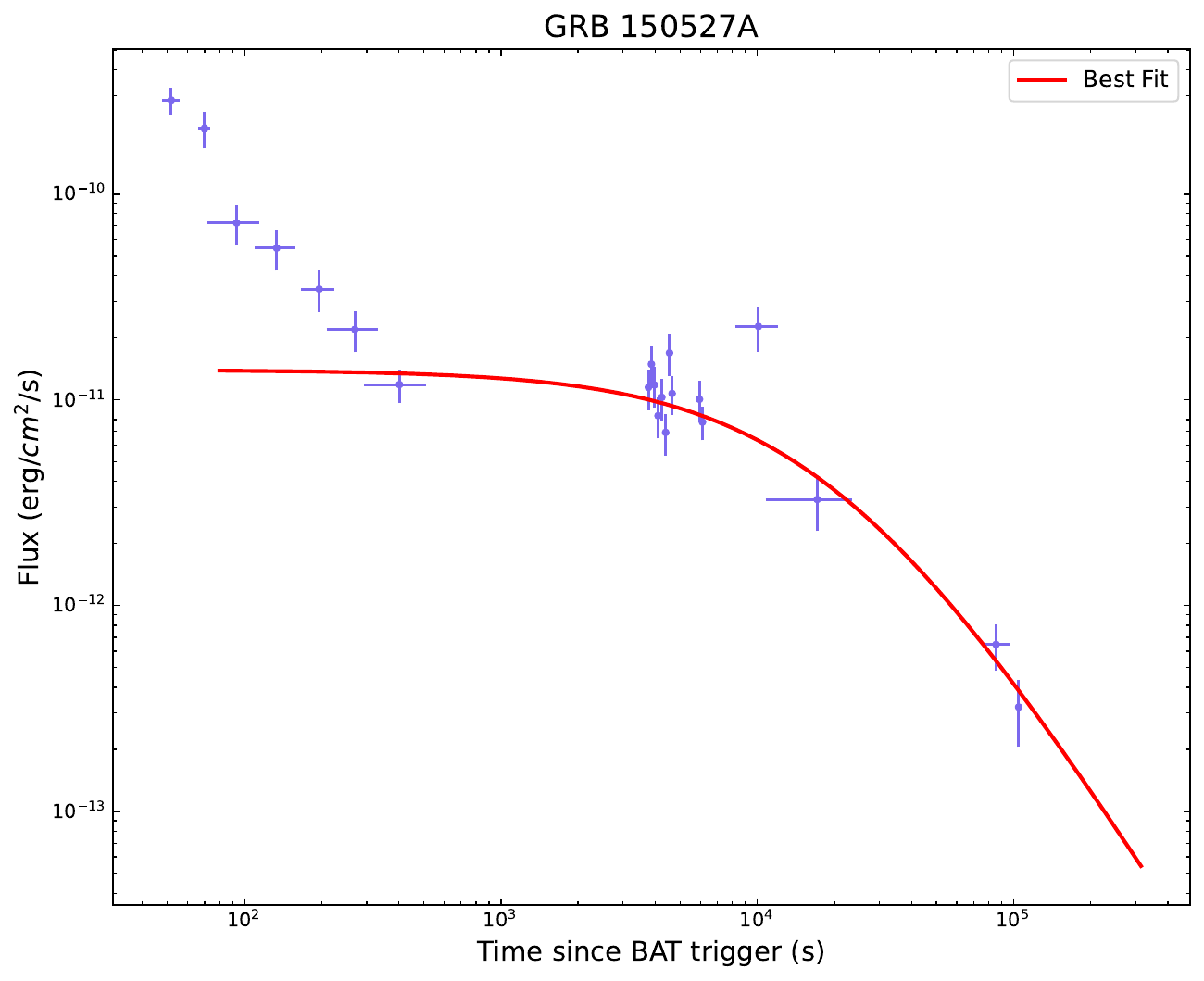}}%
\resizebox{55mm}{!}{\includegraphics[]{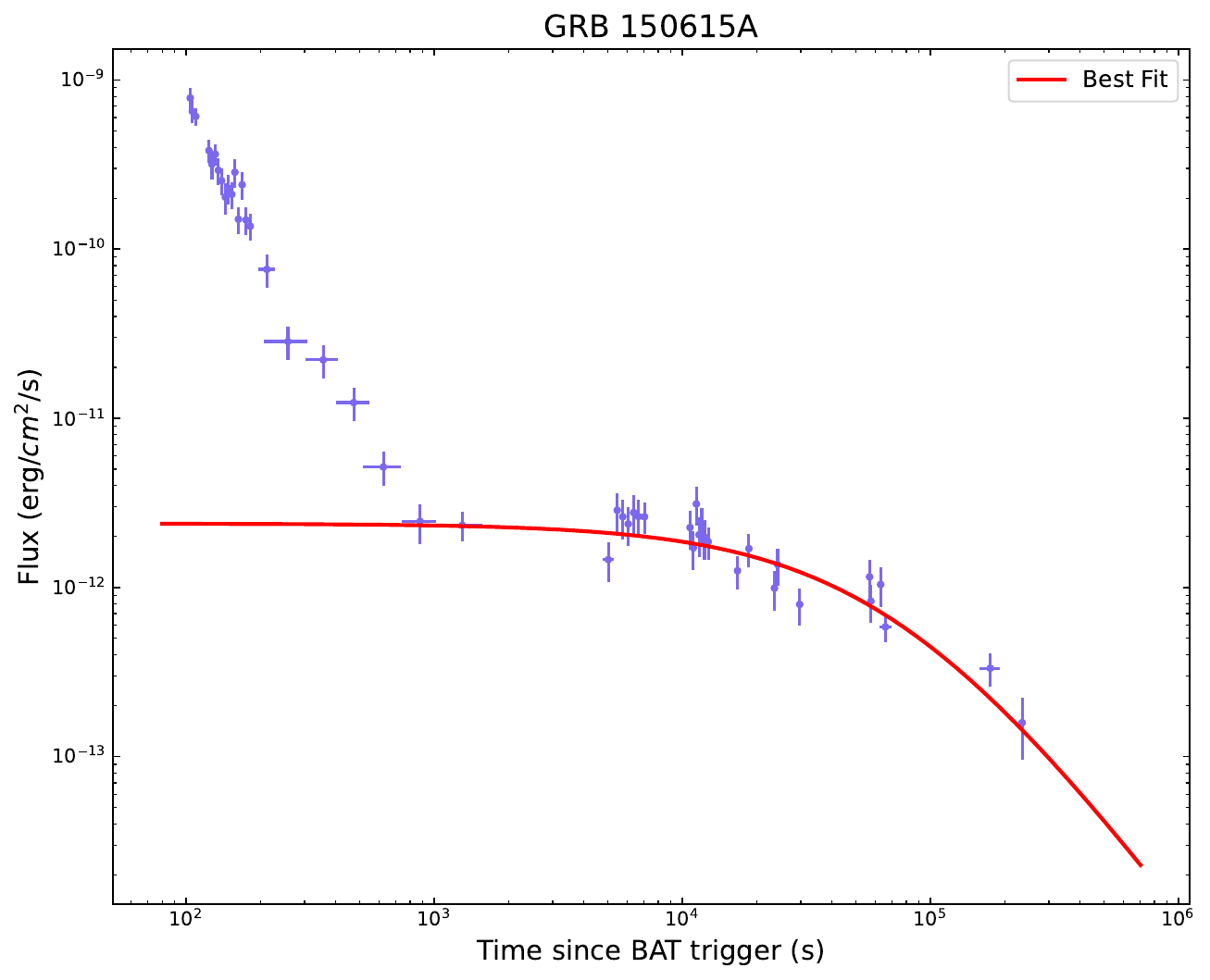}}%
\resizebox{55mm}{!}{\includegraphics[]{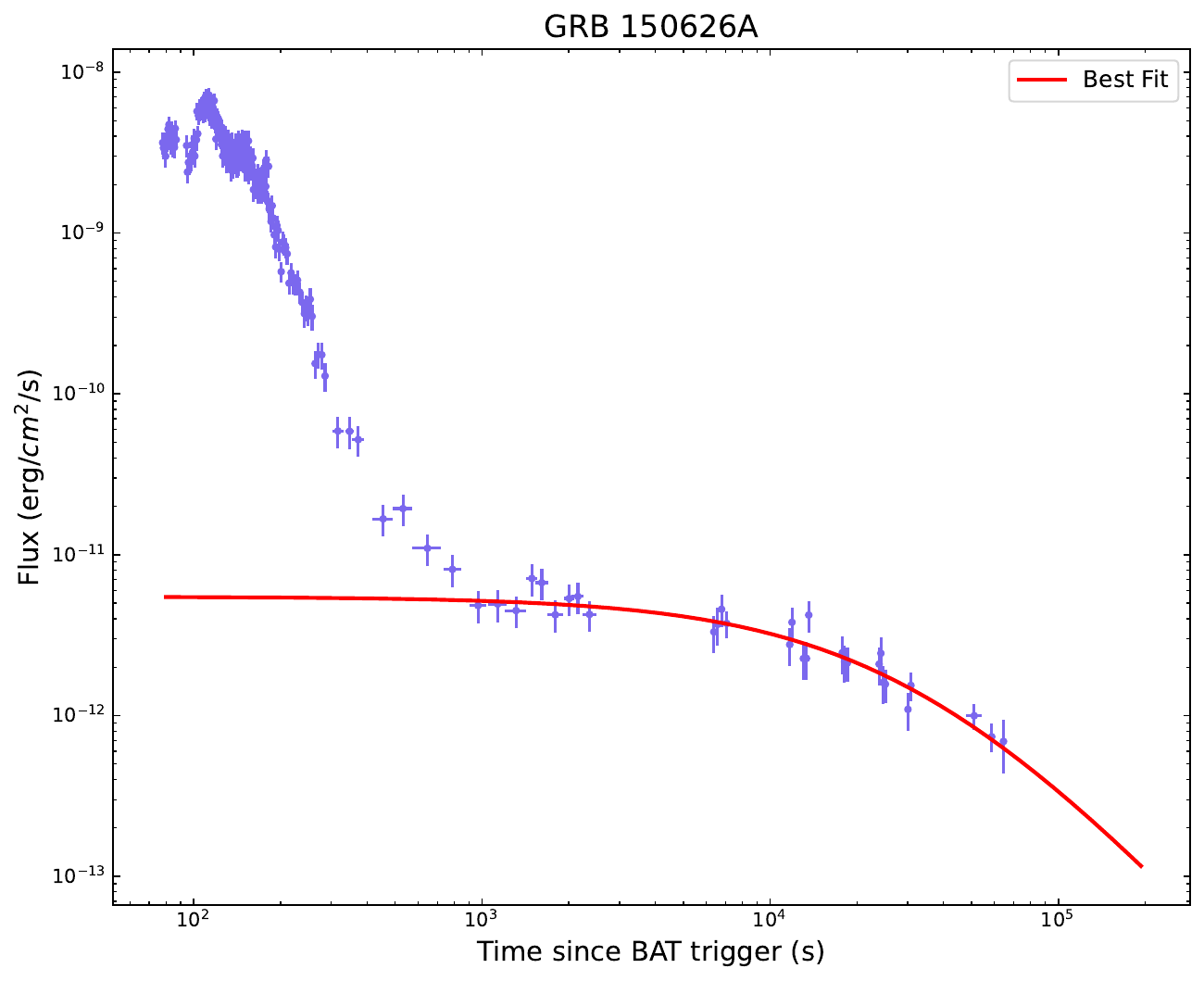}}%

\noindent
\resizebox{55mm}{!}{\includegraphics[]{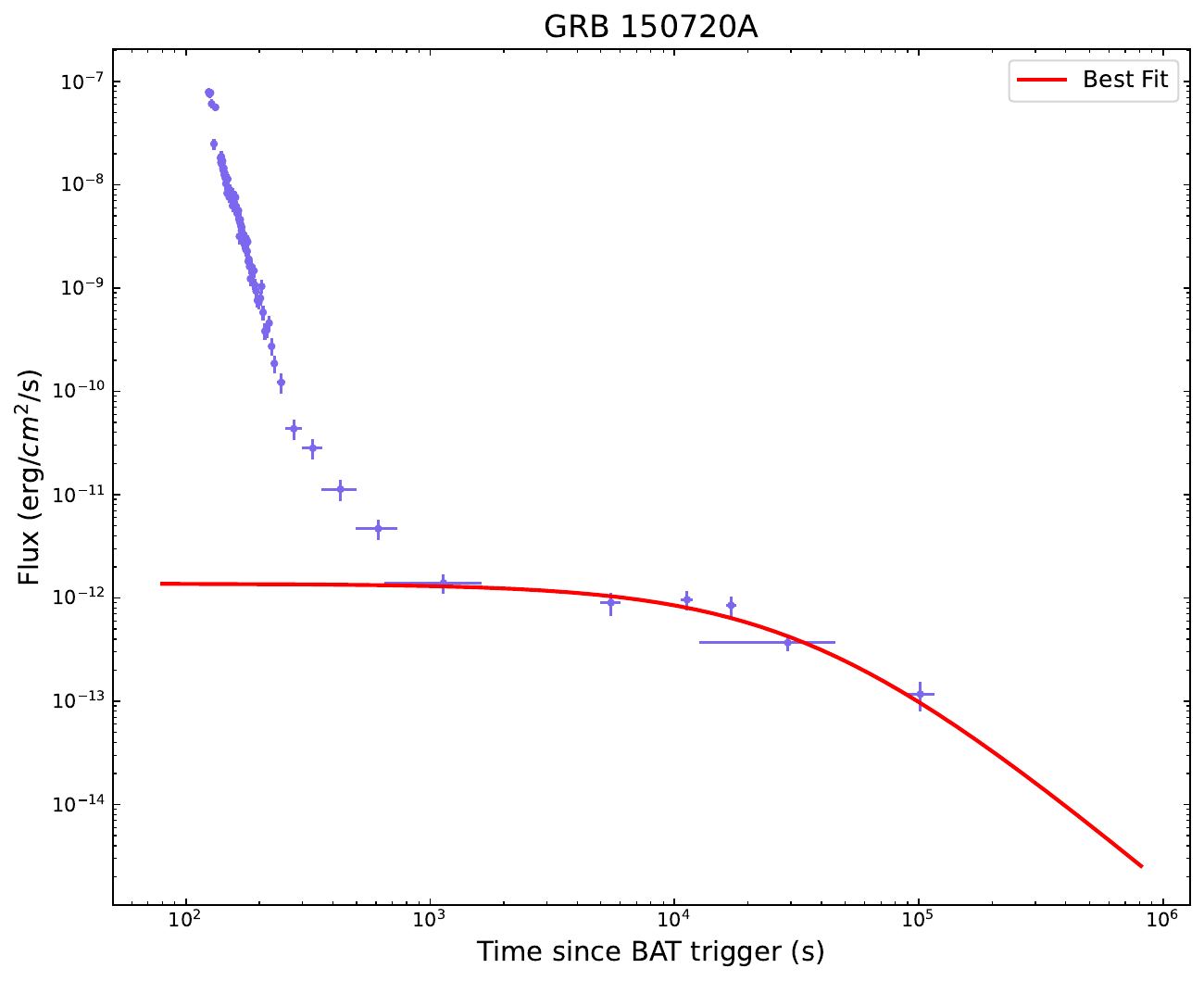}}%
\resizebox{55mm}{!}{\includegraphics[]{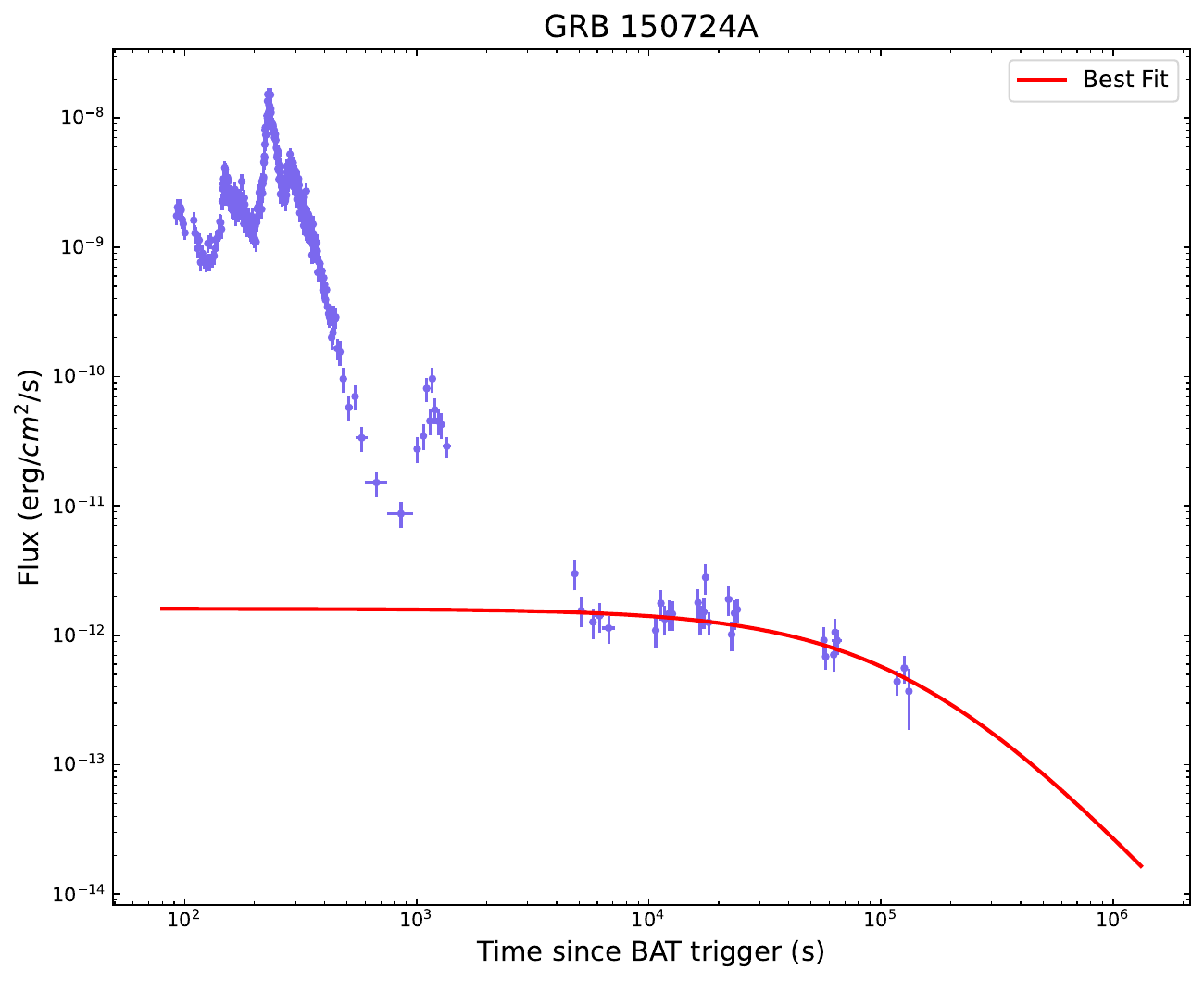}}%
\resizebox{55mm}{!}{\includegraphics[]{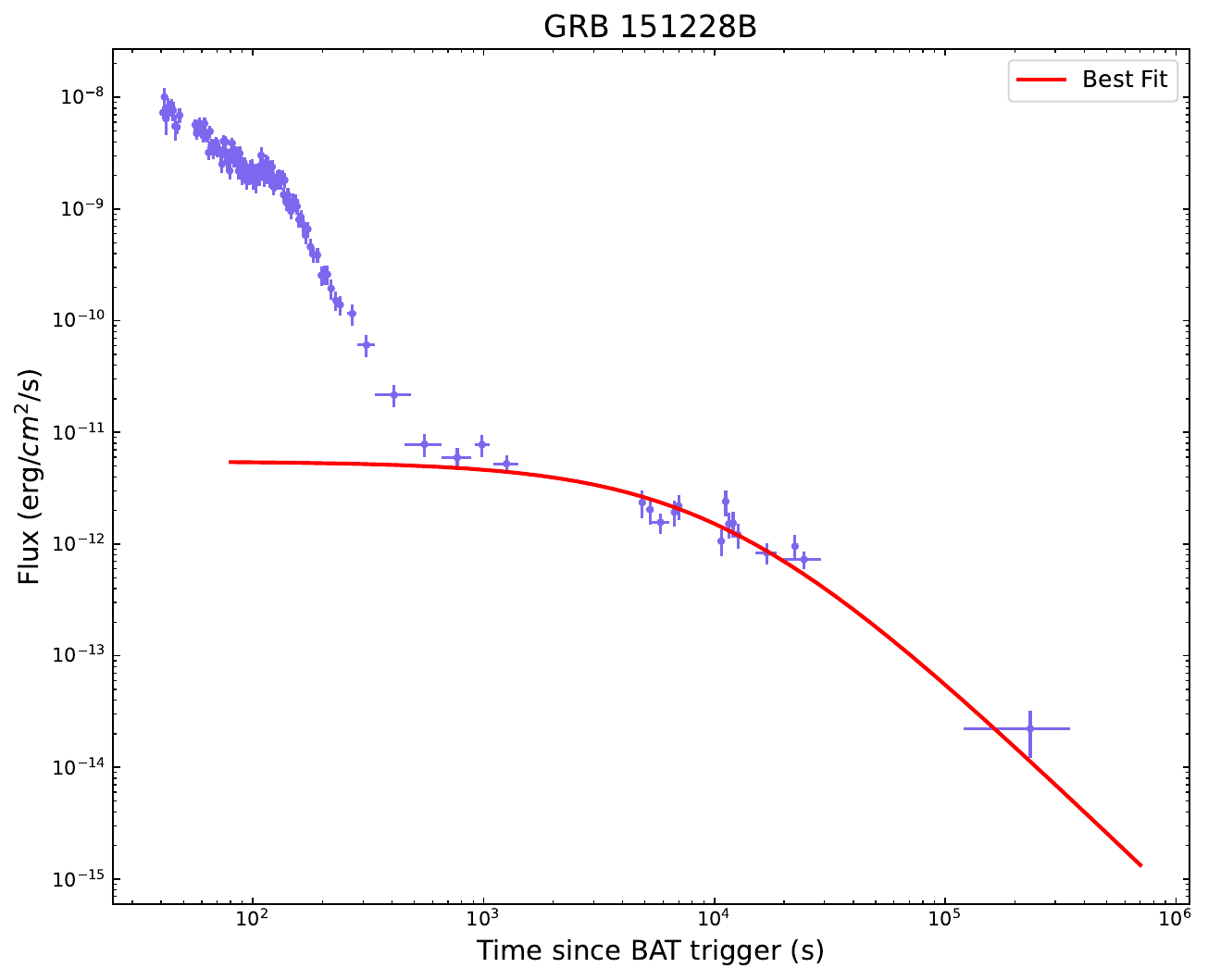}}%

\noindent
\resizebox{55mm}{!}{\includegraphics[]{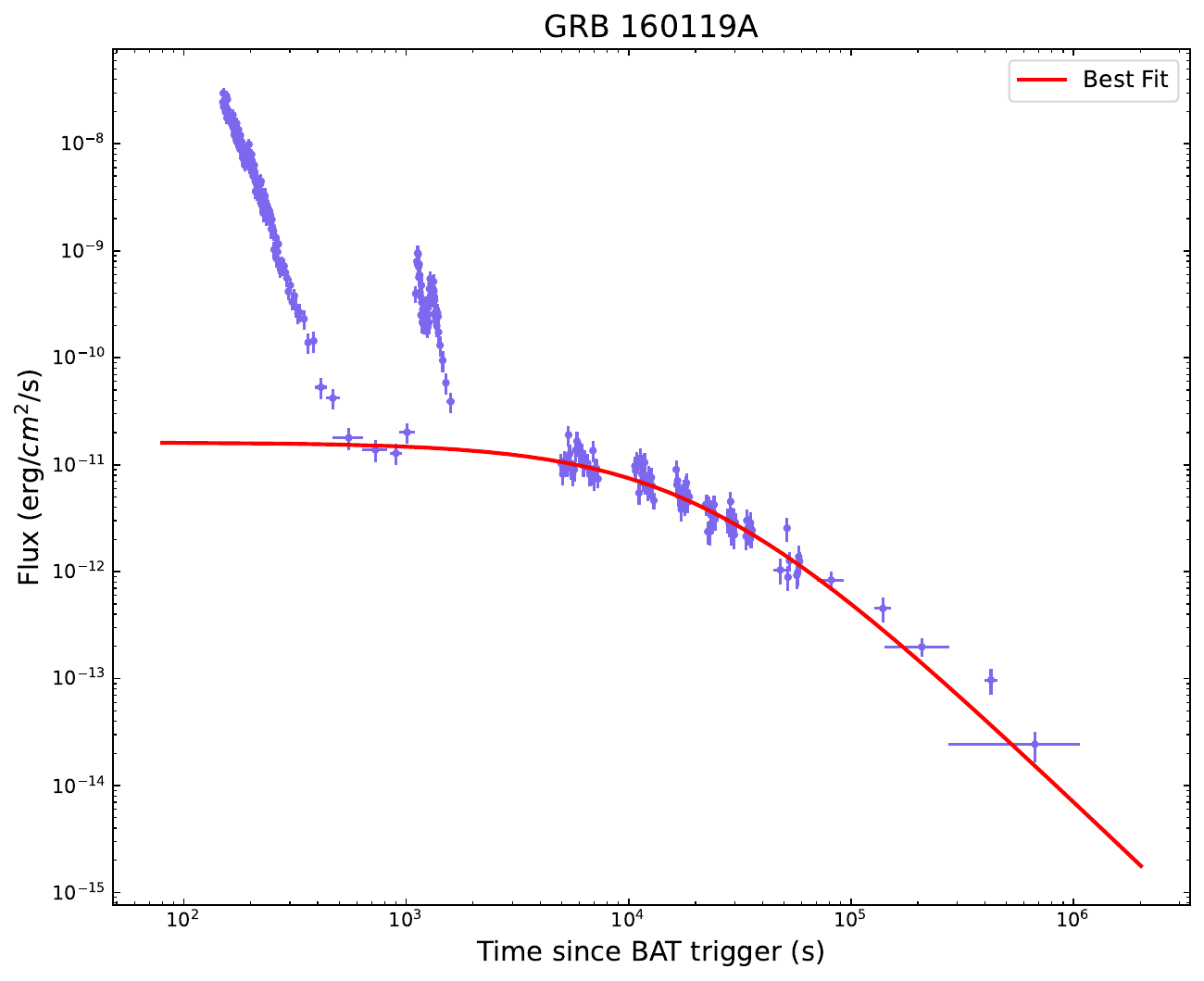}}%
\resizebox{55mm}{!}{\includegraphics[]{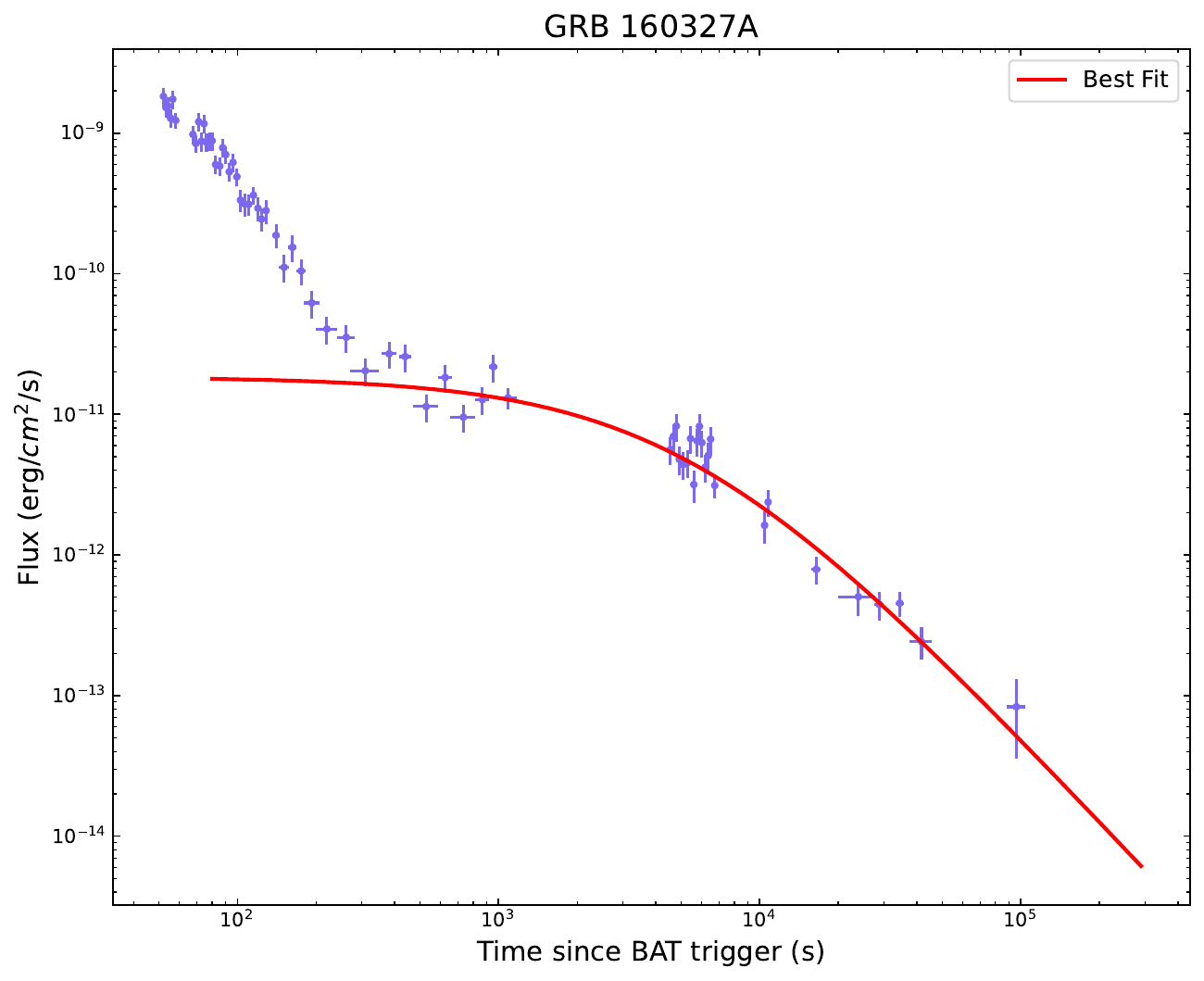}}%
\resizebox{55mm}{!}{\includegraphics[]{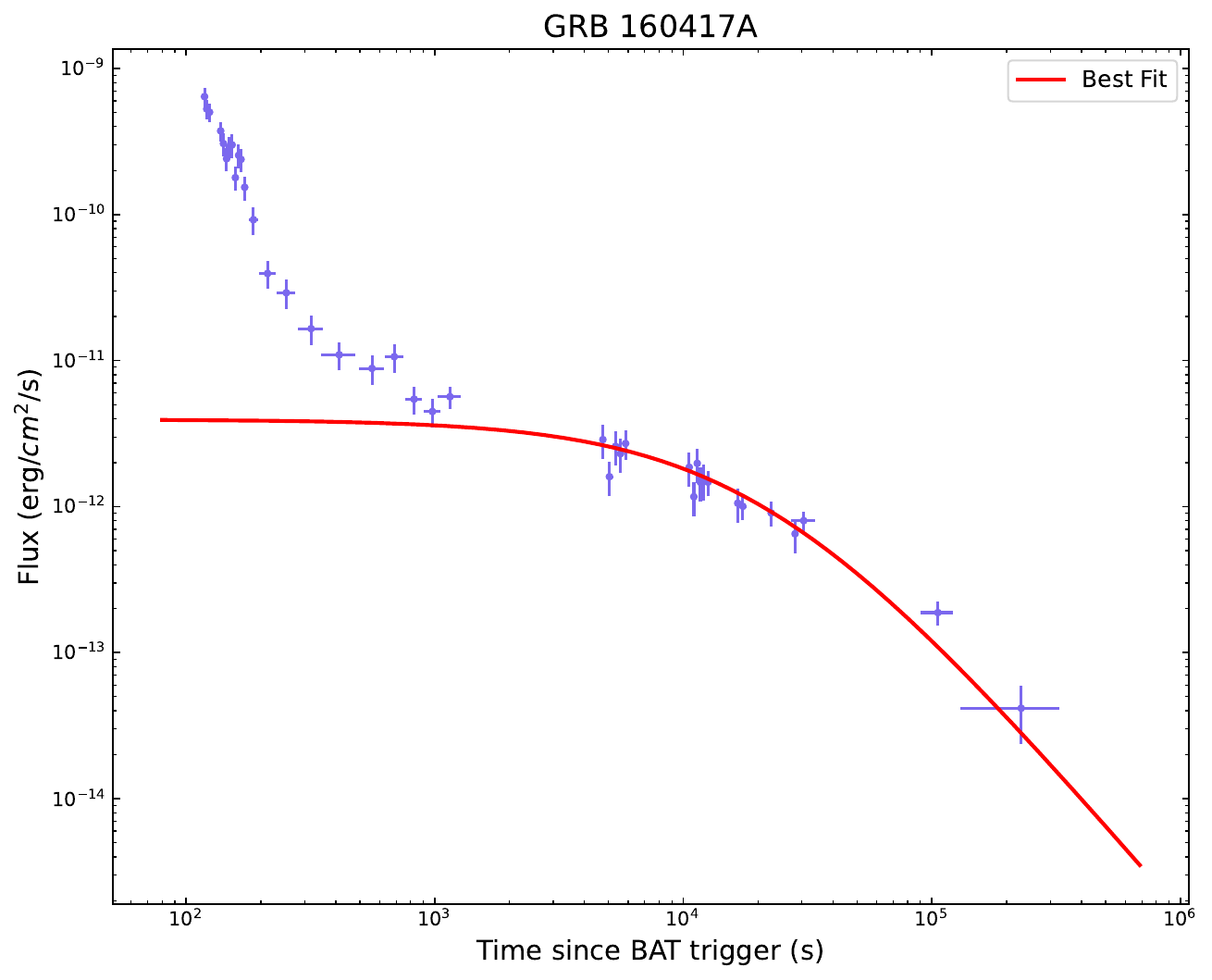}}%
\caption{(Continued)}
\end{figure*}

\addtocounter{figure}{-1}
\begin{figure*}[ht!]

\noindent
\resizebox{55mm}{!}{\includegraphics[]{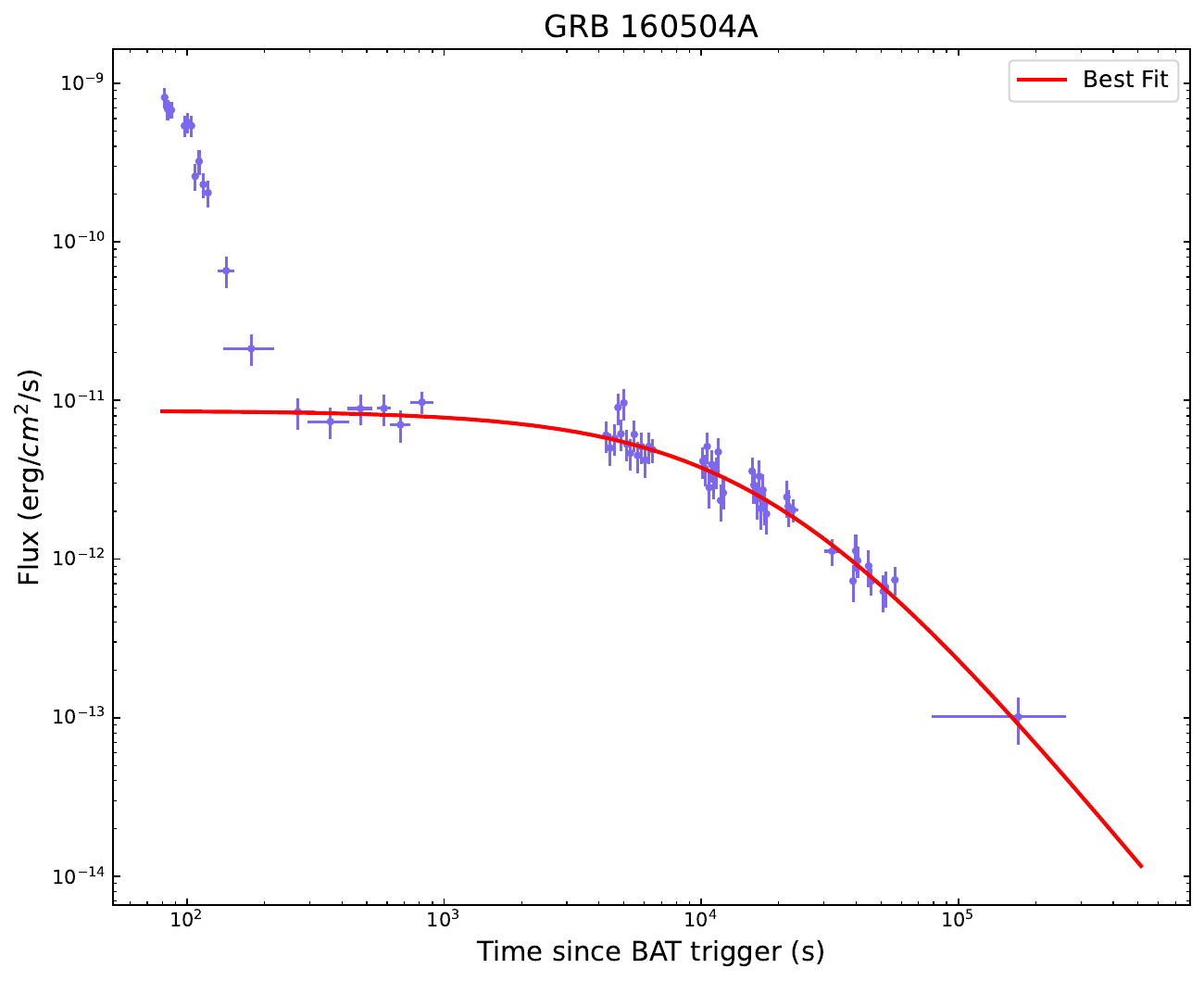}}%
\resizebox{55mm}{!}{\includegraphics[]{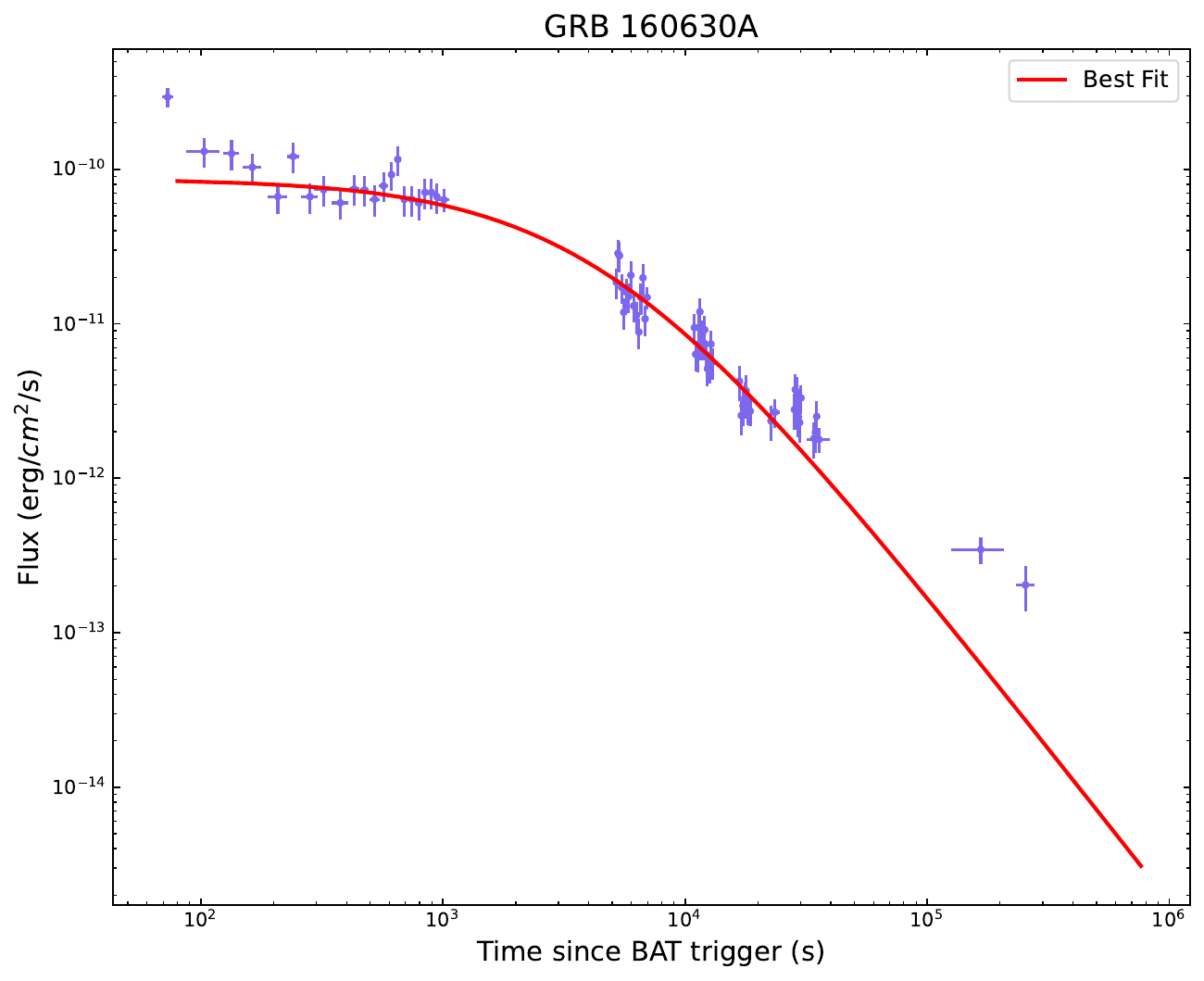}}%
\resizebox{55mm}{!}{\includegraphics[]{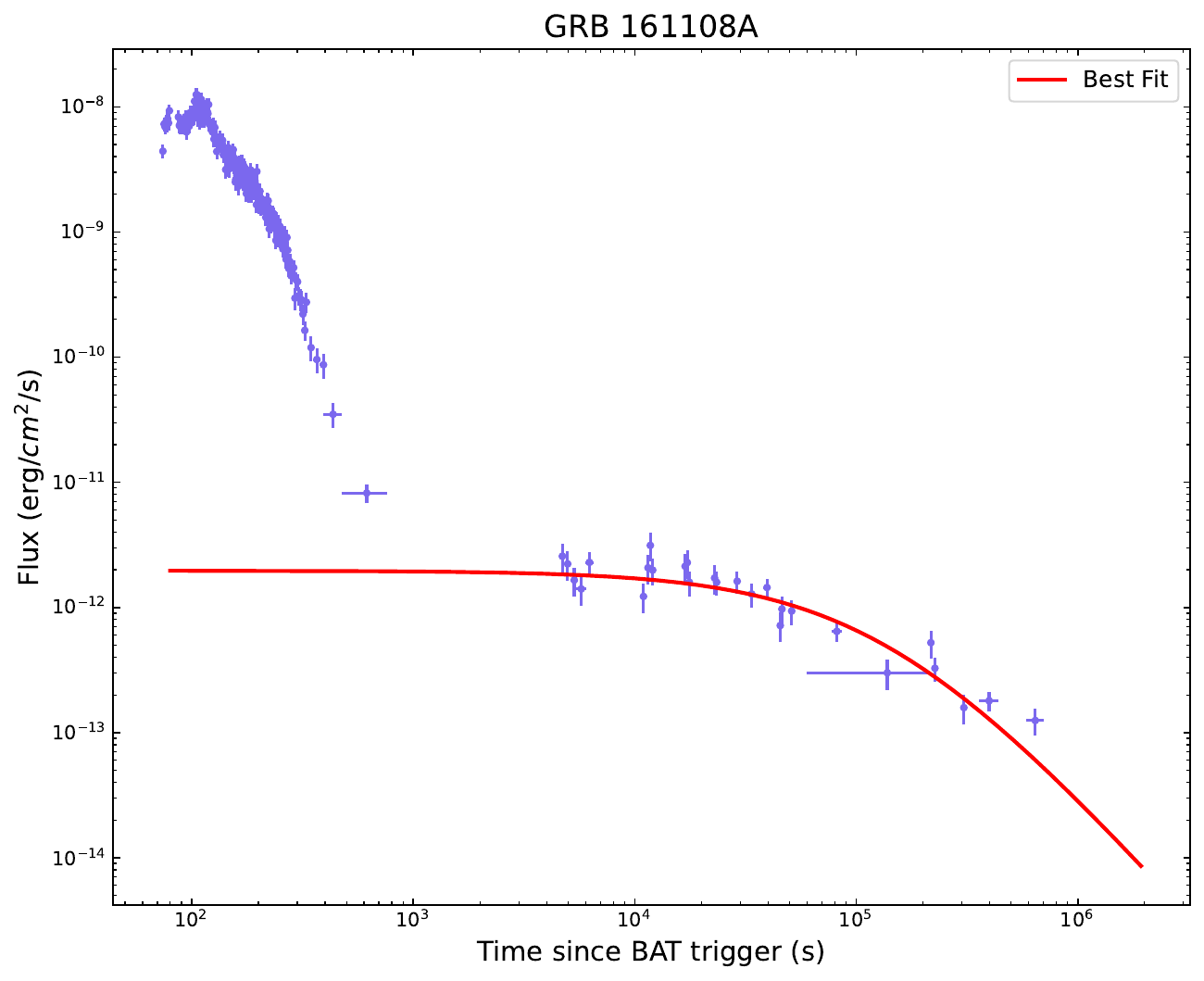}}%

\noindent
\resizebox{55mm}{!}{\includegraphics[]{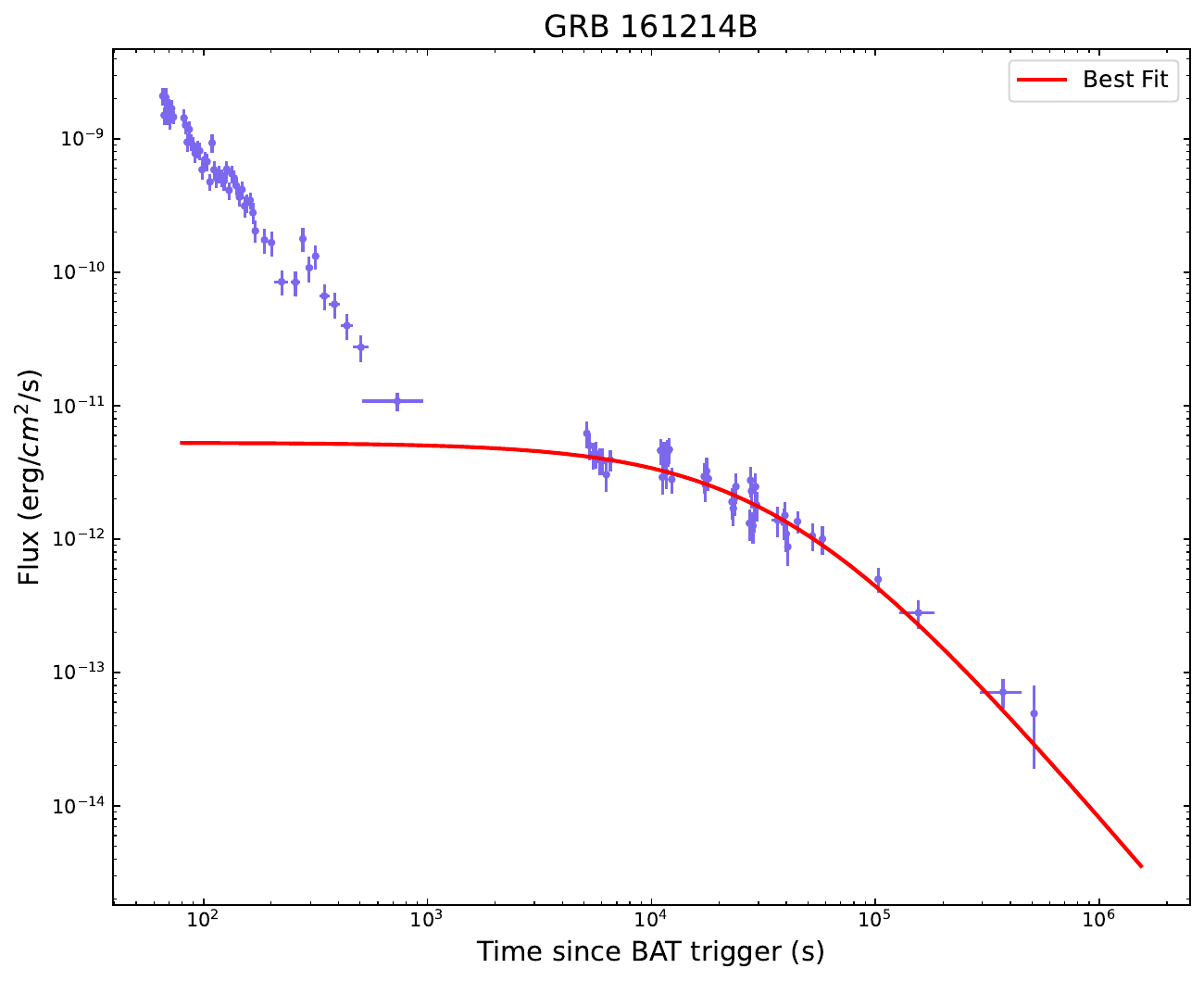}}%
\resizebox{55mm}{!}{\includegraphics[]{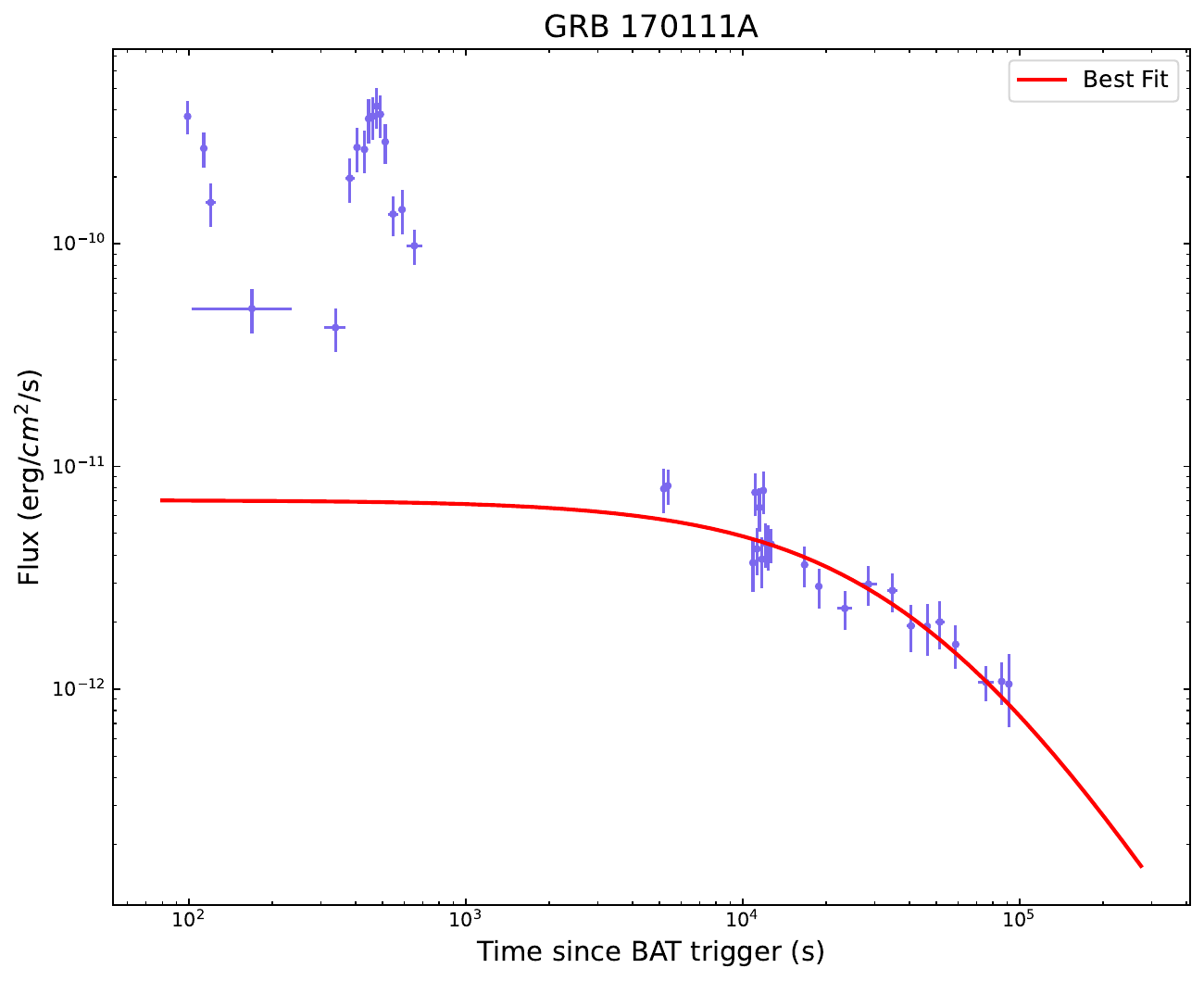}}%
\resizebox{55mm}{!}{\includegraphics[]{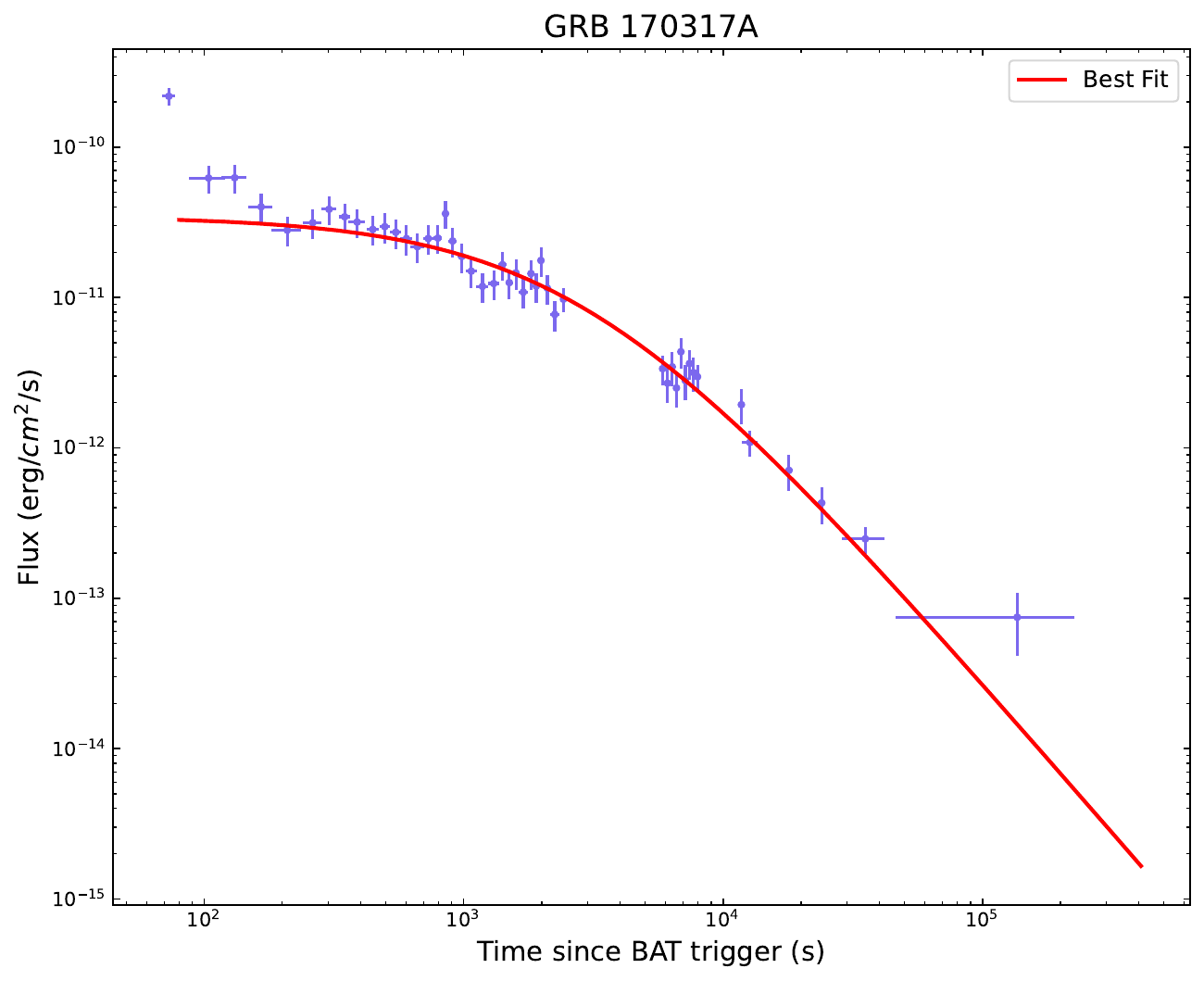}}%

\noindent
\resizebox{55mm}{!}{\includegraphics[]{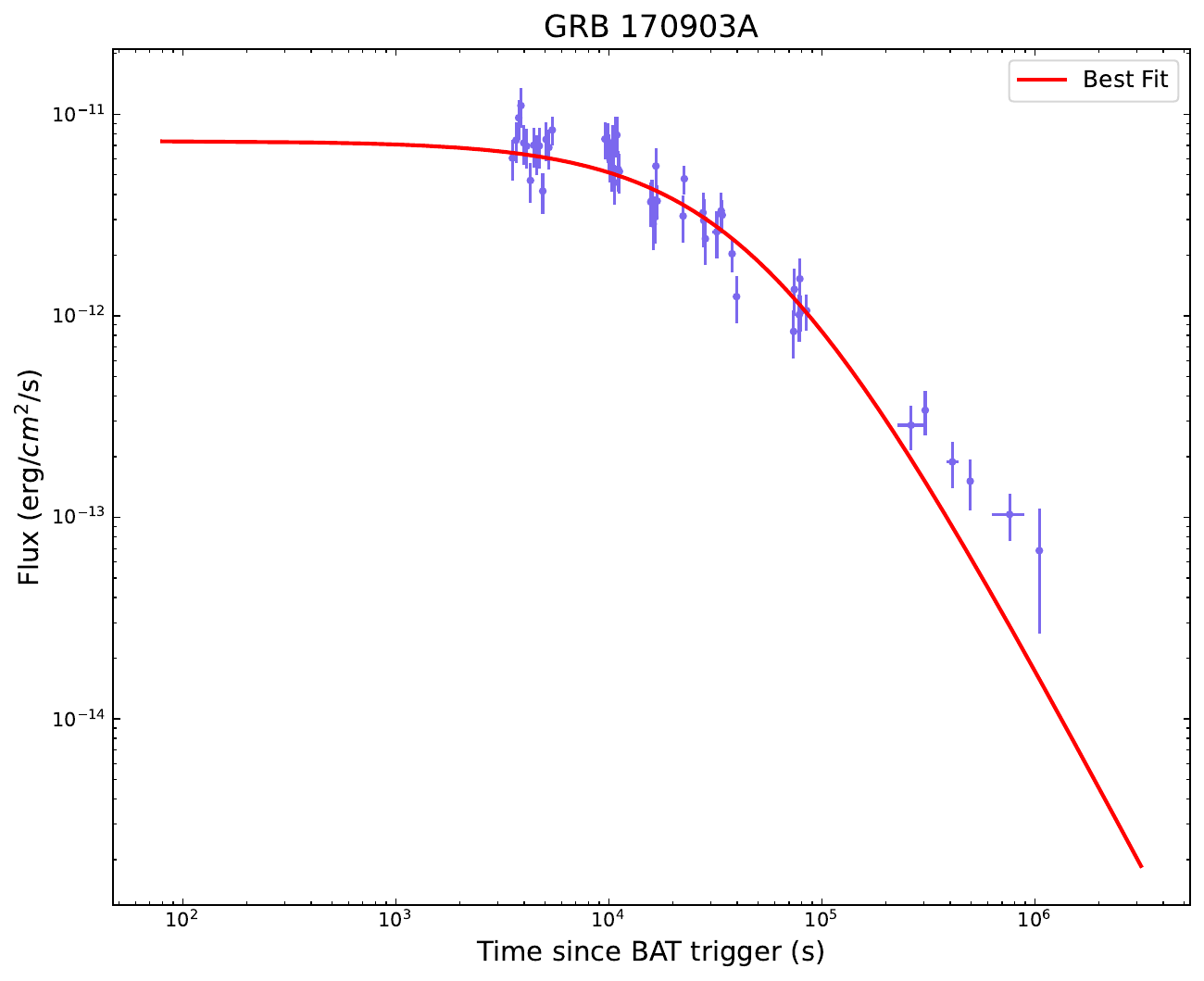}}%
\resizebox{55mm}{!}{\includegraphics[]{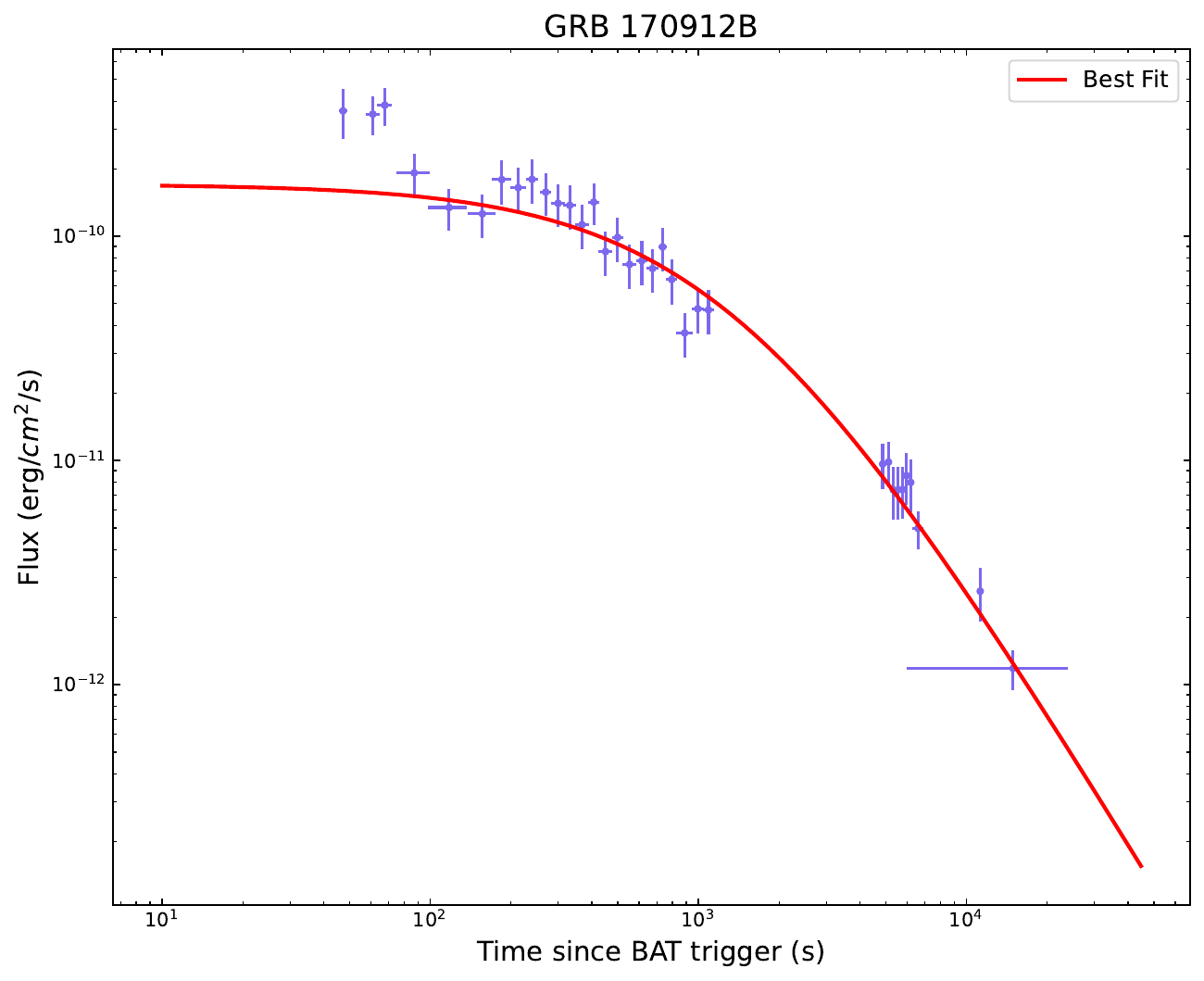}}%
\resizebox{55mm}{!}{\includegraphics[]{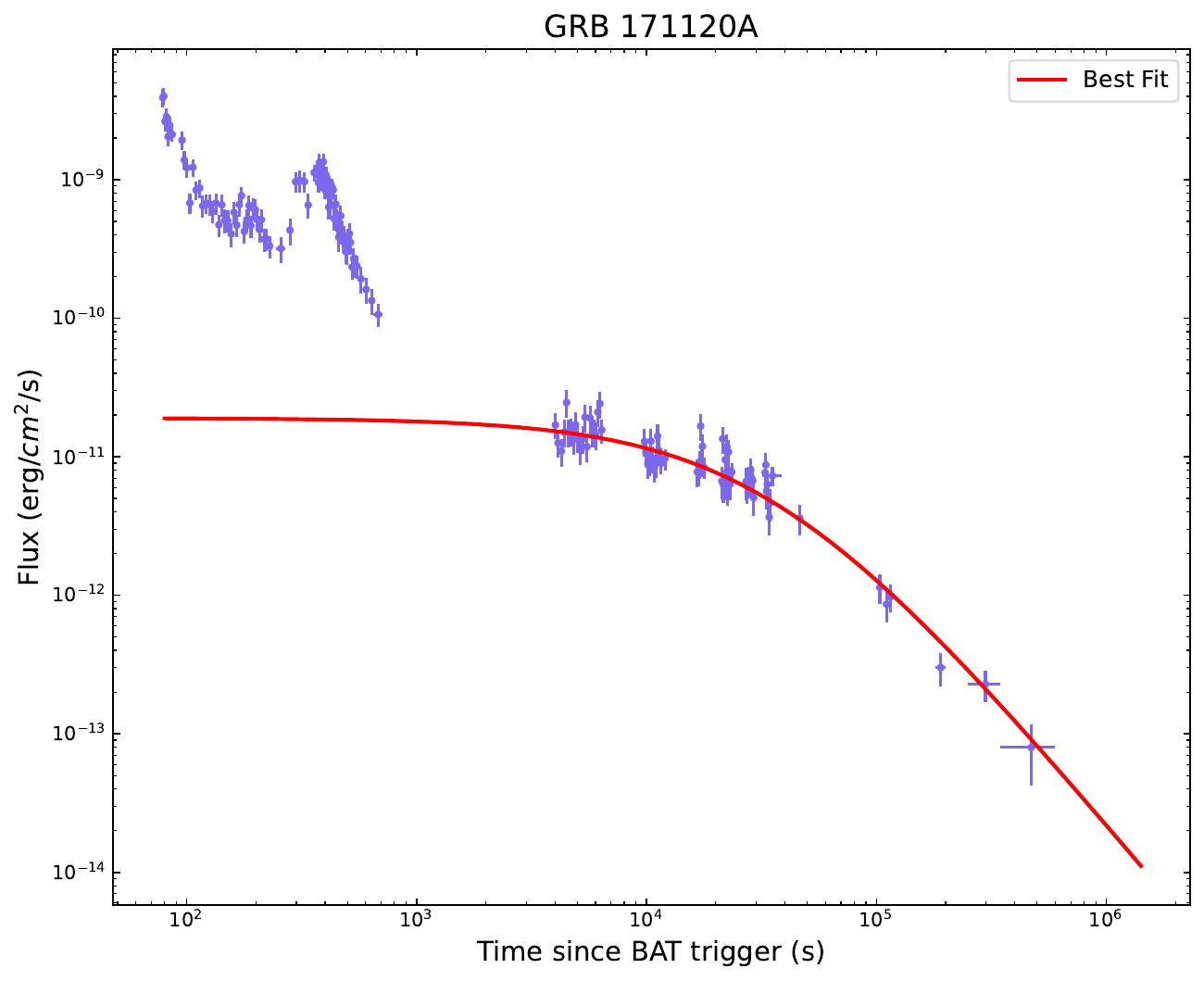}}%

\noindent
\resizebox{55mm}{!}{\includegraphics[]{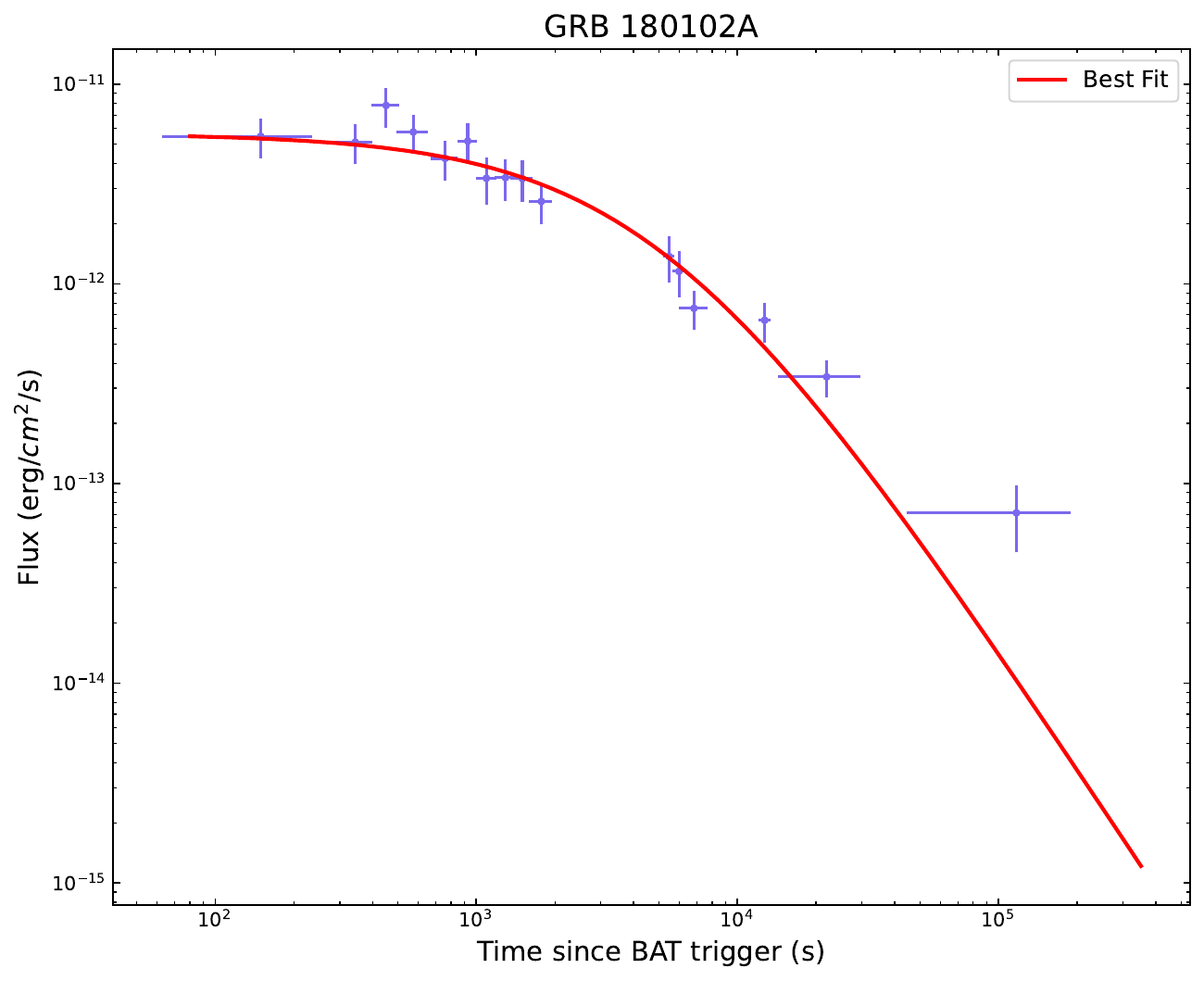}}%
\resizebox{55mm}{!}{\includegraphics[]{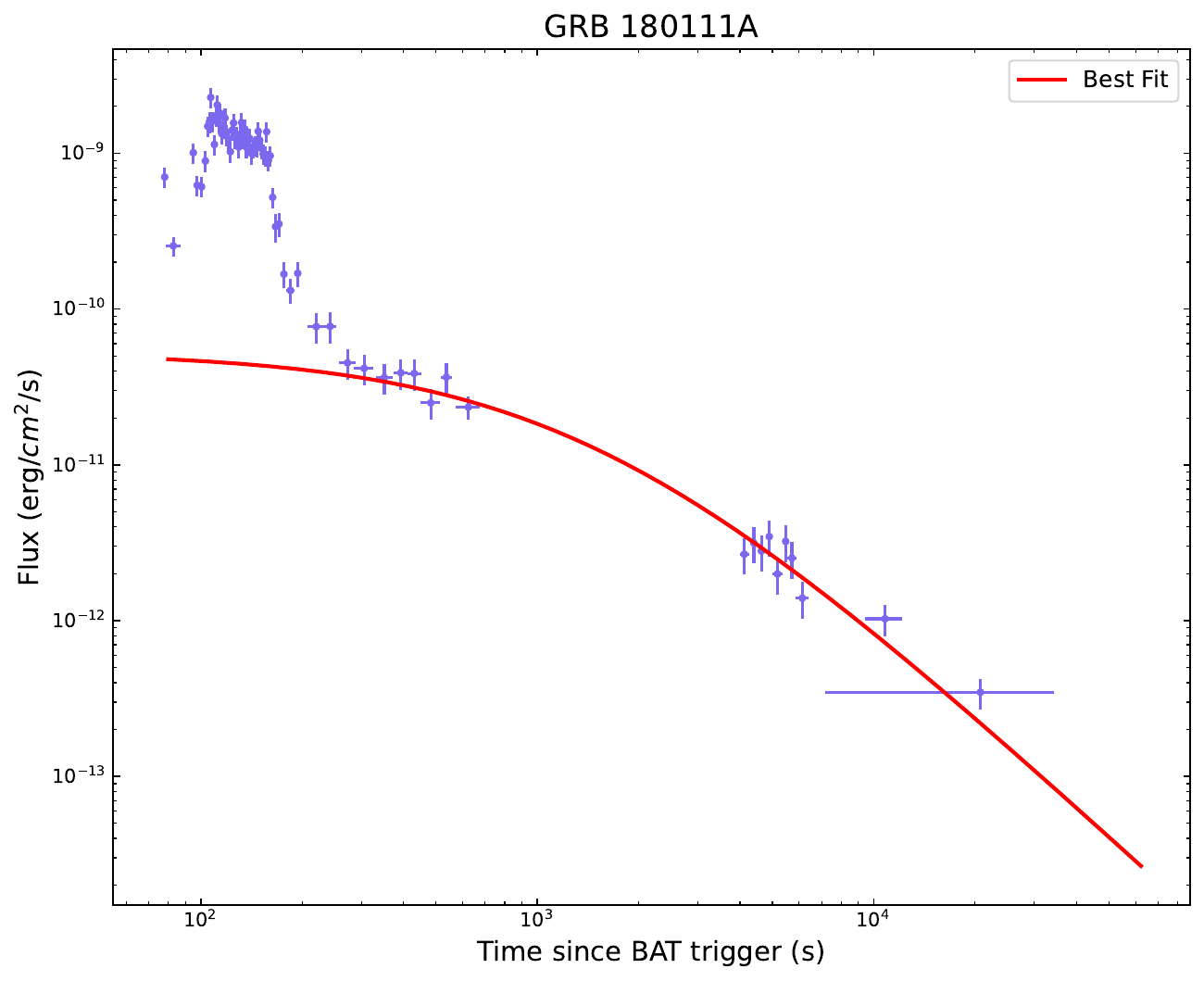}}%
\resizebox{55mm}{!}{\includegraphics[]{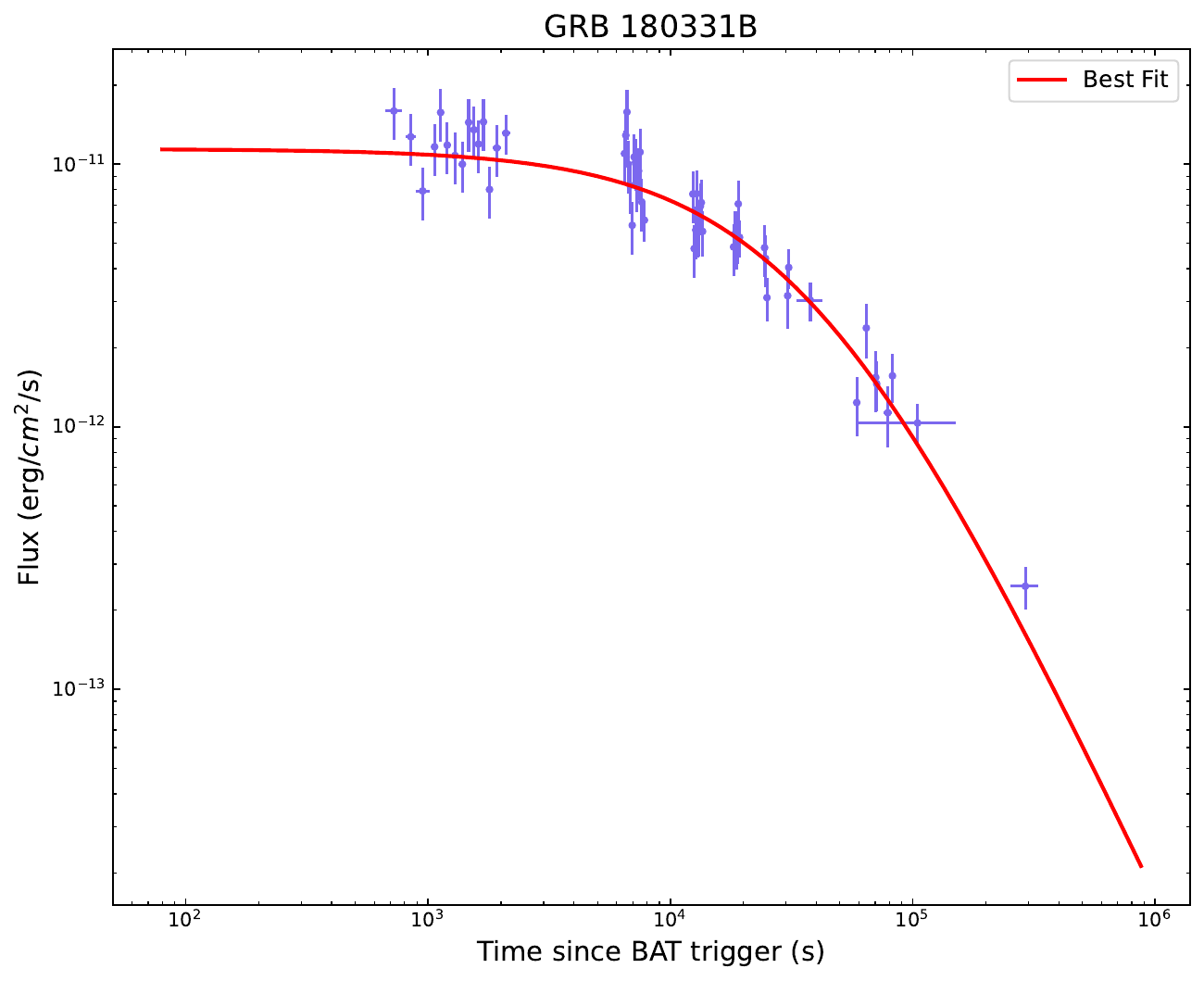}}%

\noindent
\resizebox{55mm}{!}{\includegraphics[]{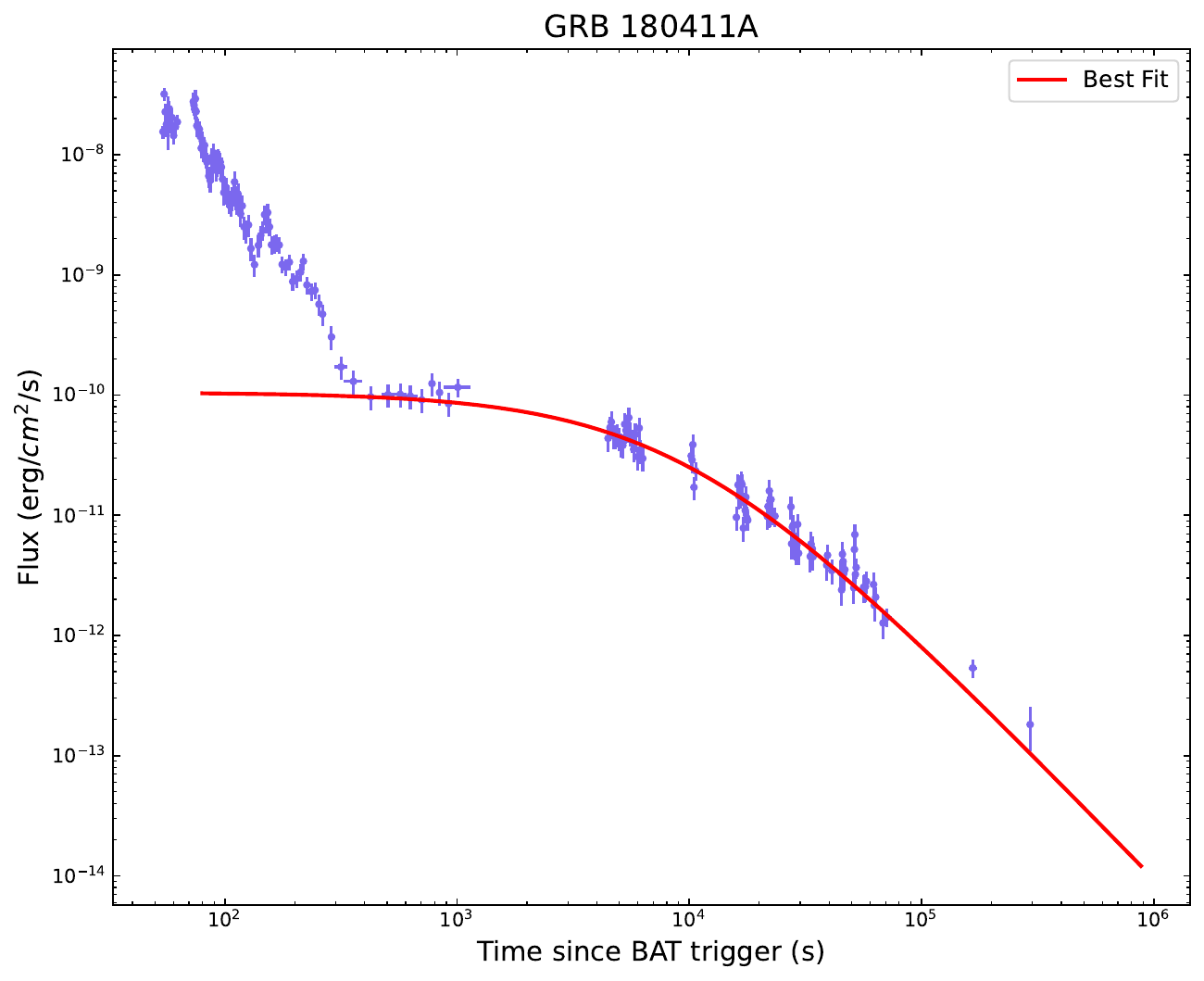}}%
\resizebox{55mm}{!}{\includegraphics[]{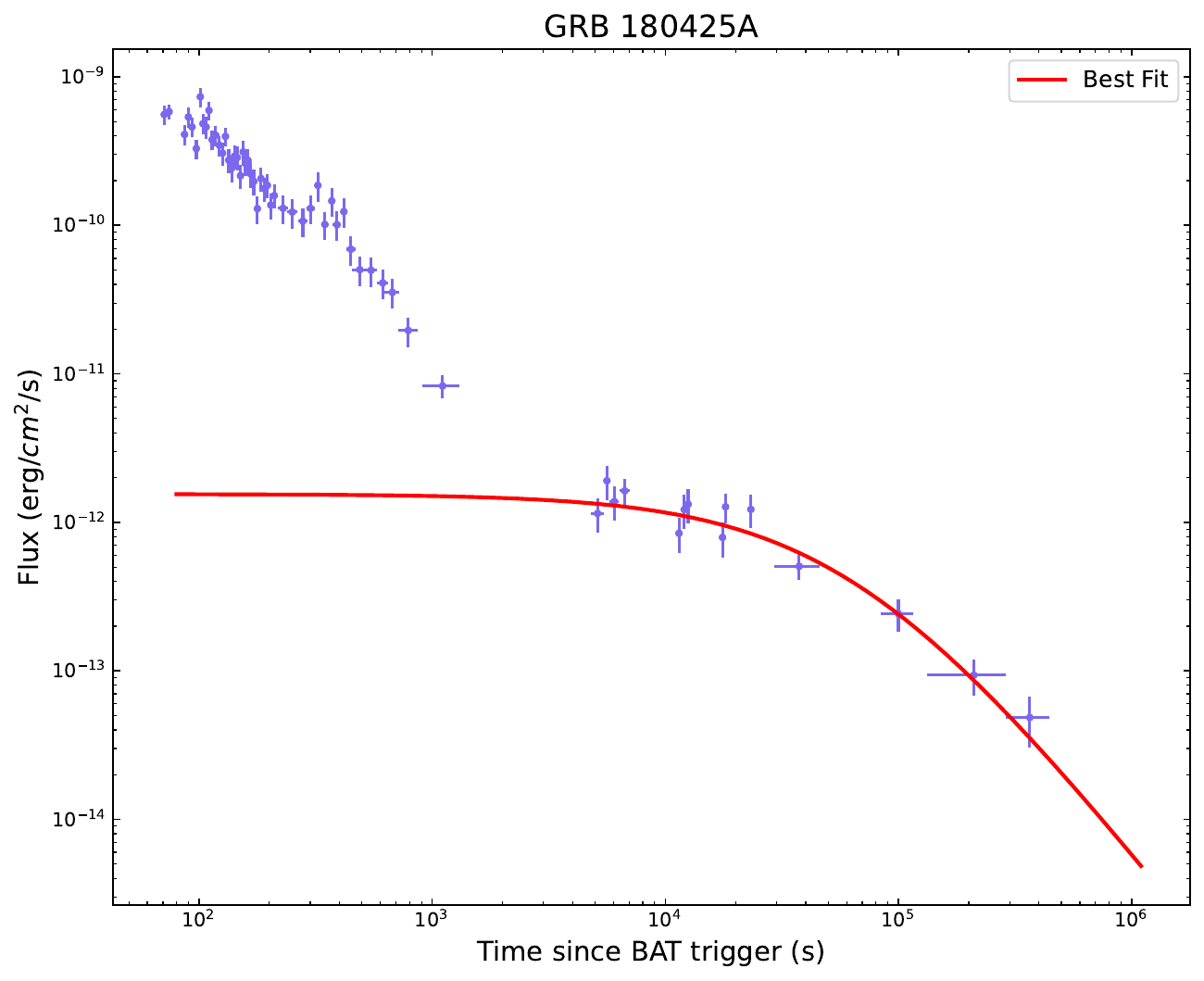}}%
\resizebox{55mm}{!}{\includegraphics[]{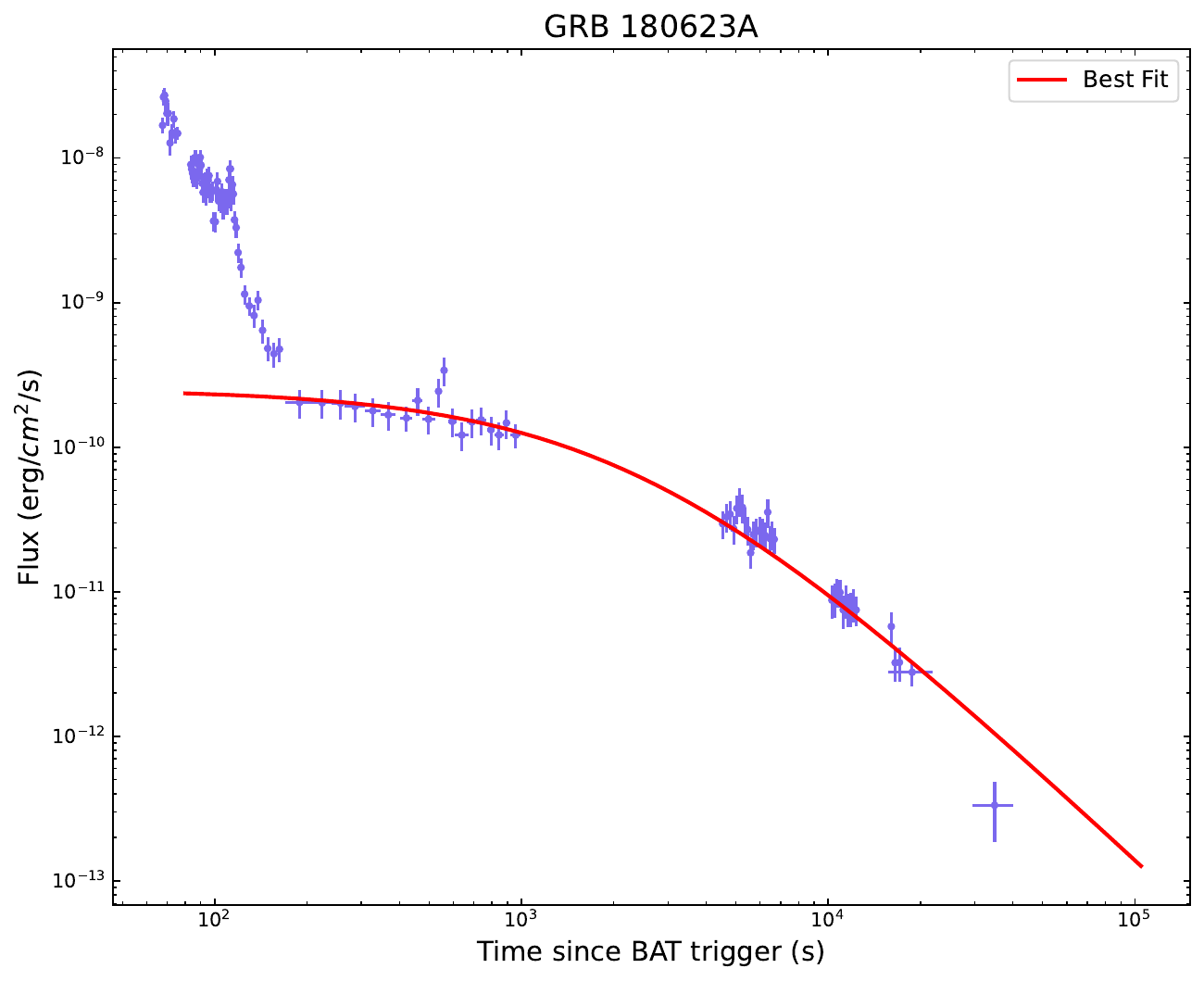}}%
\caption{(Continued)}
\end{figure*}

\addtocounter{figure}{-1}
\begin{figure*}[ht!]

\noindent
\resizebox{55mm}{!}{\includegraphics[]{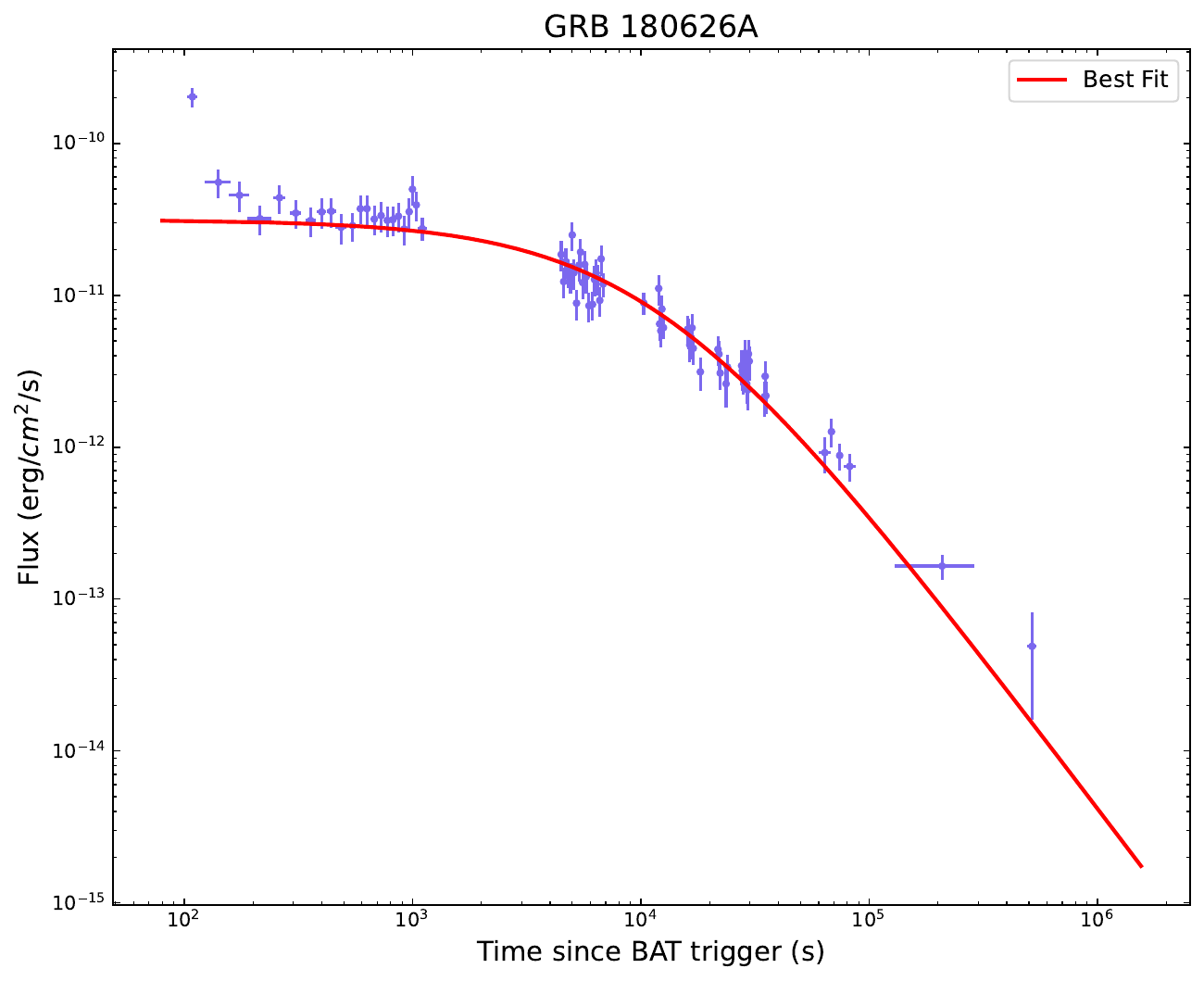}}%
\resizebox{55mm}{!}{\includegraphics[]{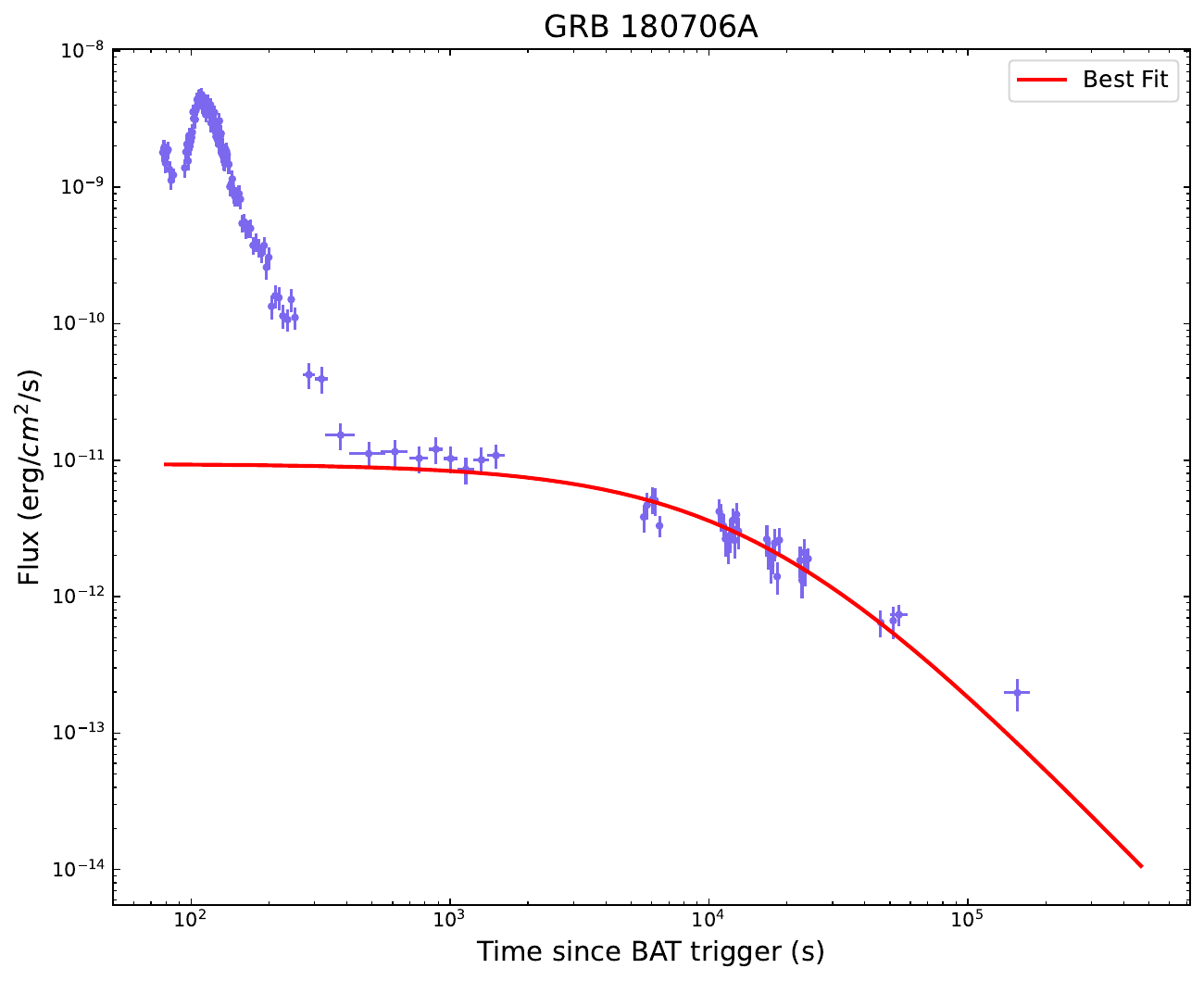}}%
\resizebox{55mm}{!}{\includegraphics[]{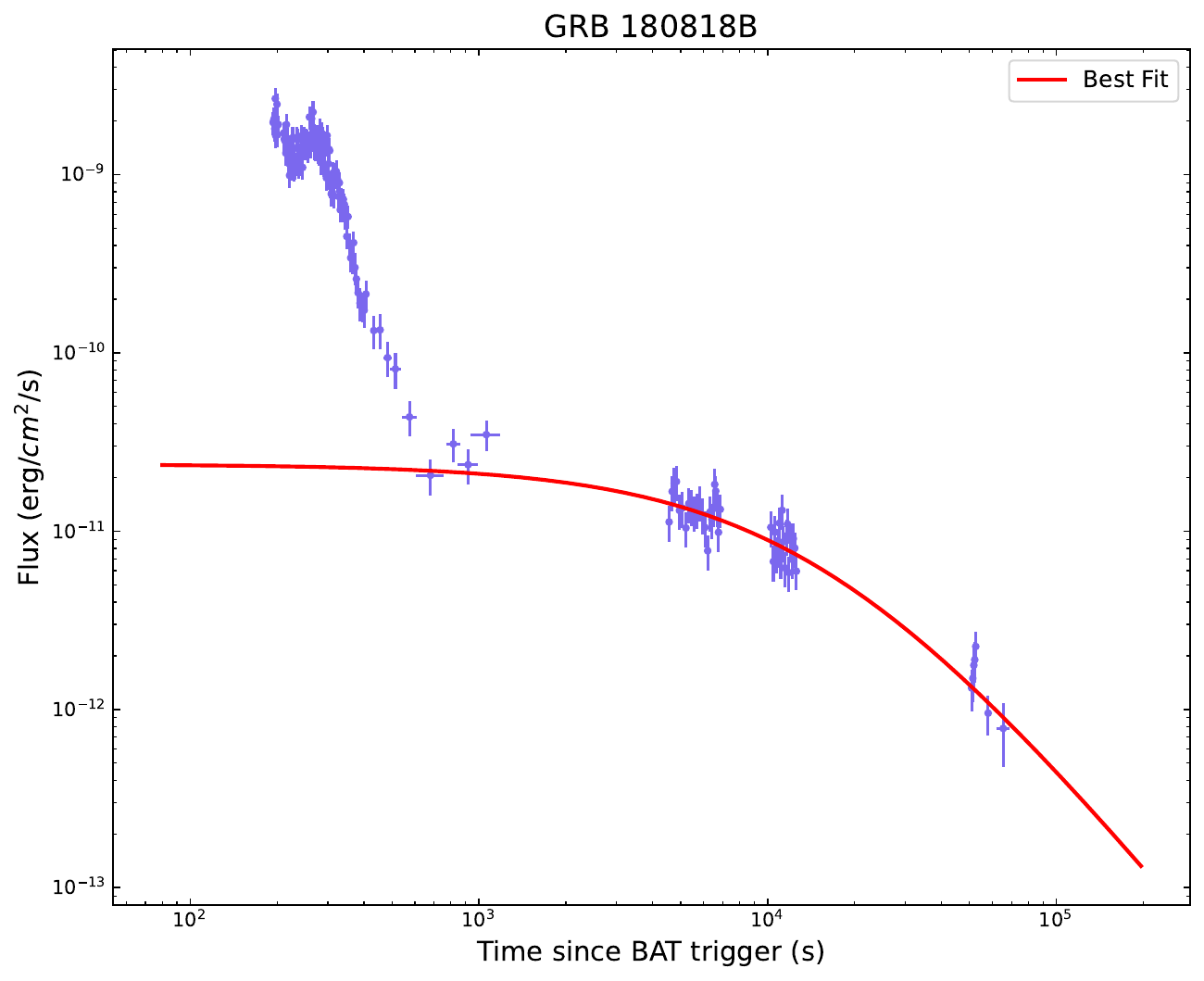}}%

\noindent
\resizebox{55mm}{!}{\includegraphics[]{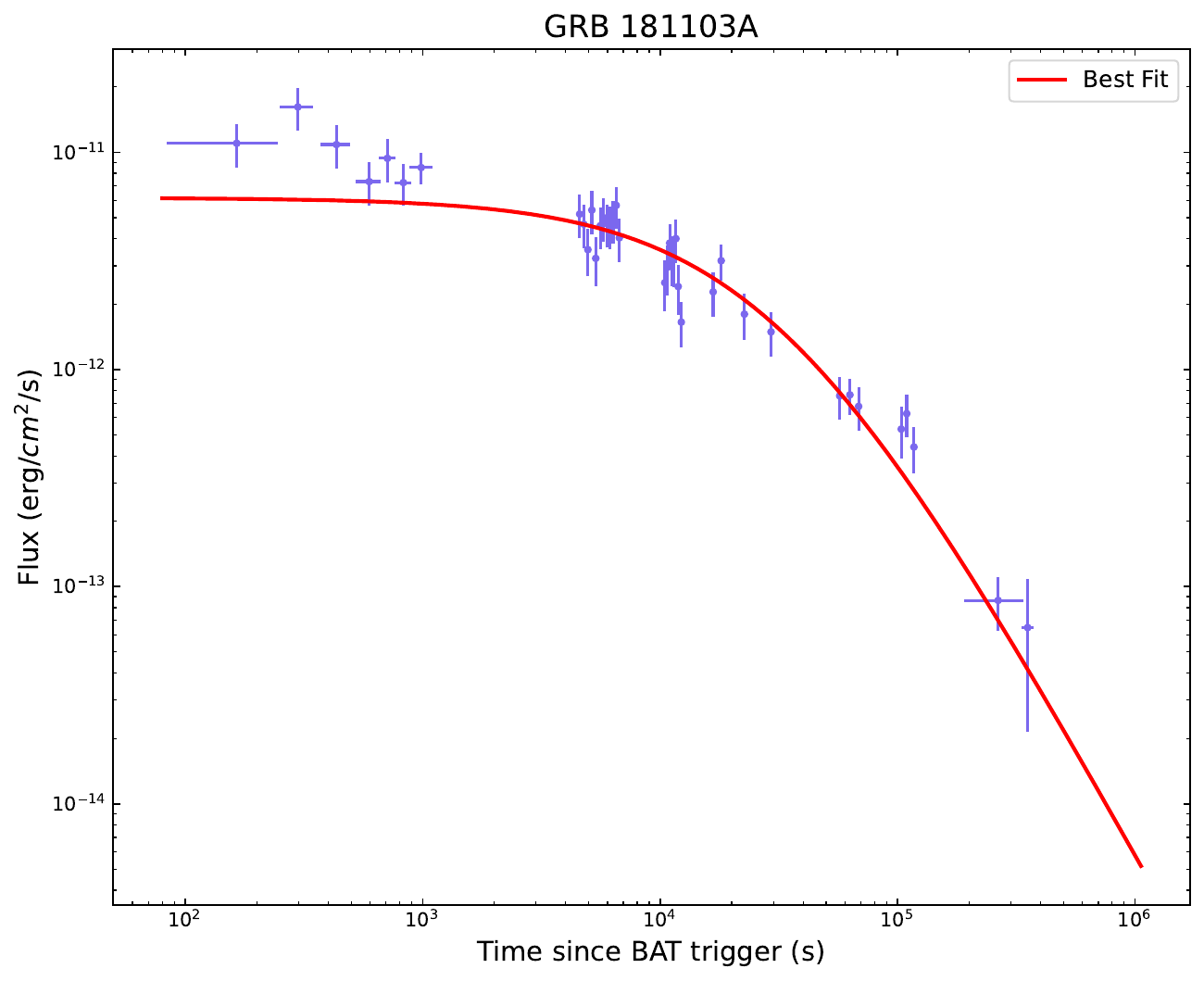}}%
\resizebox{55mm}{!}{\includegraphics[]{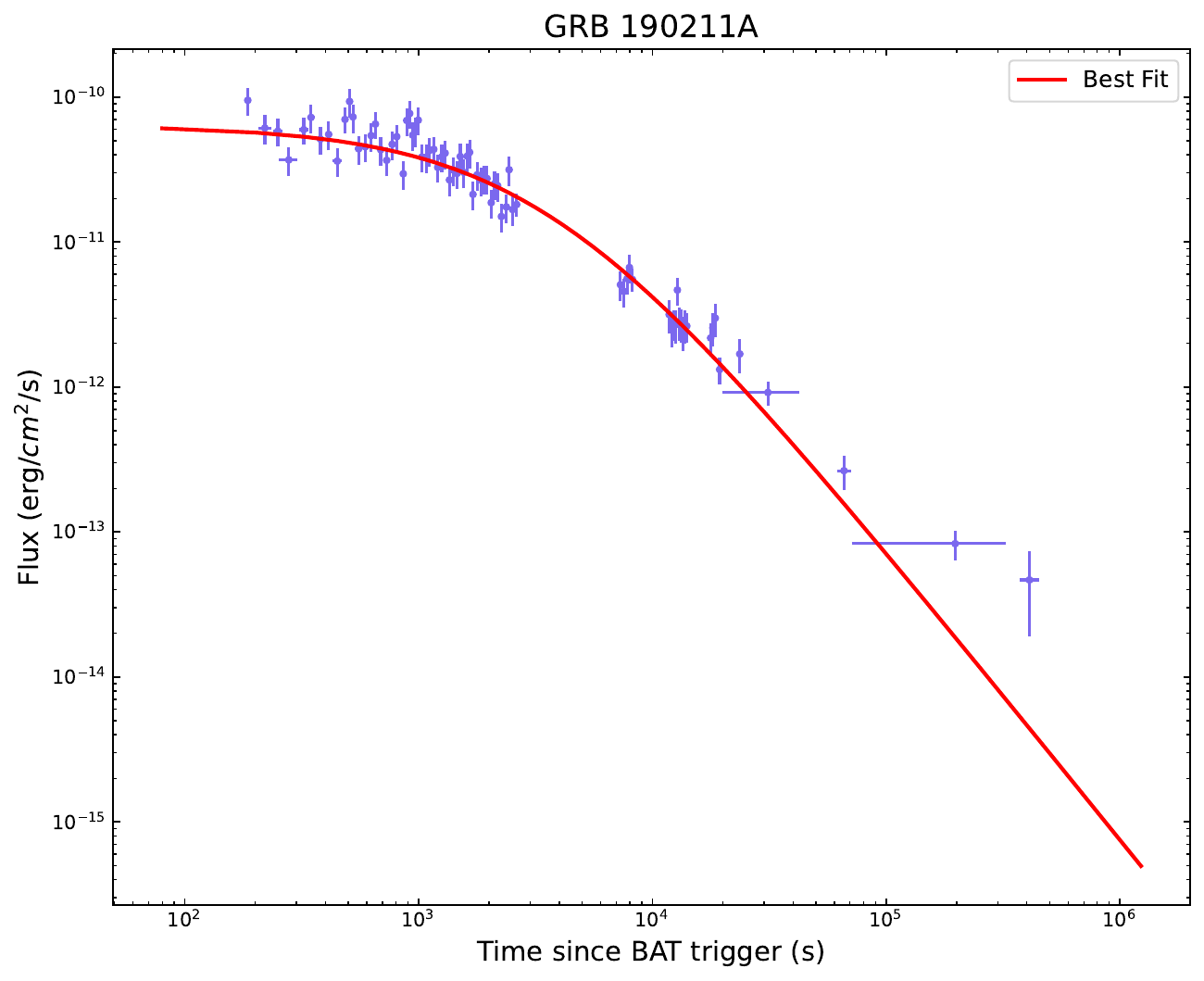}}%
\resizebox{55mm}{!}{\includegraphics[]{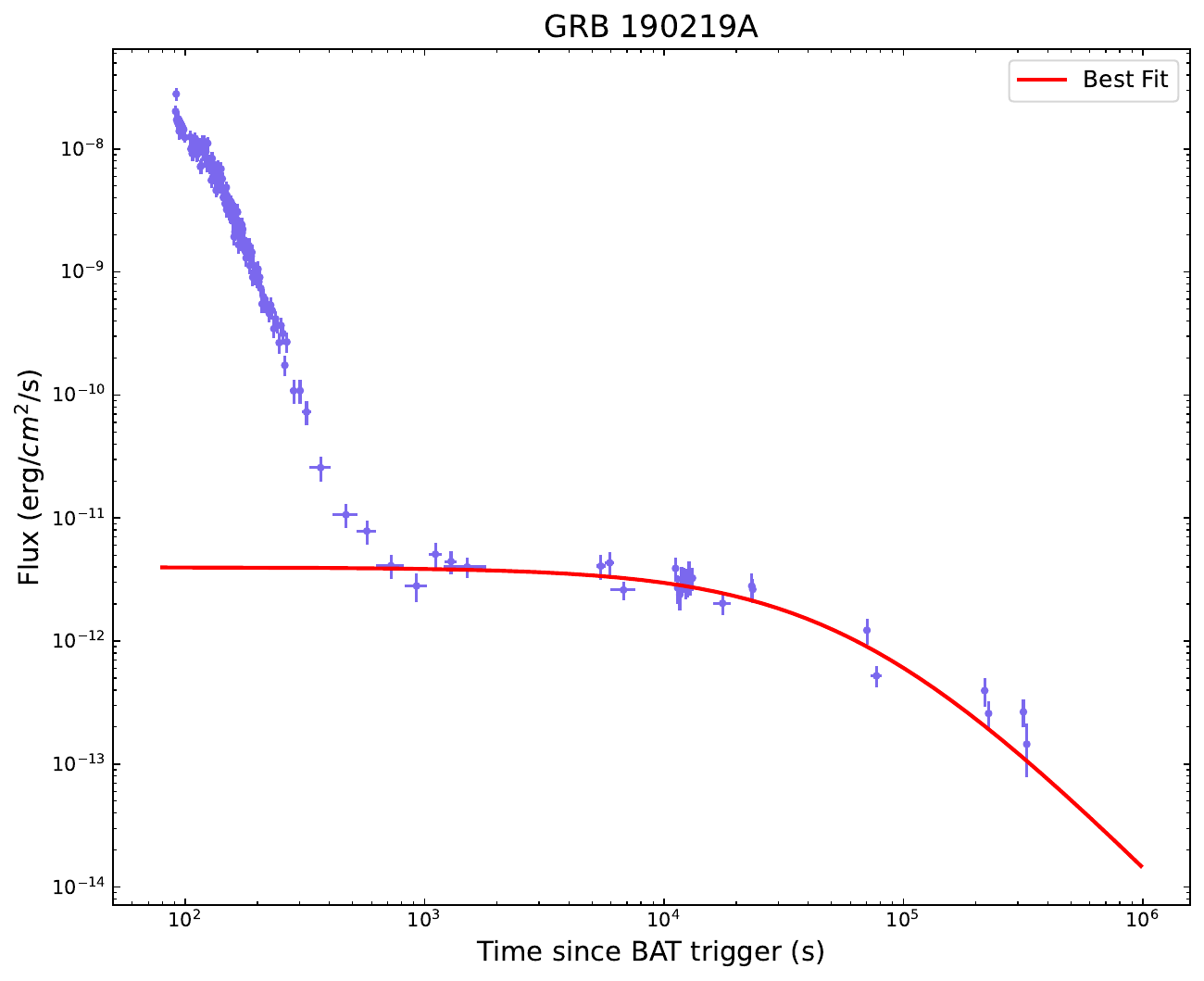}}%

\noindent
\resizebox{55mm}{!}{\includegraphics[]{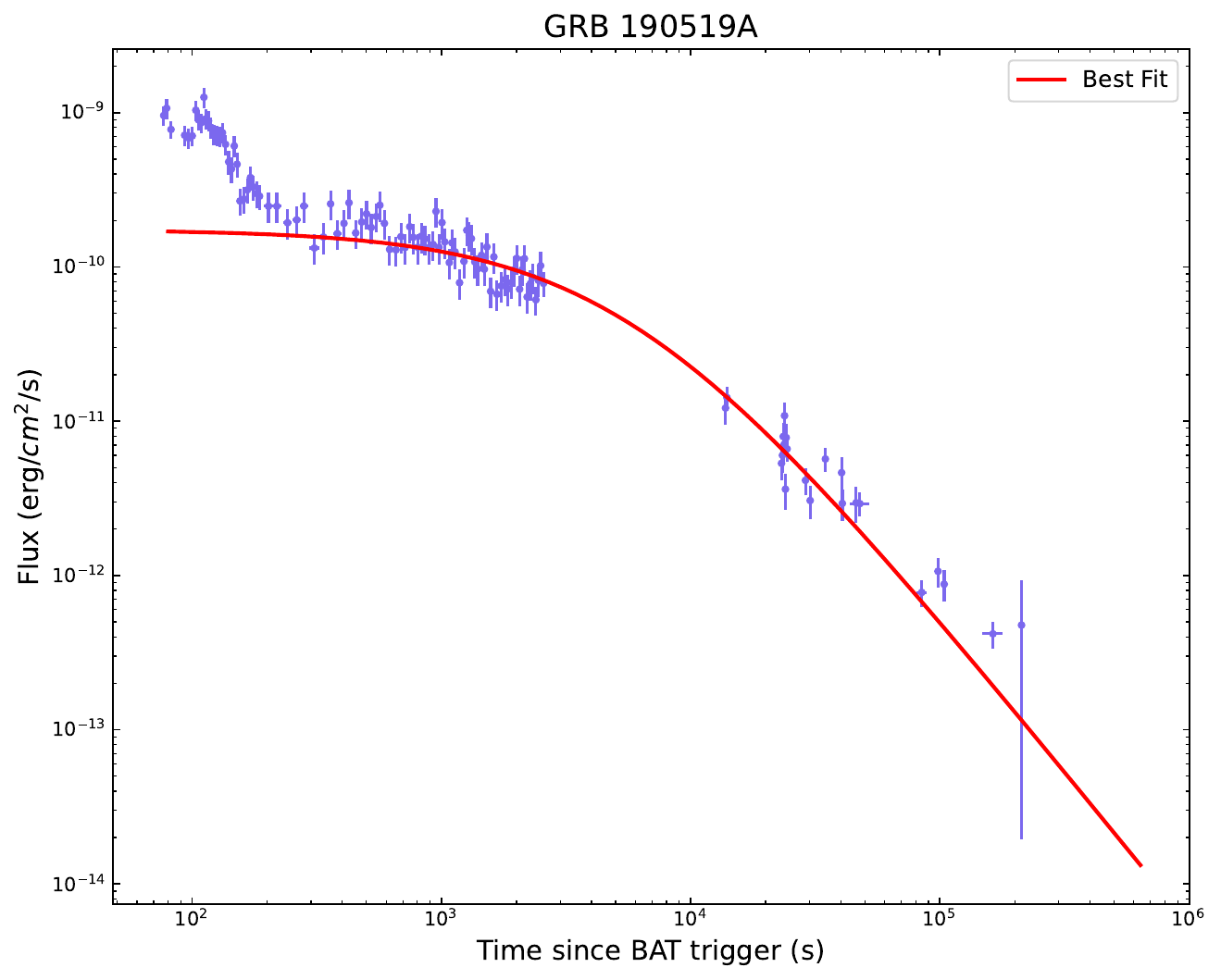}}%
\resizebox{55mm}{!}{\includegraphics[]{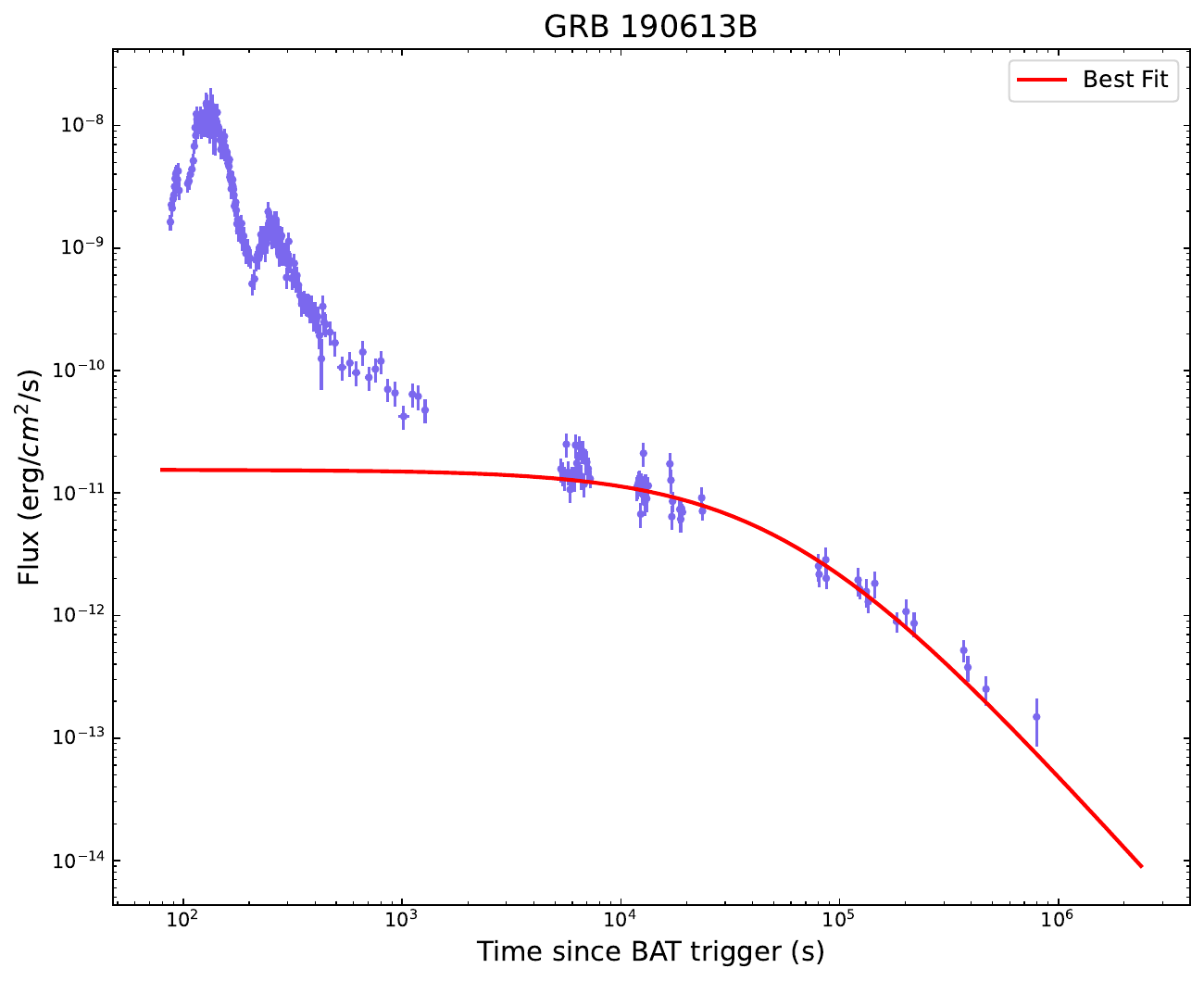}}%
\resizebox{55mm}{!}{\includegraphics[]{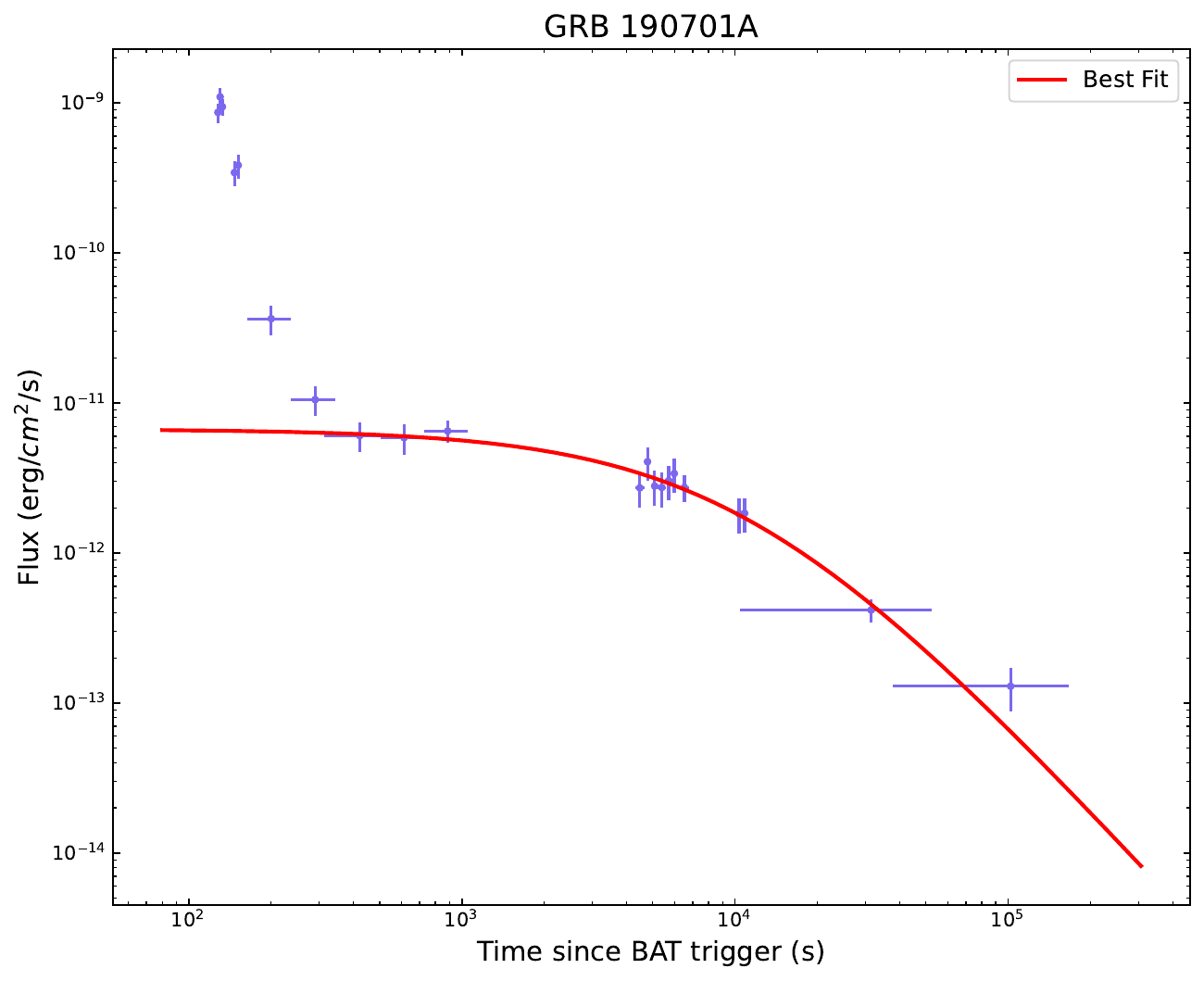}}%

\noindent
\resizebox{55mm}{!}{\includegraphics[]{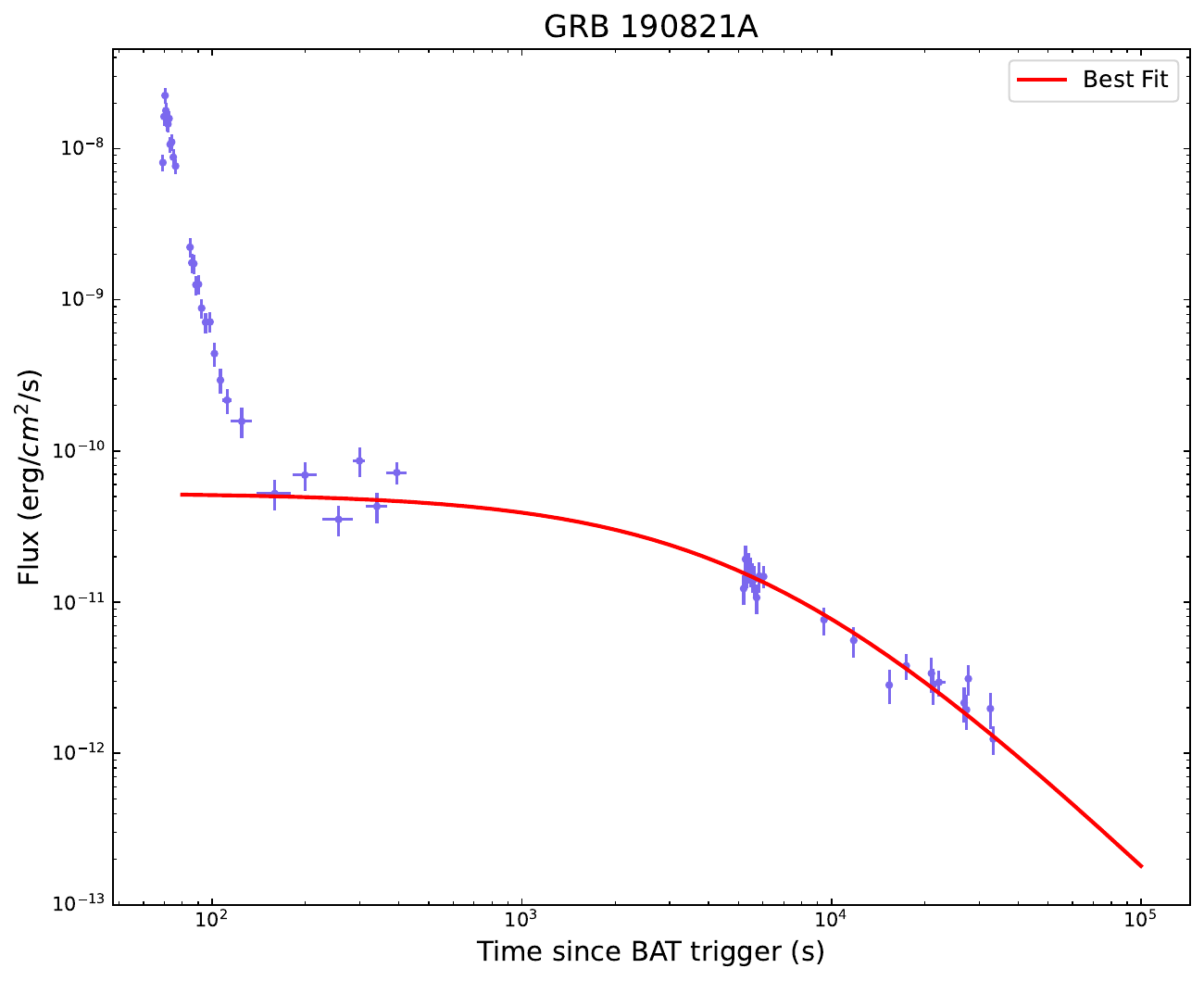}}%
\resizebox{55mm}{!}{\includegraphics[]{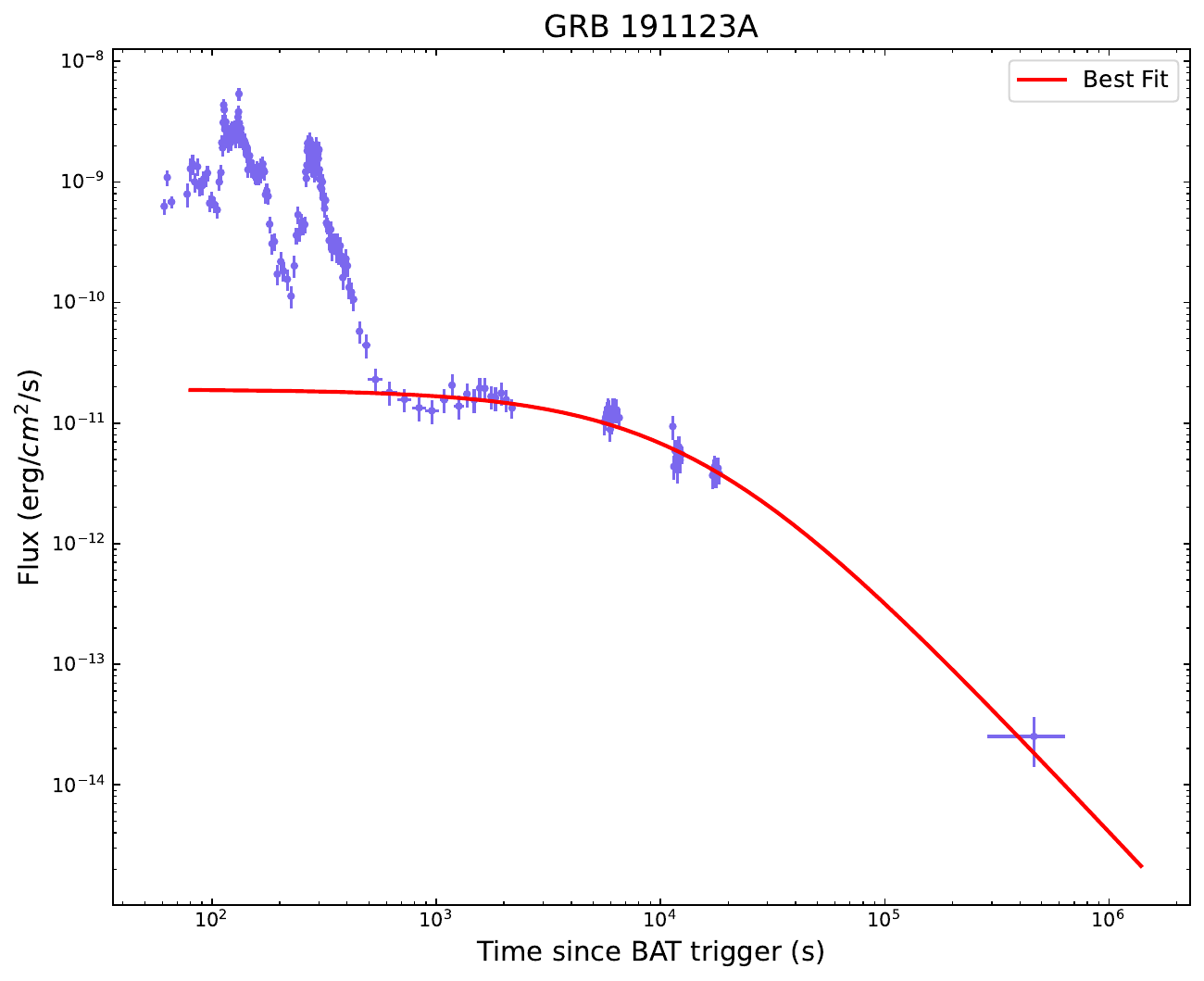}}%
\resizebox{55mm}{!}{\includegraphics[]{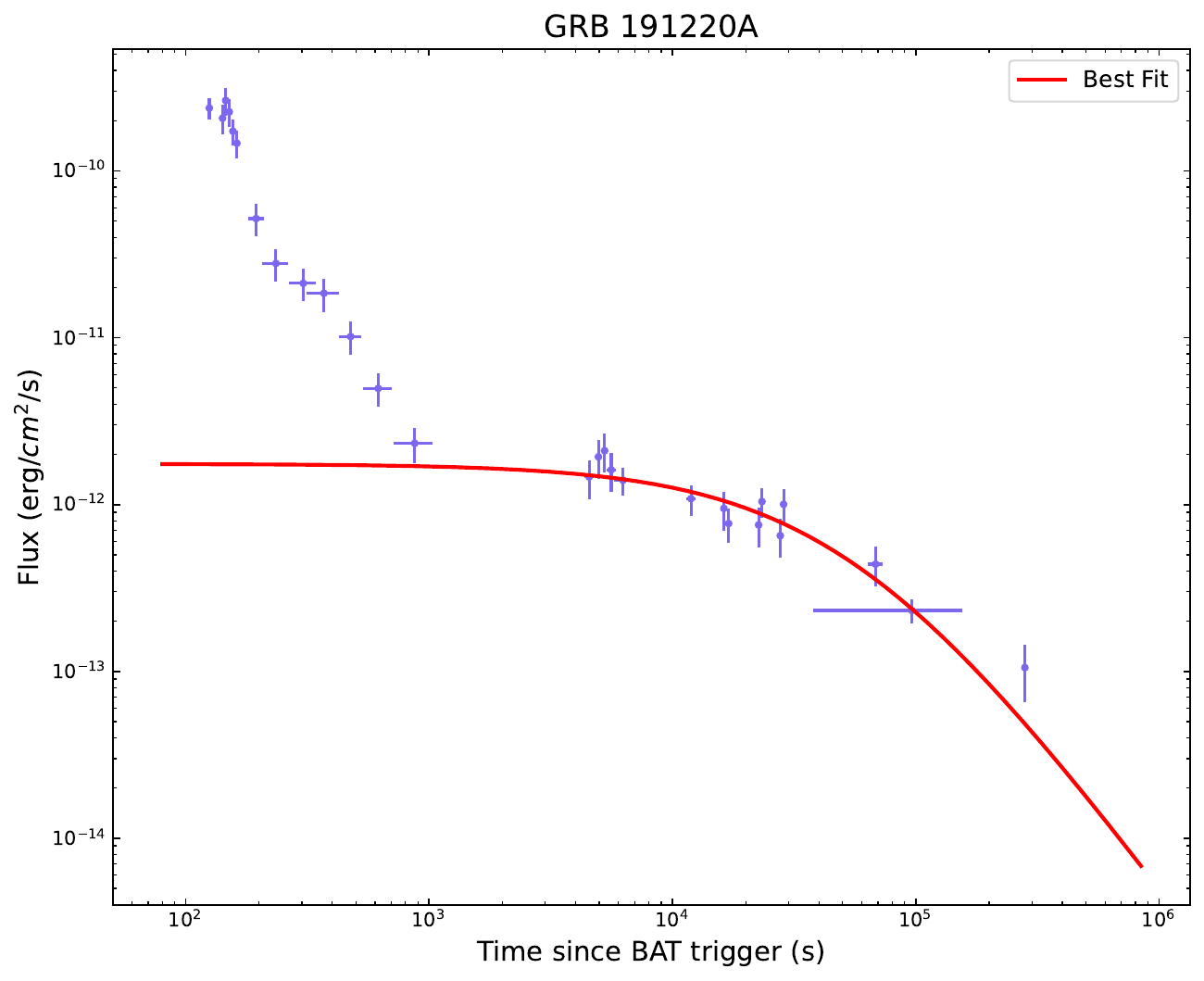}}%

\noindent
\resizebox{55mm}{!}{\includegraphics[]{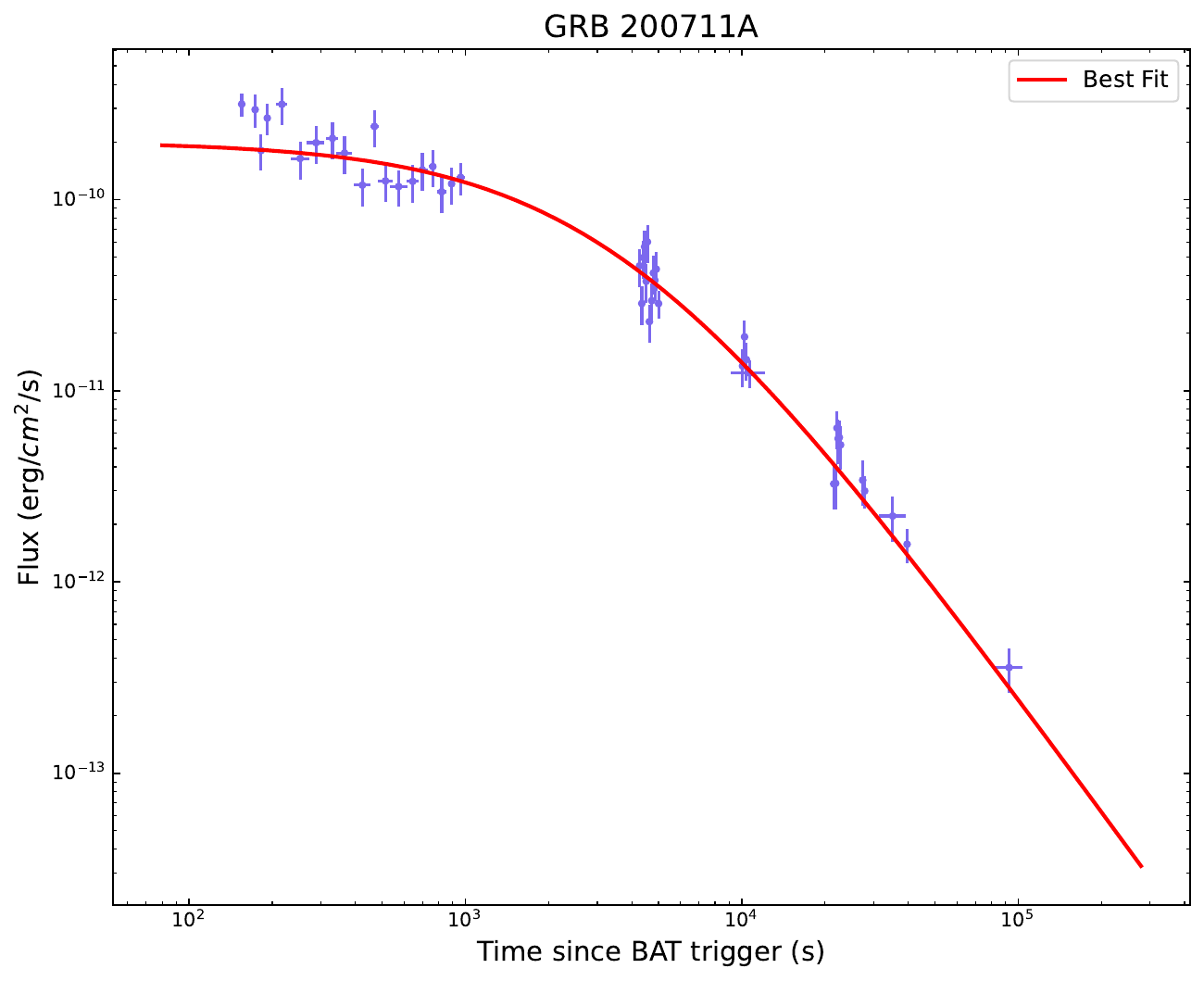}}%
\resizebox{55mm}{!}{\includegraphics[]{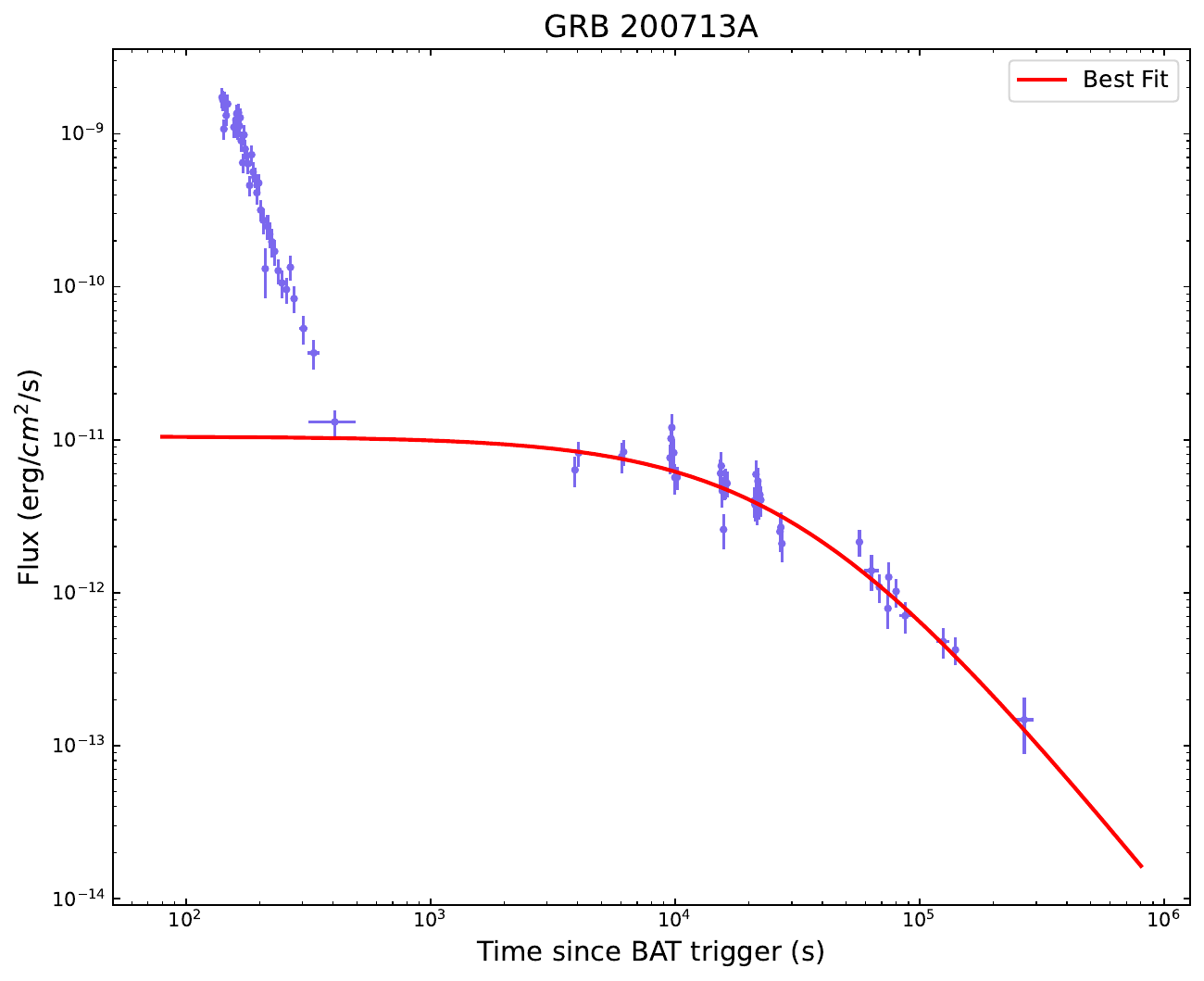}}%
\resizebox{55mm}{!}{\includegraphics[]{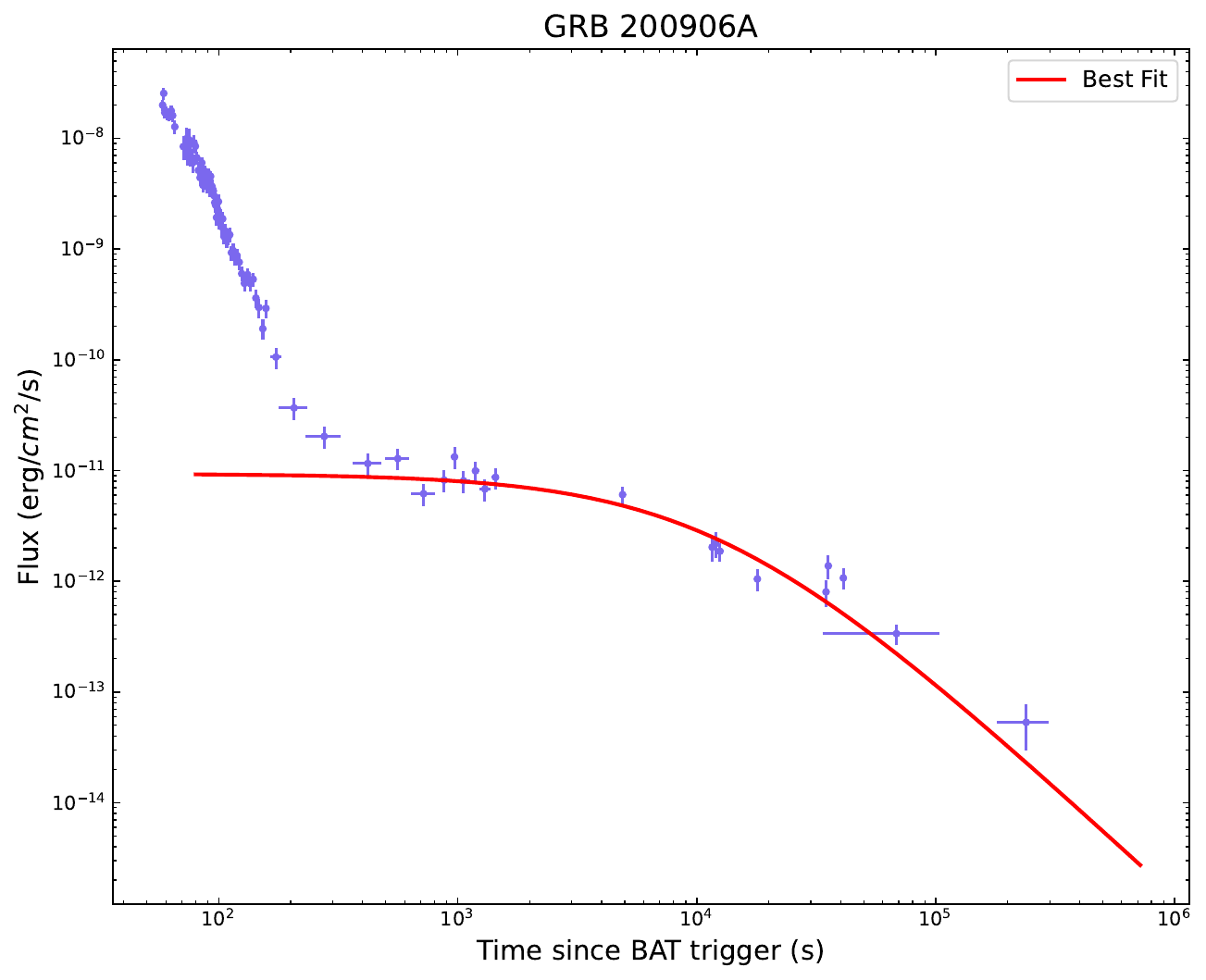}}%
\caption{(Continued)}
\end{figure*}

\addtocounter{figure}{-1}
\begin{figure*}[ht!]

\noindent
\resizebox{55mm}{!}{\includegraphics[]{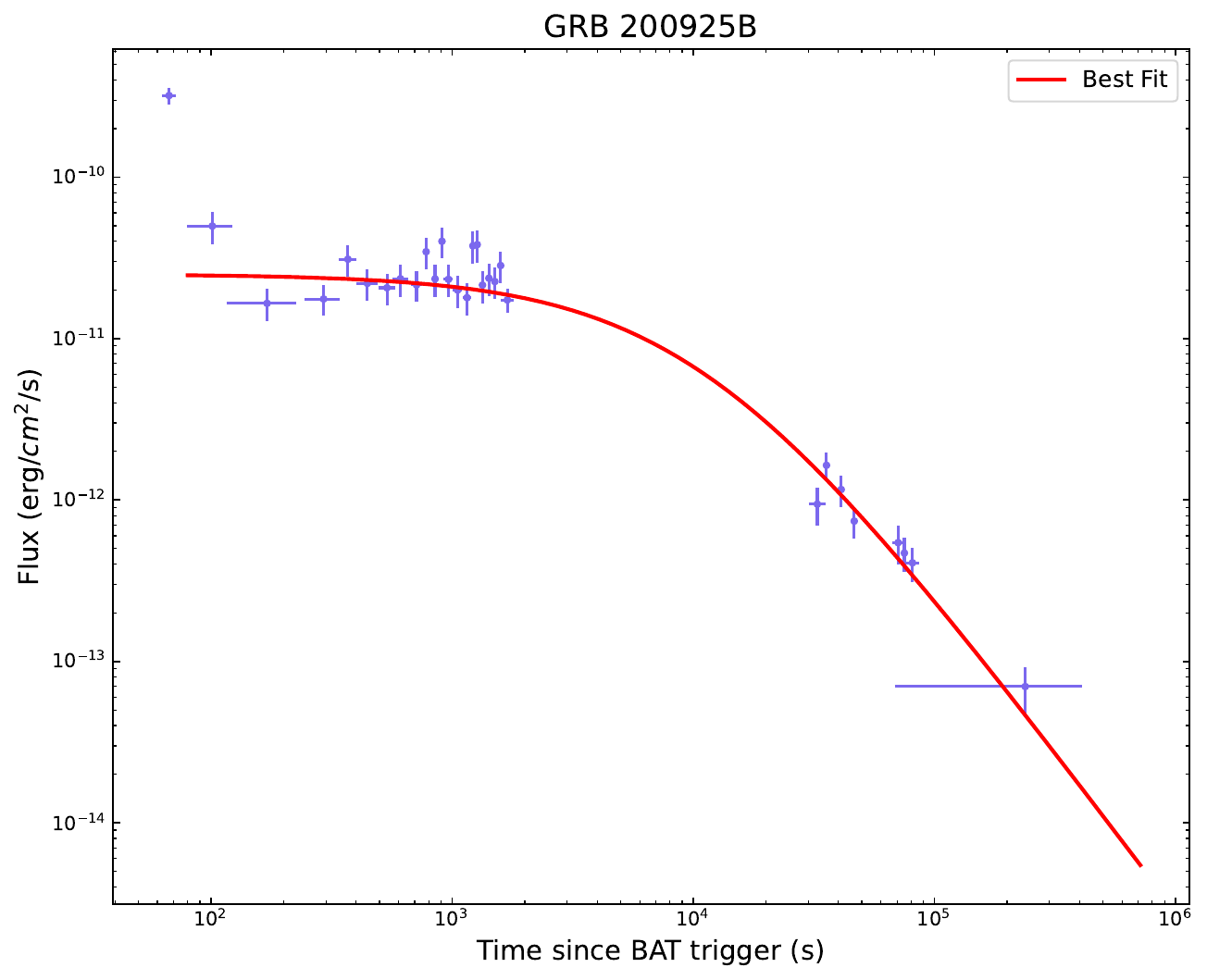}}%
\resizebox{55mm}{!}{\includegraphics[]{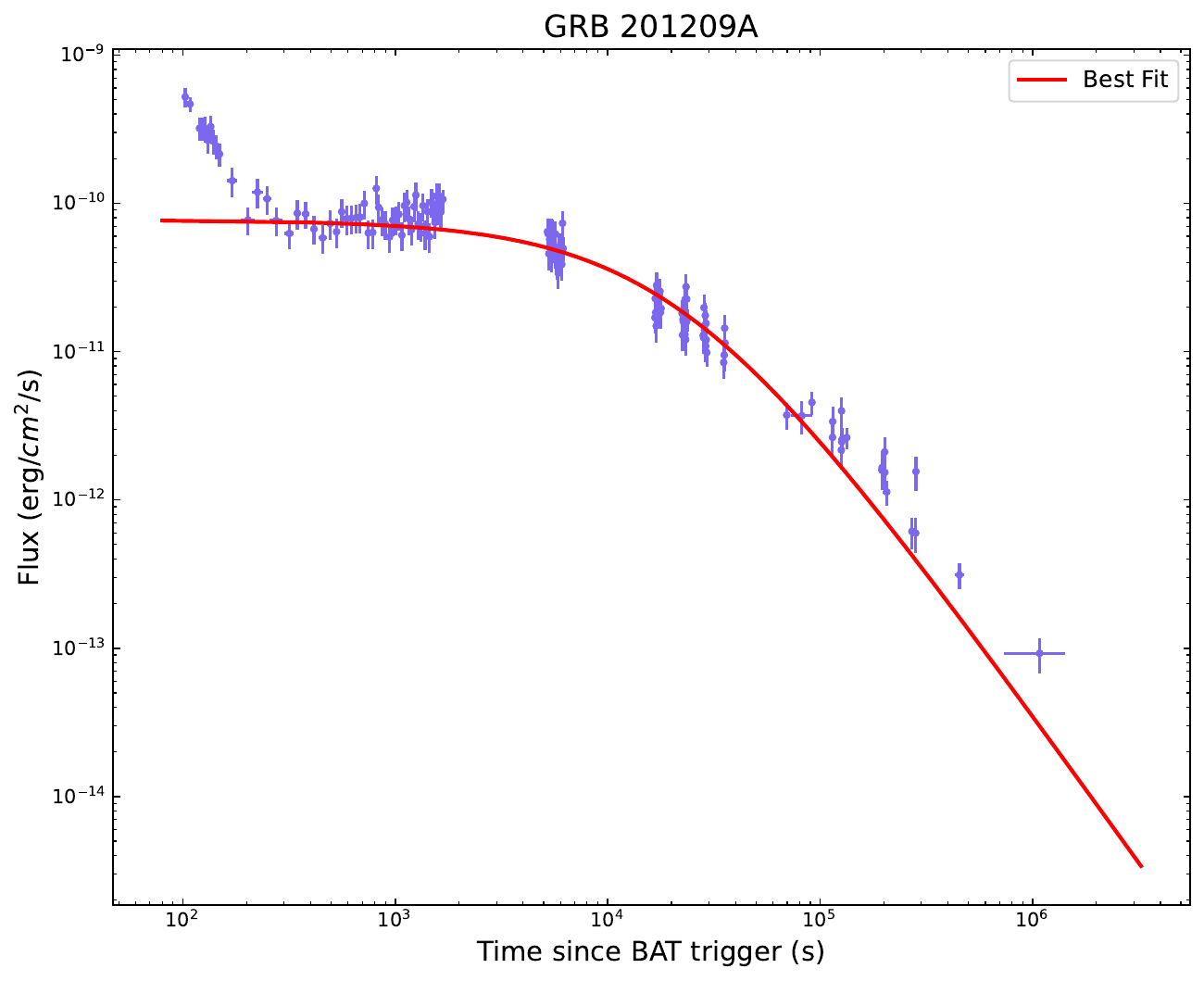}}%
\resizebox{55mm}{!}{\includegraphics[]{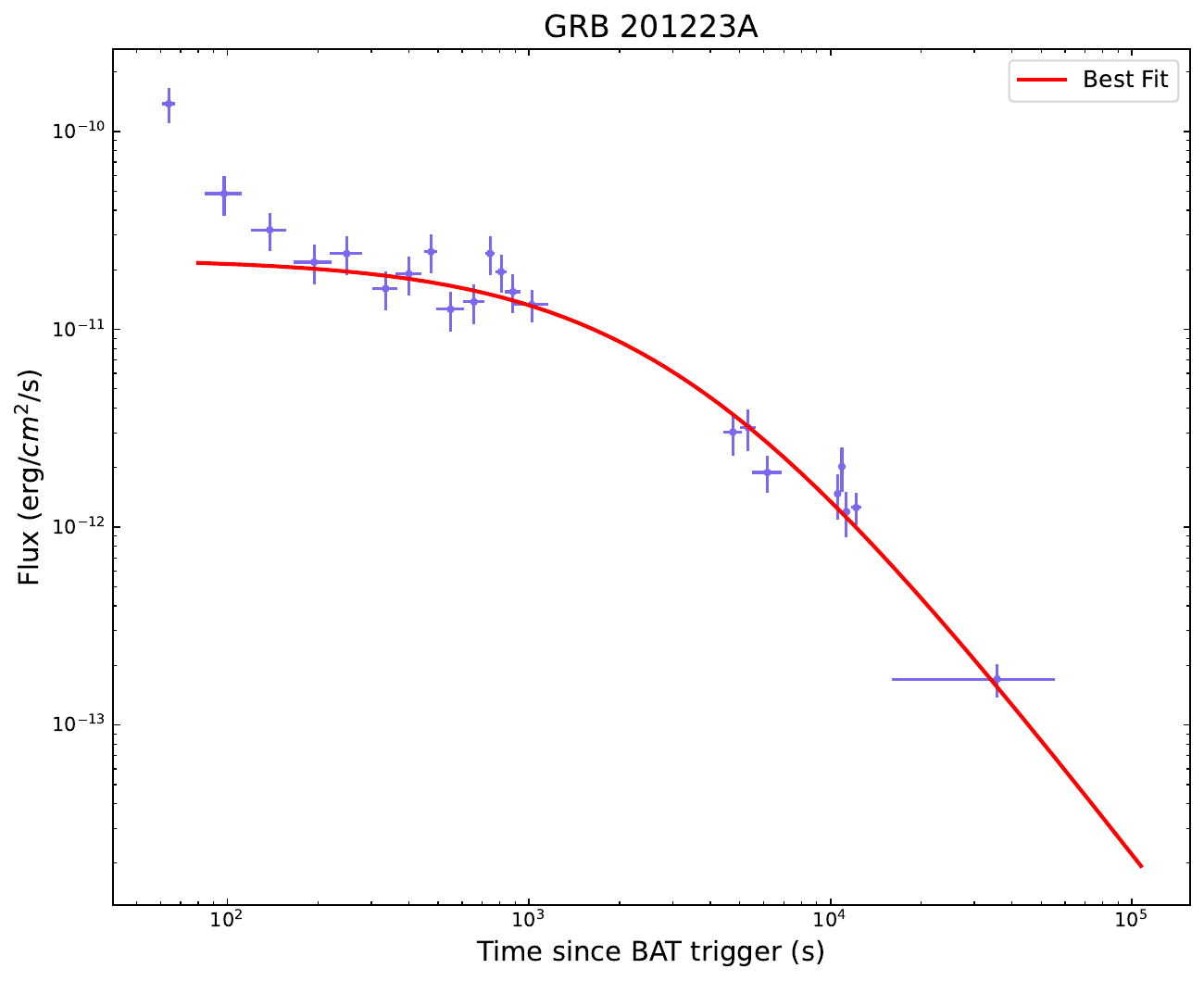}}%

\noindent
\resizebox{55mm}{!}{\includegraphics[]{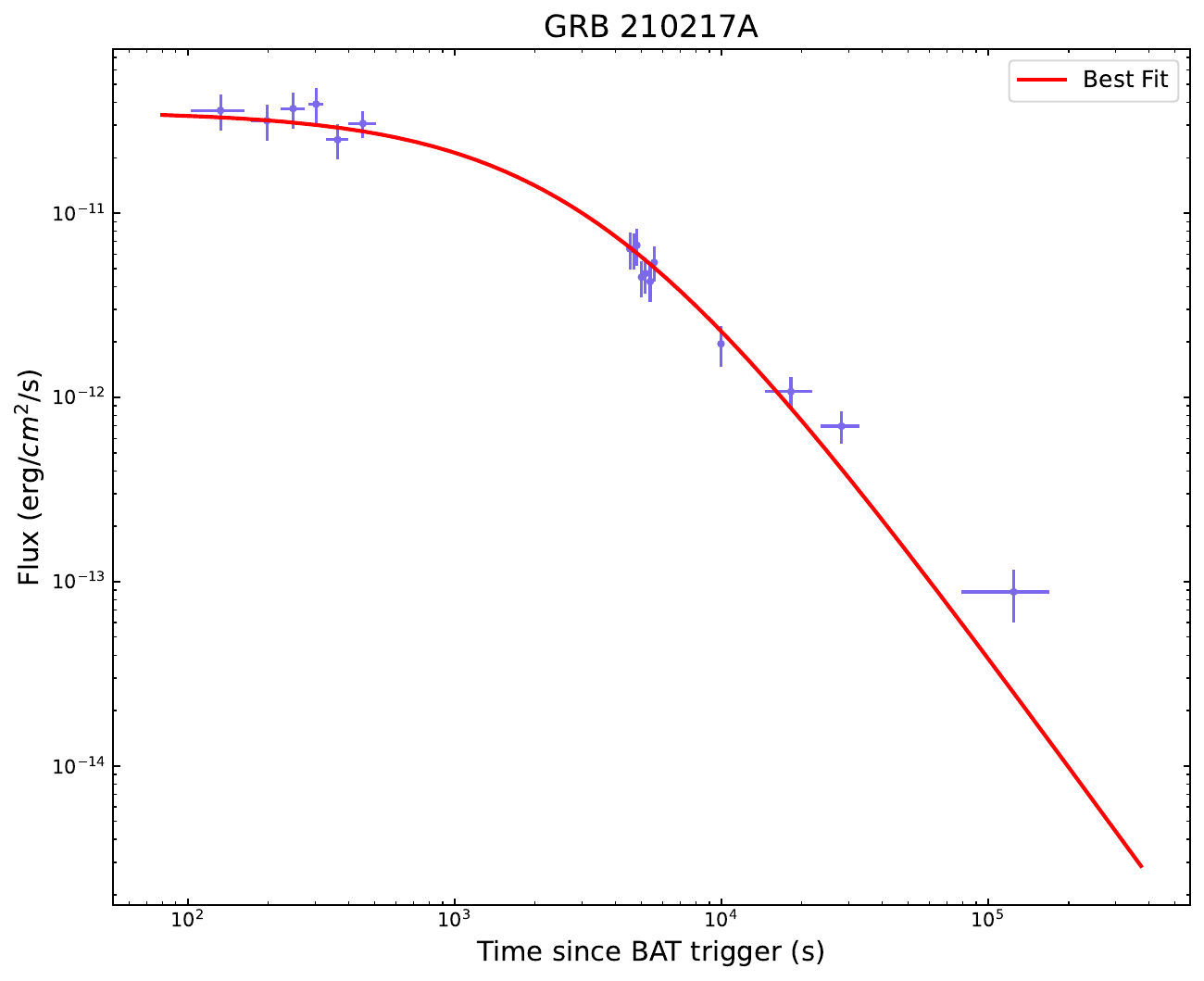}}%
\resizebox{55mm}{!}{\includegraphics[]{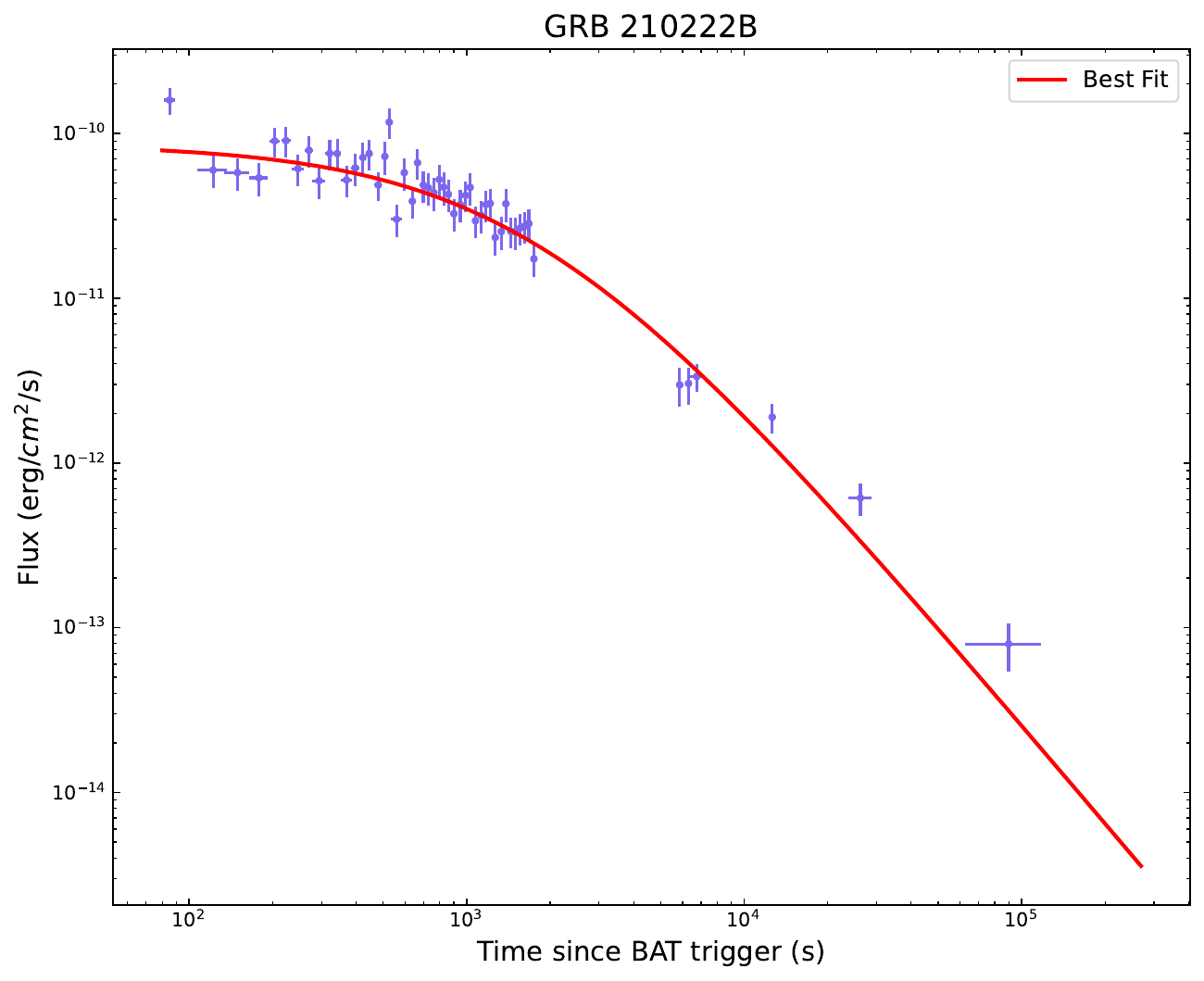}}%
\resizebox{55mm}{!}{\includegraphics[]{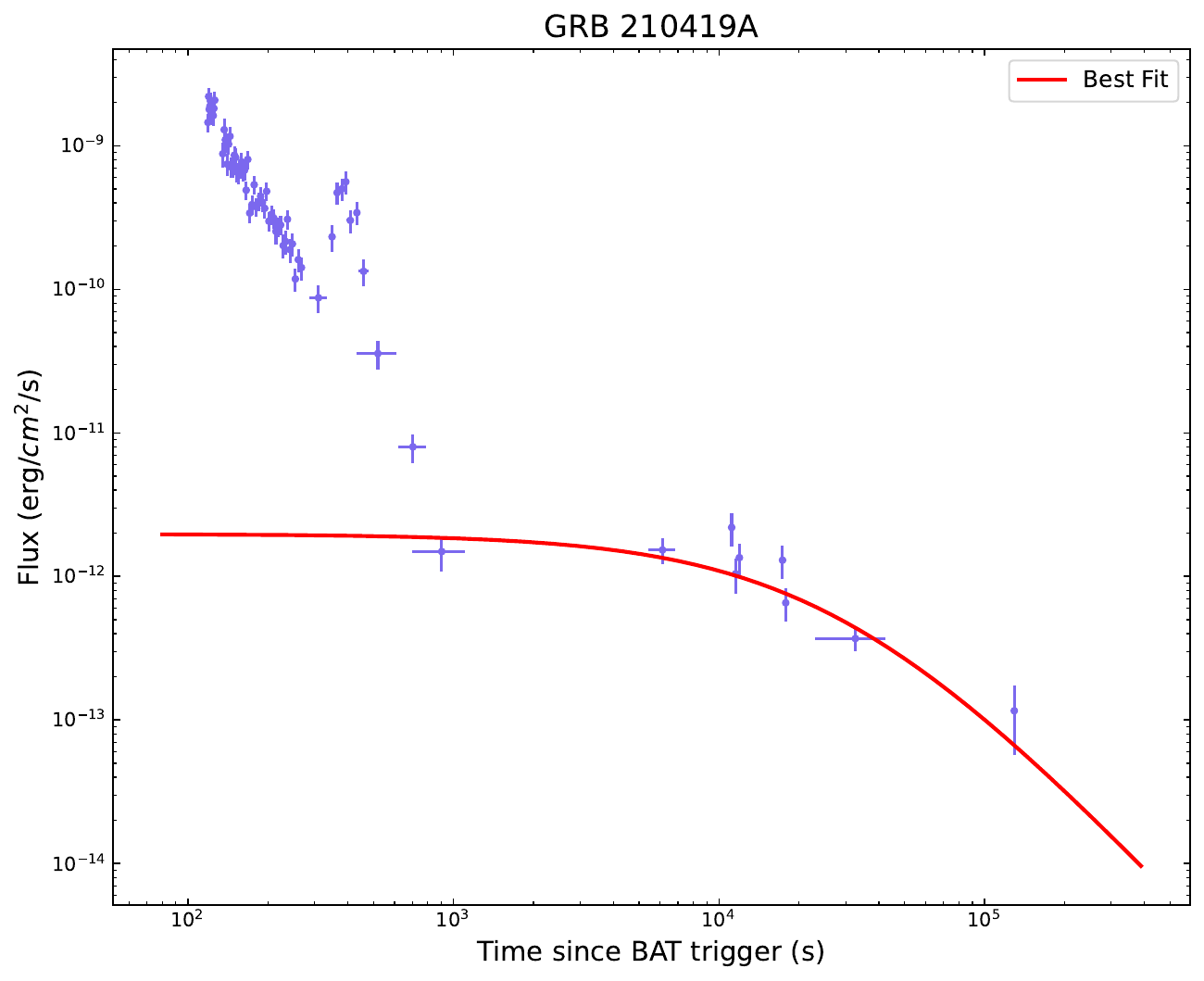}}%

\noindent
\resizebox{55mm}{!}{\includegraphics[]{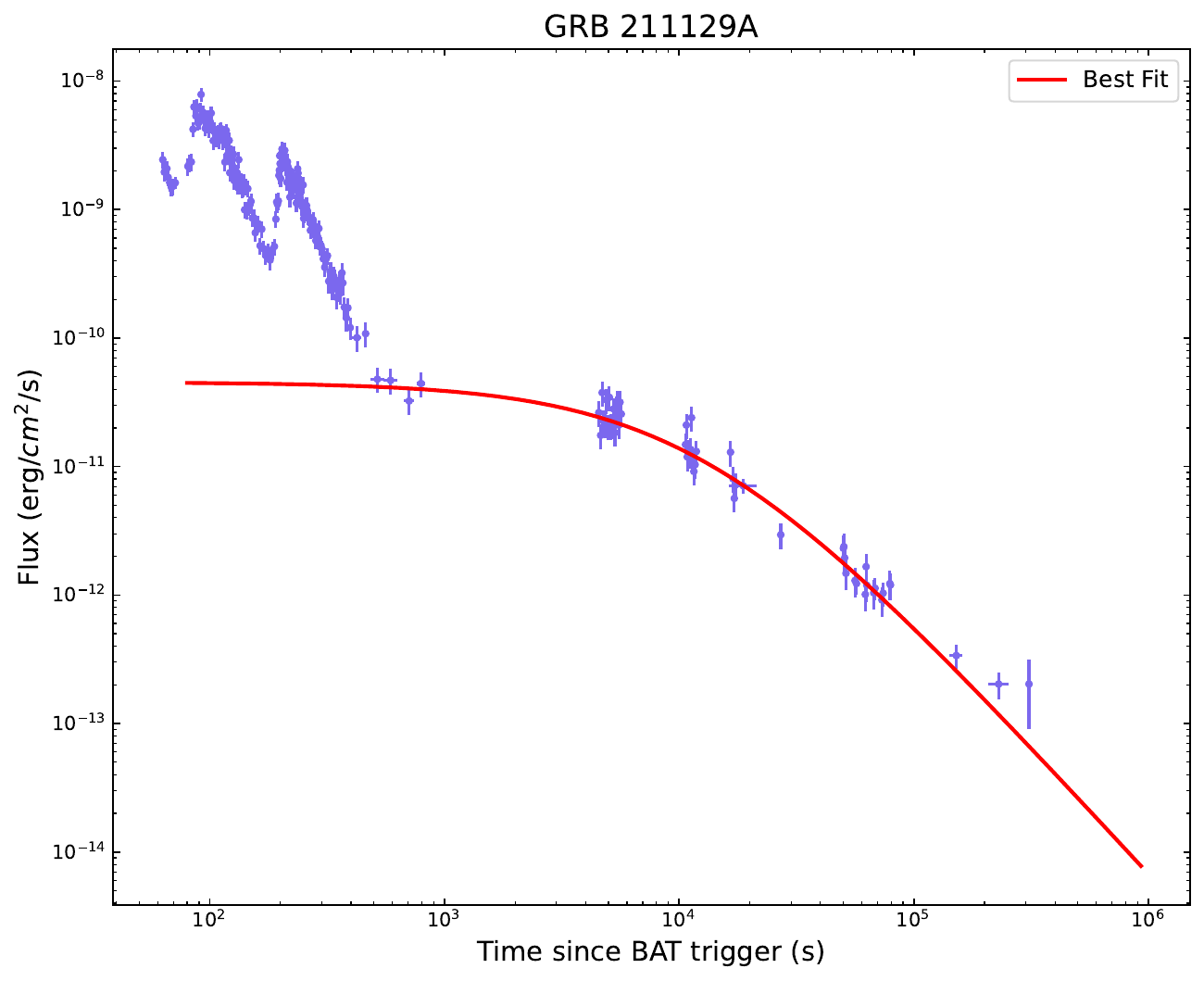}}%
\resizebox{55mm}{!}{\includegraphics[]{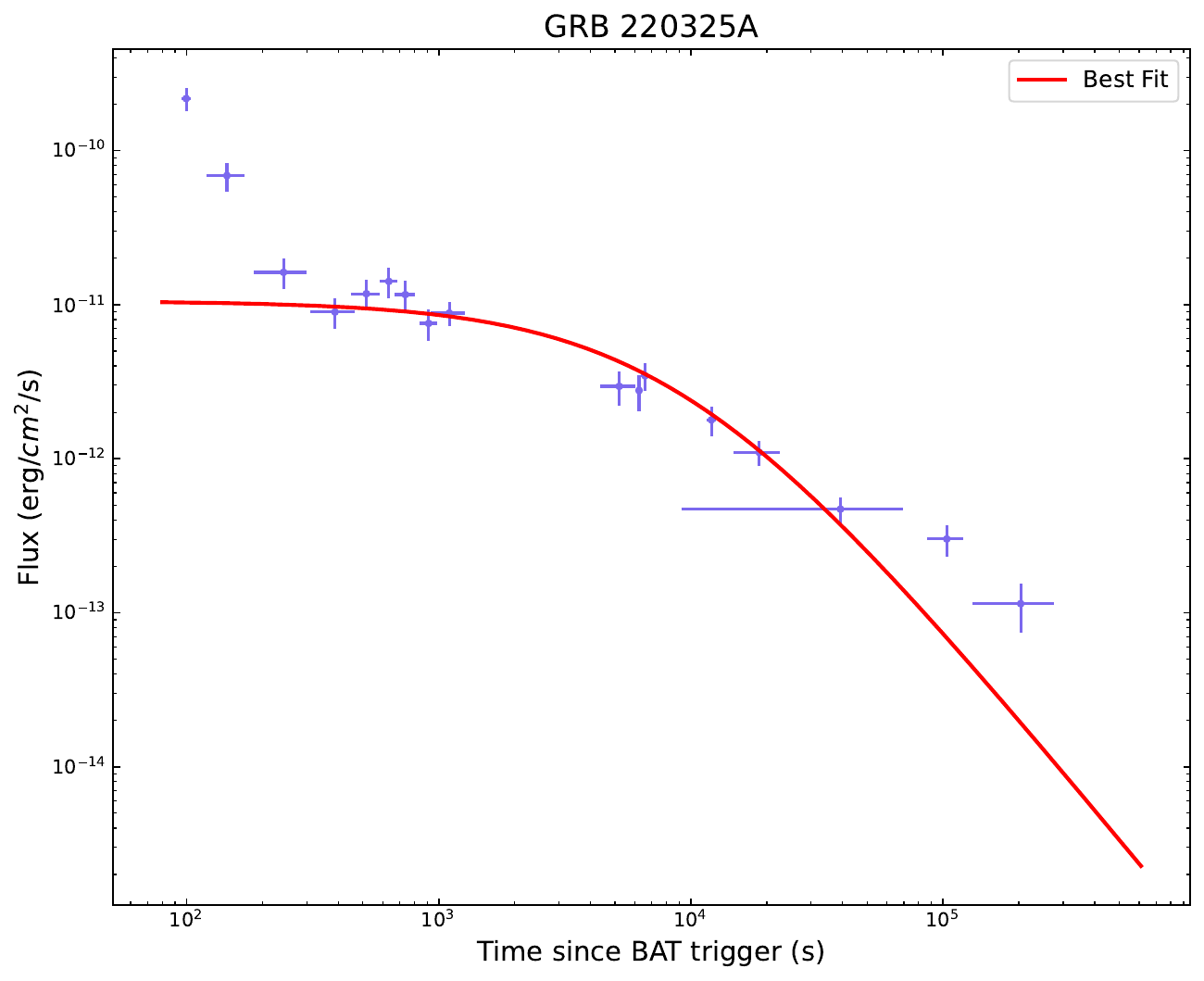}}%
\resizebox{55mm}{!}{\includegraphics[]{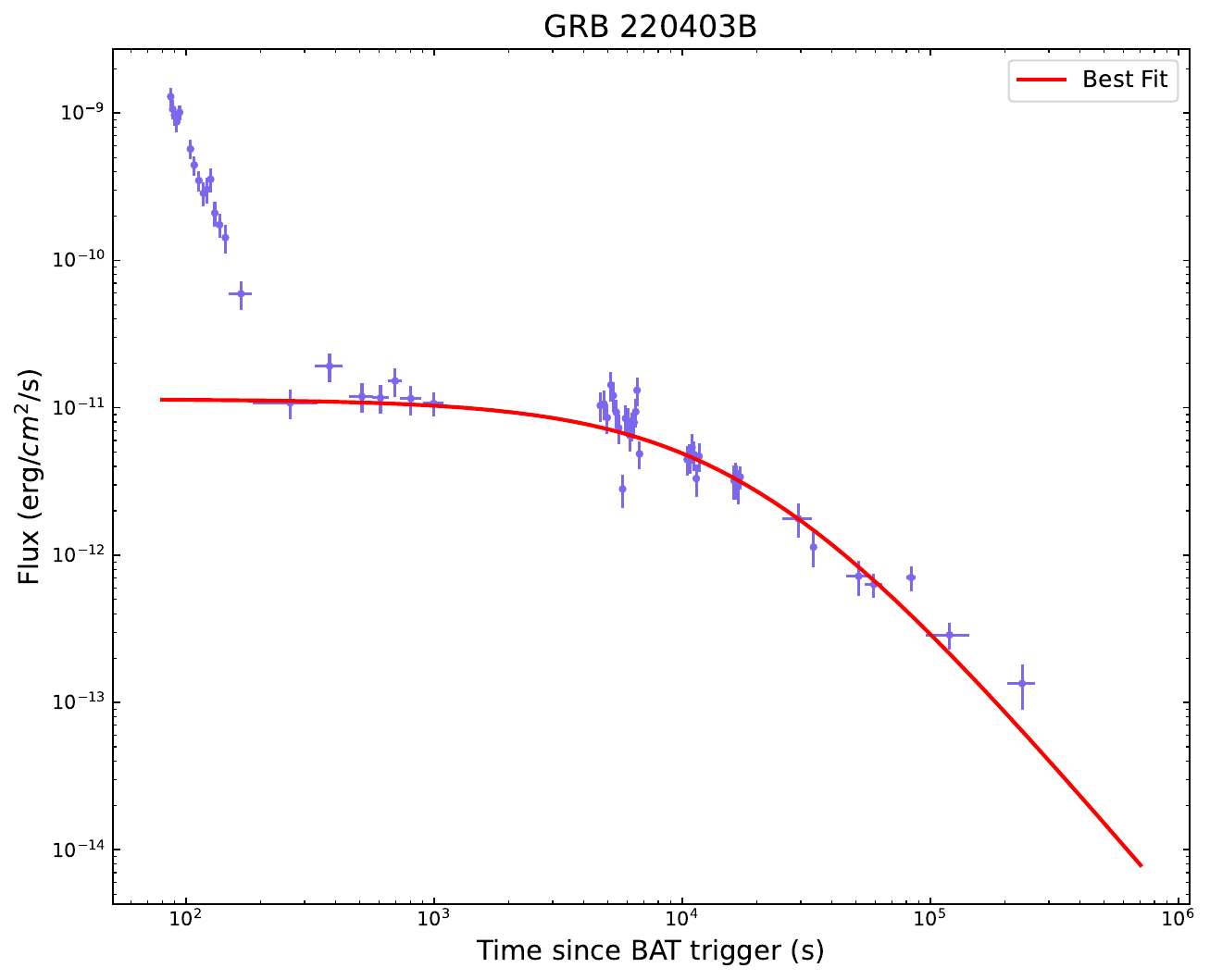}}%

\noindent
\resizebox{55mm}{!}{\includegraphics[]{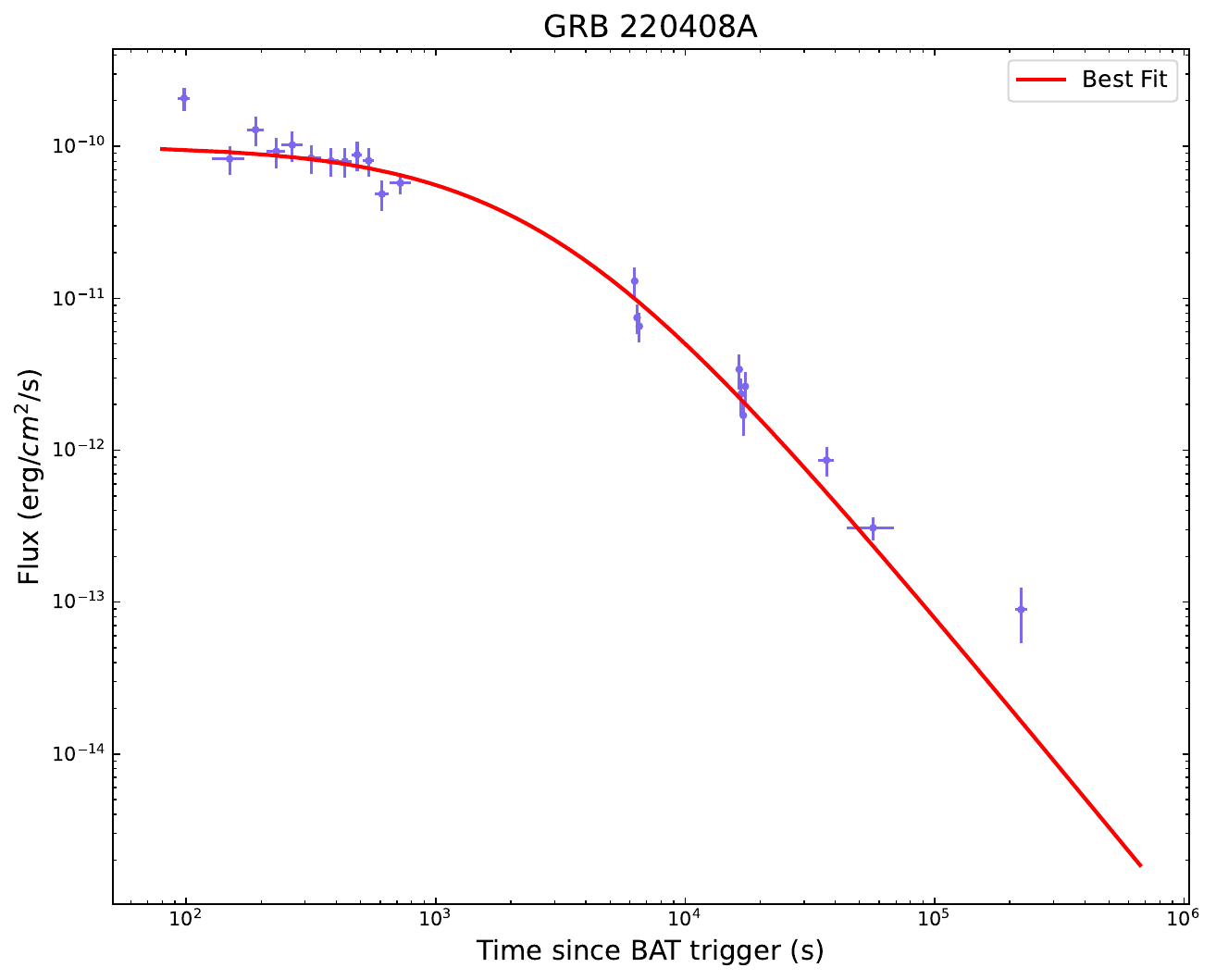}}%
\resizebox{55mm}{!}{\includegraphics[]{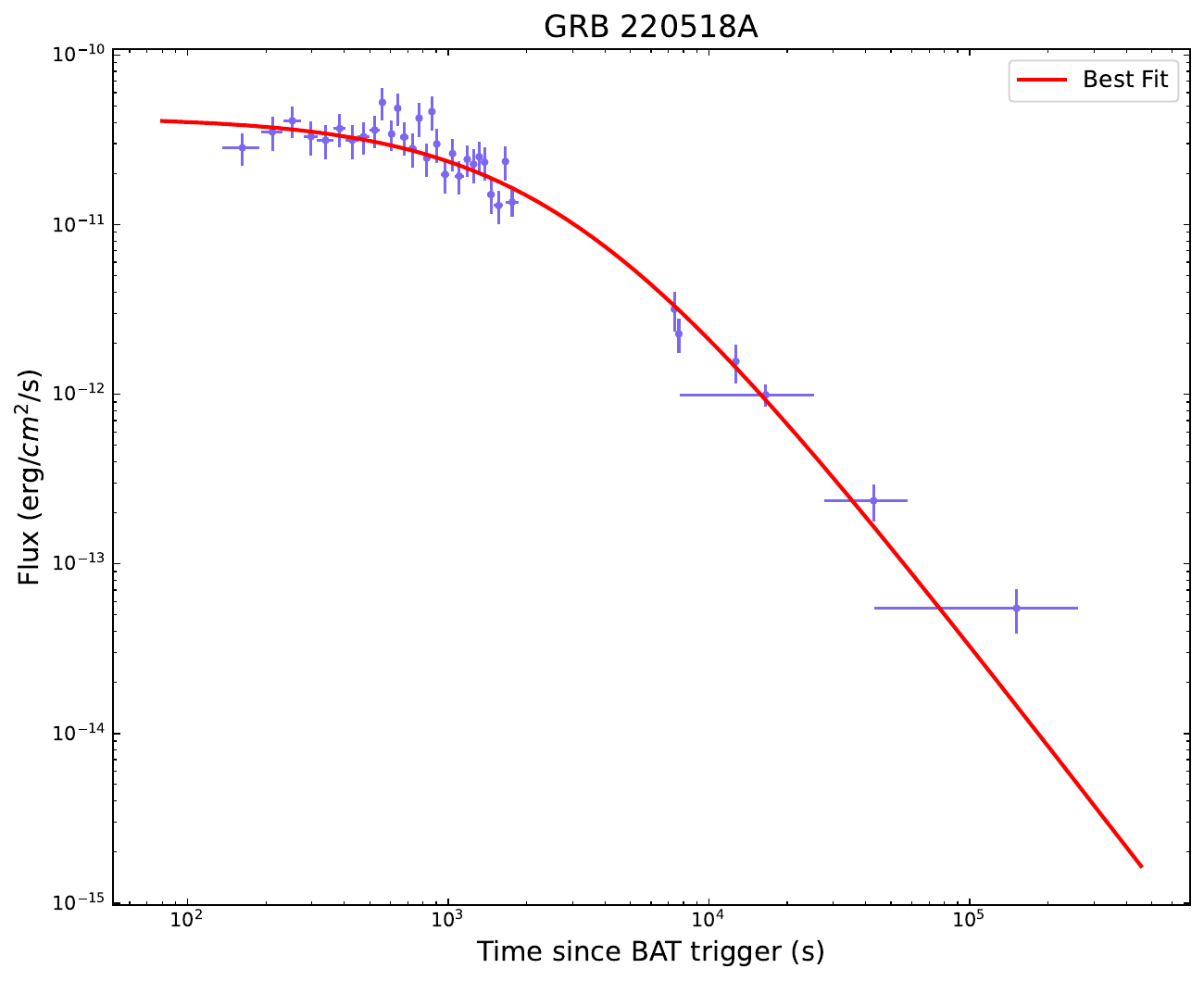}}%
\resizebox{55mm}{!}{\includegraphics[]{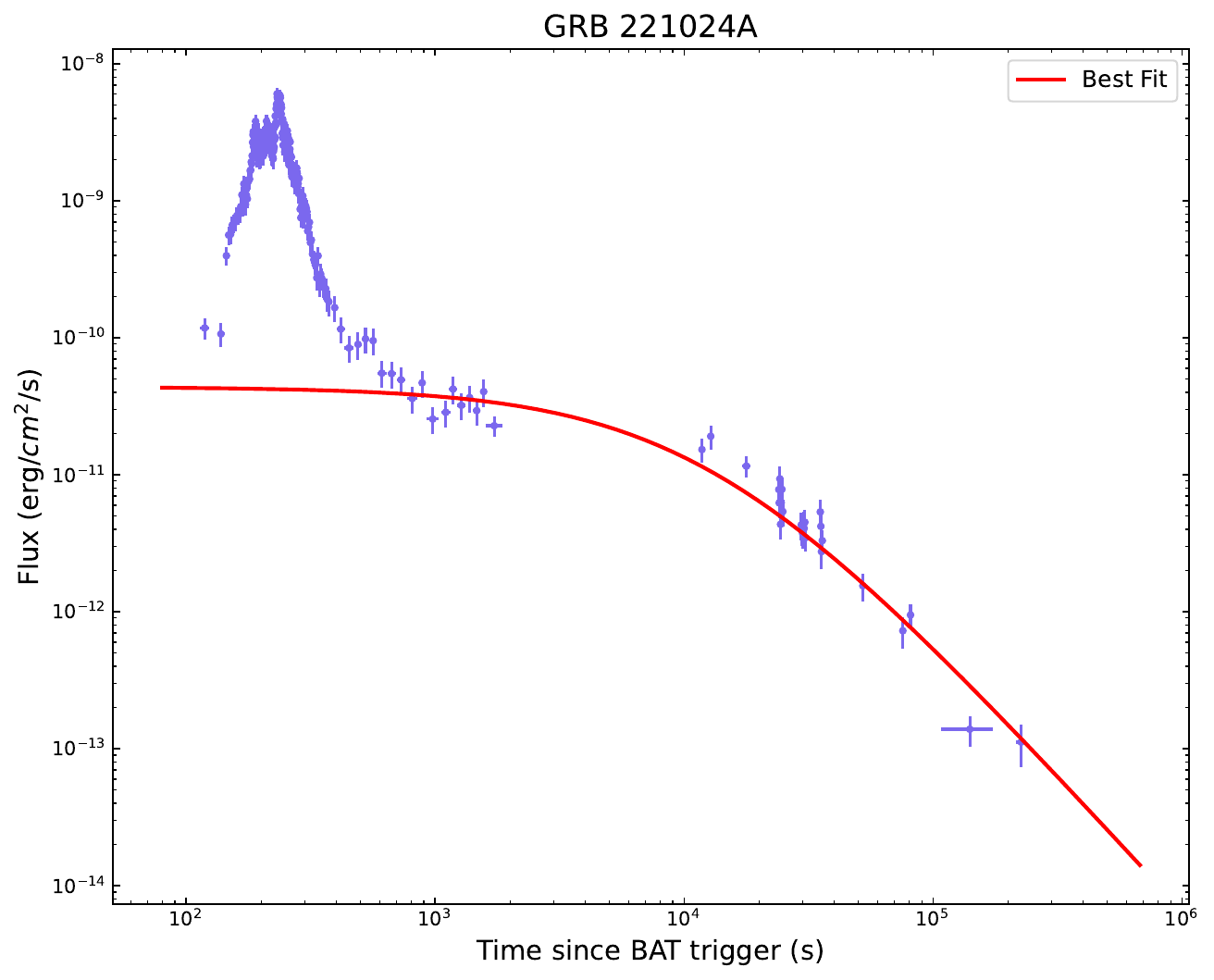}}%

\noindent
\resizebox{55mm}{!}{\includegraphics[]{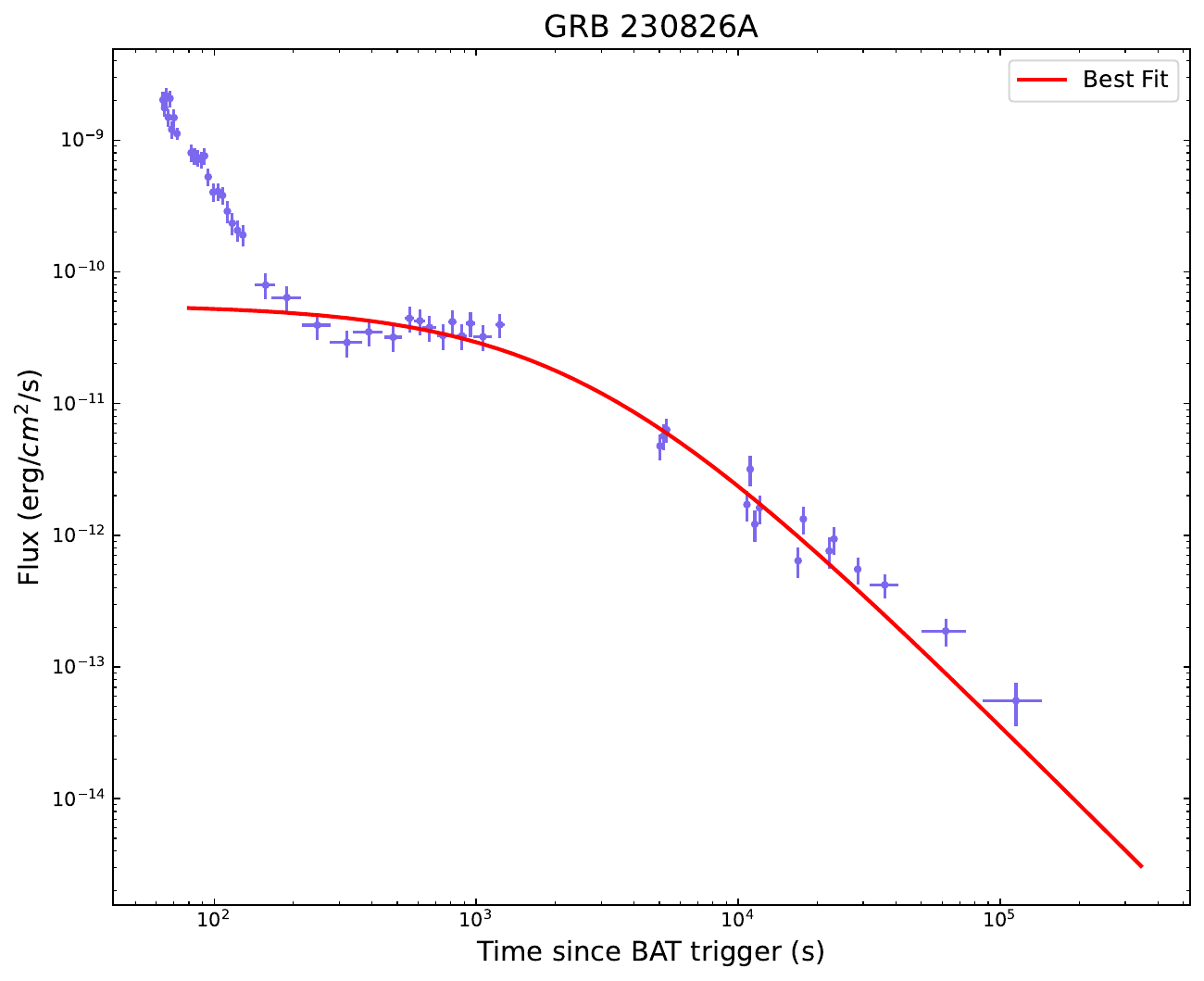}}%
\resizebox{55mm}{!}{\includegraphics[]{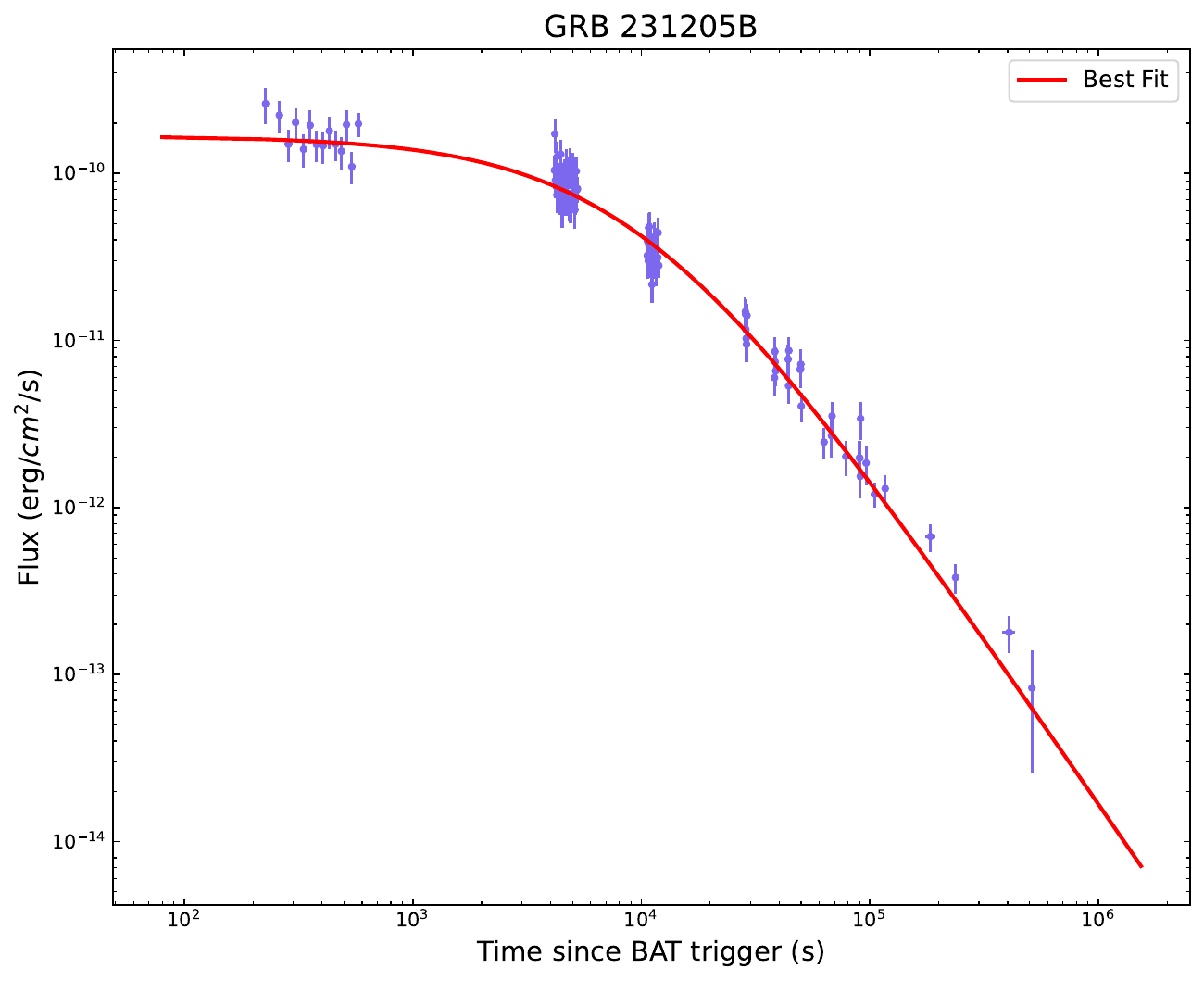}}%
\resizebox{55mm}{!}{\includegraphics[]{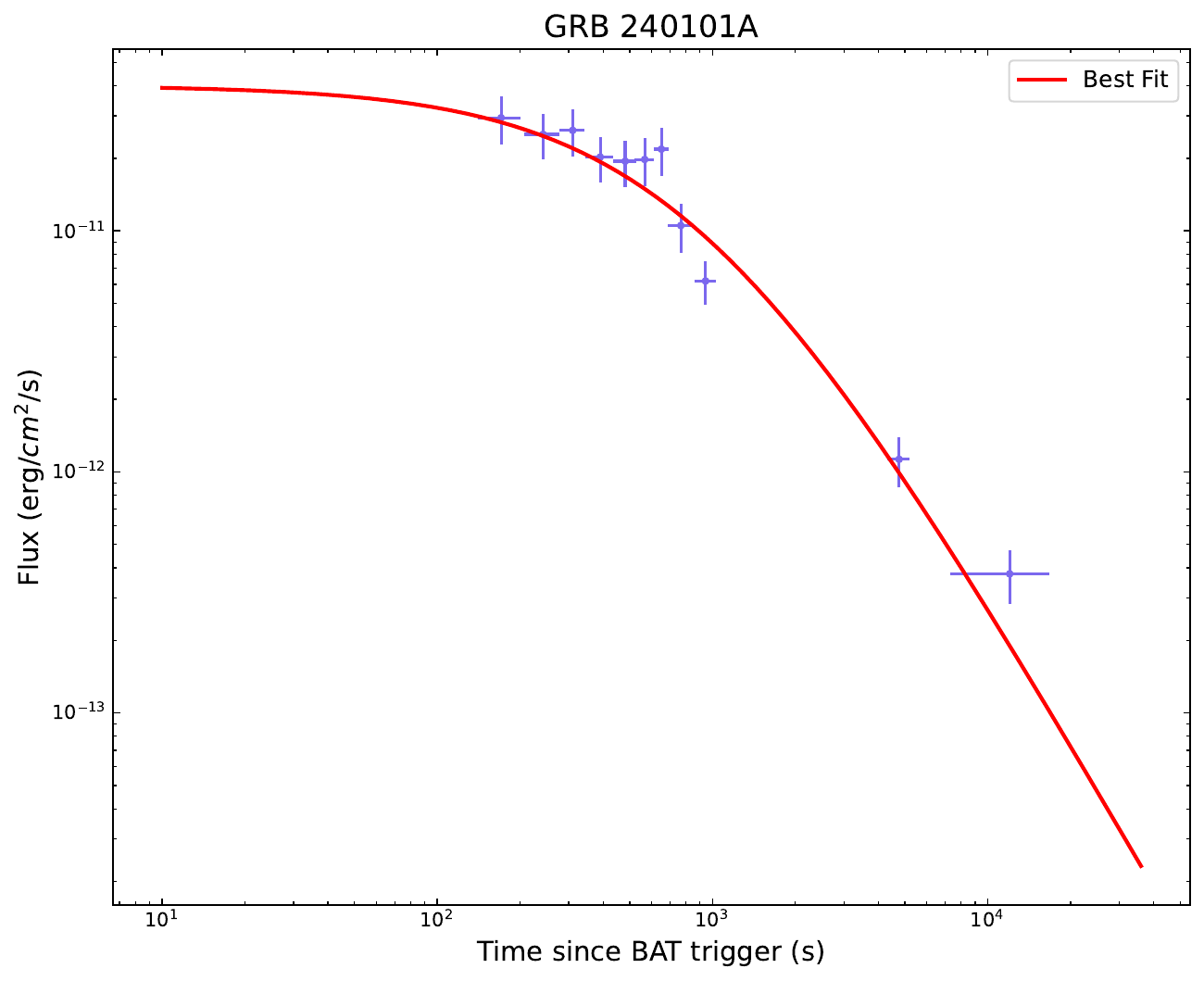}}%
\caption{(Continued)}
\end{figure*}

\addtocounter{figure}{-1}
\begin{figure*}[ht!]

\noindent
\resizebox{55mm}{!}{\includegraphics[]{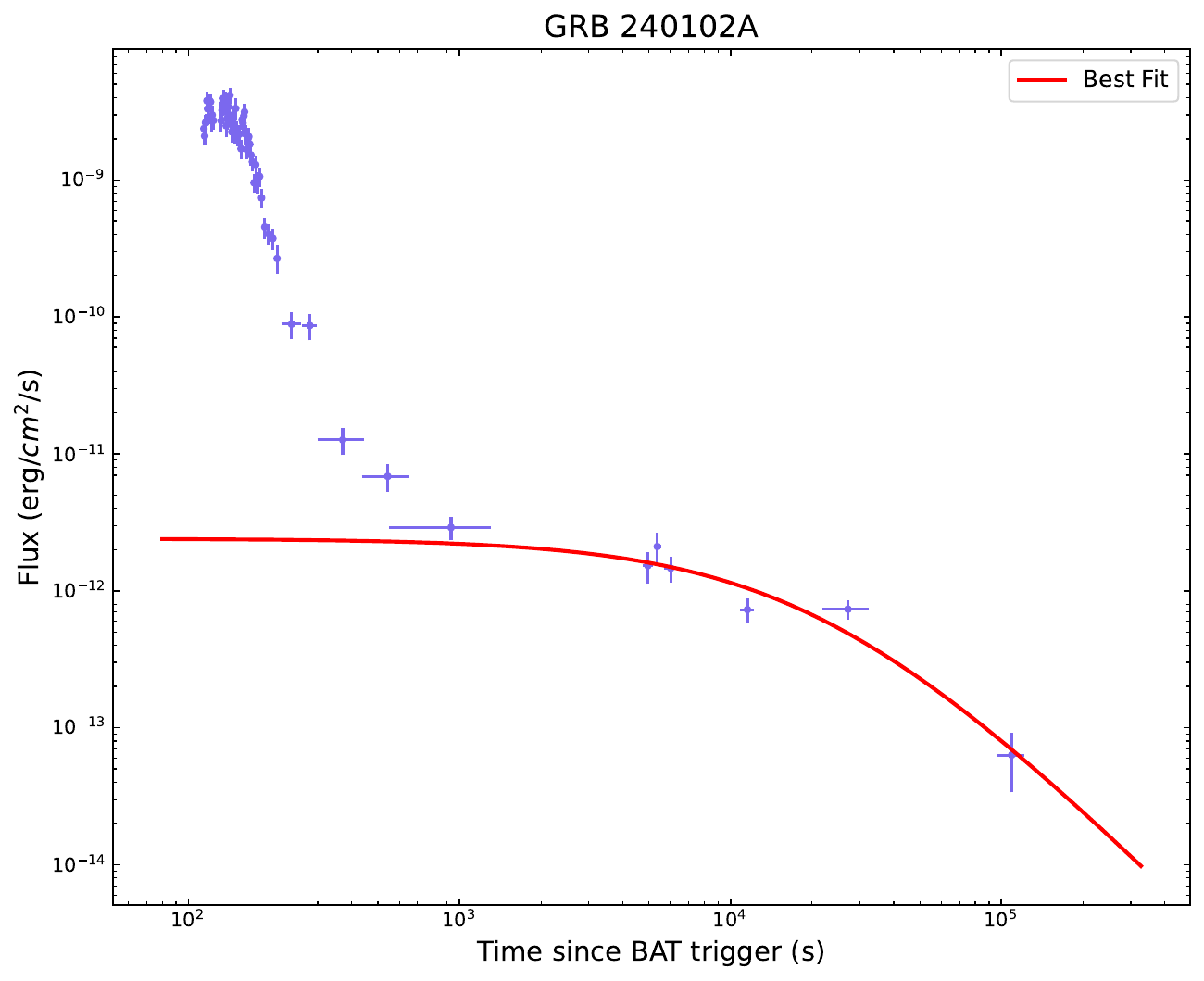}}%
\resizebox{55mm}{!}{\includegraphics[]{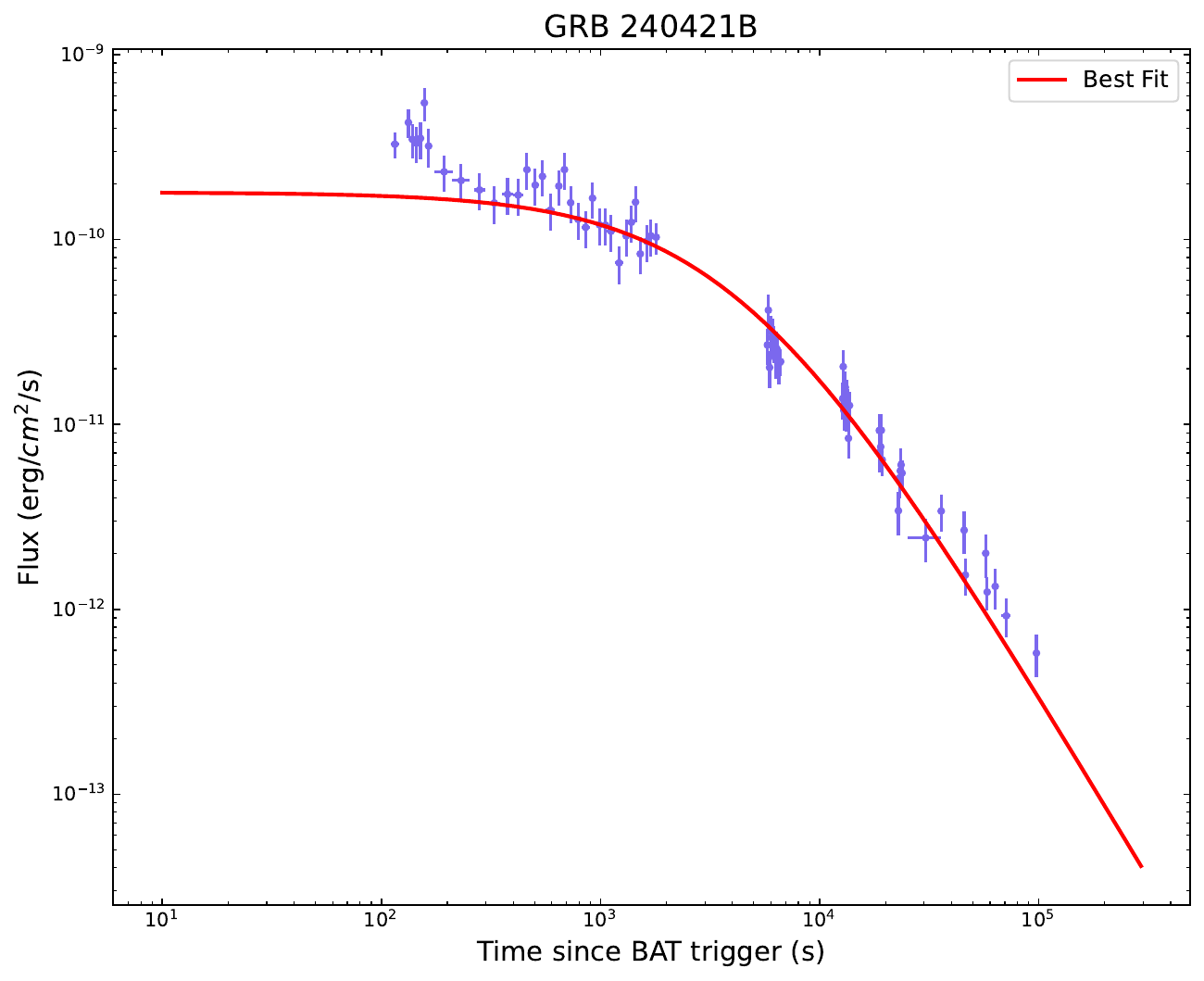}}%
\resizebox{55mm}{!}{\includegraphics[]{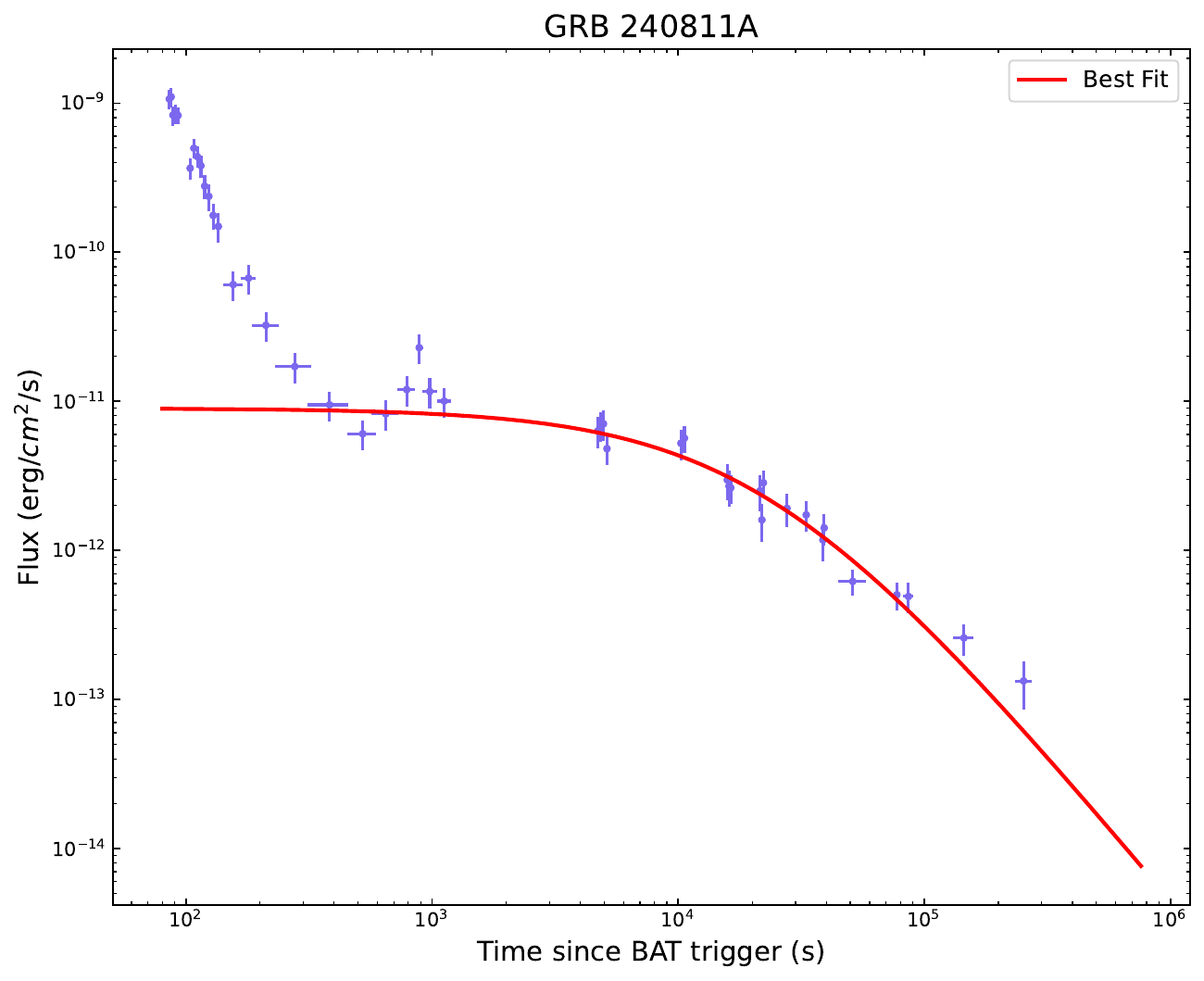}}%

\noindent
\resizebox{55mm}{!}{\includegraphics[]{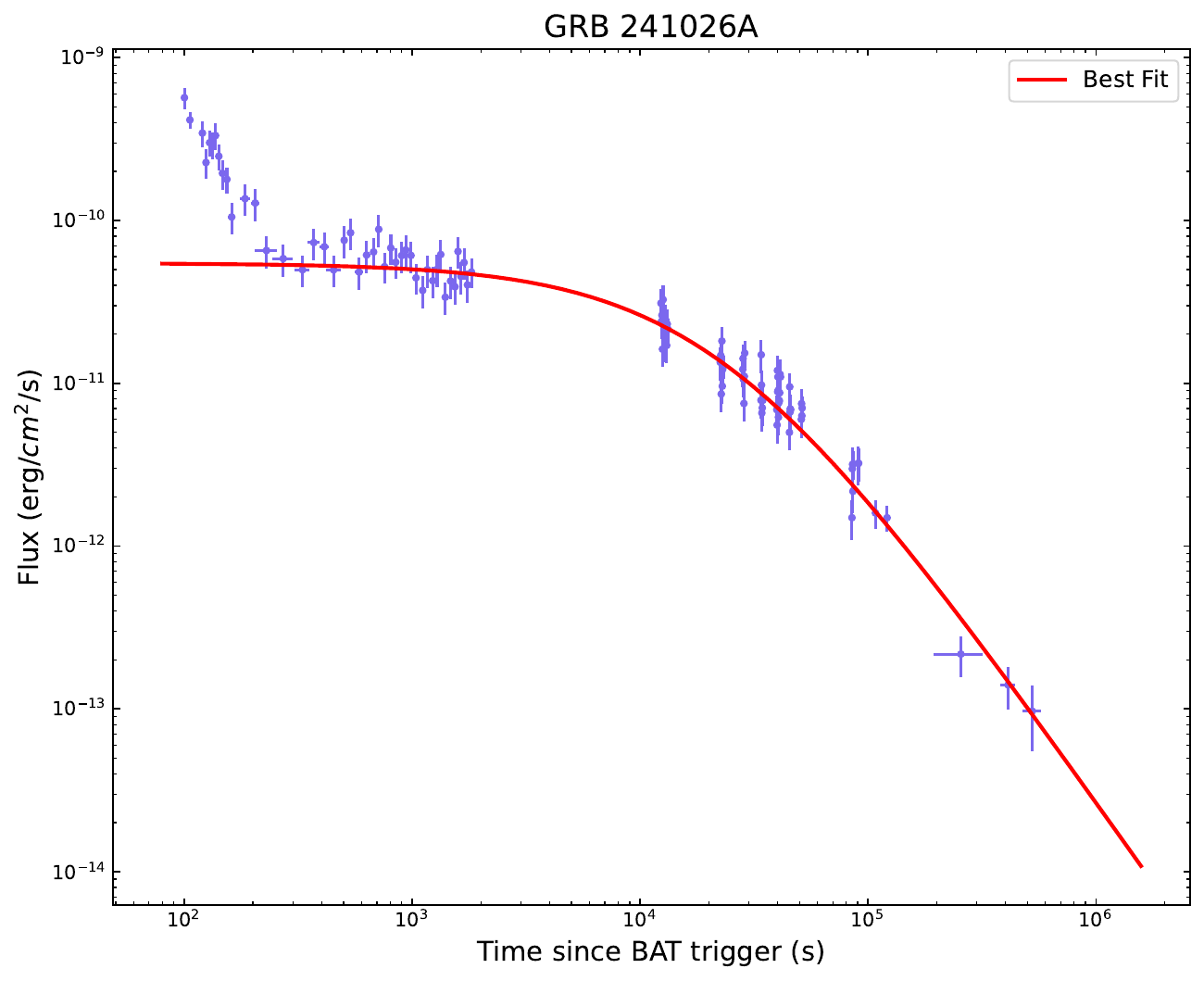}}%
\caption{(Continued)}
\end{figure*}

\begin{figure*}[ht!]
\centering
\resizebox{180mm}{!}{\includegraphics[]{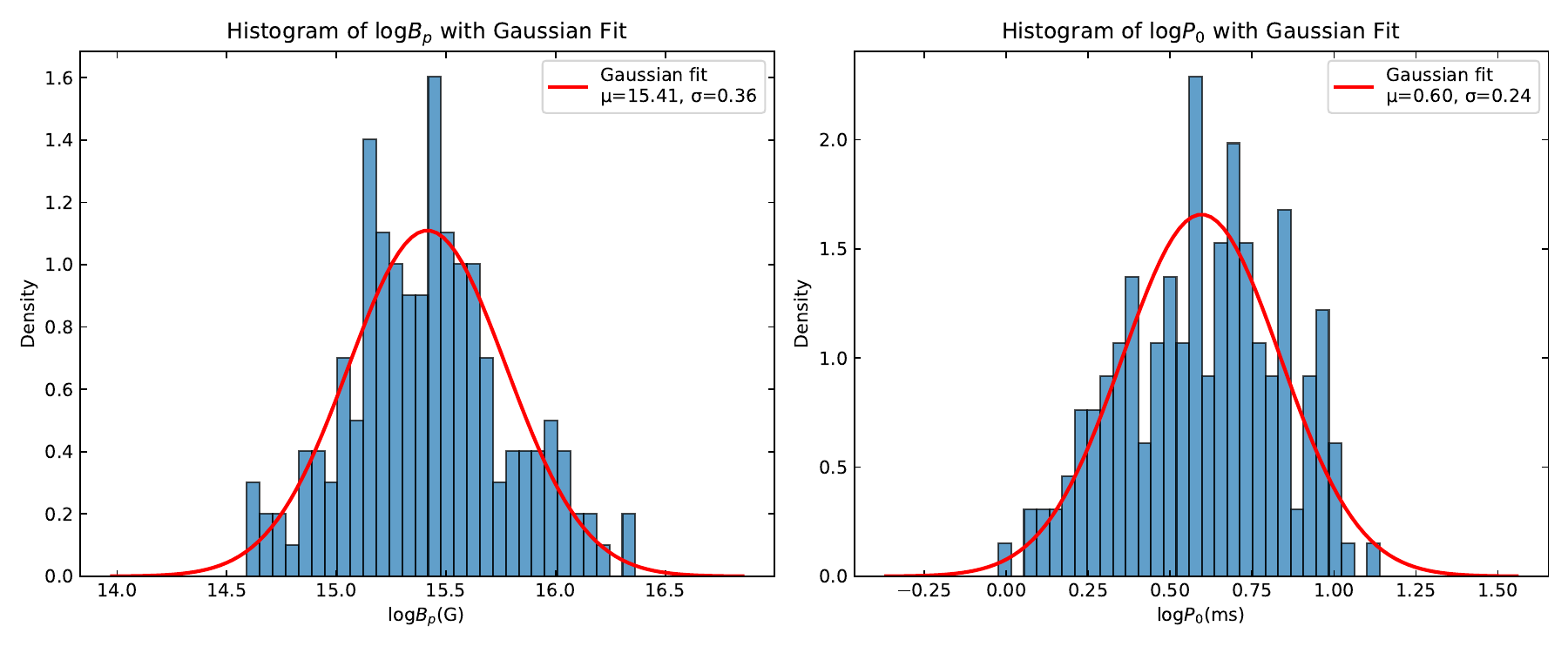}}\\
\caption{Distribution histograms of the surface magnetic field ($\log B_p$) and the initial rotation period ($\log P_0$) for magnetars, with Gaussian fits, under the assumption of an X-ray conversion efficiency of 0.5.}
\label{fig:distribution}
\end{figure*}

\begin{figure*}[ht!]
\centering
\resizebox{100mm}{!}{\includegraphics[]{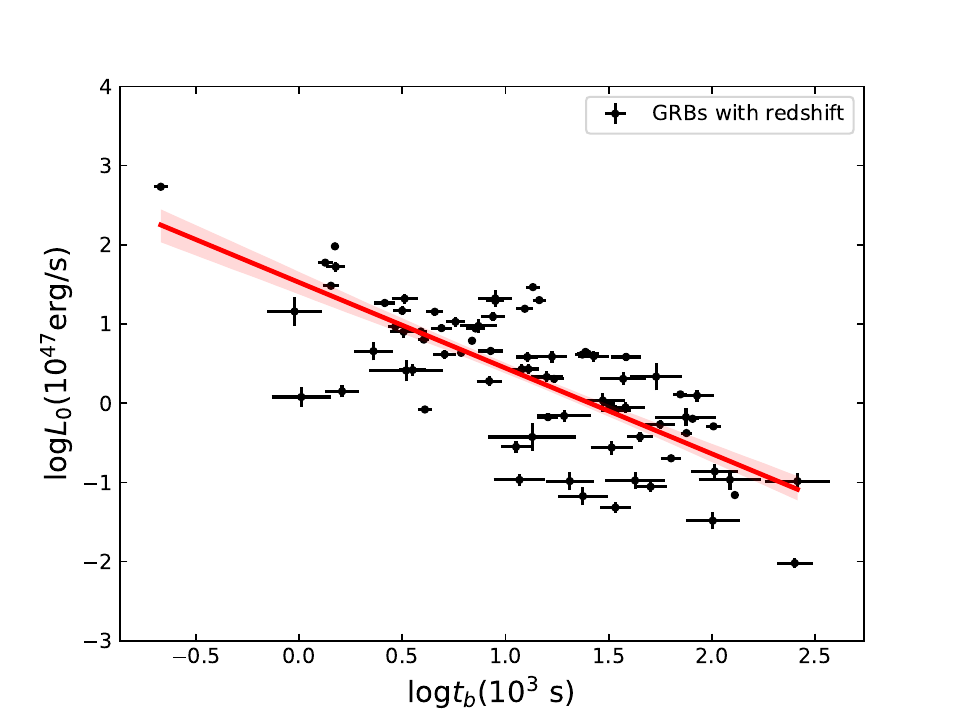}}%
\resizebox{90mm}{!}{\includegraphics[]{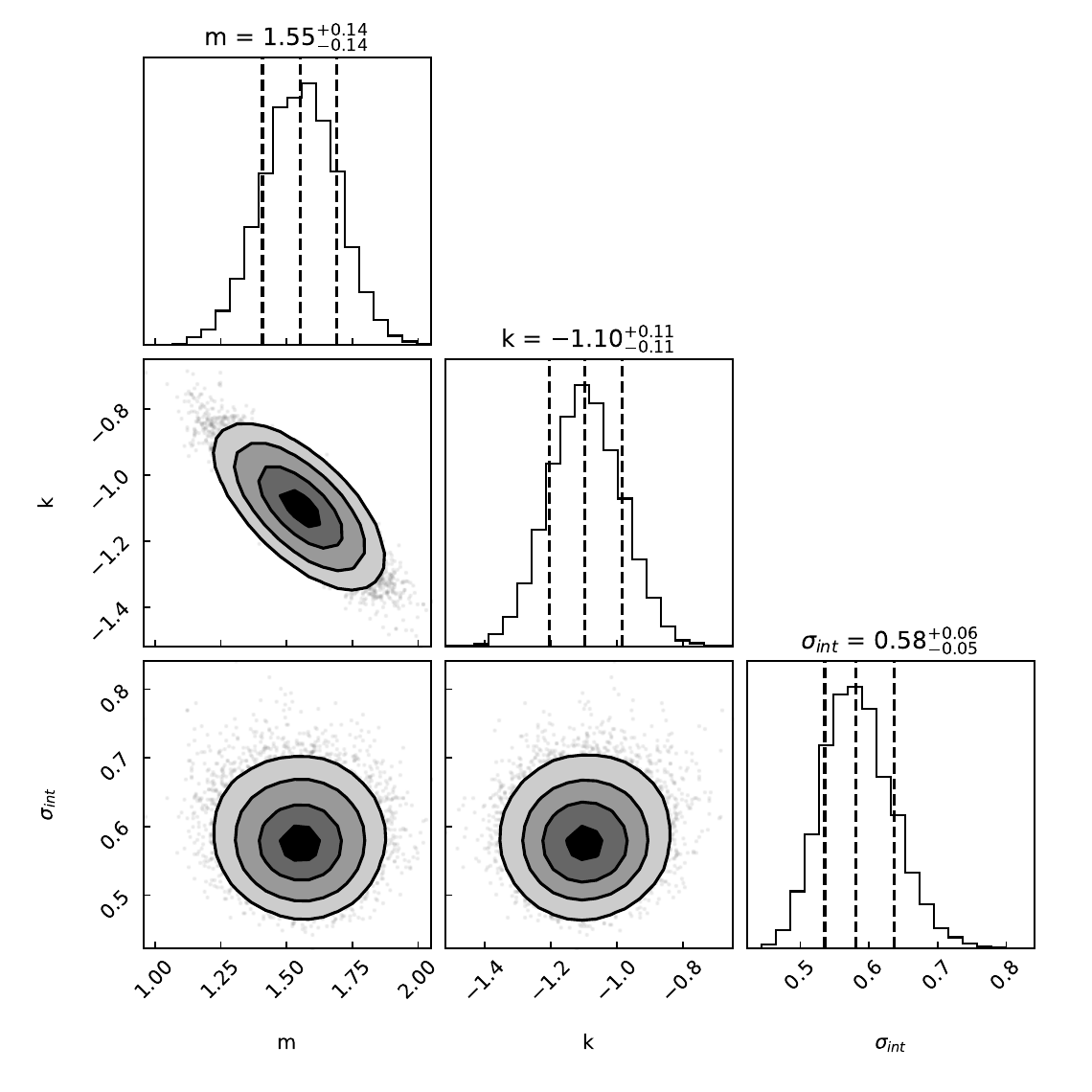}}\\
\resizebox{100mm}{!}{\includegraphics[]{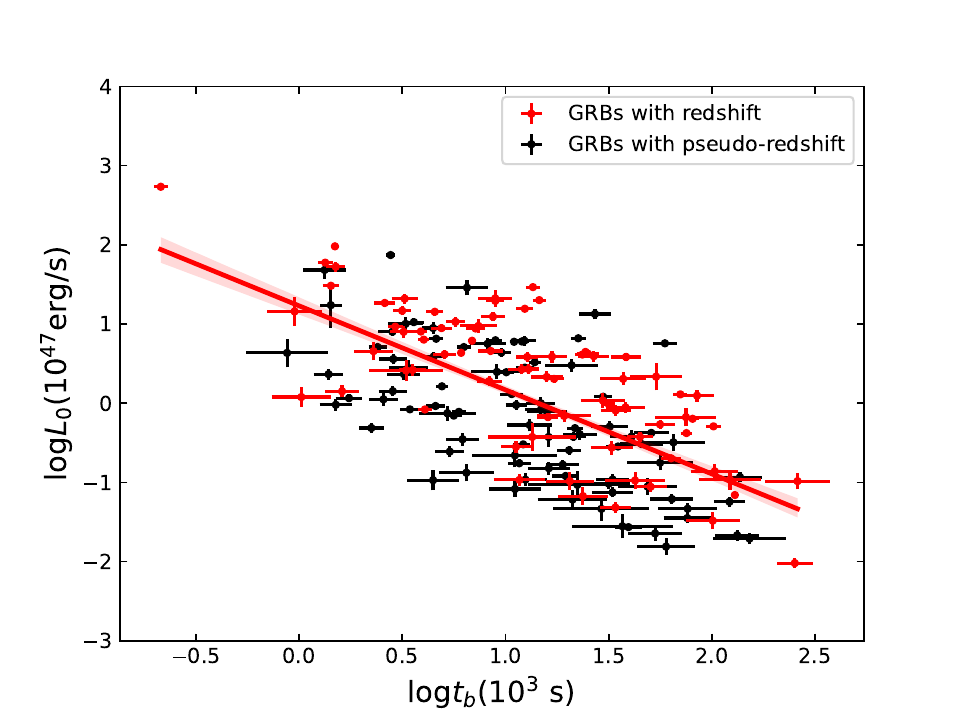}}%
\resizebox{90mm}{!}{\includegraphics[]{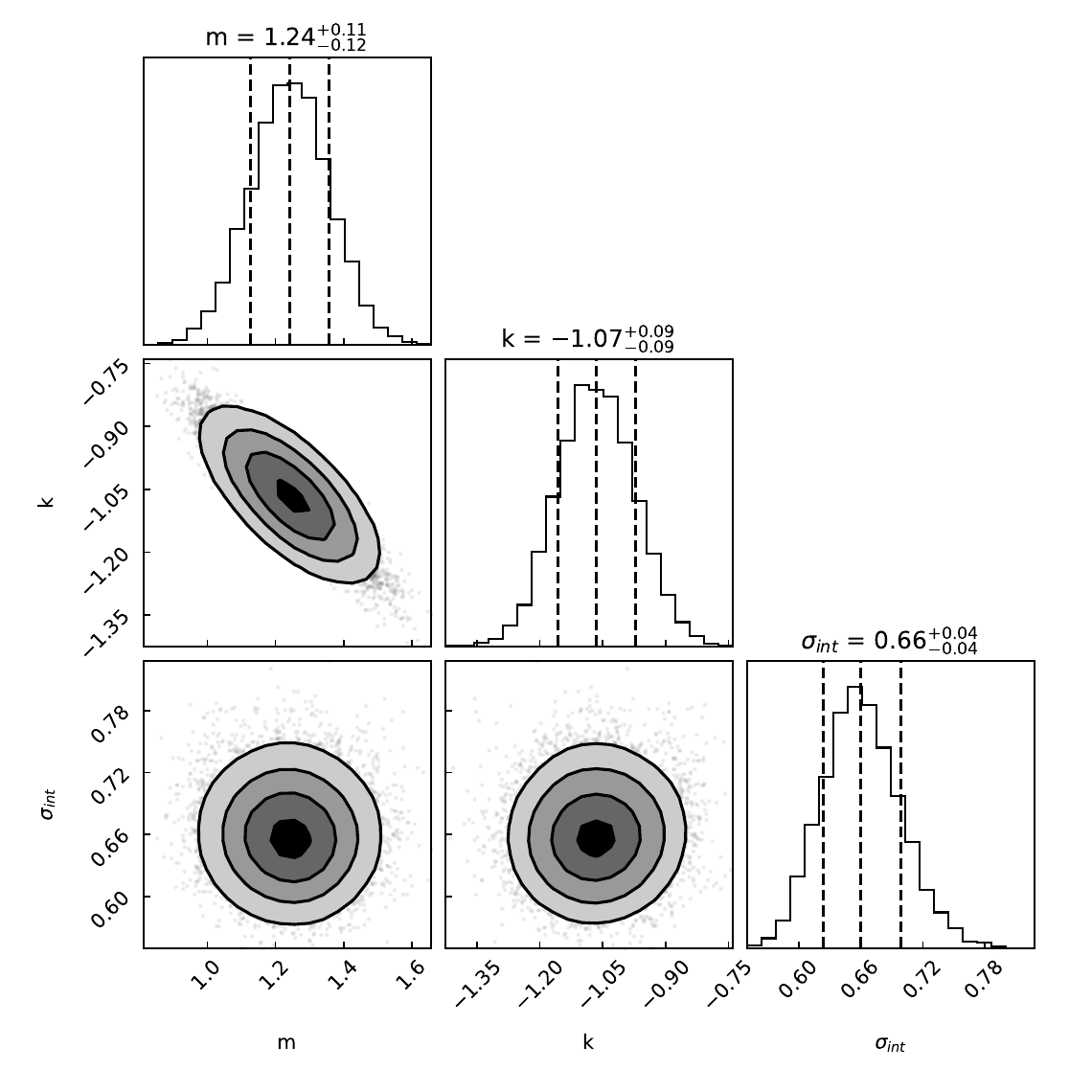}}\\

\caption{The Dainotti correlation. The best-fit curve is represented by the red solid line, with its corresponding \(1\sigma\) confidence interval indicated by the shaded region. Top: Dainotti correlation (Left) and its corner plot (Right) for GRBs with redshifts. Bottom: Dainotti correlation (Left) and its corner plot (Right) for all GRBs.}
\label{fig:Dainotti}
\end{figure*}

\begin{figure*}[ht!]
\centering
\resizebox{100mm}{!}{\includegraphics[]{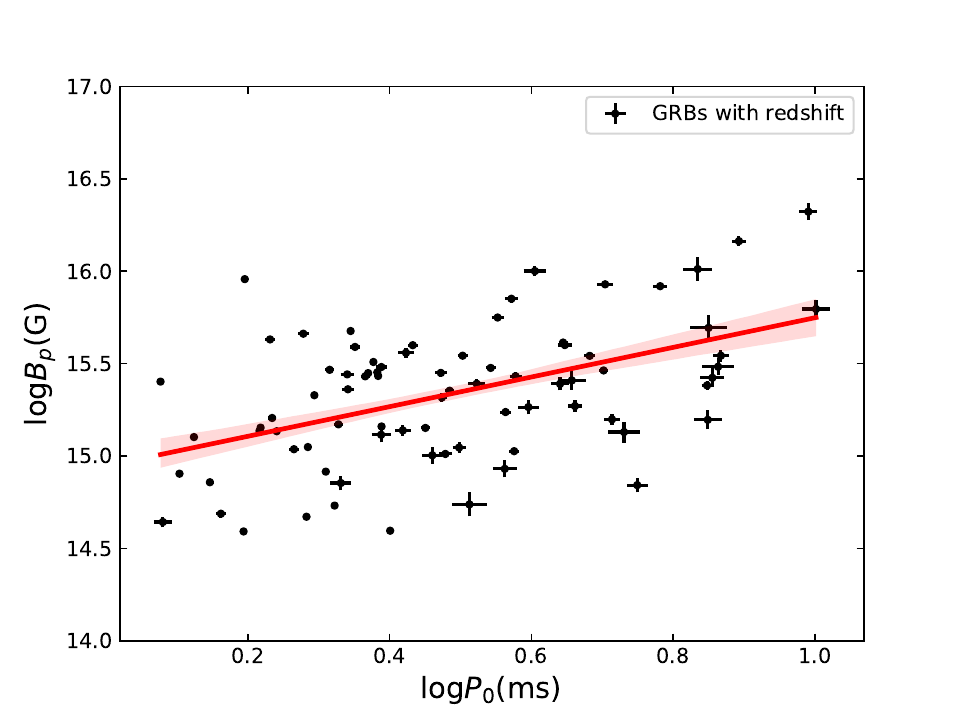}}%
\resizebox{90mm}{!}{\includegraphics[]{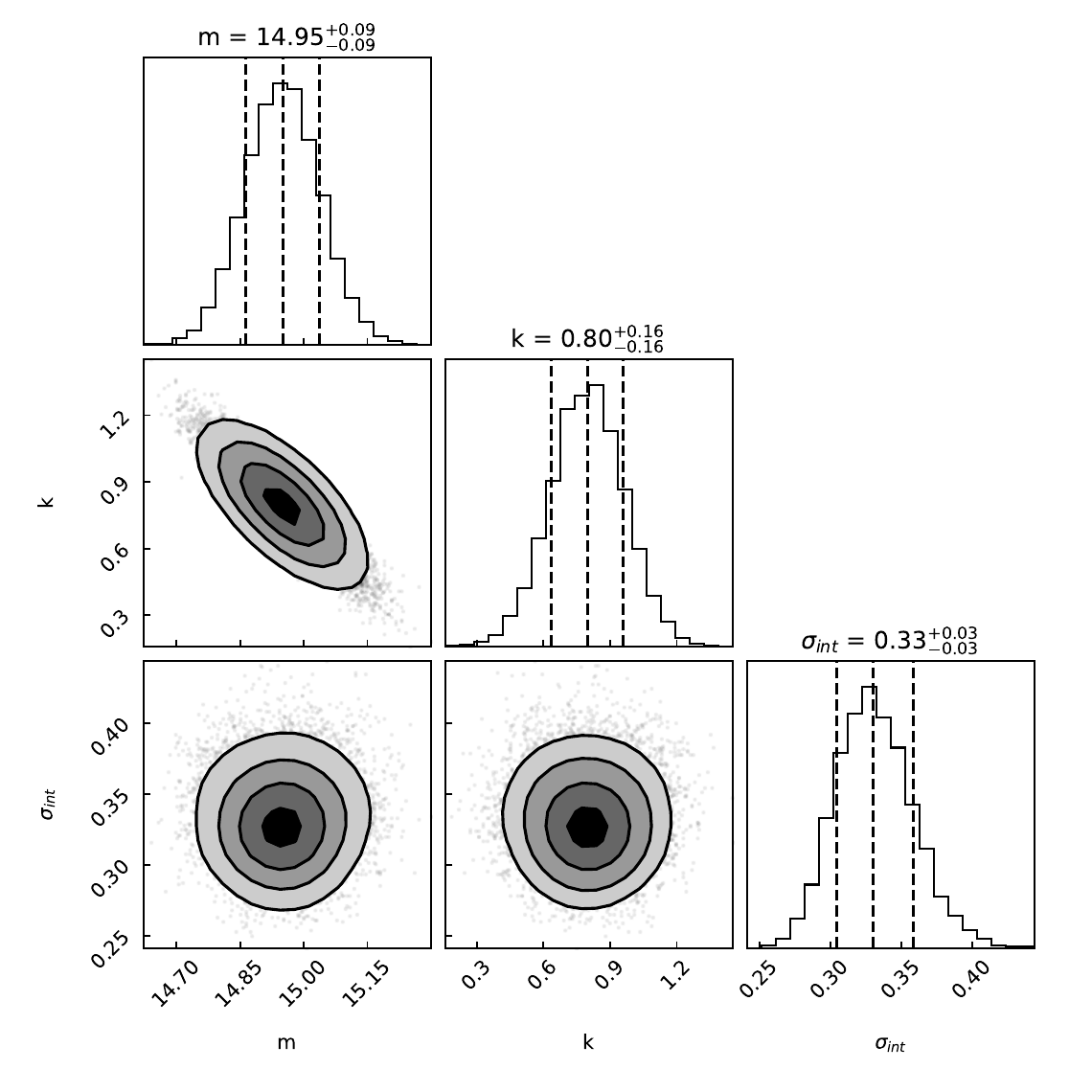}}\\
\resizebox{100mm}{!}{\includegraphics[]{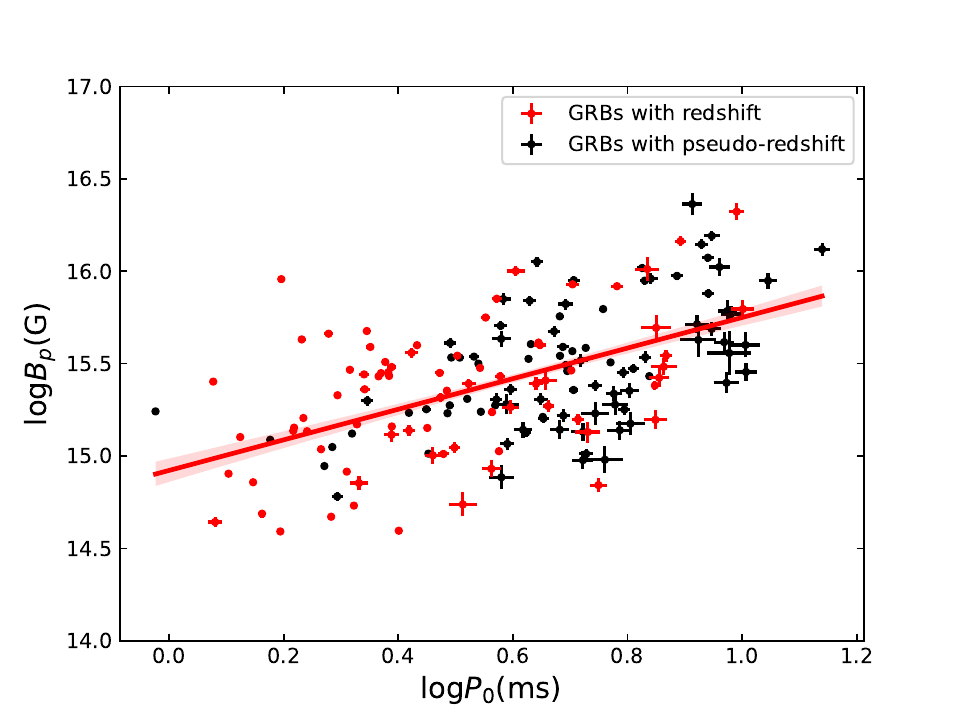}}%
\resizebox{90mm}{!}{\includegraphics[]{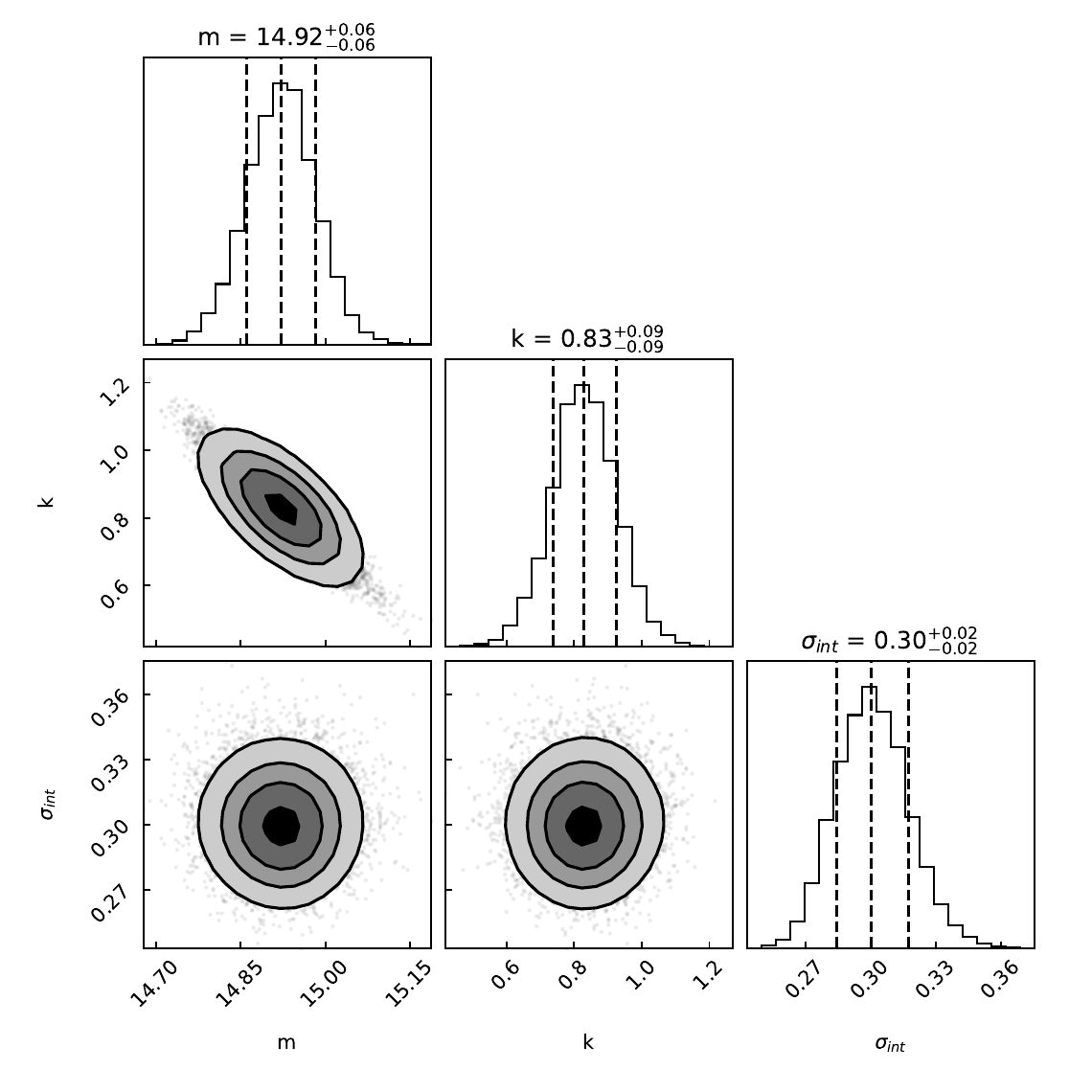}}\\

\caption{The $B_p-P_0$ correlation for GRB sample (assuming $\eta_x = 0.5$). The best-fit curve is represented by the red solid line, with its corresponding \(1\sigma\) confidence interval indicated by the shaded region. Top: $B_p-P_0$ correlation (Left) and its corner plot (Right) for GRBs with redshifts. Bottom: $B_p-P_0$ correlation (Left) and its corner plot (Right) for all GRBs.}
\label{fig:BP}
\end{figure*}

\begin{figure*}[ht!]
\centering
\resizebox{120mm}{!}{\includegraphics[]{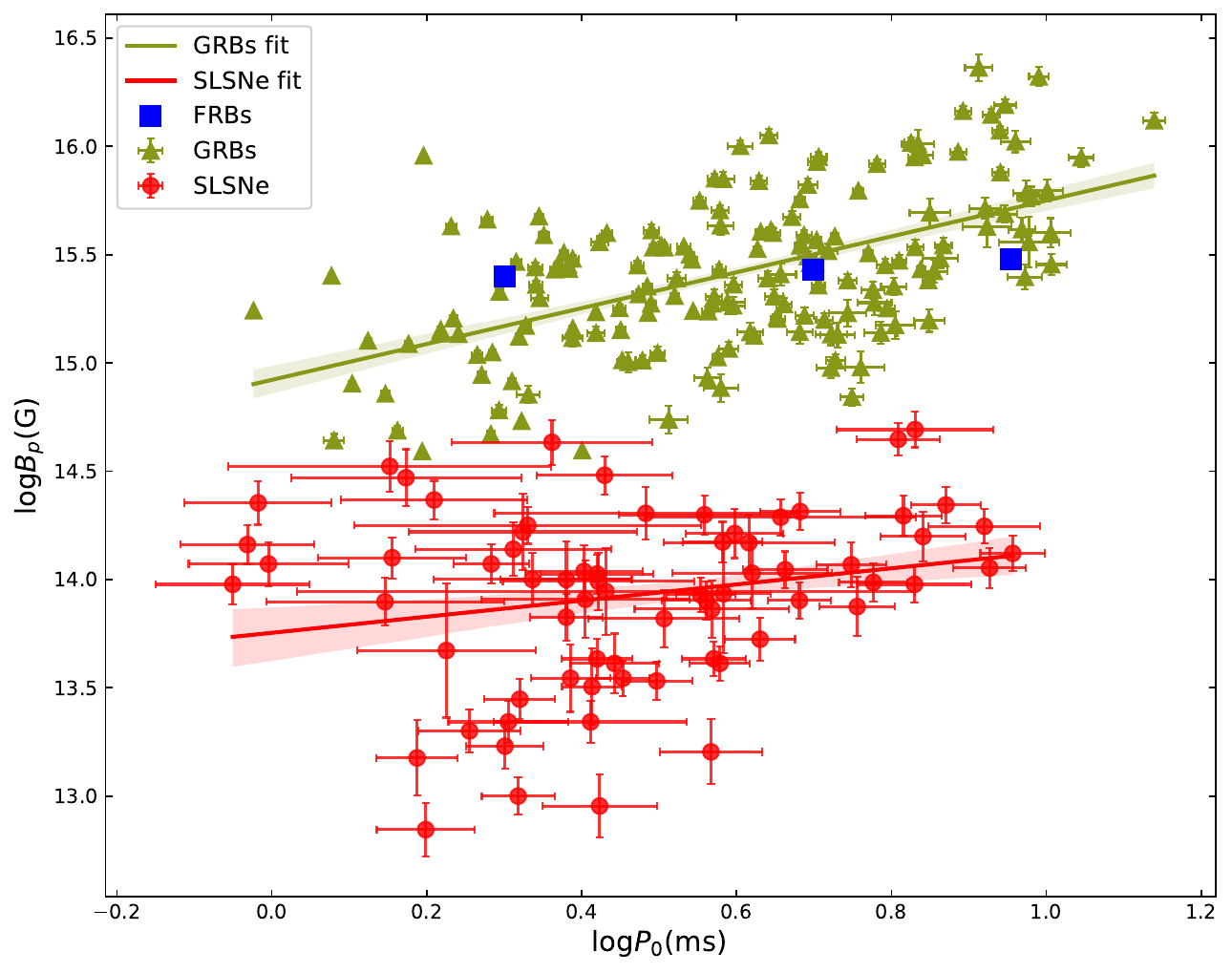}}\\

\caption{Comparison of the magnetar parameters among GRBs (assuming $\eta_x = 0.5$), SLSNe, and FRBs. The populations of GRBs and SLSNe can be roughly separated by the line $B_p = 10^{14.5} \, \mathrm{G}$. The blue rectangles mark the inferred magnetar parameter regions for three FRBs (FRB 20121102A (middle), FRB 20190520B (right), and FRB 20240114A (left)). The polar magnetic field \( B_p \) shown here is approximated by roughly taking one-tenth of the internal magnetic field $B_{\mathrm{int}}$ obtained from modelling the spectra and light curves of the persistent radio sources (i.e., $B_p \sim B_{\mathrm{int}} / 10$; Zhao et al., in prep.). The initial spin period, which lacks direct observational constraints, is assumed to be uniformly distributed between 1 and 10 ms. The magnetic field strengths of these FRB sources are notably similar to those of the GRB sample, suggesting that both populations may be powered by a similar type of strongly magnetized magnetar engine.}
\label{fig:SLSN}
\end{figure*}

\clearpage

\begin{table*}[htbp]
  \centering
  \caption{GRB Data Summary}
    \begin{tabular}{lccccccc}
    \hline
    \hline
    GRB & $z$ & T$_{90}$ & $\gamma$ & $F_0$ & $t_b$ & $P_0$ & $B_p$ \\
         &    & (s)       &   &$10^{-11}$(erg/cm$^2$/s) & $10^3$(s) & (ms)&$10^{15}$ (G) \\
    \hline
    050315 & 1.95 & 95.60 & 1.89 & 0.80$\pm$0.06 & 101.58$\pm$8.22 & 1.92$\pm$0.02 & 0.47$\pm$0.01 \\
    \hline
    050319 & 3.24 & 152.50 & 1.85 & 2.15$\pm$0.20 & 25.06$\pm$3.10 & 1.84$\pm$0.03 & 1.09$\pm$0.05 \\
    \hline
    050505 & 4.27 & 58.90 & 2.09 & 3.67$\pm$0.33 & 14.57$\pm$1.10 & 1.65$\pm$0.01 & 1.42$\pm$0.03 \\
    \hline
    050802 & 1.71 & 13.00 & 1.80 & 14.95$\pm$1.02 & 4.02$\pm$0.24 & 2.41$\pm$0.02 & 2.84$\pm$0.05 \\
    \hline
    050803 & 4.30 & 85.00 & 1.88 & 3.99$\pm$0.43 & 12.39$\pm$1.20 & 1.72$\pm$0.02 & 1.61$\pm$0.04 \\
    \hline
    050814 & 5.30 & 150.90 & 1.97 & 0.53$\pm$0.08 & 26.73$\pm$4.61 & 2.98$\pm$0.05 & 2.07$\pm$0.11 \\
    \hline
    050824 & 0.83 & 25.00 & 1.86 & 0.12$\pm$0.02 & 251.62$\pm$49.96 & 5.61$\pm$0.19 & 0.70$\pm$0.06 \\
    \hline
    051008 & 2.77 & $>$32.0 & 1.95 & 6.81$\pm$1.02 & 5.72$\pm$0.62 & 2.32$\pm$0.03 & 2.70$\pm$0.06 \\
    \hline
    051016B & 0.94 & 4.00 & 1.80 & 0.47$\pm$0.07 & 34.03$\pm$5.75 & 7.04$\pm$0.12 & 2.41$\pm$0.13 \\
    \hline
    060109* & 1.94 & 115.40 & 2.10 & 1.20$\pm$0.16 & 11.25$\pm$1.32 & 4.81$\pm$0.08 & 3.48$\pm$0.12 \\
    \hline
    060115 & 3.53 & 142.00 & 2.12 & 0.30$\pm$0.07 & 29.59$\pm$7.35 & 4.37$\pm$0.13 & 2.46$\pm$0.20 \\
    \hline
    060502A & 1.51 & 33.00 & 1.98 & 1.00$\pm$0.14 & 44.78$\pm$6.53 & 3.01$\pm$0.06 & 1.03$\pm$0.05 \\
    \hline
    060510A* & 1.37 & 20.40 & 1.76 & 24.43$\pm$1.52 & 10.99$\pm$0.56 & 1.93$\pm$0.02 & 1.12$\pm$0.02 \\
    \hline
    060526 & 3.22 & 298.20 & 1.89 & 1.05$\pm$0.17 & 15.82$\pm$2.79 & 3.33$\pm$0.09 & 2.46$\pm$0.16 \\
    \hline
    060604 & 2.68 & 10.00 & 2.08 & 0.47$\pm$0.07 & 34.13$\pm$5.77 & 3.66$\pm$0.06 & 1.73$\pm$0.09 \\
    \hline
    060605 & 3.78 & 79.10 & 1.89 & 4.82$\pm$0.60 & 4.54$\pm$0.42 & 2.71$\pm$0.04 & 3.98$\pm$0.10 \\
    \hline
    060707 & 3.43 & 68.00 & 2.00 & 0.44$\pm$0.08 & 84.57$\pm$16.13 & 2.14$\pm$0.07 & 0.71$\pm$0.07 \\
    \hline
    060714 & 2.71 & 115.00 & 2.00 & 2.88$\pm$0.37 & 8.47$\pm$1.21 & 2.97$\pm$0.06 & 2.82$\pm$0.13 \\
    \hline
    060729 & 0.54 & 116.00 & 1.92 & 2.33$\pm$0.09 & 128.92$\pm$4.71 & 2.52$\pm$0.01 & 0.39$\pm$0.00 \\
    \hline
    060807* & 0.98 & 54.00 & 1.97 & 2.36$\pm$0.19 & 12.17$\pm$0.93 & 4.95$\pm$0.07 & 2.88$\pm$0.08 \\
    \hline
    060814 & 1.92 & 146.00 & 1.99 & 2.97$\pm$0.37 & 17.18$\pm$1.81 & 2.45$\pm$0.02 & 1.44$\pm$0.04 \\
    \hline
    060906 & 3.69 & 43.50 & 2.10 & 0.69$\pm$0.10 & 11.99$\pm$1.87 & 4.44$\pm$0.10 & 3.98$\pm$0.21 \\
    \hline
    060908 & 1.88 & 19.30 & 2.07 & 20.35$\pm$8.73 & 0.95$\pm$0.29 & 4.03$\pm$0.15 & 10.02$\pm$0.62 \\
    \hline
    061121 & 1.31 & 81.00 & 1.83 & 26.11$\pm$1.11 & 6.88$\pm$0.32 & 1.65$\pm$0.01 & 1.37$\pm$0.02 \\
    \hline
    061222A & 2.09 & 71.40 & 1.84 & 6.29$\pm$0.11 & 24.40$\pm$0.62 & 1.27$\pm$0.01 & 0.80$\pm$0.02 \\
    \hline
    070103 & 2.26 & 19.00 & 2.22 & 1.98$\pm$0.33 & 3.55$\pm$0.52 & 6.05$\pm$0.14 & 8.30$\pm$0.38 \\
    \hline
    070129 & 2.34 & 460.00 & 2.07 & 0.45$\pm$0.06 & 56.00$\pm$9.58 & 3.15$\pm$0.07 & 1.11$\pm$0.07 \\
    \hline
    070306 & 1.50 & 209.50 & 1.80 & 3.12$\pm$0.22 & 31.57$\pm$2.07 & 2.04$\pm$0.02 & 0.82$\pm$0.02 \\
    \hline
    070420* & 0.74 & 76.50 & 1.85 & 12.78$\pm$1.03 & 5.94$\pm$0.47 & 5.33$\pm$0.06 & 3.85$\pm$0.10 \\
    \hline
    070521 & 2.09 & 37.90 & 1.85 & 20.66$\pm$2.83 & 3.16$\pm$0.32 & 2.07$\pm$0.03 & 2.93$\pm$0.07 \\
    \hline
    070721B & 3.63 & 259.20 & 1.46 & 37.93$\pm$5.48 & 1.51$\pm$0.16 & 1.70$\pm$0.03 & 4.27$\pm$0.11 \\
    \hline
    070810A & 2.17 & 11.00 & 2.06 & 4.57$\pm$1.23 & 2.29$\pm$0.49 & 5.06$\pm$0.13 & 8.49$\pm$0.38 \\
    \hline
    080229A* & 0.99 & 64.00 & 1.66 & 48.05$\pm$3.05 & 6.28$\pm$0.48 & 1.50$\pm$0.01 & 1.22$\pm$0.03 \\
    \hline
    080310 & 2.43 & 365.00 & 2.09 & 1.98$\pm$0.27 & 12.90$\pm$1.54 & 3.05$\pm$0.04 & 2.26$\pm$0.07 \\
    \hline
    080516* & 3.72 & 5.80 & 2.29 & 1.06$\pm$0.19 & 8.19$\pm$1.64 & 4.70$\pm$0.12 & 4.72$\pm$0.31 \\
    \hline
    080707 & 1.23 & 27.10 & 1.92 & 0.48$\pm$0.12 & 20.46$\pm$5.52 & 7.36$\pm$0.18 & 3.49$\pm$0.27 \\
    \hline
    081008 & 1.97 & 185.50 & 1.98 & 5.78$\pm$0.79 & 5.07$\pm$0.66 & 3.18$\pm$0.05 & 3.49$\pm$0.13 \\
    \hline
    081126* & 2.22 & 54.00 & 1.87 & 10.61$\pm$1.78 & 4.46$\pm$0.57 & 3.40$\pm$0.07 & 3.45$\pm$0.12 \\
    \hline
    090407 & 1.45 & 310.00 & 2.22 & 0.48$\pm$0.04 & 63.33$\pm$7.22 & 3.76$\pm$0.06 & 1.06$\pm$0.04 \\
    \hline
    090418A & 1.61 & 56.00 & 1.91 & 22.67$\pm$2.37 & 2.92$\pm$0.24 & 2.38$\pm$0.03 & 3.23$\pm$0.07 \\
    \hline
    090516 & 4.11 & 140.00 & 2.03 & 4.36$\pm$0.58 & 8.92$\pm$0.90 & 1.97$\pm$0.02 & 2.13$\pm$0.05 \\
    \hline
    090529 & 2.62 & $>$ 100 & 2.12 & 0.06$\pm$0.02 & 122.10$\pm$42.81 & 5.38$\pm$0.27 & 1.35$\pm$0.18 \\
    \hline
    090530 & 1.27 & 48.00 & 1.92 & 0.38$\pm$0.06 & 50.42$\pm$9.29 & 5.17$\pm$0.13 & 1.58$\pm$0.11 \\
    \hline
    090728* & 0.99 & 59.00 & 1.79 & 4.15$\pm$0.62 & 2.24$\pm$0.30 & 8.71$\pm$0.21 & 11.83$\pm$0.56 \\
    \hline
    100424A & 2.47 & 104.00 & 1.97 & 444.31$\pm$41.96 & 0.21$\pm$0.02 & 1.57$\pm$0.01 & 9.06$\pm$0.17 \\
    \hline
    100425A & 1.75 & 37.00 & 2.20 & 0.16$\pm$0.04 & 42.37$\pm$13.99 & 7.16$\pm$0.28 & 2.65$\pm$0.32 \\
    \hline
    100614A* & 1.52 & 225.00 & 1.86 & 0.17$\pm$0.03 & 121.45$\pm$21.08 & 5.34$\pm$0.15 & 1.03$\pm$0.07 \\
    \hline
    \end{tabular}%
    \begin{tablenotes}
    \footnotesize
    \item \textbf{Note.} \\
    1.GRB names with * indicate data from specific selection criteria.\\
    2.The results of $B_p$ and $P_0$ were obtained under the condition of $\eta_x=0.5$.
    \end{tablenotes}
  \label{tab:grb_data_1}%
\end{table*}%

\addtocounter{table}{-1}

\begin{table*}[htbp]
  \centering
  \caption{GRB Data Summary}
    \begin{tabular}{lccccccc}
    \hline
    GRB & $z$ & T$_{90}$ & $\gamma$ & $F_0$ & $t_b$ & $P_0$ & $B_p$ \\
         &    & (s)       &   &$10^{-11}$(erg/cm$^2$/s) & $10^3$(s) & (ms)&$10^{15}$ (G) \\
    \hline
    100725B* & 0.96 & 200.00 & 2.56 & 1.40$\pm$0.18 & 20.30$\pm$2.68 & 5.08$\pm$0.10 & 2.28$\pm$0.10 \\
    \hline
    100727A* & 1.70 & 84.00 & 2.08 & 0.55$\pm$0.07 & 45.57$\pm$10.89 & 4.15$\pm$0.16 & 1.39$\pm$0.14 \\
    \hline
    100805A* & 9.00 & 15.00 & 1.87 & 0.68$\pm$0.11 & 27.09$\pm$4.79 & 2.22$\pm$0.05 & 1.99$\pm$0.12 \\
    \hline
    100814A & 1.44 & 174.50 & 1.82 & 2.20$\pm$0.15 & 80.47$\pm$5.57 & 1.56$\pm$0.01 & 0.39$\pm$0.01 \\
    \hline
    100901A & 1.41 & 439.00 & 1.97 & 1.34$\pm$0.08 & 75.16$\pm$4.82 & 2.10$\pm$0.02 & 0.54$\pm$0.01 \\
    \hline
    100906A & 1.73 & 114.40 & 1.93 & 16.23$\pm$1.63 & 3.89$\pm$0.27 & 2.34$\pm$0.03 & 2.81$\pm$0.05 \\
    \hline
    110102A* & 1.53 & 264.00 & 2.03 & 8.14$\pm$0.65 & 13.84$\pm$0.95 & 3.50$\pm$0.03 & 1.74$\pm$0.03 \\
    \hline
    110208A* & 0.97 & 37.40 & 1.93 & 0.86$\pm$0.26 & 4.46$\pm$1.29 & 13.79$\pm$0.45 & 13.16$\pm$1.06 \\
    \hline
    110210A* & 0.95 & 233.00 & 1.92 & 0.19$\pm$0.03 & 132.31$\pm$32.71 & 5.27$\pm$0.22 & 0.95$\pm$0.11 \\
    \hline
    110808A & 1.35 & 48.00 & 2.22 & 0.09$\pm$0.02 & 100.78$\pm$30.35 & 7.06$\pm$0.32 & 1.57$\pm$0.19 \\
    \hline
    110820A* & 0.98 & 256.00 & 1.77 & 0.14$\pm$0.03 & 59.88$\pm$19.27 & 9.39$\pm$0.48 & 2.49$\pm$0.33 \\
    \hline
    111008A & 4.99 & 63.46 & 1.87 & 5.45$\pm$0.47 & 13.56$\pm$1.05 & 1.33$\pm$0.01 & 1.27$\pm$0.03 \\
    \hline
    111123A & 3.15 & 290.00 & 2.55 & 0.78$\pm$0.14 & 16.79$\pm$2.74 & 3.78$\pm$0.07 & 2.70$\pm$0.13 \\
    \hline
    120308A* & 2.65 & 60.60 & 1.54 & 7.88$\pm$0.67 & 4.61$\pm$0.35 & 3.11$\pm$0.04 & 3.41$\pm$0.08 \\
    \hline
    120324A* & 1.88 & 118.00 & 2.05 & 8.70$\pm$0.82 & 12.00$\pm$1.00 & 3.06$\pm$0.03 & 1.70$\pm$0.04 \\
    \hline
    120327A & 2.81 & 62.90 & 1.71 & 15.80$\pm$1.82 & 2.60$\pm$0.30 & 2.25$\pm$0.04 & 3.89$\pm$0.15 \\
    \hline
    120404A & 2.88 & 38.70 & 1.90 & 4.97$\pm$0.91 & 3.21$\pm$0.49 & 3.57$\pm$0.07 & 5.62$\pm$0.23 \\
    \hline
    120521B* & 9.57 & 31.40 & 2.00 & 0.95$\pm$0.21 & 6.50$\pm$1.55 & 3.83$\pm$0.12 & 7.07$\pm$0.55 \\
    \hline
    120521C* & 2.87 & 26.70 & 1.82 & 0.26$\pm$0.07 & 16.15$\pm$5.05 & 8.34$\pm$0.39 & 5.16$\pm$0.58 \\
    \hline
    120811C & 2.67 & 26.80 & 1.81 & 7.41$\pm$0.84 & 4.89$\pm$0.55 & 2.45$\pm$0.05 & 3.03$\pm$0.12 \\
    \hline
    121128A & 2.20 & 23.30 & 1.85 & 37.76$\pm$3.55 & 1.43$\pm$0.13 & 2.21$\pm$0.03 & 4.75$\pm$0.14 \\
    \hline
    121217A* & 1.55 & 778.00 & 1.83 & 3.48$\pm$0.36 & 29.69$\pm$3.01 & 2.83$\pm$0.03 & 1.03$\pm$0.03 \\
    \hline
    130327A* & 1.03 & 9.00 & 2.25 & 0.13$\pm$0.03 & 53.00$\pm$15.88 & 10.16$\pm$0.45 & 2.85$\pm$0.32 \\
    \hline
    130504A* & 2.29 & 50.00 & 1.91 & 10.62$\pm$2.05 & 3.28$\pm$0.65 & 3.09$\pm$0.07 & 4.09$\pm$0.25 \\
    \hline
    130511A & 1.30 & 5.43 & 1.87 & 5.02$\pm$1.44 & 1.03$\pm$0.34 & 9.78$\pm$0.29 & 20.99$\pm$2.18 \\
    \hline
    130606A & 5.91 & 276.60 & 1.86 & 2.75$\pm$0.65 & 8.91$\pm$1.68 & 2.19$\pm$0.05 & 2.76$\pm$0.13 \\
    \hline
    130609B* & 9.95 & 210.60 & 1.95 & 67.34$\pm$3.20 & 2.77$\pm$0.14 & 0.95$\pm$0.01 & 1.74$\pm$0.03 \\
    \hline
    131024A* & 1.69 & 112.00 & 1.92 & 1.14$\pm$0.21 & 13.09$\pm$3.30 & 5.23$\pm$0.15 & 3.29$\pm$0.28 \\
    \hline
    140108A* & 1.58 & 94.00 & 2.00 & 14.63$\pm$1.07 & 8.94$\pm$0.62 & 2.62$\pm$0.02 & 1.71$\pm$0.04 \\
    \hline
    140114A & 3.00 & 139.70 & 2.20 & 0.33$\pm$0.05 & 38.09$\pm$8.46 & 3.94$\pm$0.13 & 1.84$\pm$0.16 \\
    \hline
    140323A* & 0.74 & 104.90 & 2.00 & 24.31$\pm$1.78 & 4.92$\pm$0.34 & 4.24$\pm$0.05 & 3.36$\pm$0.08 \\
    \hline
    140430A & 1.60 & 173.60 & 2.13 & 0.55$\pm$0.11 & 32.64$\pm$7.62 & 4.59$\pm$0.10 & 1.86$\pm$0.14 \\
    \hline
    140518A & 4.71 & 60.50 & 2.09 & 3.02$\pm$0.43 & 3.25$\pm$0.47 & 3.73$\pm$0.08 & 7.09$\pm$0.35 \\
    \hline
    140709A* & 1.22 & 98.60 & 1.97 & 4.48$\pm$0.81 & 14.81$\pm$2.39 & 3.71$\pm$0.05 & 1.88$\pm$0.07 \\
    \hline
    140719A* & 2.28 & 48.00 & 2.14 & 0.41$\pm$0.06 & 31.97$\pm$6.61 & 4.45$\pm$0.12 & 2.03$\pm$0.15 \\
    \hline
    140730A* & 1.02 & 41.30 & 2.01 & 0.90$\pm$0.22 & 6.46$\pm$1.98 & 11.09$\pm$0.42 & 8.88$\pm$0.89 \\
    \hline
    141004A* & 6.05 & 3.92 & 1.77 & 7.10$\pm$1.81 & 1.33$\pm$0.32 & 4.38$\pm$0.11 & 11.23$\pm$0.74 \\
    \hline
    141026A & 3.35 & 146.00 & 1.92 & 0.28$\pm$0.08 & 74.74$\pm$25.44 & 2.89$\pm$0.09 & 1.01$\pm$0.11 \\
    \hline
    150202A* & 1.02 & 25.70 & 2.03 & 6.50$\pm$1.17 & 1.51$\pm$0.27 & 8.49$\pm$0.21 & 13.98$\pm$0.86 \\
    \hline
    150428B* & 0.94 & 130.90 & 1.89 & 0.32$\pm$0.05 & 75.84$\pm$19.99 & 6.11$\pm$0.30 & 1.38$\pm$0.16 \\
    \hline
    150527A* & 3.27 & 112.00 & 1.90 & 1.39$\pm$0.26 & 20.88$\pm$6.12 & 3.88$\pm$0.20 & 1.91$\pm$0.23 \\
    \hline
    150615A* & 1.01 & 27.60 & 2.47 & 0.24$\pm$0.04 & 76.18$\pm$25.16 & 6.38$\pm$0.37 & 1.50$\pm$0.22 \\
    \hline
    150626A* & 0.97 & 144.00 & 2.08 & 0.55$\pm$0.08 & 32.97$\pm$7.58 & 6.36$\pm$0.24 & 2.25$\pm$0.21 \\
    \hline
    150720A* & 1.13 & 151.00 & 2.10 & 0.14$\pm$0.05 & 36.81$\pm$20.58 & 9.49$\pm$0.85 & 3.61$\pm$0.98 \\
    \hline
    150724A* & 0.90 & 280.00 & 2.22 & 0.16$\pm$0.03 & 151.83$\pm$61.08 & 5.76$\pm$0.41 & 0.95$\pm$0.16 \\
    \hline
    151027A & 0.81 & 129.69 & 1.98 & 53.04$\pm$1.29 & 6.09$\pm$0.19 & 1.74$\pm$0.01 & 1.36$\pm$0.02 \\
    \hline
    151027B & 4.06 & 80.00 & 1.79 & 0.68$\pm$0.13 & 37.10$\pm$9.45 & 2.44$\pm$0.07 & 1.30$\pm$0.12 \\
    \hline
    \end{tabular}%
    \begin{tablenotes}
    \footnotesize
    \item \textbf{Note.} \\
    1.GRB names with * indicate data from specific selection criteria.\\
    2.The results of $B_p$ and $P_0$ were obtained under the condition of $\eta_x=0.5$.
    \end{tablenotes}
  \label{tab:grb_data_2}%
\end{table*}%

\addtocounter{table}{-1}

\begin{table*}[htbp]
  \centering
  \caption{GRB Data Summary}
    \begin{tabular}{lccccccc}
    \hline
    \hline
    GRB & $z$ & T$_{90}$ & $\gamma$ & $F_0$ & $t_b$ & $P_0$ & $B_p$ \\
         &    & (s)       &   &$10^{-11}$(erg/cm$^2$/s) & $10^3$(s) & (ms)&$10^{15}$ (G) \\
    \hline
    151031A & 1.17 & 5.00 & 2.11 & 0.49$\pm$0.09 & 11.67$\pm$3.38 & 10.03$\pm$0.46 & 6.24$\pm$0.74 \\
    \hline
    151112A & 4.10 & 19.30 & 2.28 & 0.57$\pm$0.07 & 38.33$\pm$6.57 & 2.62$\pm$0.07 & 1.38$\pm$0.09 \\
    \hline
    151215A & 2.59 & 17.80 & 2.17 & 1.47$\pm$0.46 & 3.30$\pm$1.36 & 6.83$\pm$0.32 & 10.27$\pm$1.51 \\
    \hline
    151228B* & 1.60 & 48.00 & 1.90 & 0.55$\pm$0.19 & 11.09$\pm$5.26 & 9.42$\pm$0.35 & 6.09$\pm$0.82 \\
    \hline
    160119A* & 1.21 & 116.00 & 2.07 & 1.62$\pm$0.21 & 21.30$\pm$2.91 & 6.89$\pm$0.09 & 2.71$\pm$0.11 \\
    \hline
    160121A & 1.96 & 12.00 & 2.21 & 0.76$\pm$0.14 & 19.21$\pm$5.82 & 4.54$\pm$0.21 & 2.56$\pm$0.30 \\
    \hline
    160227A & 2.38 & 316.50 & 1.67 & 1.67$\pm$0.08 & 70.23$\pm$5.49 & 1.45$\pm$0.02 & 0.49$\pm$0.02 \\
    \hline
    160327A* & 1.01 & 28.00 & 1.91 & 1.84$\pm$0.31 & 5.37$\pm$0.87 & 8.72$\pm$0.21 & 7.58$\pm$0.41 \\
    \hline
    160417A* & 1.03 & 15.00 & 2.04 & 0.40$\pm$0.11 & 21.10$\pm$8.22 & 9.30$\pm$0.36 & 4.13$\pm$0.53 \\
    \hline
    160504A* & 1.01 & 53.90 & 1.98 & 0.86$\pm$0.11 & 19.55$\pm$2.97 & 6.45$\pm$0.14 & 2.96$\pm$0.17 \\
    \hline
    160630A* & 0.96 & 29.50 & 1.75 & 8.68$\pm$1.03 & 4.58$\pm$0.48 & 4.28$\pm$0.06 & 4.03$\pm$0.12 \\
    \hline
    161014A & 2.82 & 18.30 & 1.83 & 42.79$\pm$4.77 & 1.34$\pm$0.11 & 1.90$\pm$0.03 & 4.59$\pm$0.12 \\
    \hline
    161017A & 2.01 & 216.30 & 1.99 & 5.04$\pm$0.76 & 12.73$\pm$1.57 & 2.13$\pm$0.03 & 1.48$\pm$0.05 \\
    \hline
    161108A* & 1.84 & 105.10 & 1.96 & 0.20$\pm$0.03 & 136.50$\pm$33.33 & 3.80$\pm$0.20 & 0.77$\pm$0.11 \\
    \hline
    161214B* & 1.93 & 24.80 & 2.03 & 0.53$\pm$0.10 & 40.78$\pm$9.63 & 4.88$\pm$0.13 & 1.65$\pm$0.14 \\
    \hline
    170111A* & 1.04 & 29.90 & 1.72 & 0.71$\pm$0.17 & 48.70$\pm$16.15 & 4.81$\pm$0.19 & 1.39$\pm$0.16 \\
    \hline
    170202A & 3.65 & 46.20 & 1.98 & 3.96$\pm$0.51 & 8.69$\pm$1.16 & 2.19$\pm$0.04 & 2.30$\pm$0.10 \\
    \hline
    170317A* & 1.53 & 11.94 & 2.04 & 3.47$\pm$0.50 & 2.84$\pm$0.45 & 6.76$\pm$0.13 & 8.89$\pm$0.45 \\
    \hline
    170519A & 0.82 & 216.40 & 1.97 & 3.39$\pm$0.61 & 11.24$\pm$1.84 & 5.03$\pm$0.07 & 2.90$\pm$0.13 \\
    \hline
    170903A* & 1.70 & 29.20 & 2.12 & 0.74$\pm$0.09 & 50.79$\pm$10.08 & 3.89$\pm$0.11 & 1.16$\pm$0.09 \\
    \hline
    170912B* & 1.11 & 19.60 & 1.64 & 17.07$\pm$2.71 & 1.39$\pm$0.23 & 5.08$\pm$0.12 & 8.94$\pm$0.50 \\
    \hline
    171120A* & 1.11 & 64.00 & 1.77 & 1.90$\pm$0.18 & 35.04$\pm$4.54 & 4.20$\pm$0.08 & 1.34$\pm$0.06 \\
    \hline
    171222A & 2.41 & 174.80 & 1.99 & 0.09$\pm$0.02 & 259.33$\pm$94.44 & 3.26$\pm$0.18 & 0.55$\pm$0.08 \\
    \hline
    180102A* & 2.30 & 10.80 & 2.18 & 0.57$\pm$0.13 & 5.23$\pm$1.66 & 9.13$\pm$0.40 & 10.54$\pm$1.21 \\
    \hline
    180111A* & 4.82 & 50.60 & 1.63 & 5.31$\pm$3.52 & 1.42$\pm$0.17 & 8.85$\pm$0.29 & 15.52$\pm$1.01 \\
    \hline
    180325A & 2.25 & 94.10 & 1.79 & 121.07$\pm$6.79 & 1.49$\pm$0.07 & 1.19$\pm$0.01 & 2.53$\pm$0.04 \\
    \hline
    180329B & 2.00 & 210.00 & 1.87 & 2.89$\pm$0.43 & 8.33$\pm$1.16 & 3.49$\pm$0.06 & 3.00$\pm$0.12 \\
    \hline
    180331B* & 0.50 & 147.00 & 1.77 & 1.14$\pm$0.11 & 39.35$\pm$6.04 & 6.23$\pm$0.15 & 1.78$\pm$0.11 \\
    \hline
    180404A & 1.00 & 35.20 & 1.76 & 0.57$\pm$0.15 & 23.69$\pm$6.55 & 7.30$\pm$0.38 & 3.04$\pm$0.32 \\
    \hline
    180411A* & 1.65 & 77.50 & 1.86 & 10.52$\pm$1.36 & 9.54$\pm$1.00 & 3.31$\pm$0.04 & 2.04$\pm$0.05 \\
    \hline
    180425A* & 3.04 & 11.20 & 1.99 & 0.16$\pm$0.04 & 64.99$\pm$22.95 & 5.54$\pm$0.31 & 1.70$\pm$0.24 \\
    \hline
    180623A* & 1.17 & 114.90 & 1.99 & 25.17$\pm$3.11 & 2.40$\pm$0.24 & 4.81$\pm$0.08 & 5.70$\pm$0.17 \\
    \hline
    180626A* & 0.69 & 30.07 & 1.97 & 3.13$\pm$0.35 & 11.66$\pm$1.50 & 5.89$\pm$0.09 & 3.21$\pm$0.13 \\
    \hline
    180706A* & 1.07 & 42.70 & 1.99 & 0.94$\pm$0.18 & 16.20$\pm$3.81 & 6.77$\pm$0.16 & 3.42$\pm$0.26 \\
    \hline
    180818B* & 1.55 & 134.40 & 1.78 & 2.37$\pm$0.42 & 15.86$\pm$3.55 & 3.73$\pm$0.09 & 2.02$\pm$0.16 \\
    \hline
    181103A* & 0.96 & 8.38 & 2.32 & 0.62$\pm$0.10 & 31.56$\pm$7.56 & 5.97$\pm$0.20 & 2.17$\pm$0.20 \\
    \hline
    190106A & 1.86 & 76.80 & 1.95 & 6.81$\pm$0.47 & 23.57$\pm$1.85 & 1.40$\pm$0.02 & 0.72$\pm$0.02 \\
    \hline
    190114A & 3.38 & 66.60 & 1.83 & 4.19$\pm$0.46 & 7.16$\pm$0.77 & 2.42$\pm$0.03 & 2.71$\pm$0.09 \\
    \hline
    190211A* & 0.99 & 12.48 & 1.95 & 6.37$\pm$0.63 & 3.44$\pm$0.37 & 5.72$\pm$0.09 & 6.24$\pm$0.23 \\
    \hline
    190219A* & 1.01 & 167.80 & 2.13 & 0.40$\pm$0.05 & 63.86$\pm$15.24 & 5.27$\pm$0.25 & 1.35$\pm$0.15 \\
    \hline
    190519A* & 0.62 & 45.58 & 1.81 & 17.45$\pm$1.35 & 5.61$\pm$0.52 & 5.05$\pm$0.07 & 3.69$\pm$0.13 \\
    \hline
    190613B* & 3.93 & 160.84 & 1.97 & 1.55$\pm$0.16 & 59.11$\pm$8.08 & 1.97$\pm$0.04 & 0.60$\pm$0.03 \\
    \hline
    190701A* & 1.00 & 52.40 & 1.85 & 0.67$\pm$0.15 & 11.11$\pm$3.17 & 9.56$\pm$0.39 & 5.87$\pm$0.61 \\
    \hline
    190719C & 2.47 & 185.70 & 1.54 & 3.02$\pm$1.21 & 53.59$\pm$15.44 & 1.20$\pm$0.04 & 0.44$\pm$0.03 \\
    \hline
    190821A* & 0.79 & 57.10 & 1.73 & 5.27$\pm$1.03 & 6.19$\pm$1.12 & 4.87$\pm$0.12 & 3.89$\pm$0.21 \\
    \hline
    191011A & 1.72 & 7.37 & 1.72 & 3.54$\pm$0.63 & 1.61$\pm$0.30 & 7.81$\pm$0.18 & 14.55$\pm$0.85 \\
    \hline
    191123A* & 1.58 & 275.10 & 2.02 & 1.90$\pm$0.24 & 14.82$\pm$2.63 & 3.95$\pm$0.10 & 2.30$\pm$0.16 \\
    \hline
    \end{tabular}%
    \begin{tablenotes}
    \footnotesize
    \item \textbf{Note.} \\
    1.GRB names with * indicate data from specific selection criteria.\\
    2.The results of $B_p$ and $P_0$ were obtained under the condition of $\eta_x=0.5$.
    \end{tablenotes}
  \label{tab:grb_data_3}%
\end{table*}%

\addtocounter{table}{-1}

\begin{table*}[htbp]
  \centering
  \caption{GRB Data Summary}
    \begin{tabular}{lccccccc}
    \hline
    \hline
    GRB & $z$ & T$_{90}$ & $\gamma$ & $F_0$ & $t_b$ & $P_0$ & $B_p$ \\
         &    & (s)       &   &$10^{-11}$(erg/cm$^2$/s) & $10^3$(s) & (ms)&$10^{15}$ (G) \\
    \hline
    191220A* & 2.23 & 175.55 & 2.02 & 0.18$\pm$0.04 & 56.01$\pm$20.52 & 6.00$\pm$0.30 & 1.90$\pm$0.26 \\
    \hline
    200711A* & 1.75 & 29.39 & 1.96 & 20.08$\pm$2.44 & 3.60$\pm$0.42 & 3.21$\pm$0.05 & 3.41$\pm$0.13 \\
    \hline
    200713A* & 0.86 & 48.98 & 2.12 & 1.05$\pm$0.17 & 32.98$\pm$5.97 & 4.50$\pm$0.09 & 1.60$\pm$0.09 \\
    \hline
    200906A* & 0.87 & 70.90 & 2.25 & 0.94$\pm$0.17 & 12.47$\pm$3.04 & 8.84$\pm$0.33 & 4.87$\pm$0.45 \\
    \hline
    200925B* & 1.66 & 18.25 & 2.08 & 2.50$\pm$0.28 & 10.71$\pm$1.40 & 4.92$\pm$0.13 & 3.10$\pm$0.17 \\
    \hline
    201209A* & 0.74 & 48.00 & 1.92 & 7.70$\pm$0.54 & 21.69$\pm$1.93 & 4.49$\pm$0.05 & 1.63$\pm$0.05 \\
    \hline
    201221A & 5.70 & 44.50 & 1.57 & 2.32$\pm$0.50 & 7.38$\pm$1.51 & 2.65$\pm$0.07 & 3.62$\pm$0.22 \\
    \hline
    201223A* & 2.34 & 29.00 & 1.93 & 2.28$\pm$0.40 & 3.22$\pm$0.59 & 6.92$\pm$0.22 & 9.11$\pm$0.62 \\
    \hline
    210210A & 0.71 & 6.60 & 1.82 & 15.09$\pm$1.11 & 4.07$\pm$0.31 & 4.41$\pm$0.06 & 4.11$\pm$0.12 \\
    \hline
    210217A* & 2.09 & 4.22 & 1.93 & 3.57$\pm$0.77 & 3.38$\pm$0.73 & 4.92$\pm$0.14 & 6.64$\pm$0.45 \\
    \hline
    210222B* & 1.00 & 12.82 & 1.96 & 8.62$\pm$1.03 & 1.74$\pm$0.26 & 6.69$\pm$0.13 & 10.40$\pm$0.55 \\
    \hline
    210419A* & 1.02 & 64.43 & 2.67 & 0.20$\pm$0.07 & 29.12$\pm$15.60 & 10.14$\pm$0.59 & 3.99$\pm$0.62 \\
    \hline
    210504A & 2.08 & 135.06 & 2.08 & 0.41$\pm$0.17 & 13.46$\pm$6.60 & 7.08$\pm$0.43 & 4.94$\pm$0.76 \\
    \hline
    211129A* & 2.19 & 113.01 & 2.31 & 4.54$\pm$0.64 & 12.33$\pm$1.57 & 3.09$\pm$0.05 & 1.88$\pm$0.07 \\
    \hline
    220325A* & 4.25 & 3.50 & 1.59 & 1.06$\pm$0.23 & 9.08$\pm$2.63 & 3.80$\pm$0.14 & 4.30$\pm$0.42 \\
    \hline
    220403B* & 1.06 & 27.00 & 1.88 & 1.15$\pm$0.17 & 18.92$\pm$3.50 & 6.20$\pm$0.15 & 2.83$\pm$0.19 \\
    \hline
    220408A* & 1.52 & 17.25 & 1.91 & 10.15$\pm$1.58 & 2.86$\pm$0.45 & 3.79$\pm$0.10 & 5.06$\pm$0.28 \\
    \hline
    220518A* & 3.20 & 12.29 & 1.83 & 4.32$\pm$0.56 & 2.83$\pm$0.46 & 4.26$\pm$0.10 & 6.92$\pm$0.42 \\
    \hline
    221024A* & 1.90 & 120.00 & 1.96 & 4.38$\pm$0.51 & 12.39$\pm$1.24 & 2.81$\pm$0.05 & 1.79$\pm$0.06 \\
    \hline
    230325A & 1.66 & 38.05 & 2.08 & 0.26$\pm$0.05 & 103.06$\pm$26.87 & 3.65$\pm$0.14 & 0.85$\pm$0.09 \\
    \hline
    230826A* & 1.10 & 41.07 & 2.16 & 5.65$\pm$1.09 & 2.56$\pm$0.42 & 7.69$\pm$0.19 & 9.43$\pm$0.43 \\
    \hline
    231205B* & 1.01 & 46.94 & 2.06 & 16.75$\pm$1.57 & 10.08$\pm$0.87 & 2.09$\pm$0.02 & 1.32$\pm$0.03 \\
    \hline
    240101A* & 2.19 & 15.80 & 2.10 & 4.08$\pm$1.70 & 0.87$\pm$0.40 & 8.18$\pm$0.33 & 23.08$\pm$3.25 \\
    \hline
    240102A* & 1.63 & 119.00 & 1.82 & 0.24$\pm$0.09 & 22.27$\pm$12.24 & 8.39$\pm$0.61 & 4.25$\pm$0.91 \\
    \hline
    240421B* & 1.25 & 24.00 & 1.87 & 17.95$\pm$2.31 & 4.49$\pm$0.50 & 3.47$\pm$0.05 & 3.16$\pm$0.10 \\
    \hline
    240811A* & 1.63 & 66.32 & 1.99 & 0.90$\pm$0.15 & 22.84$\pm$4.30 & 5.54$\pm$0.15 & 2.40$\pm$0.16 \\
    \hline
    241010A & 0.98 & 30.86 & 1.82 & 5.77$\pm$0.66 & 16.04$\pm$1.88 & 2.82$\pm$0.04 & 1.42$\pm$0.05 \\
    \hline
    241026A* & 2.61 & 25.20 & 1.86 & 5.47$\pm$0.45 & 22.49$\pm$1.85 & 1.87$\pm$0.02 & 0.88$\pm$0.02 \\
    \hline
    \end{tabular}%
    \begin{tablenotes}
    \footnotesize
    \item \textbf{Note.} \\
    1.GRB names with * indicate data from specific selection criteria.\\
    2.The results of $B_p$ and $P_0$ were obtained under the condition of $\eta_x=0.5$.
    \end{tablenotes}
  \label{tab:grb_data_4}%
\end{table*}%

\begin{table*}[htbp]
    \centering
    \caption{Fitting Results for Correlations in the Redshift-measured and Full GRB Samples. }
    \begin{tabular}{lcccl}
    \hline
        correlation & sample & $k$ & $m$ & $\sigma_{\mathrm{int}}$ \\ \hline
          $L_0-t_b$ & redshift-measured  & $-1.10 \pm 0.11$ & $1.55 \pm 0.14$ & 0.58 \\
          (Dainotti) & full & $-1.07 \pm 0.09$ & $1.24 \pm 0.12$ & 0.66 \\ \hline
        \multirow{2}{*}{$B_p-P_0$}
          & redshift-measured  & $0.80 \pm 0.16$ & $14.95 \pm 0.09$ & 0.33 \\
          & full & $0.83 \pm 0.09$ & $14.92 \pm 0.06$ & 0.30  \\ \hline
    \end{tabular}
    \label{tab:addlabe2}%
\end{table*}

\clearpage


\begin{thebibliography}{}
\expandafter\ifx\csname natexlab\endcsname\relax\def\natexlab#1{#1}\fi
\providecommand{\url}[1]{\href{#1}{#1}}
\providecommand{\dodoi}[1]{doi:~\href{http://doi.org/#1}{\nolinkurl{#1}}}
\providecommand{\doeprint}[1]{\href{http://ascl.net/#1}{\nolinkurl{http://ascl.net/#1}}}
\providecommand{\doarXiv}[1]{\href{https://arxiv.org/abs/#1}{\nolinkurl{https://arxiv.org/abs/#1}}}

\bibitem[Abbott et al.(2017)]{2017ApJ...848L..12A} Abbott, B.~P., Abbott, R., Abbott, T.~D., et al.\ 2017, \apjl, 848, 2, L12.
\bibitem[Amati et al.(2002)]{2002A&A...390...81A} Amati, L., Frontera, F., Tavani, M., et al.\ 2002, \aap, 390, 81.
\bibitem[Amati(2006)]{2006MNRAS.372..233A} Amati, L.\ 2006, \mnras, 372, 1, 233.
\bibitem[Amati et al.(2019)]{2019MNRAS.486L..46A} Amati, L., D'Agostino, R., Luongo, O., et al.\ 2019, \mnras, 486, 1, L46.
\bibitem[Ascenzi et al.(2020)]{2020A&A...641A..61A} Ascenzi, S., Oganesyan, G., Salafia, O.~S., et al.\ 2020, \aap, 641, A61.
\bibitem[Atteia(2003)]{2003A&A...407L...1A} Atteia, J.-L.\ 2003, \aap, 407, L1.
\bibitem[Band et al.(1993)]{1993ApJ...413..281B} Band, D., Matteson, J., Ford, L., et al.\ 1993, \apj, 413, 281.
\bibitem[Bargiacchi et al. (2025)]{Bargiacchi2025} Bargiacchi, G, Dainotti, M \& Capozziello, S., 2025, New Astronomy Reviews, 100, 101712
\bibitem[Berger et al.(2013)]{2013ApJ...774L..23B} Berger, E., Fong, W., \& Chornock, R.\ 2013, \apjl, 774, 2, L23.
\bibitem[Bhattacharya et al.(2025)]{2025MNRAS.tmp.2057B} Bhattacharya, M., Murase, K., \& Kashiyama, K.\ 2025, \mnras.
\bibitem[Blandford \& Znajek(1977)]{1977MNRAS.179..433B} Blandford, R.~D. \& Znajek, R.~L.\ 1977, \mnras, 179, 433.
\bibitem[Bloom et al.(2002)]{2002AJ....123.1111B} Bloom, J.~S., Kulkarni, S.~R., \& Djorgovski, S.~G.\ 2002, \aj, 123, 3, 1111.
\bibitem[Bloom et al.(2001)]{2001AJ....121.2879B} Bloom, J.~S., Frail, D.~A., \& Sari, R.\ 2001, \aj, 121, 6, 2879.
\bibitem[Bochenek et al.(2020)]{2020Natur.587...59B} Bochenek, C.~D., Ravi, V., Belov, K.~V., et al.\ 2020, \nat, 587, 7832, 59.
\bibitem[Cao et al.(2022)]{2022MNRAS.516.1386C} Cao, S., Dainotti, M., \& Ratra, B.\ 2022, \mnras, 516, 1, 1386.
\bibitem[Chen \& Beloborodov(2007)]{2007ApJ...657..383C} Chen, W.-X. \& Beloborodov, A.~M.\ 2007, \apj, 657, 1, 383.
\bibitem[Chen et al.(2023)]{2023ApJ...943...42C} Chen, Z.~H., Yan, L., Kangas, T., et al.\ 2023, \apj, 943, 1, 42.
\bibitem[CHIME/FRB Collaboration et al.(2020)]{2020Natur.587...54C} CHIME/FRB Collaboration, Andersen, B.~C., Bandura, K.~M., et al.\ 2020, \nat, 587, 7832, 54.
\bibitem[Cook et al.(1994)]{1994ApJ...424..823C} Cook, G.~B., Shapiro, S.~L., \& Teukolsky, S.~A.\ 1994, \apj, 424, 823.
\bibitem[Corsi \& M{\'e}sz{\'a}ros(2009)]{2009ApJ...702.1171C} Corsi, A. \& M{\'e}sz{\'a}ros, P.\ 2009, \apj, 702, 2, 1171.
\bibitem[Cucchiara et al.(2011)]{2011ApJ...736....7C} Cucchiara, A., Levan, A.~J., Fox, D.~B., et al.\ 2011, \apj, 736, 1, 7.
\bibitem[D'Avanzo et al.(2010)]{2010A&A...522A..20D} D'Avanzo, P., Perri, M., Fugazza, D., et al.\ 2010, \aap, 522, A20.
\bibitem[Dai \& Lu(1998b)]{1998PhRvL..81.4301D} Dai, Z.~G. \& Lu, T.\ 1998, \prl, 81, 20, 4301.
\bibitem[Dai \& Lu(1998a)]{1998A&A...333L..87D} Dai, Z.~G. \& Lu, T.\ 1998, \aap, 333, L87.
\bibitem[Dainotti et al.(2013)]{2013MNRAS.436...82D} Dainotti, M.~G., Cardone, V.~F., Piedipalumbo, E., et al.\ 2013, \mnras, 436, 1, 82.
\bibitem[Dainotti et al.(2008)]{2008MNRAS.391L..79D} Dainotti, M.~G., Cardone, V.~F., \& Capozziello, S.\ 2008, \mnras, 391, 1, L79.
\bibitem[Dainotti et al.(2017)]{2017ApJ...848...88D} Dainotti, M.~G., Hernandez, X., Postnikov, S., et al.\ 2017, \apj, 848, 2, 88.
\bibitem[Dainotti et al.(2020)]{2020ApJ...904...97D} Dainotti, M.~G., Lenart, A. {\L}., Sarracino, G., et al.\ 2020, \apj, 904, 2, 97.
\bibitem[Dainotti et al.(2010)]{2010ApJ...722L.215D} Dainotti, M.~G., Willingale, R., Capozziello, S., et al.\ 2010, \apjl, 722, 2, L215.
\bibitem[Dainotti et al.(2022)]{2022MNRAS.514.1828D} Dainotti, M.~G., Nielson, V., Sarracino, G., et al.\ 2022, \mnras, 514, 2, 1828.
\bibitem[Dainotti et al.(2023a)]{2023ApJ...951...63D} Dainotti, M.~G., Bargiacchi, G., Bogdan, M., et al.\ 2023, \apj, 951, 1, 63.
\bibitem[Dainotti et al.(2023b)]{2023MNRAS.518.2201D} Dainotti, M.~G., Lenart, A. {\L}., Chraya, A., et al.\ 2023, \mnras, 518, 2, 2201.
\bibitem[Dainotti et al.(2024a)]{2024ApJ...967L..30D} Dainotti, M.~G., Narendra, A., Pollo, A., et al.\ 2024, \apjl, 967, 2, L30.
\bibitem[Dainotti et al.(2024b)]{2024ApJS..271...22D} Dainotti, M.~G., Taira, E., Wang, E., et al.\ 2024, \apjs, 271, 1, 22.
\bibitem[Dainotti et al.(2026)]{2026arXiv260318223D} Dainotti, M.~G., {\L}ukasz Lenart, A., De Simone, B., et al.\ 2026, arXiv:2603.18223.
\bibitem[Dall'Osso et al.(2011)]{2011A&A...526A.121D} Dall'Osso, S., Stratta, G., Guetta, D., et al.\ 2011, \aap, 526, A121.
\bibitem[Dereli-B{\'e}gu{\'e} et al.(2022)]{2022NatCo..13.5611D} Dereli-B{\'e}gu{\'e}, H., Pe'er, A., Ryde, F., et al.\ 2022, Nature Communications, 13, 5611.
\bibitem[Du et al.(2021)]{2021ApJ...908..242D} Du, M., Yi, S.-X., Liu, T., et al.\ 2021, \apj, 908, 2, 242.
\bibitem[Eichler et al.(1989)]{1989Natur.340..126E} Eichler, D., Livio, M., Piran, T., et al.\ 1989, \nat, 340, 6229, 126.
\bibitem[Evans et al.(2007)]{2007A&A...469..379E} Evans, P.~A., Beardmore, A.~P., Page, K.~L., et al.\ 2007, \aap, 469, 1, 379.
\bibitem[Evans et al.(2009)]{2009MNRAS.397.1177E} Evans, P.~A., Beardmore, A.~P., Page, K.~L., et al.\ 2009, \mnras, 397, 3, 1177.
\bibitem[Evans et al.(2010)]{2010A&A...519A.102E} Evans, P.~A., Willingale, R., Osborne, J.~P., et al.\ 2010, \aap, 519, A102.
\bibitem[Favale et al.(2024)]{2024JHEAp..44..323F} Favale, A., Dainotti, M.~G., G{\'o}mez-Valent, A., et al.\ 2024, Journal of High Energy Astrophysics, 44, 323.
\bibitem[Foreman-Mackey et al.(2013)]{2013PASP..125..306F} Foreman-Mackey, D., Hogg, D.~W., Lang, D., et al.\ 2013, \pasp, 125, 925, 306.
\bibitem[Fox et al.(2005)]{2005Natur.437..845F} Fox, D.~B., Frail, D.~A., Price, P.~A., et al.\ 2005, \nat, 437, 7060, 845.
\bibitem[Gao et al.(2016)]{2016PhRvD..93d4065G} Gao, H., Zhang, B., \& L{\"u}, H.-J.\ 2016, \prd, 93, 4, 044065.
\bibitem[Gehrels et al.(2009)]{2009ARA&A..47..567G} Gehrels, N., Ramirez-Ruiz, E., \& Fox, D.~B.\ 2009, \araa, 47, 1, 567.
\bibitem[Ghirlanda et al.(2004)]{2004ApJ...616..331G} Ghirlanda, G., Ghisellini, G., \& Lazzati, D.\ 2004, \apj, 616, 1, 331.
\bibitem[Hjorth et al.(2003)]{2003Natur.423..847H} Hjorth, J., Sollerman, J., M{\o}ller, P., et al.\ 2003, \nat, 423, 6942, 847.
\bibitem[Kasen \& Bildsten (2010)]{Kasen2010} Kasen, D. \& Bildsten, L., 2010, ApJ, 717, 245
\bibitem[Khadka et al.(2021)]{2021JCAP...09..042K} Khadka, N., Luongo, O., Muccino, M., et al.\ 2021, \jcap, 2021, 9, 042.
\bibitem[Kouveliotou et al.(1993)]{1993ApJ...413L.101K} Kouveliotou, C., Meegan, C.~A., Fishman, G.~J., et al.\ 1993, \apjl, 413, L101.
\bibitem[Kumar \& Zhang(2015)]{2015PhR...561....1K} Kumar, P. \& Zhang, B.\ 2015, \physrep, 561, 1.
\bibitem[Lan et al.(2025a)]{2025ApJS..280...45L} Lan, L., Gao, H., Ai, S., et al.\ 2025, \apjs, 280, 1, 45.
\bibitem[Lan et al.(2025b)]{2025arXiv251122149L} Lan, L., Gao, H., Zhao, L., et al.\ 2025, , arXiv:2511.22149.
\bibitem[Lasky et al.(2017)]{2017ApJ...843L...1L} Lasky, P.~D., Leris, C., Rowlinson, A., et al.\ 2017, \apjl, 843, 1, L1.
\bibitem[Lattimer \& Prakash(2001)]{2001ApJ...550..426L} Lattimer, J.~M. \& Prakash, M.\ 2001, \apj, 550, 1, 426.
\bibitem[Lei et al.(2009)]{2009ApJ...700.1970L} Lei, W.~H., Wang, D.~X., Zhang, L., et al.\ 2009, \apj, 700, 2, 1970.
\bibitem[Lei et al.(2013)]{2013ApJ...765..125L} Lei, W.-H., Zhang, B., \& Liang, E.-W.\ 2013, \apj, 765, 2, 125.
\bibitem[Lenart et al.(2025)]{2025JHEAp..4700384L} Lenart, A. {\L}., Dainotti, M.~G., Khatiya, N., et al.\ 2025, Journal of High Energy Astrophysics, 47, 100384.
\bibitem[Levine et al.(2022)]{2022ApJ...925...15L} Levine, D., Dainotti, M., Zvonarek, K.~J., et al.\ 2022, \apj, 925, 1, 15.
\bibitem[Li et al.(2023)]{2023ApJ...953...58L} Li, J.-L., Yang, Y.-P., Yi, ., et al.\ 2023, \apj, 953, 1, 58.
\bibitem[Li et al.(2026)]{2026arXiv260101586L} Li, X.-Y., Liu, T., Huang, B.-Q., et al.\ 2026, , arXiv:2601.01586.
\bibitem[Liu et al.(2017)]{2017NewAR..79....1L} Liu, T., Gu, W.-M., \& Zhang, B.\ 2017, \nar, 79, 1.
\bibitem[Lyons et al.(2010)]{2010MNRAS.402..705L} Lyons, N., O'Brien, P.~T., Zhang, B., et al.\ 2010, \mnras, 402, 2, 705.
\bibitem[L{\"u} et al.(2019)]{2019ApJ...871...54L} L{\"u}, H.-J., Lan, L., \& Liang, E.-W.\ 2019, \apj, 871, 1, 54.
\bibitem[L{\"u} et al.(2015)]{2015ApJ...805...89L} L{\"u}, H.-J., Zhang, B., Lei, W.-H., et al.\ 2015, \apj, 805, 2, 89.
\bibitem[L{\"u} \& Zhang(2014)]{2014ApJ...785...74L} L{\"u}, H.-J. \& Zhang, B.\ 2014, \apj, 785, 1, 74.
\bibitem[L{\"u} et al.(2018)]{2018MNRAS.480.4402L} L{\"u}, H.-J., Zou, L., Lan, L., et al.\ 2018, \mnras, 480, 4, 4402.
\bibitem[MacFadyen \& Woosley(1999)]{1999ApJ...524..262M} MacFadyen, A.~I. \& Woosley, S.~E.\ 1999, \apj, 524, 1, 262.
\bibitem[Margalit \& Metzger(2018)]{2018ApJ...868L...4M} Margalit, B. \& Metzger, B.~D.\ 2018, \apjl, 868, 1, L4.
\bibitem[Metzger et al.(2011)]{2011MNRAS.413.2031M} Metzger, B.~D., Giannios, D., Thompson, T.~A., et al.\ 2011, \mnras, 413, 3, 2031.
\bibitem[Metzger et al.(2008)]{2008MNRAS.385.1455M} Metzger, B.~D., Quataert, E., \& Thompson, T.~A.\ 2008, \mnras, 385, 3, 1455.
\bibitem[Metzger \& Piro(2014)]{2014MNRAS.439.3916M} Metzger, B.~D. \& Piro, A.~L.\ 2014, \mnras, 439, 4, 3916.
\bibitem[Murase et al. (2016)]{Murase2016} Murase, K., Kashiyama, K. \& Meszaros, P., 2016, MNRAS, 461, 1498
\bibitem[Narayan et al.(1992)]{1992ApJ...395L..83N} Narayan, R., Paczynski, B., \& Piran, T.\ 1992, \apjl, 395, L83.
\bibitem[Narendra et al.(2025)]{2025A&A...698A..92N} Narendra, A., Dainotti, M.~G., Sarkar, M., et al.\ 2025, \aap, 698, A92.
\bibitem[Nicholl et al.(2017)]{2017ApJ...843...84N} Nicholl, M., Williams, P.~K.~G., Berger, E., et al.\ 2017, \apj, 843, 2, 84.
\bibitem[Nousek et al.(2006)]{2006ApJ...642..389N} Nousek, J.~A., Kouveliotou, C., Grupe, D., et al.\ 2006, \apj, 642, 1, 389.
\bibitem[O'Connor et al.(2023)]{2023SciA....9I1405O} O'Connor, B., Troja, E., Ryan, G., et al.\ 2023, Science Advances, 9, 23, eadi1405.
\bibitem[Oganesyan et al.(2020)]{2020ApJ...893...88O} Oganesyan, G., Ascenzi, S., Branchesi, M., et al.\ 2020, \apj, 893, 2, 88.
\bibitem[Piran(2004)]{2004RvMP...76.1143P} Piran, T.\ 2004, Reviews of Modern Physics, 76, 4, 1143.
\bibitem[Planck Collaboration et al.(2020)]{2020A&A...641A...6P} Planck Collaboration, Aghanim, N., Akrami, Y., et al.\ 2020, \aap, 641, A6.
\bibitem[Platts et al.(2019)]{2019PhR...821....1P} Platts, E., Weltman, A., Walters, A., et al.\ 2019, \physrep, 821, 1.
\bibitem[Popham et al.(1999)]{1999ApJ...518..356P} Popham, R., Woosley, S.~E., \& Fryer, C.\ 1999, \apj, 518, 1, 356.
\bibitem[Racusin et al.(2009)]{2009ApJ...698...43R} Racusin, J.~L., Liang, E.~W., Burrows, D.~N., et al.\ 2009, \apj, 698, 1, 43.
\bibitem[Rahaman et al.(2025)]{2025ApJ...988..276R} Rahaman, S.~M., Acharya, S.~K., Beniamini, P., et al.\ 2025, \apj, 988, 2, 276.
\bibitem[Rea et al.(2015)]{2015ApJ...813...92R} Rea, N., Gull{\'o}n, M., Pons, J.~A., et al.\ 2015, \apj, 813, 2, 92.
\bibitem[Rowlinson et al.(2010)]{2010MNRAS.409..531R} Rowlinson, A., O'Brien, P.~T., Tanvir, N.~R., et al.\ 2010, \mnras, 409, 2, 531.
\bibitem[Rowlinson et al.(2014)]{2014MNRAS.443.1779R} Rowlinson, A., Gompertz, B.~P., Dainotti, M., et al.\ 2014, \mnras, 443, 2, 1779.
\bibitem[Rowlinson et al.(2013)]{2013MNRAS.430.1061R} Rowlinson, A., O'Brien, P.~T., Metzger, B.~D., et al.\ 2013, \mnras, 430, 2, 1061.
\bibitem[Ruffert et al.(1997)]{1997A&A...319..122R} Ruffert, M., Janka, H.-T., Takahashi, K., et al.\ 1997, \aap, 319, 122.
\bibitem[Sarin et al.(2020)]{2020MNRAS.499.5986S} Sarin, N., Lasky, P.~D., \& Ashton, G.\ 2020, \mnras, 499, 4, 5986.
\bibitem[Schaefer(2007)]{2007ApJ...660...16S} Schaefer, B.~E.\ 2007, \apj, 660, 1, 16.
\bibitem[Si et al.(2018)]{2018ApJ...863...50S} Si, S.-K., Qi, Y.-Q., Xue, F.-X., et al.\ 2018, \apj, 863, 1, 50.
\bibitem[Stratta et al.(2018)]{2018ApJ...869..155S} Stratta, G., Dainotti, M.~G., Dall'Osso, S., et al.\ 2018, \apj, 869, 2, 155.
\bibitem[Tang et al.(2019)]{2019ApJS..245....1T} Tang, C.-H., Huang, Y.-F., Geng, J.-J., et al.\ 2019, \apjs, 245, 1, 1.
\bibitem[Thompson(1994)]{1994MNRAS.270..480T} Thompson, C.\ 1994, \mnras, 270, 480.
\bibitem[Tian et al.(2023)]{2023ApJ...958...74T} Tian, X., Li, J.-L., Yi, S.-X., et al.\ 2023, \apj, 958, 1, 74.
\bibitem[Troja et al.(2007)]{2007ApJ...665..599T} Troja, E., Cusumano, G., O'Brien, P.~T., et al.\ 2007, \apj, 665, 1, 599.
\bibitem[Usov(1992)]{1992Natur.357..472U} Usov, V.~V.\ 1992, \nat, 357, 6378, 472.
\bibitem[Wang \& Dai (2013)]{Wang2013} Wang, F.~Y. \& Dai, Z.~G., 2013, Nature Physics, 9, 463
\bibitem[Wang et al.(2015)]{2015NewAR..67....1W} Wang, F.~Y., Dai, Z.~G., \& Liang, E.~W.\ 2015, \nar, 67, 1.
\bibitem[Wang et al.(2022a)]{2022ApJ...924...97W} Wang, F.~Y., Hu, J.~P., Zhang, G.~Q., et al.\ 2022, \apj, 924, 2, 97.
\bibitem[Wang et al. (2022b)]{Wang2022} Wang, F. Y., Zhang, G. Q., Dai, Z. G. \& Cheng, K. S., 2022, Nature Communications, 13, 4382
\bibitem[Wang et al. (2011)]{Wang2011} Wang, F. Y., Qi, S. \& Dai, Z. G., 2011, \mnras, 415, 3423
\bibitem[Wang et al. (2016)]{Wang2016} Wang, J. S., Wang, F. Y., Cheng, K. S. \& Dai, Z. G., 2016, A\&A, 585, A68
\bibitem[Wei et al. (2016)] {Wei2016} Wei, J., et al., 2016, arXiv: 1610.06892
\bibitem[Willingale et al.(2007)]{2007ApJ...662.1093W} Willingale, R., O'Brien, P.~T., Osborne, J.~P., et al.\ 2007, \apj, 662, 2, 1093.
\bibitem[Woosley \& Bloom(2006)]{2006ARA&A..44..507W} Woosley, S.~E. \& Bloom, J.~S.\ 2006, \araa, 44, 1, 507.
\bibitem[Woosley(1993)]{1993ApJ...405..273W} Woosley, S.~E.\ 1993, \apj, 405, 273.
\bibitem[Woosley (2010)]{Woosley2010} Woosley, S. E., 2010, ApJL, 719, L204
\bibitem[Xiao \& Dai(2019)]{2019ApJ...878...62X} Xiao, D. \& Dai, Z.-G.\ 2019, \apj, 878, 1, 62.
\bibitem[Xie et al.(2022)]{2022ApJ...934..125X} Xie, L., Wei, D.-M., Wang, Y., et al.\ 2022, \apj, 934, 2, 125.
\bibitem[Yi et al.(2013)]{2013ApJ...776..120Y} Yi, S.-X., Wu, X.-F., \& Dai, Z.-G.\ 2013, \apj, 776, 2, 120.
\bibitem[Yi et al.(2017)]{2017JHEAp..13....1Y} Yi, S.-X., Lei, W.-H., Zhang, B., et al.\ 2017, Journal of High Energy Astrophysics, 13, 1.
\bibitem[Yi et al.(2016)]{2016ApJS..224...20Y} Yi, S.-X., Xi, S.-Q., Yu, H., et al.\ 2016, \apjs, 224, 2, 20.
\bibitem[Yi et al.(2021)]{2021MNRAS.507.1047Y} Yi, S.-X., Xie, W., Ma, S.-B., et al.\ 2021, \mnras, 507, 1, 1047.
\bibitem[Yonetoku et al.(2004)]{2004ApJ...609..935Y} Yonetoku, D., Murakami, T., Nakamura, T., et al.\ 2004, \apj, 609, 2, 935.
\bibitem[Yu et al.(2020)]{2020ChA&A..44..269Y} Yu, S.-. jing ., Gonzalez, F., Wei, J.-. yan ., et al.\ 2020, \caa, 44, 2, 269.
\bibitem[Yu et al.(2017)]{2017ApJ...840...12Y} Yu, Y.-W., Zhu, J.-P., Li, S.-Z., et al.\ 2017, \apj, 840, 1, 12.
\bibitem[Yuan et al.(2015)]{2015arXiv150607735Y} Yuan, W., Zhang, C., Feng, H., et al.\ 2015, , arXiv:1506.07735.
\bibitem[Zhang et al.(2007)]{2007ApJ...666.1002Z} Zhang, B.-B., Liang, E.-W., \& Zhang, B.\ 2007, \apj, 666, 2, 1002.
\bibitem[Zhang \& M{\'e}sz{\'a}ros(2001)]{2001ApJ...552L..35Z} Zhang, B. \& M{\'e}sz{\'a}ros, P.\ 2001, \apjl, 552, 1, L35.
\bibitem[Zhang et al.(2006)]{2006ApJ...642..354Z} Zhang, B., Fan, Y.~Z., Dyks, J., et al.\ 2006, \apj, 642, 1, 354.
\bibitem[Zhang \& Yan(2011)]{2011ApJ...726...90Z} Zhang, B. \& Yan, H.\ 2011, \apj, 726, 2, 90.
\bibitem[Zhang(2020)]{2020Natur.587...45Z} Zhang, B.\ 2020, \nat, 587, 7832, 45.
\bibitem[Zhao \& Wang(2021)]{2021ApJ...923L..17Z} Zhao, Z.~Y. \& Wang, F.~Y.\ 2021, \apjl, 923, 1, L17.
\bibitem[Zhong et al.(2024)]{2024ApJ...963L..26Z} Zhong, S.-Q., Li, L., Xiao, D., et al.\ 2024, \apjl, 963, 1, L26.
\bibitem[Zhou et al.(2025)]{2025ApJ...991..145Z} Zhou, Y.-Q., Yi, S.-X., Yang, Y.-P., et al.\ 2025, \apj, 991, 2, 145.

\end{thebibliography}
\end{document}